\documentclass[prd,preprint,tightenlines,floatfix,
preprintnumbers,nofootinbib,eqsecnum,superscriptaddress]{revtex4}

\usepackage[T1]{fontenc}		

\usepackage{amsmath,amsfonts,amssymb,amstext,mathrsfs}
\usepackage{mathpazo}

\usepackage[dvips]{graphicx}
\usepackage{epsf,float}
\usepackage{revsymb}

\usepackage{dcolumn}
\usepackage{braket}
\usepackage{color,xcolor}
\usepackage{graphicx}
\usepackage{subfigure}
\usepackage{multirow}
\usepackage{tabularx}
\usepackage{pstricks}
\usepackage[section]{placeins}
\usepackage{booktabs}
\usepackage{array}

\usepackage{hyperref}

\newcommand{\Pom}{\mathbb{P}}
\newcommand{\Ode}{\mathbb{O}}
\newcommand{\Reg}{\mathbb{R}}

\renewcommand\slash[1]{\not \! #1}

\newcommand{\bdPt}{\mbox{\boldmath $dP_{t}$}}
\newcommand{\bqta}{\mbox{\boldmath $q_{t,1}$}}
\newcommand{\bqtb}{\mbox{\boldmath $q_{t,2}$}}
\newcommand{\bpta}{\mbox{\boldmath $p_{t,1}$}}
\newcommand{\bptb}{\mbox{\boldmath $p_{t,2}$}}

\newcommand{\bpa}{\mbox{\boldmath $p_{a}$}}
\newcommand{\bpb}{\mbox{\boldmath $p_{b}$}}

\newcommand{\bhpa}{\mbox{\boldmath $\hat{p}_{a}$}}
\newcommand{\bhpb}{\mbox{\boldmath $\hat{p}_{b}$}}
\newcommand{\bhpc}{\mbox{\boldmath $\hat{p}_{3}$}}

\newcommand{\bea}{\mbox{\boldmath $e_{1}$}}
\newcommand{\beb}{\mbox{\boldmath $e_{2}$}}
\newcommand{\bec}{\mbox{\boldmath $e_{3}$}}

\begin{document}

\title{Searching for the odderon in \boldmath{$pp \to pp K^{+}K^{-}$}
and \boldmath{$pp \to pp \mu^{+}\mu^{-}$} reactions
in the \boldmath{$\phi(1020)$} resonance region at the LHC}

\vspace{0.6cm}

\author{Piotr Lebiedowicz}
\email{Piotr.Lebiedowicz@ifj.edu.pl}
\affiliation{Institute of Nuclear Physics Polish Academy of Sciences, Radzikowskiego 152, PL-31342 Krak\'ow, Poland}

\author{Otto Nachtmann}
\email{O.Nachtmann@thphys.uni-heidelberg.de}
\affiliation{Institut f\"ur Theoretische Physik, Universit\"at Heidelberg,
Philosophenweg 16, D-69120 Heidelberg, Germany}

\author{Antoni Szczurek
\footnote{Also at \textit{College of Natural Sciences, 
Institute of Physics, University of Rzesz\'ow, 
Pigonia 1, PL-35310 Rzesz\'ow, Poland}.}}
\email{Antoni.Szczurek@ifj.edu.pl}
\affiliation{Institute of Nuclear Physics Polish Academy of Sciences, Radzikowskiego 152, PL-31342 Krak\'ow, Poland}

\begin{abstract}
We explore the possibility of observing odderon exchange in the $pp \to pp K^{+}K^{-}$ and $pp \to pp \mu^{+}\mu^{-}$ reactions at the LHC. We consider the central exclusive production (CEP) of the $\phi(1020)$ resonance decaying into $K^{+} K^{-}$ and $\mu^{+}\mu^{-}$. We compare the purely diffractive contribution (odderon-pomeron fusion) to the photoproduction contribution (photon-pomeron fusion). The theoretical results are calculated within the tensor-pomeron and vector-odderon model for soft reactions. We include absorptive corrections at the amplitude level. In order to fix the coupling constants for the photon-pomeron fusion contribution we discuss the reactions $\gamma p \to \omega p$ and $\gamma p \to \phi p$ including $\phi$-$\omega$ mixing. We compare our results for these reactions with the available data, especially those from HERA. Our coupling constants for the pomeron-odderon-$\phi$ vertex are taken from an analysis of the WA102 data for the $p p \to p p \phi$ reaction. We show that the odderon-exchange contribution significantly improves the description of the $pp$ azimuthal correlations and the ${\rm dP_{t}}$ ``glueball-filter variable'' dependence of $\phi$ CEP measured by WA102. To describe the low-energy data more accurately we consider also subleading processes with reggeized vector-meson exchanges. However, they do not play a significant role at the LHC. We present predictions for two possible types of measurements: at midrapidity and with forward measurement of protons (relevant for ATLAS-ALFA or CMS-TOTEM), and at forward rapidities and without measurement of protons (relevant for LHCb). We discuss the influence of experimental cuts on the integrated cross sections and on various differential distributions. With the corresponding LHC data one should be able to get a decisive answer concerning the presence of an odderon-pomeron fusion contribution in single $\phi$ CEP.
\end{abstract}


\maketitle

\section{Introduction}
\label{sec:intro}

So far there is no unambiguous experimental evidence for the odderon ($\Ode$),
the charge conjugation $C = -1$ counterpart of the $C= +1$ pomeron ($\Pom$),
introduced on theoretical grounds in \cite{Lukaszuk:1973nt, Joynson:1975az}
and predicted in QCD as the exchange of a colourless
$C$-odd three-gluon compound state
\cite{Kwiecinski:1980wb,Bartels:1980pe,Jaroszewicz:1981jz,Janik:1998xj,Bartels:1999yt}.
A hint of the odderon was seen in ISR results \cite{Breakstone:1985pe}
as a small difference between the differential cross sections
of elastic proton-proton ($pp$) and proton-antiproton ($p \bar{p}$) scattering
in the diffractive dip region at $\sqrt{s} = 53$~GeV.
The interpretation of this difference is, however, 
complicated due to non-negligible contributions from secondary reggeons.
Recently the TOTEM Collaboration has published data from high-energy elastic
proton-proton scattering experiments at the LHC.
In \cite{Antchev:2017yns} results were given for the $\rho$ parameter,
the ratio of real to imaginary part of the forward scattering amplitude.
This is a measurement at $t = 0$.
In \cite{Antchev:2018rec} the differential cross section $d\sigma/dt$
was measured for 0.36~GeV$^{2} < |t| < 0.74$~GeV$^{2}$.
The interpretation of these results is controversial at the moment.
Some authors claim for instance that the $\rho$ measurements
show that there must be an odderon effect at $t = 0$
\cite{Martynov:2017zjz,Martynov:2018nyb}. 
But other authors find that no odderon contribution is needed
at $t = 0$ \cite{Khoze:2017swe,Khoze:2018bus,Khoze:2018kna,Broilo:2018qqs,Donnachie:2019ciz}.
For a general analysis of $pp$ and $p \bar{p}$ 
elastic scattering see, e.g.,~\cite{Csorgo:2018uyp,Bence:2018ain}.


As was discussed in \cite{Schafer:1991na} 
exclusive diffractive $J/\psi$ and $\phi$ production from the pomeron-odderon fusion
in high-energy $pp$ and $p\bar{p}$ collisions 
is a direct probe for a possible odderon exchange.
The photoproduction mechanism (i.e., pomeron-photon fusion) constitutes a background 
for pomeron-odderon exchanges in these reactions.
Other sources of background involve secondary reggeon exchanges, for instance
pomeron-($\phi_{\Reg}$-reggeon) exchanges.
Exclusive production of heavy vector mesons,
$J/\psi$ and $\Upsilon$, from the pomeron-odderon and the pomeron-photon fusion
in the pQCD $k_{t}$-factorization approach was discussed in \cite{Bzdak:2007cz}.
The exclusive $p p \to p p \phi$ reaction via the (pQCD-pomeron)-photon fusion 
in the high-energy corner was studied in \cite{Cisek:2010jk};
see also \cite{Cisek:2014ala} for the exclusive photoproduction of charmonia $J/\psi$ and $\psi'$
and \cite{Cisek:2011vt} for the exclusive $\omega$ production.

A possible probe of the odderon is photoproduction of $C = +1$ mesons 
\cite{Schafer:1992pq,Barakhovsky:1991ra}.
At sufficiently high energies only odderon and photon exchange contribute to these reactions.
Photoproduction of the pseudoscalars $\pi^{0}$, $\eta$, $\eta'$, $\eta_{c}$, 
and of the tensor $f_{2}(1270)$ in $ep$ scattering at high energies 
was discussed in \cite{Kilian:1997ew,Berger:1999ca,Berger:2000wt,Donnachie:2005bu,Ewerz:2006gd}.
For exclusive $\eta_{c}$ photoproduction within the high-energy 
framework of eikonal dipole scattering see \cite{Dumitru:2019qec}.
In \cite{Motyka:1998kb,Braunewell:2004pf} a probe of the perturbative odderon
in the quasidiffractive process $\gamma^{*} \gamma^{*} \to \eta_{c} \eta_{c}$ was studied.

Another interesting possibility is to study the charge asymmetry caused by
the interference between pomeron and odderon exchange.
This was discussed in diffractive $c \bar{c}$ pair photoproduction
\cite{Brodsky:1999mz},
in diffractive $\pi^{+} \pi^{-}$ pair photoproduction
\cite{Hagler:2002nh,Hagler:2002nf,Ginzburg:2002zd,Bolz:2014mya},
and in the production of two pion pairs in photon-photon collisions \cite{Pire:2008xe}. 
However, so far in no one of the exclusive reactions 
a clear identification of the odderon was found experimentally.
For a more detailed review of the phenomenological and 
theoretical status of the odderon we refer the reader to \cite{Ewerz:2003xi,Ewerz:2005rg}.
In this context we would also like to mention the EMMI workshop on
``Central exclusive production at the LHC'' which was held
in Heidelberg in February 2019. There, questions of odderon searches
were extensively discussed. Corresponding remarks and the link to the talks
presented at this workshop can be found in~\cite{Ewerz:2019arb}.

Recently, the possibility of probing the odderon 
in ultraperipheral proton-ion collisions was considered
\cite{Goncalves:2018pbr,Harland-Lang:2018ytk}.
In \cite{Goncalves:2015hra} the measurement of the exclusive $\eta_{c}$ production
in nuclear collisions was discussed.
The situation of the odderon in this context is also not obvious and requires further studies.

In \cite{Ewerz:2013kda} the tensor-pomeron and vector-odderon concept 
was introduced for soft reactions.
In this approach, the $C = +1$ pomeron and the reggeons 
$\Reg_{+} = f_{2 \Reg}, a_{2 \Reg}$ are treated as effective
rank-2 symmetric tensor exchanges
while the $C = -1$ odderon and the reggeons 
$\Reg_{-} = \omega_{\Reg}, \rho_{\Reg}$ are treated 
as effective vector exchanges.
For these effective exchanges a number of propagators and vertices, 
respecting the standard rules of quantum field theory, 
were derived from comparisons with experiments.
This allows for an easy construction of amplitudes for specific processes.
In \cite{Ewerz:2016onn} the helicity structure of small-$|t|$ proton-proton elastic scattering
was considered in three models for the pomeron: tensor, vector, and scalar.
Only the tensor ansatz for the pomeron was found to be compatible with 
the high-energy experiment on polarised $pp$ elastic scattering \cite{Adamczyk:2012kn}.
In \cite{Britzger:2019lvc} the authors, 
using combinations of two tensor-type pomerons (a soft one and a hard one) 
and the $\Reg_{+}$-reggeon exchange,
successfully described low-$x$ deep-inelastic lepton-nucleon scattering and photoproduction.

Applications of the tensor-pomeron and vector-odderon ansatz
were given for photoproduction of pion pairs in \cite{Bolz:2014mya}
and for a number of central-exclusive-production (CEP) reactions in proton-proton collisions 
in \cite{Lebiedowicz:2013ika,Lebiedowicz:2014bea,Lebiedowicz:2016ioh,
Lebiedowicz:2016zka,Lebiedowicz:2016ryp,Lebiedowicz:2018sdt,
Lebiedowicz:2018eui,Lebiedowicz:2019jru,Lebiedowicz:2019por}.
Also contributions from the subleading exchanges, $\Reg_{+}$ and $\Reg_{-}$, 
were discussed in these works.
As an example, for the $pp \to pp p\bar{p}$ reaction \cite{Lebiedowicz:2018sdt}
the contributions involving the odderon are expected to be small 
since its coupling to the proton is very small.
We have predicted asymmetries in the (pseudo)rapidity distributions 
of the centrally produced antiproton and proton.
The asymmetry is caused by interference effects of the dominant ($\Pom, \Pom$)
with the subdominant ($\Ode + \Reg_{-}$, $\Pom + \Reg_{+}$) 
and ($\Pom + \Reg_{+}$, $\Ode + \Reg_{-}$) exchanges.
We find for the odderon only very small effects,
roughly a factor 10 smaller than the effects due to reggeons.

In this paper we consider the possibility of observing odderon exchange
in the $p p \to p p \phi$, $pp \to pp (\phi \to K^{+} K^{-})$,
and $p p \to p p (\phi \to \mu^{+} \mu^{-})$ reactions
in the light of our recent analysis of the $p p \to p p \phi \phi$ reaction \cite{Lebiedowicz:2019jru}.
In the diffractive production of $\phi$ meson pairs 
it is possible to have pomeron-pomeron fusion 
with intermediate $\hat{t}/\hat{u}$-channel odderon exchange.
Thus, the $pp \to pp \phi \phi$ reaction
is a good candidate for the odderon-exchange searches, 
as it does not involve the coupling of the odderon to the proton.
By confronting our model results, including the odderon,
the reggeized $\phi$ exchange, and the $f_{2}(2340)$ resonance exchange contributions,
with the WA102 data from \cite{Barberis:1998bq} 
we derived an upper limit for the $\Pom \Ode \phi$ coupling.
Taking into account typical kinematic cuts for LHC experiments
in the $pp \to pp \phi \phi \to pp K^{+}K^{-}K^{+}K^{-}$ reaction
we have found that the odderon exchange contribution
should be distinguishable from other contributions
for large rapidity distance between the outgoing $\phi$ mesons 
and in the region of large four-kaon invariant masses.
At least, it should be possible to derive an upper limit 
on the odderon contribution in this reaction.

Here we will try to understand the $p p \to p p \phi$ reaction
at relatively low center-of-mass energy $\sqrt{s} = 29.1$~GeV 
by comparing our model results with the WA102 experimental data from \cite{Kirk:2000ws}.
We shall calculate the photoproduction mechanism. 
For this purpose we have to consider also low-energy photon-proton collisions 
in the $\gamma p \to \phi p$ reaction where
the corresponding mechanism is not well established yet;
see, e.g., Refs.~\cite{Laget:2000gj,Oh:2001bq,
Titov:1999eu,Titov:2003bk,Ozaki:2009mj,
Kiswandhi:2010ub,Kiswandhi:2011cq,
Ryu:2012tw,Kong:2016scm,Kim:2019kef}.
Of course, the amplitude for $\gamma p \to \phi p$ 
cannot be realised by the \mbox{$C = -1$} odderon exchange.
In addition to the $\gamma$-$\Pom$-fusion processes
we shall estimate also subleading contributions, e.g.
the $\gamma$-pseudoscalar-meson fusion,
the $\phi$-$\Pom$ fusion, the $\omega$-$\Pom$ fusion, 
the $\omega$-$f_{2 \Reg}$ fusion, and the $\rho$-$\pi^{0}$ fusion,
to determine their role in the $p p \to p p \phi$ reaction.
Our aim is to see how much room is left for the $\Ode$-$\Pom$ fusion
which is the main object of our studies.

Our paper is organized as follows.
In Sec.~\ref{sec:pp_ppKK} we consider the $p p \to p p (\phi \to K^{+} K^{-})$ reaction.
Section \ref{sec:pp_ppLL} deals with $\mu^{+} \mu^{-}$ production.
For both reactions we give analytic expressions for the resonant amplitudes.
Section~\ref{sec:results} contains the comparison of our results for 
the $pp \to pp \phi$ reaction with the WA102 data.
We discuss the role of different contributions
such as $\gamma$-$\Pom$, $\Ode$-$\Pom$, $\phi$-$\Pom$, 
$\omega$-$\Pom$,
and $\omega$-$f_{2 \Reg}$ fusion processes.
Then we turn to high energies and show numerical results for total
and differential cross sections calculated 
with typical experimental cuts for the LHC experiments.
We discuss our predictions for the $K^+ K^-$ channel for $\sqrt{s} = 13$~TeV.
In addition, we present our predictions 
for the $\mu^{+} \mu^{-}$ production
also at $\sqrt{s} = 13$~TeV which is currently under analysis by the LHCb Collaboration.
We briefly discuss and/or provide references to relevant works
for the continuum contributions.
Section~\ref{sec:conclusions} presents our conclusions and further prospects.
In Appendices~\ref{sec:appendixA} and \ref{sec:appendixB}
we discuss useful relations and properties concerning
the photoproduction of $\omega$ and $\phi$ mesons.
In Appendix~\ref{sec:subleading} 
we discuss the subleading processes
contributing to $pp \to pp (\phi \to K^{+}K^{-})$.
We have collected there some useful formulas concerning
details of the calculations.
In Appendix~\ref{sec:appendixC} we give
the definition of the Collins-Soper (CS) frame used in our paper.

In our paper we denote by $e > 0$ the proton charge.
We use the $\gamma$-matrix conventions of Bjorken and Drell \cite{Bjorken_Drell_1965}.
The totally antisymmetric Levi-Civita symbol
$\varepsilon_{\mu \nu \kappa \lambda}$ is used with the normalisation
$\varepsilon_{0123} = 1$.

\section{The $pp \to pp \phi \to pp K^{+}K^{-}$ reaction}
\label{sec:pp_ppKK}

Here we discuss the reaction
\begin{eqnarray}
p(p_{a},\lambda_{a}) + p(p_{b},\lambda_{b}) \to
p(p_{1},\lambda_{1}) + K^{+}(p_{3}) + K^{-}(p_{4}) + p(p_{2},\lambda_{2}) \,,
\label{2to4_reaction_KK}
\end{eqnarray}
where $p_{a,b}$, $p_{1,2}$ and 
$\lambda_{a,b}, \lambda_{1,2} = \pm \frac{1}{2}$ 
denote the four-momenta and helicities of the protons and
$p_{3,4}$ denote the four-momenta of the $K$ mesons, respectively.

\begin{figure}
(a)\includegraphics[width=6.5cm]{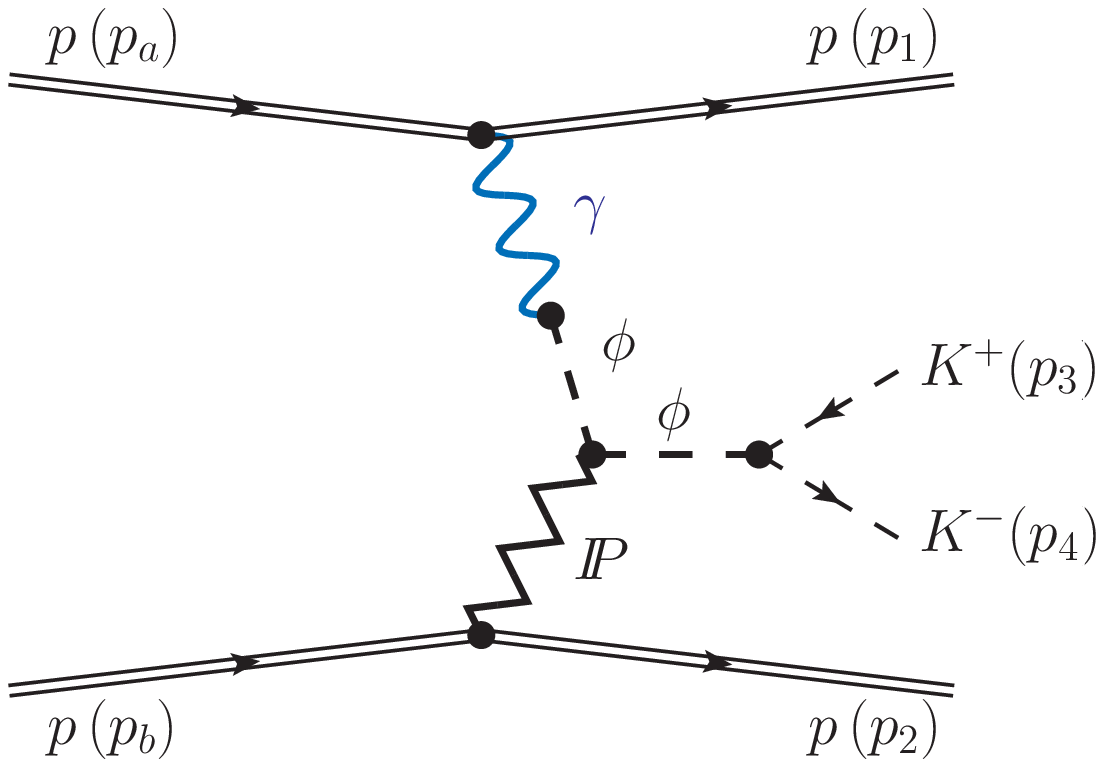}   
(b)\includegraphics[width=6.5cm]{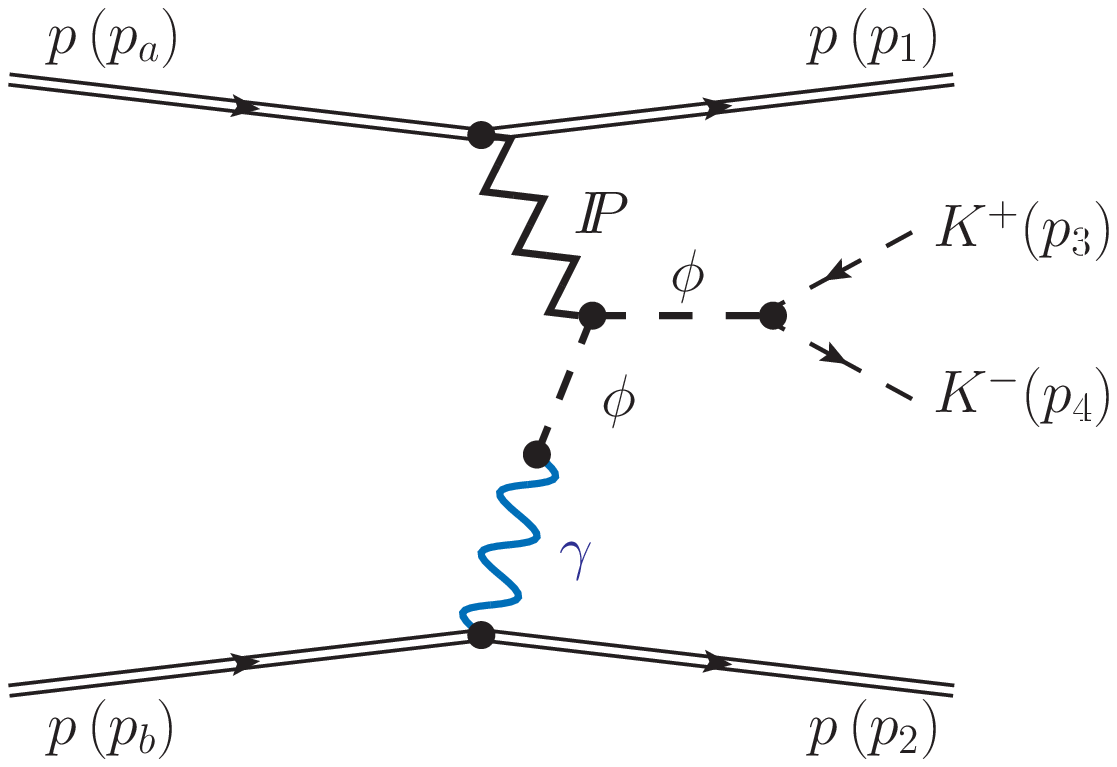}   
  \caption{\label{fig:diagrams_phi_photon}
The Born-level diagrams for central exclusive $\phi$-meson
photoproduction in proton-proton collisions
with the subsequent decay $\phi \to K^{+}K^{-}$:
(a) photon-pomeron fusion; (b) pomeron-photon fusion.}
\end{figure}
\begin{figure}
(a)\includegraphics[width=6.5cm]{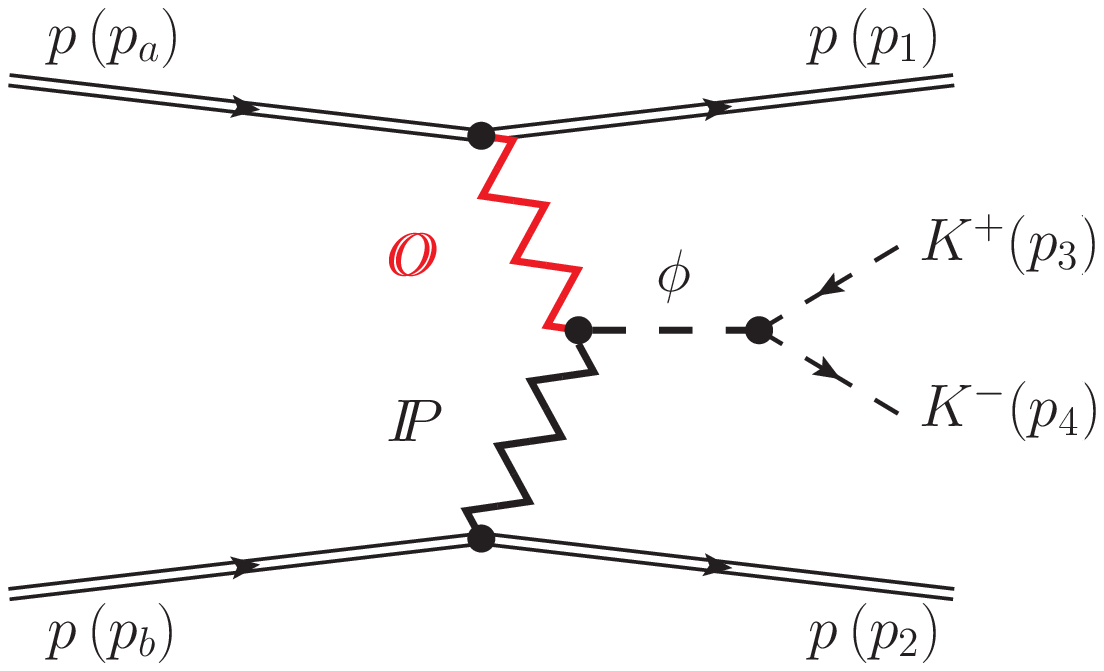}   
(b)\includegraphics[width=6.5cm]{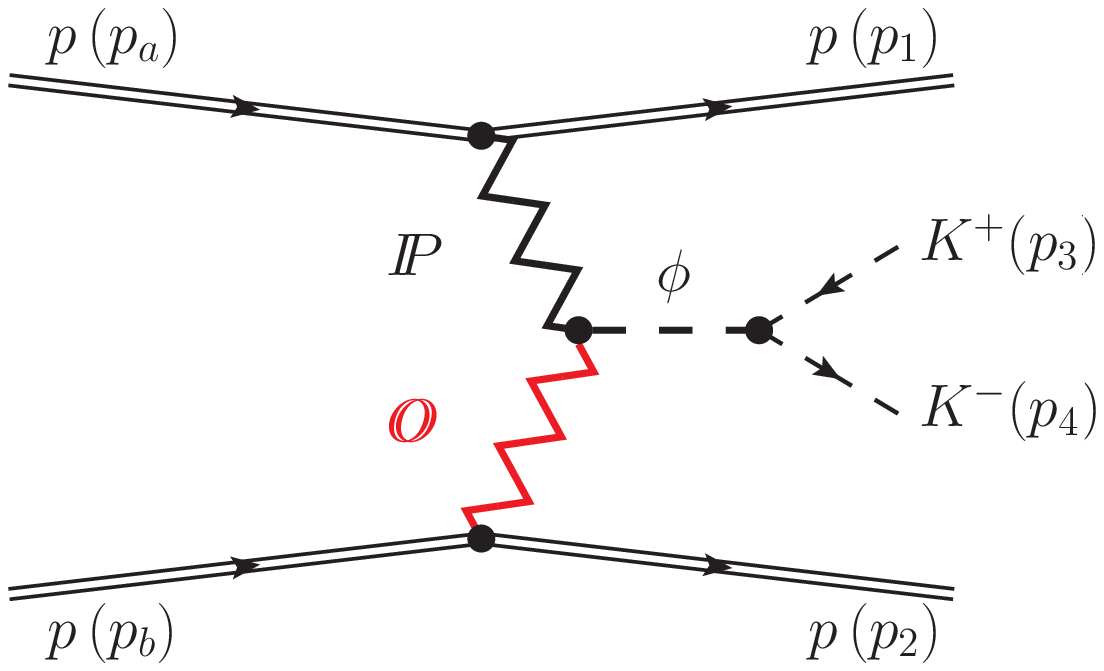}  
  \caption{\label{fig:diagrams_phi_odderon}
The Born-level diagrams for diffractive production of a $\phi$ meson
decaying to $K^+ K^-$ in proton-proton collisions with odderon exchange:
(a) odderon-pomeron fusion; (b) pomeron-odderon fusion.
}
\end{figure}

The full amplitude of the reaction (\ref{2to4_reaction_KK})
is a sum of the continuum amplitude and the amplitudes
through the $s$-channel resonances as was discussed in detail in 
\cite{Lebiedowicz:2018eui}.
Here we focus on the limited dikaon invariant mass region,
i.e., the $\phi \equiv \phi(1020)$ resonance region,
\begin{eqnarray}
1.01 \; \mathrm{GeV} < M_{K^{+}K^{-}} < 1.03\; \mathrm{GeV}\,.
\label{M43_phi_region}
\end{eqnarray}
That is, we consider the reaction
\begin{eqnarray}
p(p_{a},\lambda_{a}) + p(p_{b},\lambda_{b}) \to
p(p_{1},\lambda_{1}) + 
\left[ \phi(p_{34}) \to K^{+}(p_{3}) + K^{-}(p_{4}) \right] 
+ p(p_{2},\lambda_{2}) \,.
\label{2to4_reaction_KK_via_phi}
\end{eqnarray}
The kinematic variables are
\begin{eqnarray}
&&p_{34} = p_{3} + p_{4}\,, \quad q_1 = p_{a} - p_{1}\,, \quad q_2 = p_{b} - p_{2}\,,
\nonumber \\ 
&&s = (p_{a} + p_{b})^{2} = (p_{1} + p_{2} + p_{34})^{2}\,,
\nonumber \\ 
&&t_1 = q_{1}^{2}\,, 
\quad t_2 = q_{2}^{2}\,,
\nonumber \\
&&
s_{1} = (p_{1} + p_{34})^{2}\,, \quad
s_{2} = (p_{2} + p_{34})^{2}\,.
\label{2to4_kinematic}
\end{eqnarray}

For high energies and central $\phi$ production we 
expect the process (\ref{2to4_reaction_KK_via_phi})
to be dominated by diffractive scattering.
The corresponding diagrams are shown in Figs.~\ref{fig:diagrams_phi_photon} 
and~\ref{fig:diagrams_phi_odderon}.
That is, we consider the fusion processes $\gamma \Pom \to \phi$
and $\Ode \Pom \to \phi$.
For the first process all couplings are, in essence, known.
For the odderon-exchange process we shall use the ans{\"a}tze from \cite{Ewerz:2013kda}
and we shall try to get information on the odderon parameters 
and couplings from the reaction (\ref{2to4_reaction_KK_via_phi}).
The amplitude for (\ref{2to4_reaction_KK_via_phi}) gets the following contributions 
from these diagrams
\begin{eqnarray}
&&^{(1)}{\cal M}_{pp \to pp K^{+}K^{-}}^{(\phi \to K^{+}K^{-})} = 
{\cal M}^{(\gamma \Pom)}_{pp \to pp K^{+}K^{-}} + {\cal M}^{(\Pom \gamma)}_{pp \to pp K^{+}K^{-}}\,, \\
&&^{(2)}{\cal M}_{pp \to pp K^{+}K^{-}}^{(\phi \to K^{+}K^{-})} = 
{\cal M}^{(\Ode \Pom)}_{pp \to pp K^{+}K^{-}} + {\cal M}^{(\Pom \Ode)}_{pp \to pp K^{+}K^{-}}\,.
\label{pp_ppphi}
\end{eqnarray}

At the relatively low center-of-mass energy of the WA102 experiment, 
$\sqrt{s} = 29.1$~GeV, we have to include also 
subleading contributions with meson exchanges
discussed in Appendix~\ref{sec:subleading}.

To give the full physical amplitude, for instance, 
for the $pp \to pp K^{+}K^{-}$ process (\ref{2to4_reaction_KK})
we should include absorptive corrections to the Born amplitudes.
For the details how to include the $pp$-rescattering corrections 
in the eikonal approximation for the four-body reaction
see, e.g., Sec.~3.3 of \cite{Lebiedowicz:2014bea} and \cite{Lebiedowicz:2015eka}.

Below, in Table~\ref{tab:table2} of Sec.~\ref{predictions_LHC}, 
we give numerical values
for the gap survival factors (``soft survival probability'' factors) 
denoted as $\langle S^{2} \rangle$,
the ratios of full (including absorption) and Born cross sections.

The measurement of forward protons would be useful
to better understand absorption effects.
The {\tt GenEx} Monte Carlo generator \cite{Kycia:2014hea,Kycia:2017ota}
could be used in this context.
We refer the reader to \cite{Kycia:2017iij} where a first calculation
of four-pion continuum production in the $pp \to pp \pi^{+}\pi^{-}\pi^{+}\pi^{-}$ reaction
with the help of the {\tt GenEx} code was performed.

\subsection{$\gamma$-$\Pom$ fusion}
\label{sec:gamma_pomeron}

The Born-level amplitude for the $\gamma$-$\Pom$ exchange,
see diagram~(a) in Fig.~\ref{fig:diagrams_phi_photon},
reads 
\begin{eqnarray}
&&{\cal M}^{(\gamma \Pom)}_{pp \to pp K^{+}K^{-}} 
= (-i)
\bar{u}(p_{1}, \lambda_{1}) 
i\Gamma^{(\gamma pp)}_{\mu}(p_{1},p_{a}) 
u(p_{a}, \lambda_{a}) \nonumber \\
&&\qquad \times  
i\Delta^{(\gamma)\,\mu \sigma}(q_{1})\, 
i\Gamma^{(\gamma \to \phi)}_{\sigma \nu}(q_{1})\,
i\Delta^{(\phi)\,\nu \rho_{1}}(q_{1}) \,
i\Gamma^{(\Pom \phi \phi)}_{\rho_{2} \rho_{1} \alpha \beta}(p_{34},q_{1})\, 
i\Delta^{(\phi)\,\rho_{2} \kappa}(p_{34})\,
i\Gamma^{(\phi KK)}_{\kappa}(p_{3},p_{4})
\nonumber \\
&& \qquad \times 
i\Delta^{(\Pom)\,\alpha \beta, \delta \eta}(s_{2},t_{2}) \,
\bar{u}(p_{2}, \lambda_{2}) 
i\Gamma^{(\Pom pp)}_{\delta \eta}(p_{2},p_{b}) 
u(p_{b}, \lambda_{b}) \,.
\label{amplitude_gamma_pomeron}
\end{eqnarray}

The $\gamma pp$ vertex and the photon propagator are given in \cite{Ewerz:2013kda} 
by formulas (3.26) and (3.1), respectively.
The $\gamma \to \phi$ transition is made here through
the vector-meson-dominance (VMD) model;
see (3.23)--(3.25) of \cite{Ewerz:2013kda}.
$\Delta^{(\Pom)}$ and $\Gamma^{(\Pom pp)}$ 
denote the effective propagator and proton vertex function,
respectively, for the tensorial pomeron.
The corresponding expressions, as given in Sec.~3 of \cite{Ewerz:2013kda}, are as follows
\begin{eqnarray}
&&i \Delta^{(\Pom)}_{\mu \nu, \kappa \lambda}(s,t) = 
\frac{1}{4s} \left( g_{\mu \kappa} g_{\nu \lambda} 
                  + g_{\mu \lambda} g_{\nu \kappa}
                  - \frac{1}{2} g_{\mu \nu} g_{\kappa \lambda} \right)
(-i s \alpha'_{\Pom})^{\alpha_{\Pom}(t)-1}\,,
\label{A1}\\
&&i\Gamma_{\mu \nu}^{(\Pom pp)}(p',p)
=-i 3 \beta_{\Pom NN} F_{1}(t)
\left\lbrace 
\frac{1}{2} 
\left[ \gamma_{\mu}(p'+p)_{\nu} 
     + \gamma_{\nu}(p'+p)_{\mu} \right]
- \frac{1}{4} g_{\mu \nu} (\slash{p}' + \slash{p})
\right\rbrace, \qquad
\label{A4}
\end{eqnarray}
where $t = (p'-p)^{2}$ and $\beta_{\Pom NN} = 1.87$~GeV$^{-1}$.
For simplicity we use for the pomeron-nucleon coupling 
the electromagnetic Dirac form factor $F_{1}(t)$ of the proton.
The pomeron trajectory $\alpha_{\Pom}(t)$
is assumed to be of standard linear form, see e.g. \cite{Donnachie:1992ny,Donnachie:2002en},
\begin{eqnarray}
&&\alpha_{\Pom}(t) = \alpha_{\Pom}(0)+\alpha'_{\Pom}\,t\,,\\ 
&&
\alpha_{\Pom}(0) = 1.0808\,, \;\;
\alpha'_{\Pom} = 0.25 \; \mathrm{GeV}^{-2}\,.
\label{trajectory}
\end{eqnarray}

Our ansatz for the $\Pom \phi\phi$ vertex follows the one for the $\Pom \rho\rho$ 
in (3.47) of \cite{Ewerz:2013kda} 
with the replacements $a_{\Pom \rho \rho} \to a_{\Pom \phi \phi}$ and 
$b_{\Pom \rho \rho} \to b_{\Pom \phi \phi}$.
This was already used in Sec.~IV~B of \cite{Lebiedowicz:2018eui}.
The $\Pom \phi\phi$ vertex function is taken with the same Lorentz structure 
as for the $f_{2} \gamma \gamma$ coupling defined in (3.39) of \cite{Ewerz:2013kda}.
With $k', \mu$ and $k,\nu$ the momentum and vector index
of the outgoing and incoming $\phi$, respectively,
and $\kappa \lambda$ the pomeron indices the $\Pom \phi\phi$ vertex reads
\begin{eqnarray}
i\Gamma^{(\Pom \phi \phi)}_{\mu \nu \kappa \lambda}(k',k) &=&
i F_{M}((k'-k)^{2}) \,\tilde{F}^{(\phi)}(k'^{2})\,\tilde{F}^{(\phi)}(k^{2}) \nonumber \\ 
&&\times 
\left[2a_{\Pom \phi \phi}\,\Gamma^{(0)}_{\mu \nu \kappa \lambda}(k',-k)\,
     - b_{\Pom \phi \phi}\,\Gamma^{(2)}_{\mu \nu \kappa \lambda}(k',-k) \right]\,,
\label{vertex_pomphiphi}
\end{eqnarray}  
%
with form factors $F_{M}$ and $\tilde{F}^{(\phi)}$ and two rank-four tensor functions,
\begin{eqnarray}
\label{3.16}
&&\Gamma_{\mu\nu\kappa\lambda}^{(0)} (k_1,k_2) =
\Big[(k_1 \cdot k_2) g_{\mu\nu} - k_{2\mu} k_{1\nu}\Big] 
\Big[k_{1\kappa}k_{2\lambda} + k_{2\kappa}k_{1\lambda} - 
\frac{1}{2} (k_1 \cdot k_2) g_{\kappa\lambda}\Big] \,,\\
\label{3.17}
&&\Gamma_{\mu\nu\kappa\lambda}^{(2)} (k_1,k_2) = \,
 (k_1\cdot k_2) (g_{\mu\kappa} g_{\nu\lambda} + g_{\mu\lambda} g_{\nu\kappa} )
+ g_{\mu\nu} (k_{1\kappa} k_{2\lambda} + k_{2\kappa} k_{1\lambda} ) \nonumber \\
&& \qquad \qquad \qquad \quad - k_{1\nu} k_{2 \lambda} g_{\mu\kappa} - k_{1\nu} k_{2 \kappa} g_{\mu\lambda} 
- k_{2\mu} k_{1 \lambda} g_{\nu\kappa} - k_{2\mu} k_{1 \kappa} g_{\nu\lambda} 
\nonumber \\
&& \qquad \qquad \qquad \quad - [(k_1 \cdot k_2) g_{\mu\nu} - k_{2\mu} k_{1\nu} ] \,g_{\kappa\lambda} \,.
\end{eqnarray}
For details see Eqs.~(3.18)--(3.22) of \cite{Ewerz:2013kda}.
In (\ref{vertex_pomphiphi}) the coupling parameters 
$a_{\Pom \phi \phi}$ and $b_{\Pom \phi \phi}$ 
have dimensions GeV$^{-3}$ and GeV$^{-1}$, respectively.
In \cite{Lebiedowicz:2018eui} we have fixed the coupling parameters 
of the tensor pomeron to the $\phi$ meson based on the HERA experimental data for 
the $\gamma p \to \phi p$ reaction \cite{Derrick:1996af,Breitweg:1999jy}.
However, the $\omega$-$\phi$ mixing effect was not taken into account there.
In the calculation here we include the $\omega$-$\phi$ mixing and
we take the coupling parameters found in Appendix~\ref{sec:appendixB}.

The full form of the vector-meson propagator is given by (3.2) of \cite{Ewerz:2013kda}.
Using the properties of the tensorial functions (\ref{3.16}) and (\ref{3.17}), 
see (3.18)--(3.22) of \cite{Ewerz:2013kda}, 
we can make for the $\phi$-meson propagator the following replacement
\begin{eqnarray}
\Delta^{(\phi)}_{\mu \nu}(k) \to -g_{\mu \nu} \, 
\Delta^{(\phi)}_{T}(k^{2})\,,
\label{delta_phi}
\end{eqnarray}
where we take the simple Breit-Wigner expression,
as discussed in \cite{Lebiedowicz:2018eui},
\begin{eqnarray}
&&\Delta_T^{(\phi)}(s) = 
\frac{1}{s - m_{\phi}^2 + i \sqrt{s} \Gamma_{\phi}(s)} \,,
\label{phi_propagator}\\
&&\Gamma_{\phi}(s) = \Gamma_{\phi}
\left( \frac{s - 4 m_{K}^2}{m_{\phi}^2 - 4 m_{K}^2} \right)^{3/2}
\frac{m_{\phi}^2}{s} \, \theta(s-4 m_{K}^2)\,.
\label{phi_propagator_aux}
\end{eqnarray}
%

For the $\phi KK$ vertex we have from (4.24)--(4.26) of \cite{Lebiedowicz:2018eui}
\begin{eqnarray}
i\Gamma^{(\phi KK)}_{\kappa}(p_{3},p_{4}) = 
-\frac{i}{2} \,g_{\phi K^{+} K^{-}} \, \left( p_{3} - p_{4} \right)_{\kappa}\,
F^{(\phi K K)}(p_{34}^{2})
\label{phiKK}
\end{eqnarray}
with $g_{\phi K^{+} K^{-}} = 8.92$ and $F^{(\phi K K)}$ a form factor.

In the hadronic vertices we take into account corresponding form factors.
We insert in the $\Pom \phi \phi$ vertex (\ref{vertex_pomphiphi})
the form factor $F_{M}(k^{2})$ to take into account 
the extended nature of $\phi$ mesons and
$\tilde{F}^{(\phi)}(k^{2})$ since we are dealing with two off-shell $\phi$ mesons;
see (4.27) of \cite{Lebiedowicz:2018eui} and (B.85) of \cite{Bolz:2014mya}.
Convenient forms are 
\begin{eqnarray}
&&F_{M}(k^{2})= \frac{1}{1-k^{2}/\Lambda_{0,\,\Pom\phi\phi}^{2}}\,, 
\label{Fpion}
\\
&&\tilde{F}^{(\phi)}(k^{2}) 
= \left[1 + \dfrac{k^{2}(k^{2}-m_{\phi}^{2})}{\tilde{\Lambda}_{\phi}^{4}} \right]^{-\tilde{n}_{\phi}},
\quad \tilde{\Lambda}_{\phi} = 2\;{\rm GeV}\,,
\quad \tilde{n}_{\phi} = 0.5\,.
\label{ff_Nachtmann}
\end{eqnarray}
We have $\tilde{F}^{(\phi)}(0) = \tilde{F}^{(\phi)}(m_{\phi}) = 1$.
In (\ref{Fpion}) we take $\Lambda_{0,\,\Pom\phi\phi}^{2} = 1.0$~GeV$^{2}$ (set A) or
$\Lambda_{0,\,\Pom\phi\phi}^{2} = 4.0$~GeV$^{2}$ (set B); 
see Fig.~\ref{fig:gamp_phip_highW} of Appendix~\ref{sec:appendixB}.
In practical calculations we include also in the $\phi KK$ vertex the form factor
\mbox{[see (4.28) of \cite{Lebiedowicz:2018eui}]}
\begin{eqnarray}
F^{(\phi K K)}(k^{2}) = 
\exp{ \left( \frac{-(k^{2}-m_{\phi}^{2})^{2}}{\Lambda_{\phi}^{4}} \right)}\,,
\quad \Lambda_{\phi} = 1\;{\rm GeV}\,.
\label{FphiKK_ff}
\end{eqnarray}

Inserting all this in (\ref{amplitude_gamma_pomeron}) 
we can write the amplitude for the $\gamma \Pom$ fusion as follows
\begin{eqnarray}
&&{\cal M}^{(\gamma \Pom)}_{pp \to pp K^{+}K^{-}} 
= -i\,e^{2}\,
\bar{u}(p_{1}, \lambda_{1}) 
\left[ \gamma^{\alpha}F_{1}(t_{1}) + \frac{i}{2m_{p}} \sigma^{\alpha \alpha'} (p_{1}-p_{a})_{\alpha'} F_{2}(t_{1}) \right]
u(p_{a}, \lambda_{a})\nonumber \\
&&\qquad \times  
\frac{1}{t_{1}}\, 
\frac{(-m_{\phi}^{2})}{t_{1}-m_{\phi}^{2}}\,\frac{1}{\gamma_{\phi}} 
\Delta_{T}^{(\phi)}(p_{34}^{2})\,
\frac{g_{\phi K^{+}K^{-}}}{2}\,(p_{3}-p_{4})^{\beta}\, F^{(\phi K K)}(p_{34}^{2})
\nonumber \\
&& \qquad \times 
\left[ 2 a_{\Pom \phi \phi}\, \Gamma^{(0)}_{\beta \alpha \kappa \lambda}(p_{34},-q_{1})
- b_{\Pom \phi \phi}\,\Gamma^{(2)}_{\beta \alpha \kappa \lambda}(p_{34},-q_{1}) \right]
\tilde{F}^{(\phi)}(t_{1})\,\tilde{F}^{(\phi)}(p_{34}^{2})\,F_{M}(t_{2})
\nonumber \\
&& \qquad \times 
\frac{1}{2 s_{2}} \left( -i s_{2} \alpha'_{\Pom} \right)^{\alpha_{\Pom}(t_{2})-1}\,
3 \beta_{\Pom NN} \, F_{1}(t_{2})\, 
\bar{u}(p_{2}, \lambda_{2}) 
\left[ \gamma^{\kappa} (p_{2} + p_{b})^{\lambda} \right]
u(p_{b}, \lambda_{b})\,.
\label{amplitude_gamma_pomeron_aux}
\end{eqnarray}
Here $\gamma_{\phi}$ is the $\gamma$-$\phi$ coupling constant;
see (3.23)--(3.25) of \cite{Ewerz:2013kda}.

For the $\Pom \gamma$-exchange we have the same structure as for the above amplitude
with 
\begin{eqnarray}
( p\, (p_{a}, \lambda_{a}), p\, (p_{1}, \lambda_{1}) )
\leftrightarrow 
( p\, (p_{b}, \lambda_{b}), p\, (p_{2}, \lambda_{2}) )\,,\;
t_{1}  \leftrightarrow t_{2}\,, \;
q_{1}  \leftrightarrow q_{2}\,, \; 
s_{1}  \leftrightarrow s_{2}\,.
\label{replace}
\end{eqnarray}

In the following we shall also consider the single $\phi$ CEP
in $pp$ collisions
\begin{eqnarray}
p(p_{a},\lambda_{a}) + p(p_{b},\lambda_{b}) \to
p(p_{1},\lambda_{1}) + \phi(p_{34}, \epsilon_{(\phi)}) + p(p_{2},\lambda_{2}) \,.
\label{2to3_reaction}
\end{eqnarray}
In (\ref{2to3_reaction}) $\epsilon_{(\phi)}$ denotes the polarisation vector
of the $\phi$ and we have $p_{34}^{2} = m_{\phi}^{2}$.
The amplitude for the $\gamma \Pom$-fusion contribution 
to the reaction (\ref{2to3_reaction}) 
is obtained from (\ref{amplitude_gamma_pomeron}) 
by making the replacement
\begin{eqnarray}
i\Delta^{(\phi)\,\rho_{2} \kappa}(p_{34})\, 
i\Gamma^{(\phi KK)}_{\kappa}(p_{3},p_{4}) \to
\epsilon_{(\phi)}^{*\, \rho_{2}}\,.
\label{2to3_replacement}
\end{eqnarray}
The same replacement holds for the $\Pom \gamma$-fusion contribution.
Analogous replacements hold for all other diagrams
when going from the reaction (\ref{2to4_reaction_KK_via_phi}) to (\ref{2to3_reaction}).

\subsection{$\Ode$-$\Pom$ fusion}
\label{sec:ode_pom}

The amplitude for the diffractive production of the $\phi(1020)$ 
via odderon-pomeron fusion,
see diagram~(a) in Fig.~\ref{fig:diagrams_phi_odderon},
can be written as
\begin{eqnarray}
{\cal M}^{(\Ode \Pom)}_{pp \to pp K^{+}K^{-}} 
&=& (-i)
\bar{u}(p_{1}, \lambda_{1}) 
i\Gamma^{(\Ode pp)}_{\mu}(p_{1},p_{a}) 
u(p_{a}, \lambda_{a}) \nonumber \\
&&\times  
i\Delta^{(\Ode)\,\mu \rho_{1}}(s_{1}, t_{1}) \,
i\Gamma^{(\Pom \Ode \phi)}_{\rho_{1} \rho_{2} \alpha \beta}(-q_{1},p_{34})\, 
i\Delta^{(\phi)\,\rho_{2} \kappa}(p_{34})\,
i\Gamma^{(\phi KK)}_{\kappa}(p_{3},p_{4})
\nonumber \\
&&\times 
i\Delta^{(\Pom)\,\alpha \beta, \delta \eta}(s_{2},t_{2}) \,
\bar{u}(p_{2}, \lambda_{2}) 
i\Gamma^{(\Pom pp)}_{\delta \eta}(p_{2},p_{b}) 
u(p_{b}, \lambda_{b}) \,.
\label{amplitude_oderon_pomeron}
\end{eqnarray}

Our ansatz for the $C = -1$ odderon follows (3.16), (3.17) and (3.68), (3.69) 
of \cite{Ewerz:2013kda}:
\begin{eqnarray}
&&i \Delta^{(\Ode)}_{\mu \nu}(s,t) = 
-i g_{\mu \nu} \frac{\eta_{\Ode}}{M_{0}^{2}} \,(-i s \alpha'_{\Ode})^{\alpha_{\Ode}(t)-1}\,,
\label{A12} \\
&&i\Gamma_{\mu}^{(\Ode pp)}(p',p) 
= -i 3\beta_{\Ode pp} \,M_{0}\,F_{1} ( (p'-p)^{2} ) \,\gamma_{\mu}\,,
\label{A13}
\end{eqnarray}
where $\eta_{\Ode}$ is a parameter with value $\eta_{\Ode} = \pm 1$;
$M_{0} = 1$~GeV is inserted for dimensional reasons;
$\alpha_{\Ode}(t)$ is the odderon trajectory, assumed to be linear in~$t$:
%
\begin{eqnarray}
\alpha_{\Ode}(t) = \alpha_{\Ode}(0)+\alpha'_{\Ode}\,t\,. 
\label{A14}
\end{eqnarray}
The odderon parameters are not yet known from experiment.
In our calculations we shall choose as default values
\begin{eqnarray}
\alpha_{\Ode}(0) = 1.05\,,\;\; \alpha'_{\Ode} = 0.25 \; \mathrm{GeV}^{-2}\,.
\label{A14_aux}
\end{eqnarray}
The coupling of the odderon to the proton, $\beta_{\Ode pp}$, in (\ref{A13})
has dimension GeV$^{-1}$.
For our study here we shall assume 
\begin{eqnarray}
\beta_{\Ode pp} = 0.1 \;\beta_{\Pom NN} \simeq 0.18\;{\rm GeV}^{-1}\,,
\label{beta_ONN}
\end{eqnarray}
which is not excluded by the data of small-$t$ proton-proton 
high-energy elastic scattering from the TOTEM experiment 
\cite{Antchev:2017yns,Antchev:2018rec}.

For the $\Pom \Ode \phi$ vertex we use an ansatz analogous to the $\Pom \phi \phi$ vertex;
see (3.48)--(3.50) of \cite{Lebiedowicz:2019jru}.
We get then with $(-q_{1},\rho_{1})$ and $(p_{34},\rho_{2})$
the outgoing oriented momenta and the vector indices
of the odderon and the $\phi$ meson, respectively,
and $\alpha \beta$ the pomeron indices,
\begin{eqnarray}
i\Gamma^{(\Pom \Ode \phi)}_{\rho_{1} \rho_{2} \alpha \beta}(-q_{1},p_{34}) 
&=&
i \left[ 2\,a_{\Pom \Ode \phi}\, \Gamma^{(0)}_{\rho_{1} \rho_{2} \alpha \beta}(-q_{1},p_{34})
- b_{\Pom \Ode \phi}\,\Gamma^{(2)}_{\rho_{1} \rho_{2} \alpha \beta}(-q_{1},p_{34}) \right] \nonumber\\
&&\times F^{(\Pom \Ode \phi)}((p_{34}-q_{1})^{2},q_{1}^{2},p_{34}^{2}) \nonumber\\
&=&
i \left[ 2\,a_{\Pom \Ode \phi}\, \Gamma^{(0)}_{\rho_{2} \rho_{1} \alpha \beta}(p_{34},-q_{1})
- b_{\Pom \Ode \phi}\,\Gamma^{(2)}_{\rho_{2} \rho_{1} \alpha \beta}(p_{34},-q_{1}) \right] \nonumber\\
&&\times F^{(\Pom \Ode \phi)}(q_{2}^{2},q_{1}^{2},p_{34}^{2}) \,.
\label{A15}
\end{eqnarray}  
Here we use the relations (3.20) of \cite{Ewerz:2013kda}
and as in (3.49) of \cite{Lebiedowicz:2019jru}
we take the factorised form for the $\Pom \Ode \phi$ form factor
\begin{eqnarray}
F^{(\Pom \Ode \phi)}(q_{2}^{2},q_{1}^{2},p_{34}^{2}) = 
\tilde{F}_{M}(q_{2}^{2})\, \tilde{F}_{M}(q_{1}^{2})\, F^{(\phi)}(p_{34}^{2})
\label{Fpomodephi}
\end{eqnarray}
with the form factors $\tilde{F}_{M}(q^{2})$ as in (\ref{Fpion})~\footnote{Here we assume 
that $\tilde{F}_{M}(q_{1}^{2})$ and $\tilde{F}_{M}(q_{2}^{2})$
have the same form (\ref{Fpion})
with the same $\Lambda_{0,\,\Pom\Ode\phi}^{2}$ parameter.
In principle, we could take different form factors with different $\Lambda_{0}^{2}$ parameters.},
but with $\Lambda_{0,\,\Pom\phi\phi}^{2}$ replaced by $\Lambda_{0,\,\Pom\Ode\phi}^{2}$,
and $F^{(\phi)}(p_{34}^{2}) = F^{(\phi KK)}(p_{34}^{2})$ (\ref{FphiKK_ff}), respectively.
The coupling parameters $a_{\Pom \Ode \phi}$, $b_{\Pom \Ode \phi}$ in (\ref{A15})
and the cutoff parameter $\Lambda_{0,\,\Pom \Ode \phi}^{2}$ 
in the form factor $\tilde{F}_{M}(q^{2})$ (\ref{Fpomodephi})
could be adjusted to experimental data; 
see (\ref{parameters_ode_a})--(\ref{parameters_ode_c}) 
in Sec.~\ref{sec:comparison_WA102} below.

The amplitude for the $\Ode \Pom$ fusion can now be written as
\begin{eqnarray}
&&{\cal M}^{(\Ode \Pom)}_{pp \to pp K^{+}K^{-}} 
= -i\,3 \beta_{\Ode pp}\,M_{0}\,F_{1}(t_{1})\,
\bar{u}(p_{1}, \lambda_{1}) \gamma^{\alpha} u(p_{a}, \lambda_{a})\nonumber \\
&&\qquad \times  
\frac{\eta_{\Ode}}{M_{0}^{2}} \left(-i s_{1} \alpha'_{\Ode} \right)^{\alpha_{\Ode}(t_{1})-1}
\Delta_{T}^{(\phi)}(p_{34}^{2})\,
\frac{g_{\phi K^{+}K^{-}}}{2}\,(p_{3}-p_{4})^{\beta}\, F^{(\phi K K)}(p_{34}^{2})
\nonumber \\
&& \qquad \times 
\left[ 2 a_{\Pom \Ode \phi}\, \Gamma^{(0)}_{\beta \alpha \kappa \lambda}(p_{34},-q_{1})
- b_{\Pom \Ode \phi}\,\Gamma^{(2)}_{\beta \alpha \kappa \lambda}(p_{34},-q_{1}) \right]
F^{(\Pom \Ode \phi)} ( q_{2}^{2},q_{1}^{2},p_{34}^{2} )
\nonumber \\
&& \qquad \times 
\frac{1}{2 s_{2}} \left( -i s_{2} \alpha'_{\Pom} \right)^{\alpha_{\Pom}(t_{2})-1}\,
3 \beta_{\Pom NN} \, F_{1}(t_{2})\, 
\bar{u}(p_{2}, \lambda_{2}) 
\left[ \gamma^{\kappa} (p_{2} + p_{b})^{\lambda} \right]
u(p_{b}, \lambda_{b})\,.\quad
\label{amplitude_odderon_pomeron_aux}
\end{eqnarray}
For the $\Pom \Ode$-exchange we have the same structure as for the above amplitude
with the replacements (\ref{replace}).

\section{The $pp \to pp \phi \to pp \mu^{+}\mu^{-}$ reaction}
\label{sec:pp_ppLL}

In this section we will focus on the exclusive reaction
\begin{eqnarray}
p(p_{a},\lambda_{a}) + p(p_{b},\lambda_{b}) &\to&
p(p_{1},\lambda_{1}) + \phi(p_{34}) + p(p_{2},\lambda_{2}) \nonumber \\ 
&\to& p(p_{1},\lambda_{1}) + \mu^{+}(p_{3},\lambda_{3}) + \mu^{-}(p_{4},\lambda_{4}) + p(p_{2},\lambda_{2}) \,,
\label{2to4_reaction_mumu}
\end{eqnarray}
where 
$p_{a,b}$, $p_{1,2}$ and $\lambda_{a,b}, \lambda_{1,2} = \pm \frac{1}{2}$ 
denote the four-momenta and helicities of the protons and
$p_{3,4}$ and $\lambda_{3,4} = \pm \frac{1}{2}$ 
denote the four-momenta and helicities of the muons, respectively.

\begin{figure}
\includegraphics[width=5.7cm]{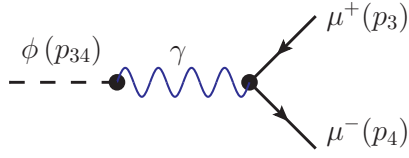}  
  \caption{\label{phi_mumu_diagram}
Decay of a $\phi$ meson to $\mu^{+}\mu^{-}$.}
\end{figure}

The amplitudes for the reaction (\ref{2to4_reaction_mumu})
through $\phi$ resonance production
can be obtained from the amplitudes discussed in Sec.~\ref{sec:pp_ppKK}
with $i\Gamma^{(\phi KK)}_{\kappa}(p_{3},p_{4})$ replaced by
$\bar{u}(p_{4}, \lambda_{4}) 
i \Gamma_{\kappa}^{(\phi \mu \mu)}(p_{3},p_{4}) 
v(p_{3}, \lambda_{3})$.
Here we describe the transition $\phi \to \gamma \to \mu^{+}\mu^{-}$,
see Fig.~\ref{phi_mumu_diagram}, by an effective vertex
\begin{eqnarray}
i\Gamma_{\kappa}^{(\phi \mu \mu)}(p_{3},p_{4}) = i g_{\phi \mu^{+} \mu^{-}} \,\gamma_{\kappa} \,.
\label{gamma_mumu}
\end{eqnarray}

The standard $\phi$-$\gamma$ coupling (see e.g. (3.23), (3.24) of \cite{Ewerz:2013kda}) gives
\begin{eqnarray}
g_{\phi \mu^{+} \mu^{-}} = -e^{2} \frac{1}{\gamma_{\phi}} \,, \quad \gamma_{\phi} < 0\,.
\label{phi_gamma_coupling_aux}
\end{eqnarray}
%

The decay rate $\phi \to \mu^{+} \mu^{-}$ is calculated from
the diagram Fig.~\ref{phi_mumu_diagram} (neglecting radiative corrections) as
\begin{eqnarray}
\Gamma(\phi \to \mu^{+} \mu^{-}) = \frac{1}{12 \pi}\, 
|g_{\phi \to \mu^{+} \mu^{-}}|^{2} \,
m_{\phi} \left( 1 + \frac{2 m_{\mu}^{2}}{m_{\phi}^{2}}\right)
\left( 1 - \frac{4 m_{\mu}^{2}}{m_{\phi}^{2}}\right)^{1/2}\,.
\label{gamma_phi_to_mumu}
\end{eqnarray}

From the experimental values \cite{Tanabashi:2018oca}
\begin{eqnarray}
&&m_{\phi} = (1019.461 \pm 0.016)\;{\rm MeV}\,, \nonumber\\
&&\Gamma(\phi \to \mu^{+} \mu^{-})/\Gamma_{\phi} = (2.86 \pm 0.19) \times 10^{-4}\,,
\nonumber\\
&&\Gamma_{\phi} = (4.249 \pm 0.013)\;{\rm MeV}\,,
\end{eqnarray}
we get 
\begin{eqnarray}
\Gamma(\phi \to \mu^{+} \mu^{-}) = (1.21 \pm 0.08)\times 10^{-3}\;{\rm MeV}
\end{eqnarray}
and using (\ref{gamma_phi_to_mumu})
\begin{eqnarray}
g_{\phi \mu^{+} \mu^{-}} = (6.71 \pm 0.22) \times 10^{-3} \,.
\label{g_phimumu_1}
\end{eqnarray}
On the other hand, using (\ref{phi_gamma_coupling_aux})
directly with the standard range for $\gamma_{\phi}$ 
quoted {\mbox{in (3.24) of \cite{Ewerz:2013kda}},
$4 \pi / \gamma_{\phi}^{2} = 0.0716 \pm 0.0017$, we get
\begin{eqnarray}
g_{\phi \mu^{+} \mu^{-}} =(6.92 \pm 0.08) \times 10^{-3}\,.
\label{g_phimumu_2}
\end{eqnarray}
Within the errors the two values obtained in (\ref{g_phimumu_1}) and (\ref{g_phimumu_2})
are compatible.
In the following we shall take (\ref{g_phimumu_2}) for our calculations.

\section{Results}
\label{sec:results}
In this section we wish to present first results 
for three cases $pp \to pp \phi(1020)$, and with $\phi$ decaying to
$K^{+}K^{-}$ or $\mu^{+}\mu^{-}$,
corresponding to the processes discussed in Secs.~\ref{sec:pp_ppKK}
and \ref{sec:pp_ppLL}.
For details how to calculate the subleading processes 
contributing to $pp \to pp (\phi \to K^{+}K^{-})$ 
we refer the reader to Appendix~\ref{sec:subleading}.

\subsection{Comparison with the WA102 data}
\label{sec:comparison_WA102}

The $\phi$-meson production in central proton-proton collisions
was studied by the WA102 Collaboration at $\sqrt{s} = 29.1$~GeV.
The experimental cross section quoted in \mbox{Table~1 of \cite{Kirk:2000ws} is}
\begin{eqnarray}
\sigma_{\rm exp} = (60 \pm 21)\,\mathrm{nb}\,.
\label{WA102_xsection}
\end{eqnarray}
In \cite{Kirk:2000ws} also the $\mathrm{dP_{t}}$ dependence of $\phi$ production
and the distribution in $\phi_{pp}$ were presented.
Here $\mathrm{dP_{t}}$ is the ``glueball-filter variable'' 
\cite{Close:1997pj,Barberis:1996iq} defined as:
\begin{eqnarray}
\bdPt = \bqta - \bqtb = \bptb - \bpta \,, \quad \mathrm{dP_{t}} = |\bdPt|\,,
\label{dPt_variable}
\end{eqnarray}
and $\phi_{pp}$ is the azimuthal angle between the transverse momentum vectors 
$\bpta$, $\bptb$ of the outgoing protons.
Both variables, $\mathrm{dP_{t}}$ 
and $\phi_{pp}$, are defined in the $pp$ center-of-mass frame.
For the kinematics see e.g. Appendix~D of \cite{Lebiedowicz:2013ika}.

In Fig.~\ref{fig:WA102_photo} (left panel) we compare our theoretical predictions 
for the $\phi_{pp}$ distribution to the WA102 experimental data 
for the $pp \to pp \phi$ reaction normalised to the central value 
of the total cross section 
$\sigma_{\rm exp} = 60$~nb from \cite{Kirk:2000ws}; see (\ref{WA102_xsection}).
We consider the two photoproduction contributions:
$\gamma \Pom$ plus $\Pom \gamma$ 
and $\gamma \widetilde{M}$ plus $\widetilde{M} \gamma$ 
with $\widetilde{M} = \pi^{0}, \eta$.
We denote, for brevity,
the coherent sum of the contributions $\gamma \Pom$ and $\Pom \gamma$
by $\gamma$-$\Pom$, the coherent sum of $\gamma \widetilde{M}$ 
and $\widetilde{M} \gamma$ by $\gamma$-$\widetilde{M}$. 
The analogous notation will be used for
these and all other contributions in the following.
For the photon-pomeron fusion we show the results 
for the two parameter sets, A and B, discussed 
in Appendix~\ref{sec:appendixB} (see Fig.~\ref{fig:gamp_phip_highW}).
For the estimation of an upper limit of the
$\gamma$-$\widetilde{M}$ contribution 
we take $\Lambda_{\widetilde{M}NN} = \Lambda_{\phi \gamma \widetilde{M}} = 1.2$~GeV
in (\ref{ff_PSNN}) and (\ref{ff_phigamPS});
see the discussion and Fig.~\ref{fig:gamp_phip} in Appendix~\ref{sec:appendixB}.
We find that the $\gamma$-$\widetilde{M}$ contribution 
is much smaller than the $\gamma$-$\Pom$ contribution. 
It constitutes about 15~$\%$ of $\gamma$-$\Pom$ in the integrated cross section.
The $\gamma$-$S$ ($S = f_{0}(500)$, $f_{0}(980)$, $a_{0}(980)$)
contribution terms are expected to be even smaller than 
the $\gamma$-$\widetilde{M}$ [$\widetilde{M} = \pi^{0}$, $\eta$] ones; 
see Fig.~\ref{fig:gamp_phip} in Appendix~\ref{sec:appendixB}.
Therefore, we neglect the $\gamma$-$\widetilde{M}$- and $\gamma$-$S$-fusion 
contributions in the further considerations.
Clearly, we see that the photoproduction mechanism is not enough to describe the WA102 data,
at least if we take the central value of $\sigma_{\rm exp}$ quoted in (\ref{WA102_xsection})
for normalising the data for the $\phi_{pp}$ distribution.

In Fig.~\ref{fig:WA102_photo} (right panel) we show 
the distributions in rapidity of the $\phi$ meson.
The photoproduction mechanisms with $\Pom$ exchange 
($\gamma \Pom$ and $\Pom \gamma$) dominate at midrapidity.
The $\gamma \widetilde{M}$ and $\widetilde{M} \gamma$ components are separated
and contribute in the backward and forward regions of ${\rm y}_{\phi}$, respectively.
The separation in rapidity means also the lack of interference effects between the
$\gamma \widetilde{M}$ and $\widetilde{M} \gamma$ components.
\begin{figure}[!ht]
\includegraphics[width=0.46\textwidth]{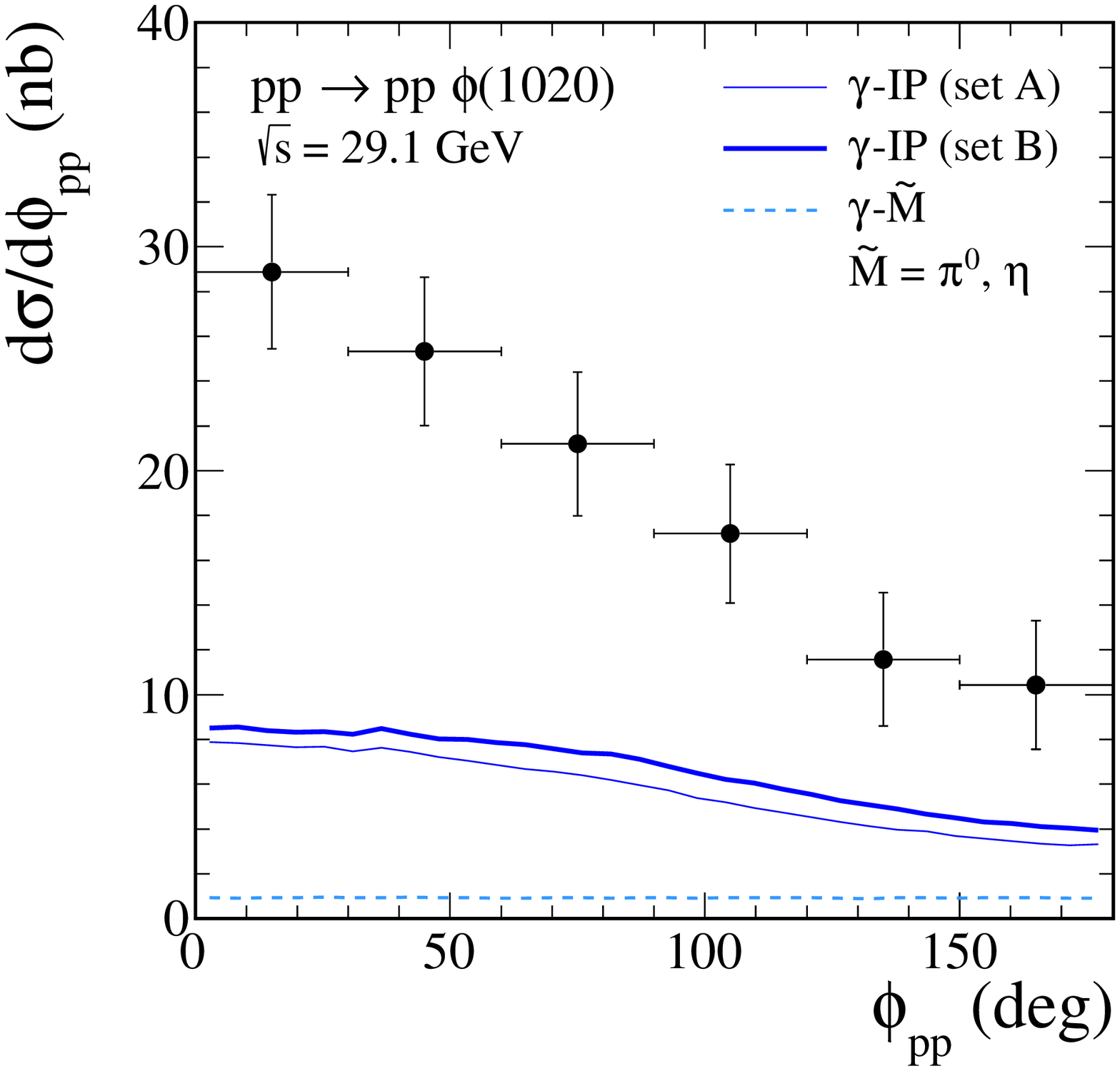}
\includegraphics[width=0.46\textwidth]{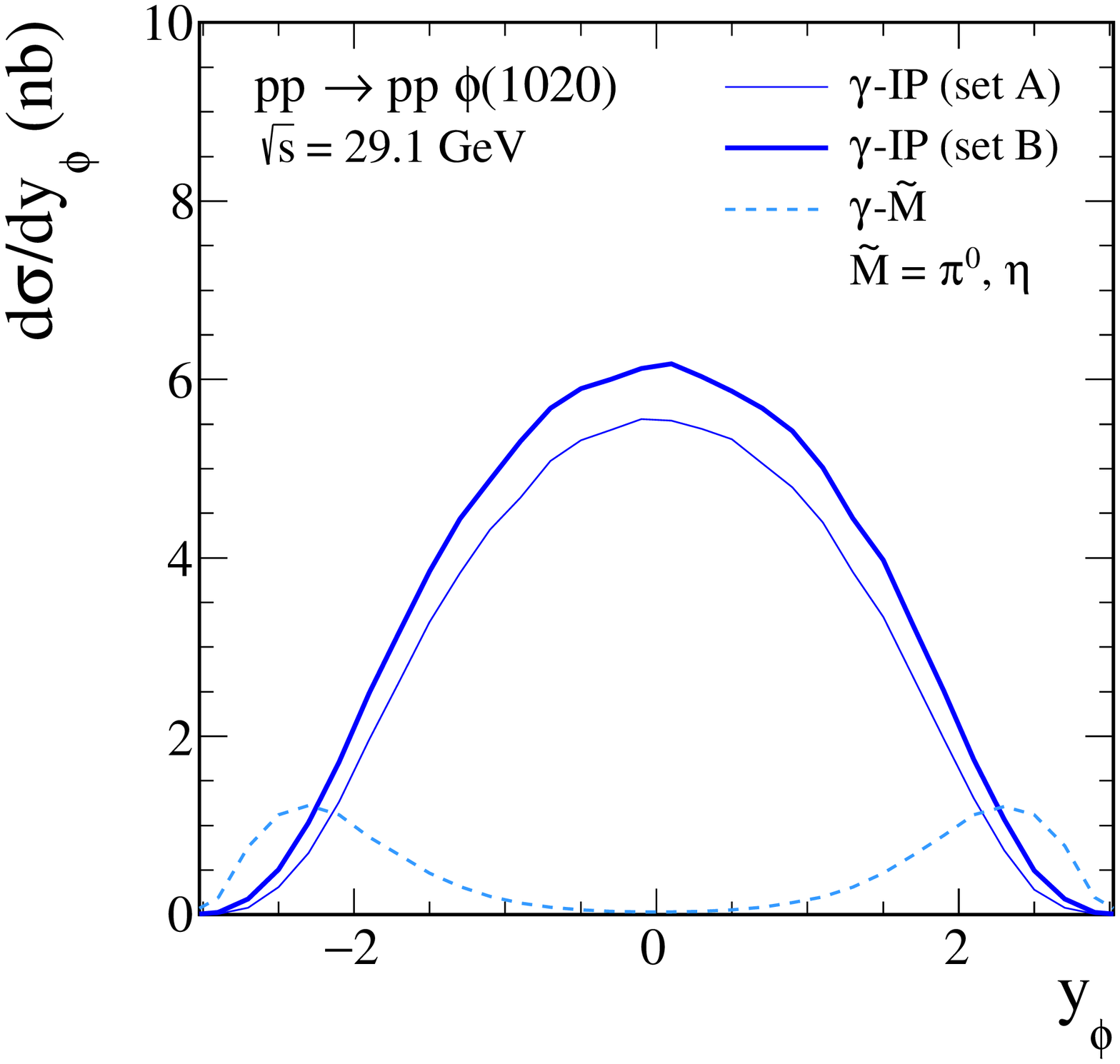}
\caption{\label{fig:WA102_photo}
The distributions in $\phi_{pp}$ and in ${\rm y}_{\phi}$
for the $\phi$ photoproduction processes
in the $pp \to pp \phi$ reaction at $\sqrt{s} = 29.1$~GeV.
The data points have been normalized to the central value for $\sigma_{\rm exp}$
(\ref{WA102_xsection}) from \cite{Kirk:2000ws}.
The results for the photon-pomeron fusion are presented
for the two parameter sets, set A and set B,
as defined in Appendix~B, see the caption of Fig.~\ref{fig:gamp_phip_highW},
(the bottom and top solid lines, respectively).
We also show the contribution from 
the $\gamma$-$\widetilde{M}$ ($\widetilde{M}= \pi^{0}, \eta$) fusion 
(the dashed lines).
The absorption effects are included here.}
\end{figure}

It is a known fact that absorption effects due to strong proton-proton interactions
have an influence on the shape of the distributions 
in $\phi_{pp}$, $\rm{dP_{t}}$, $|t_{1}|$ and $|t_{2}|$.
Thus, absorption effects should be included in realistic calculations.
In the calculations presented
we have included the absorptive corrections
in the one-channel eikonal approximation as was discussed, 
e.g., in Sec.~3.3 of \cite{Lebiedowicz:2014bea}.
The absorption effects lead to a large damping of the cross sections
for purely hadronic diffractive processes 
and a relatively small reduction of the cross section
for the photoproduction mechanism.
We obtain the ratio of full and Born cross sections $\langle S^{2} \rangle$ 
(the gap survival factor) at $\sqrt{s} = 29.1$~GeV and without any cuts included as follows
$\langle S^{2} \rangle \cong 0.8$ for the photoproduction contribution and
$\langle S^{2} \rangle \cong 0.4$ 
for the purely hadronic diffractive contributions discussed below.
However, the absorption strongly depends 
on the kinematic cuts on $|t_{1}|$ and $|t_{2}|$.
This will be discussed in detail when presenting 
our predictions for the LHC; see Sec.~\ref{predictions_LHC} below.

The question is now: what are the contributions to $\phi$ CEP
which could fill the gap between the photoproduction result
and the WA102 data in the left panel of Fig.~\ref{fig:WA102_photo}?
In the following we shall explore if this can be achieved
by the subleading fusion processes
$\omega$-$\Pom$, $\phi$-$\Pom$, $\omega$-$f_{2 \Reg}$, 
and $\rho$-$\pi^{0}$ and/or the odderon-pomeron fusion
giving a $\phi$ meson; see Appendix~\ref{sec:subleading}
and Sec.~\ref{sec:ode_pom}, respectively.

In Fig.~\ref{fig:WA102_ABS_noodd}
we show results for the $\gamma$-$\Pom$ and 
the subleading fusion processes
($\omega$-$\Pom$, $\phi$-$\Pom$, $\omega$-$f_{2 \Reg}$, and $\rho$-$\pi^{0}$).
We present results for two approaches as follows.
In the top panels (approach II) we show results 
for the reggeon-pomeron ($\phi_{\Reg}$-$\Pom$, $\omega_{\Reg}$-$\Pom$) 
and the reggeon-reggeon ($\omega_{\Reg}$-$f_{2 \Reg}$) contributions,
(\ref{omeR_to_phi_aux1})--(\ref{amplitude_omeR_pomeron_aux}),
and in the bottom panels (approach I) we show results 
for the reggeized-$\phi/\omega$-meson exchanges
(\ref{amplitude_V_pomeron_aux})--(\ref{trajectory_phi}).
The $\rho$-$\pi^{0}$ fusion contribution is calculated in the approach I,
i.e., for the reggeized $\rho^{0}$-meson exchange.
\begin{figure}[!ht]
\includegraphics[width=0.46\textwidth]{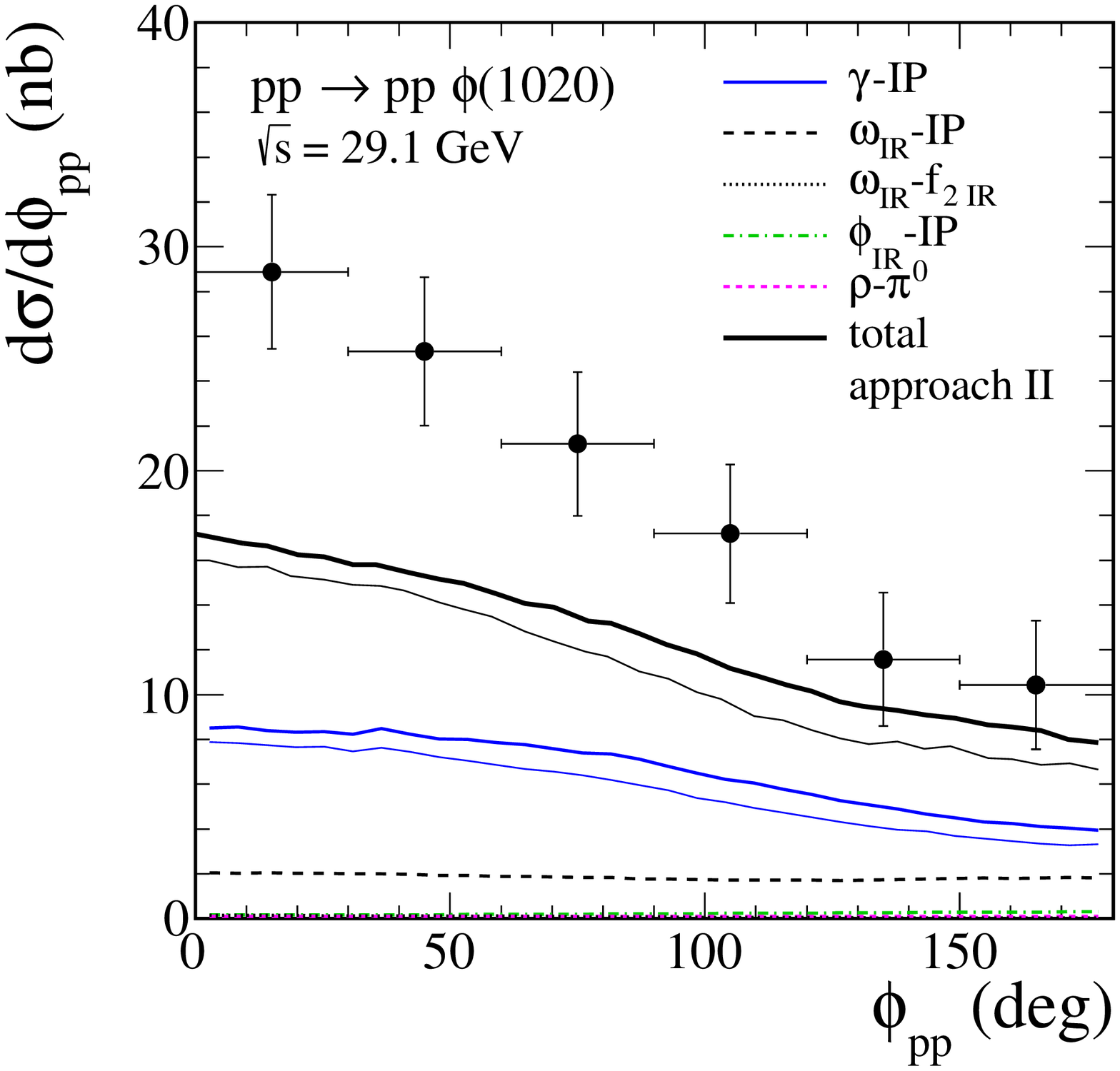}
\includegraphics[width=0.46\textwidth]{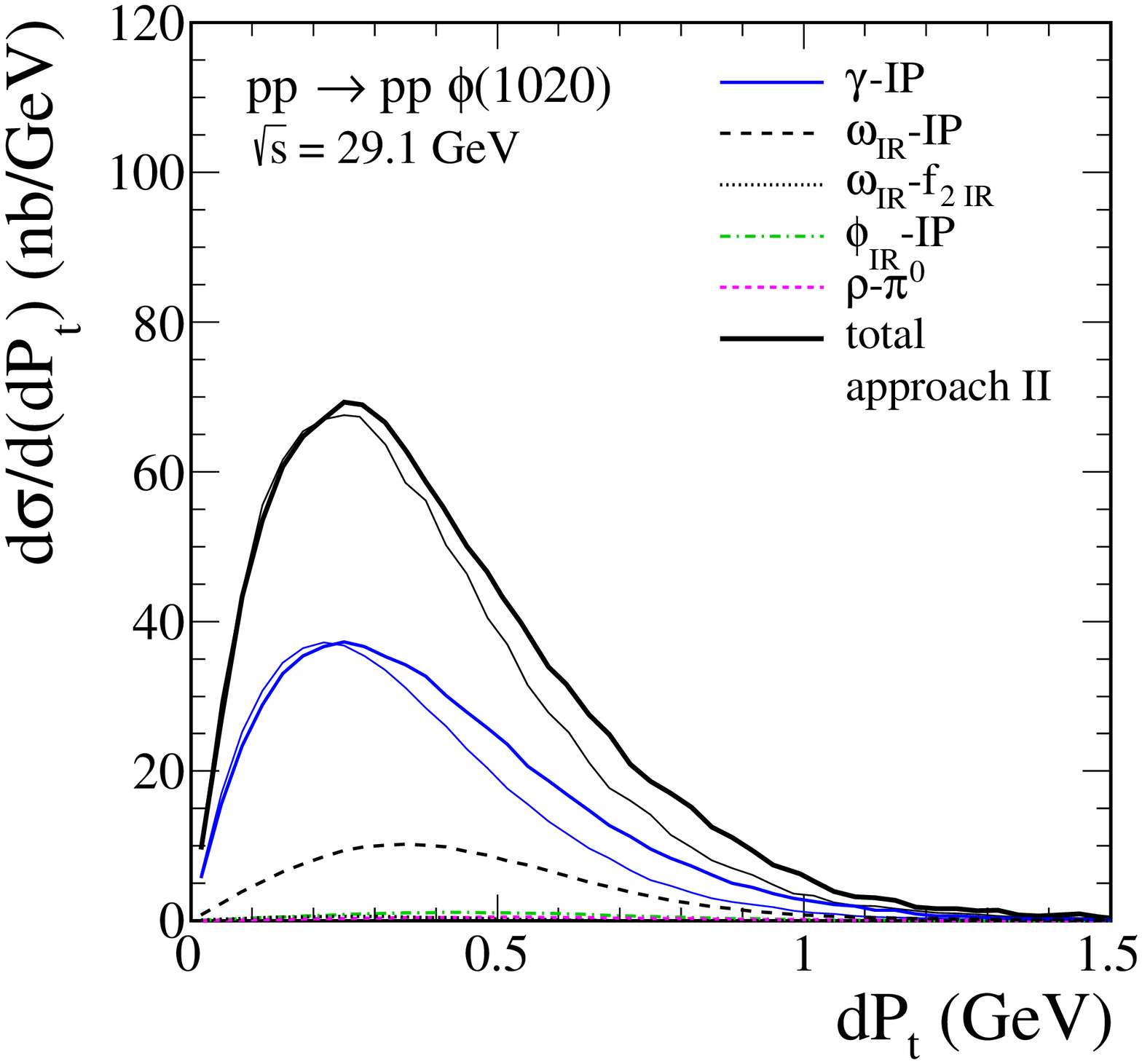}
\includegraphics[width=0.46\textwidth]{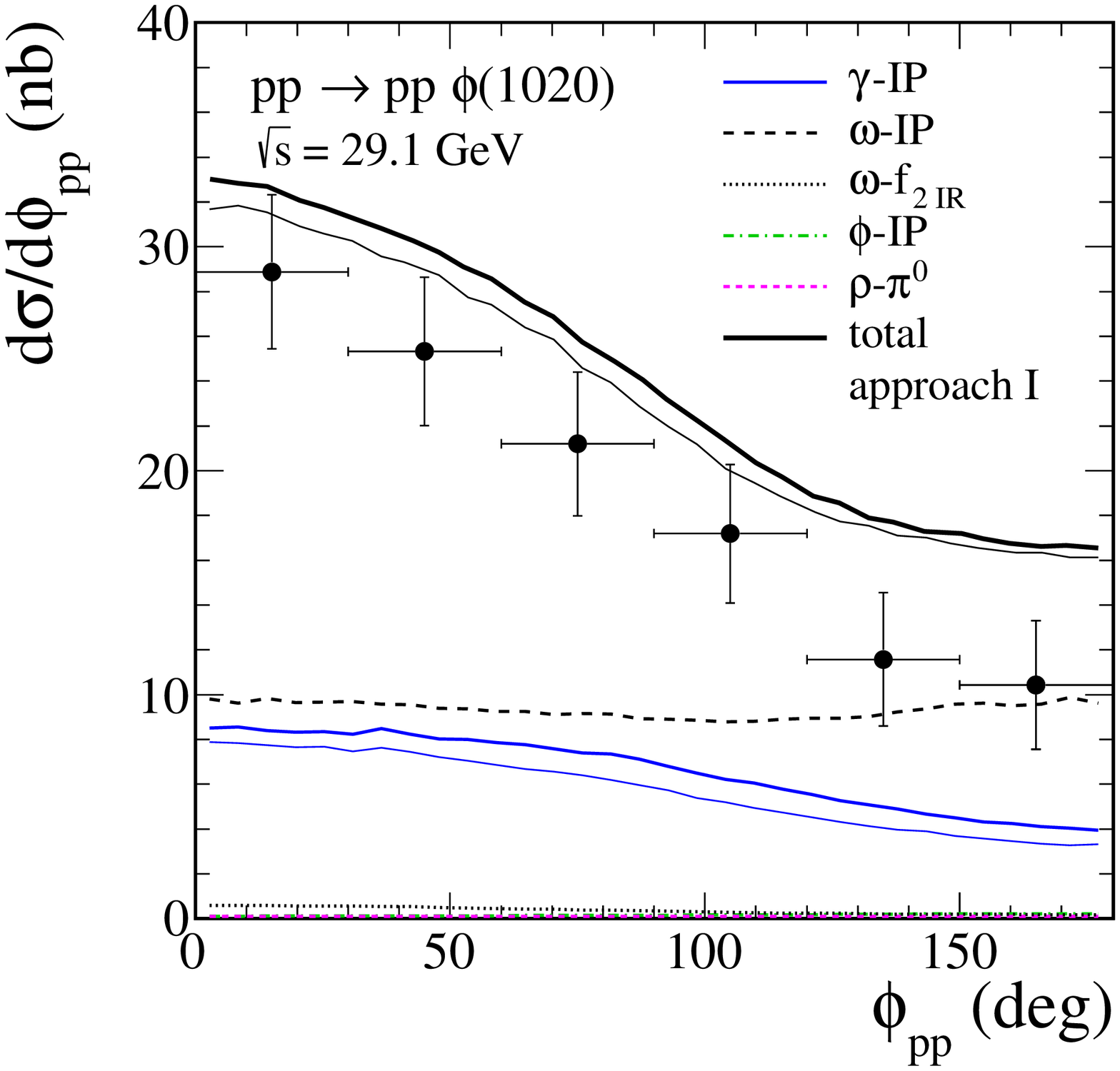}
\includegraphics[width=0.46\textwidth]{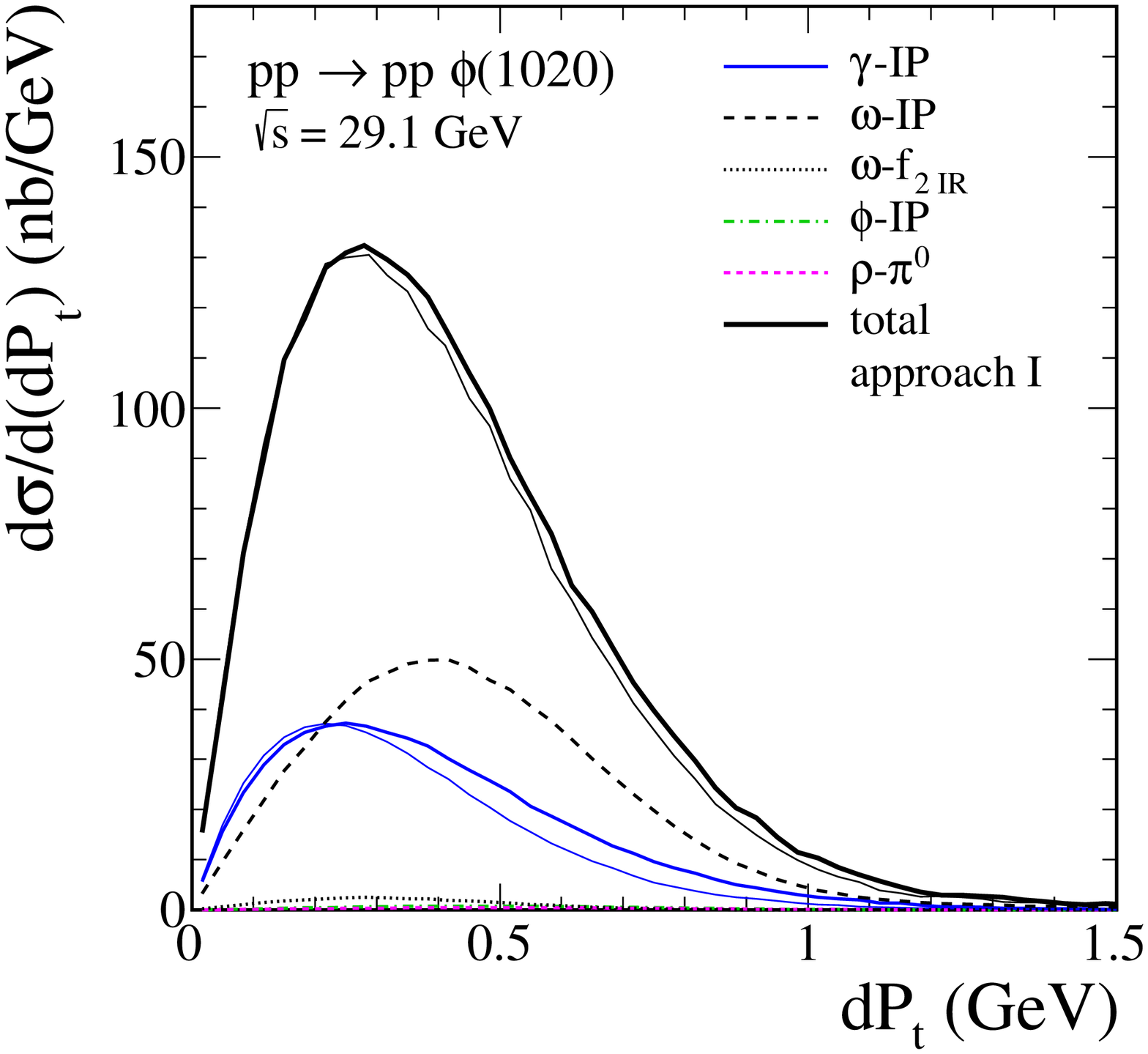}
\caption{\label{fig:WA102_ABS_noodd}
Distributions in proton-proton relative azimuthal angle $\phi_{pp}$ (left panels)
and in $\rm{dP_{t}}$ (\ref{dPt_variable}), the ``glueball filter'' variable (right panels),
for the $pp \to pp \phi$ reaction at $\sqrt{s} = 29.1$~GeV.
The data points have been normalized to the central value 
of the total cross section (\ref{WA102_xsection}) from \cite{Kirk:2000ws}.
The results for the fusion processes 
$\gamma$-$\Pom$ (the two blue solid lines), 
$\omega$-$\Pom$ (the black dashed line),
$\omega$-$f_{2 \Reg}$ (the black dotted line),
$\phi$-$\Pom$ (the green dash-dotted line), 
and $\rho$-$\pi^{0}$ (the violet dotted line) are presented.
In the top panels the $\omega$-$\Pom$, $\phi$-$\Pom$ and $\omega$-$f_{2 \Reg}$ exchanges
are treated, respectively, as reggeon-pomeron and reggeon-reggeon exchanges (approach II)
while in the bottom panels these contributions 
are calculated in the reggeized-vector-meson approach (\ref{reggeization_2}) (approach I).
The coherent sum of these contributions
is shown by the two black solid lines.
The lower blue and black solid lines are for the parameter set~A (\ref{photoproduction_setA})
and the upper lines are for the parameter set B (\ref{photoproduction_setB})
in the calculation of the $\gamma$-$\Pom$ fusion contribution.
The absorption effects are included here.}
\end{figure}

In Figs.~\ref{fig:WA102_ABS_phi12}--\ref{fig:WA102_ABS_rapidity}
we present several differential distributions 
for the $\gamma$-$\Pom$ and the $\Ode$-$\Pom$ fusion processes
corresponding to the diagrams shown 
in Figs.~\ref{fig:diagrams_phi_photon} and \ref{fig:diagrams_phi_odderon},
respectively, and for the subleading processes
$\omega$-$\Pom$, $\phi$-$\Pom$, $\omega$-$f_{2 \Reg}$ and $\rho$-$\pi^{0}$ fusion.
In the panels~(a) and (b) of Fig.~\ref{fig:WA102_ABS_phi12}
the $\omega$- and $\phi$-exchanges
are treated as reggeon exchanges (approach~II)
while in the panel~(c) as the reggeized-vector-meson exchange 
(\ref{reggeization_2}) (approach~I).
For the $\Ode$-$\Pom$ fusion contribution we take the following parameters, 
see (\ref{A12})--(\ref{Fpomodephi}),
\begin{eqnarray}
&&\eta_{\Ode} = -1\,, \;\;
\alpha_{\Ode}(0) = 1.05\,,\;\;
\alpha'_{\Ode} = 0.25 \; \mathrm{GeV}^{-2}\,, 
\label{parameters_ode} \\
&&\Lambda_{0,\,\Pom \Ode \phi}^{2} = 0.5\;{\rm GeV}^{2}\,,
\label{parameters_ode_lambda}
\end{eqnarray}
and we choose different values for $a_{\Pom \Ode \phi}$ and $b_{\Pom \Ode \phi}$:
\begin{eqnarray}
{\rm (a)}&&a_{\Pom \Ode \phi}= -0.8 \; {\rm GeV}^{-3}\,,\;\;
  b_{\Pom \Ode \phi}= 1.0 \; {\rm GeV}^{-1}\,; 
\label{parameters_ode_a}\\
{\rm (b)}&&a_{\Pom \Ode \phi}= -0.8 \; {\rm GeV}^{-3}\,,\;\;
  b_{\Pom \Ode \phi}= 1.6 \; {\rm GeV}^{-1}\,;
\label{parameters_ode_b}\\
{\rm (c)}&&a_{\Pom \Ode \phi}= -0.6 \; {\rm GeV}^{-3}\,,\;\;
  b_{\Pom \Ode \phi}= 1.6 \; {\rm GeV}^{-1}\,.
\label{parameters_ode_c}
\end{eqnarray}
The results shown in panels (a) and (b) of Fig.~\ref{fig:WA102_ABS_phi12}
correspond to the approach~II and
the $\Pom \Ode \phi$ parameters in
(\ref{parameters_ode_a}) and (\ref{parameters_ode_b}),
respectively.
The results shown in panel (c) correspond to the approach I and (\ref{parameters_ode_c}).
The coherent sum of all contributions is shown by the black solid lines.
The lower line is for the parameter set~A of photoproduction 
(\ref{photoproduction_setA})
and the upper line is for set~B (\ref{photoproduction_setB}).

We have checked that these parameters 
are compatible with our analysis of the WA102 data for
the $pp \to pp \phi \phi$ reaction discussed in \cite{Lebiedowicz:2019jru}.
Comparing the results shown in Fig.~\ref{fig:WA102_ABS_noodd} 
with those in Fig.~\ref{fig:WA102_ABS_phi12}
we can see that the complete results indicate a large interference effect
between the $\gamma$-$\Pom$, $\Ode$-$\Pom$, 
$\omega$-$\Pom$, $\omega$-$f_{2 \Reg}$, 
and $\phi$-$\Pom$ terms.
\begin{figure}[!ht]
(a)\includegraphics[width=0.41\textwidth]{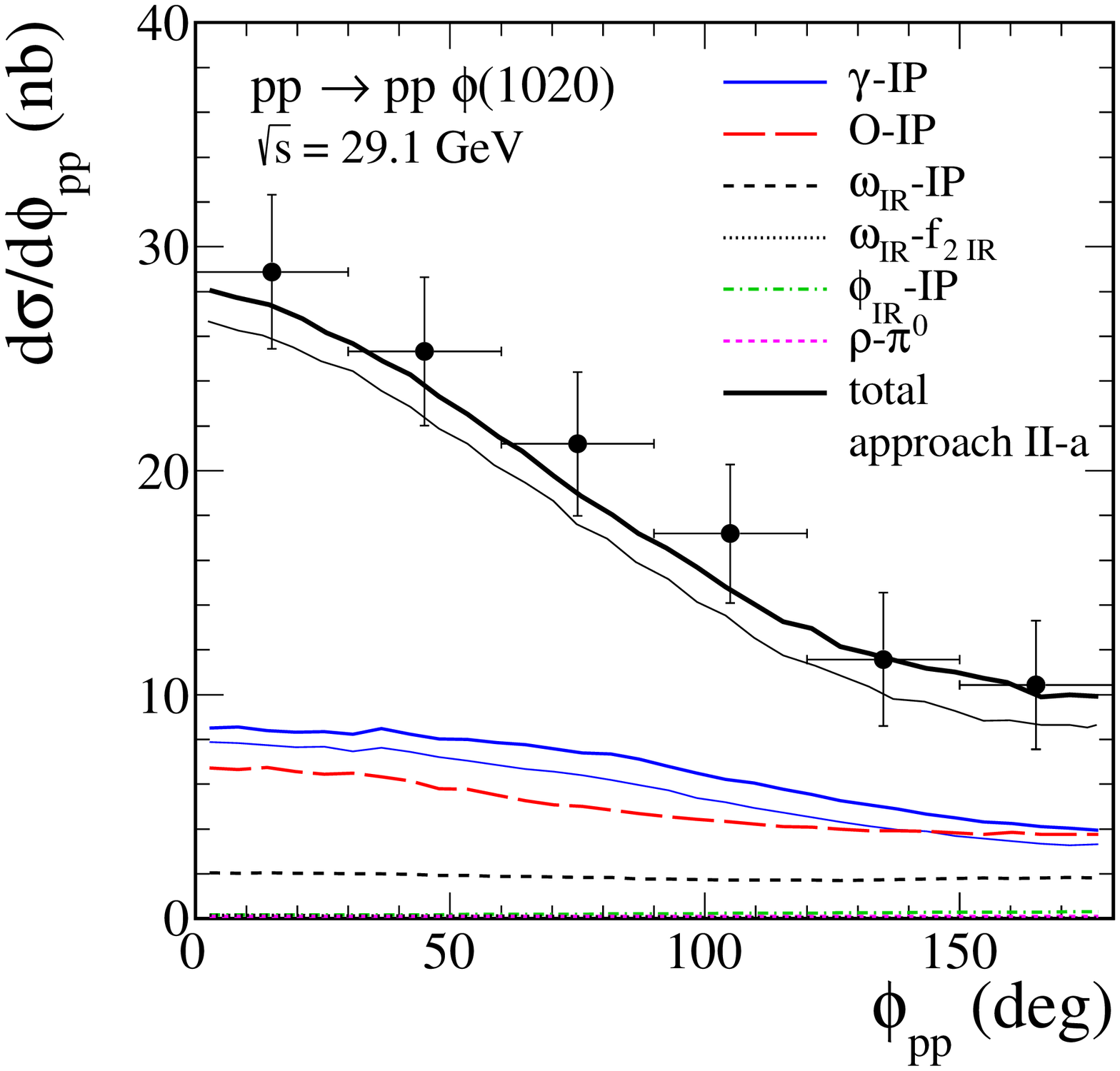}
   \includegraphics[width=0.41\textwidth]{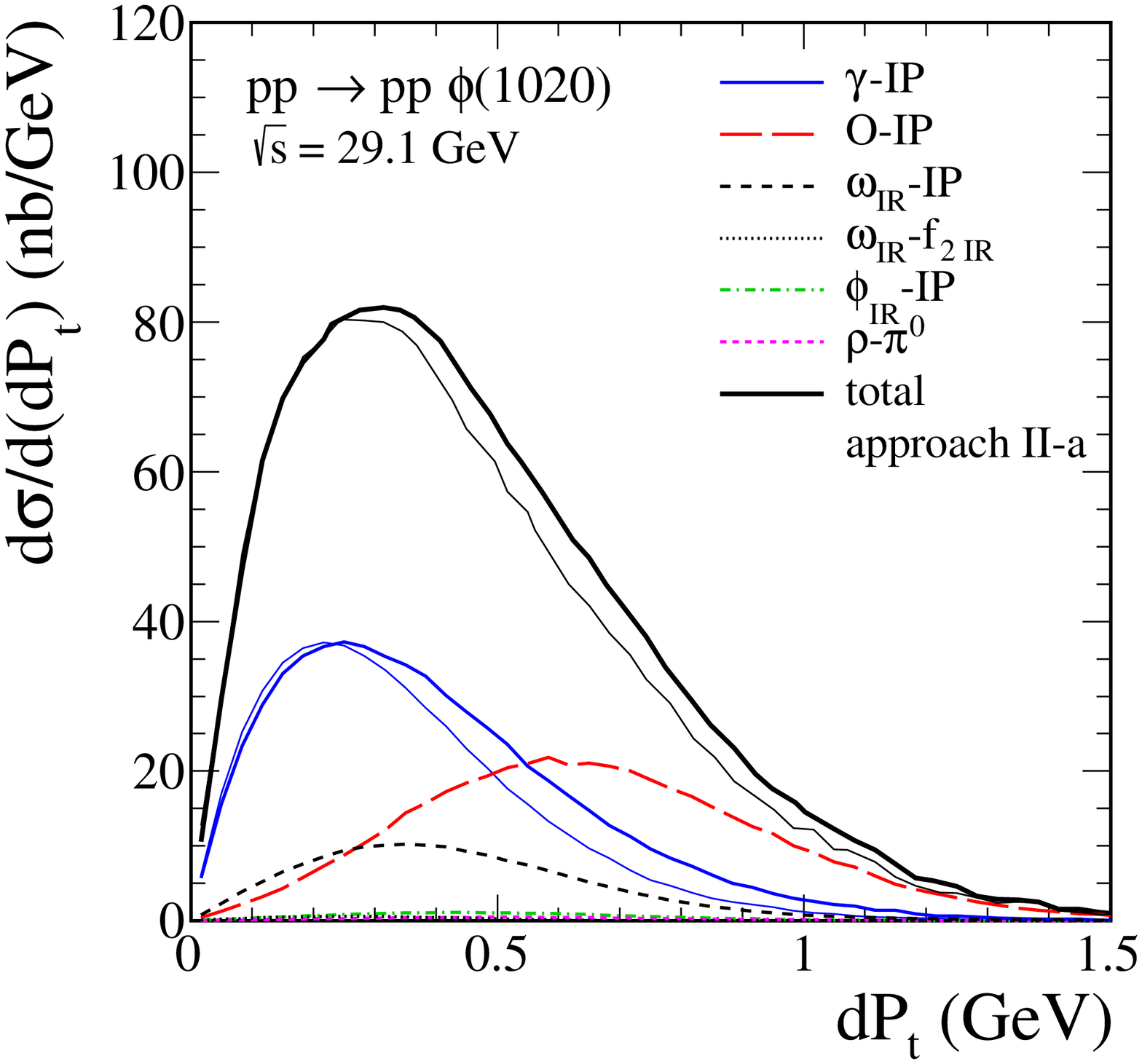}
(b)\includegraphics[width=0.41\textwidth]{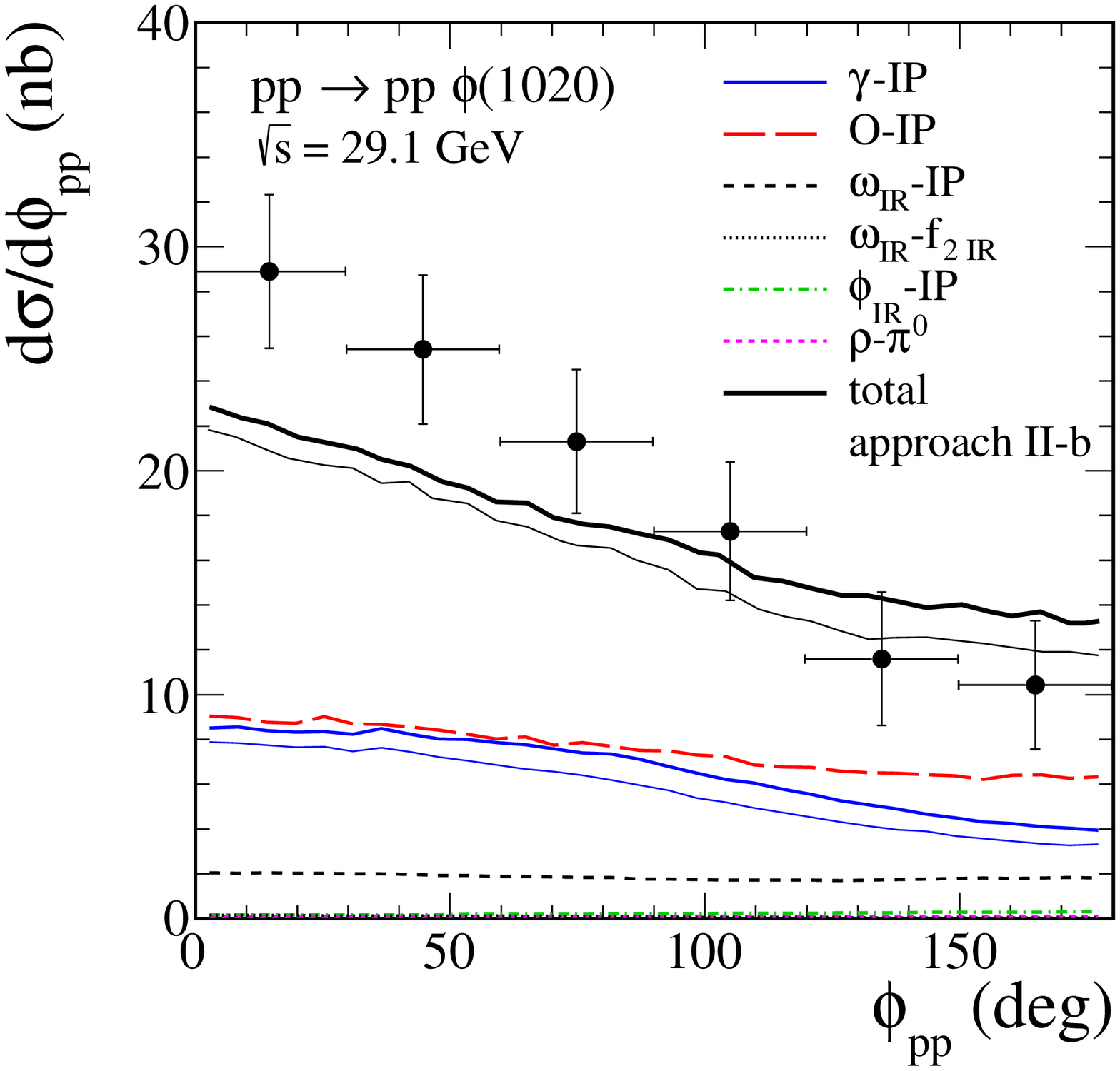}
   \includegraphics[width=0.41\textwidth]{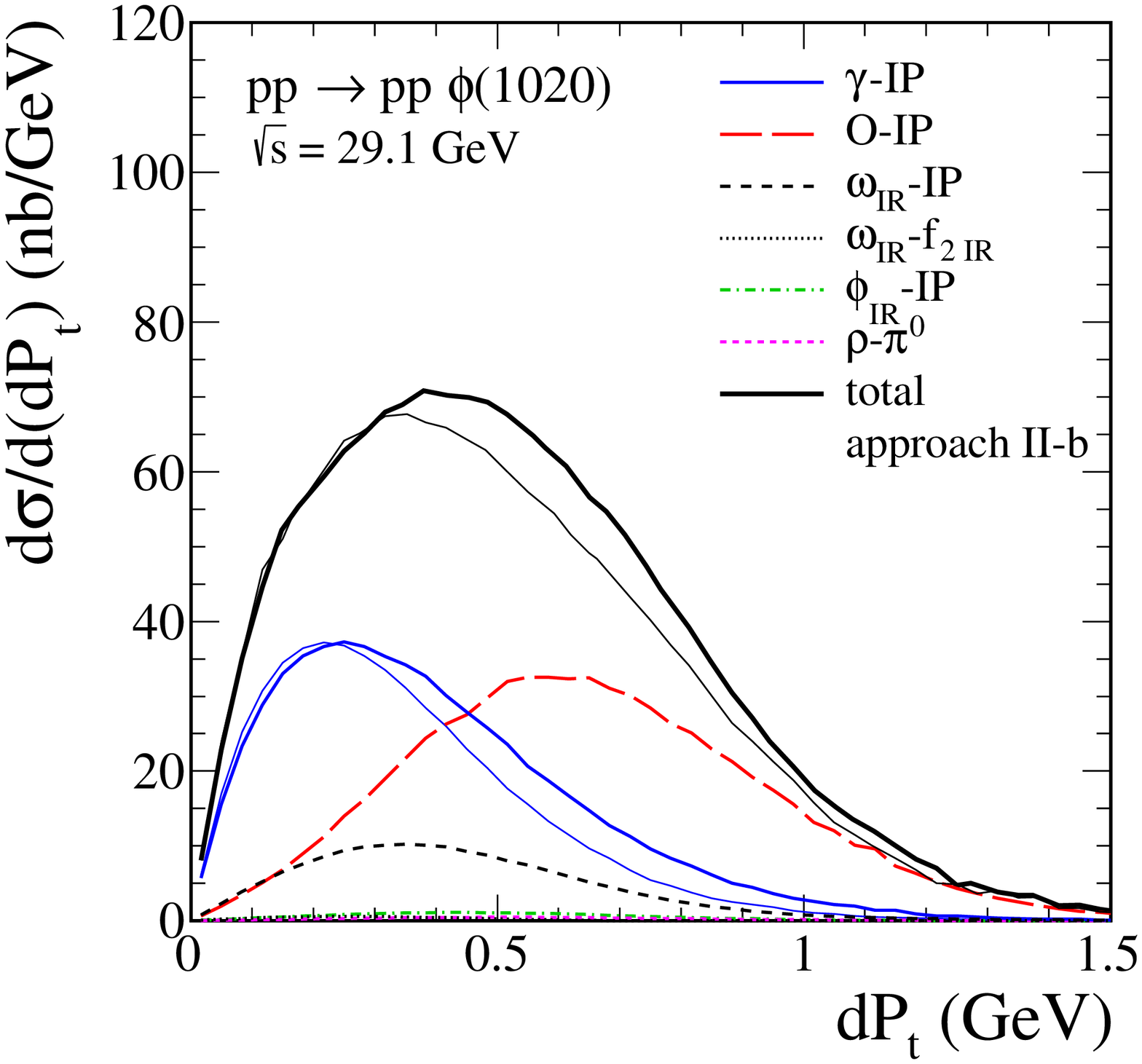}
(c)\includegraphics[width=0.41\textwidth]{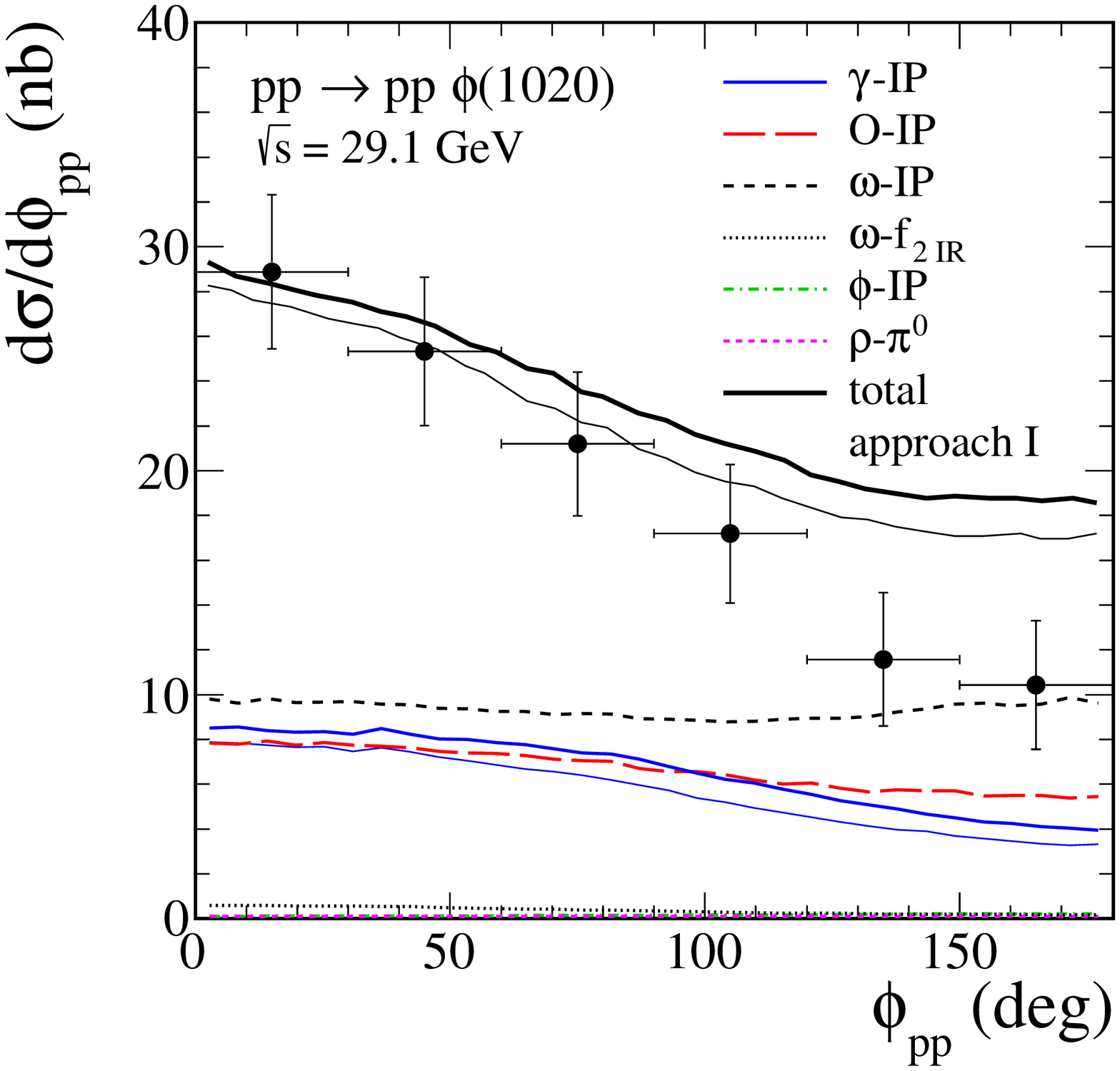}
   \includegraphics[width=0.41\textwidth]{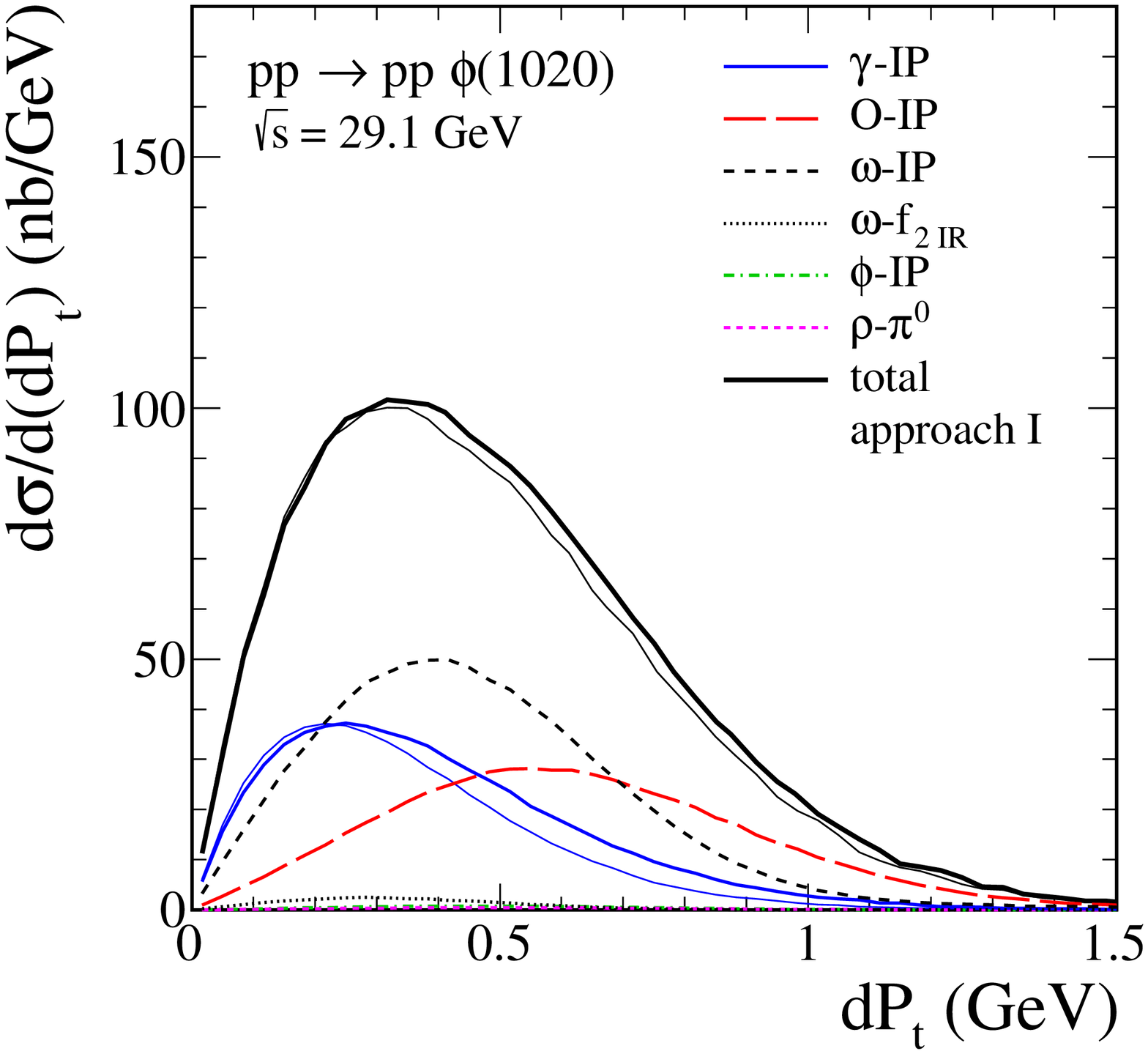}
\caption{\label{fig:WA102_ABS_phi12}
The $\phi_{pp}$ (left panels) and $\rm{dP_{t}}$ (right panels) distributions 
for the $pp \to pp \phi$ reaction at $\sqrt{s} = 29.1$~GeV.
The data points have been normalized to the central value 
of the total cross section (\ref{WA102_xsection}) from \cite{Kirk:2000ws}.
The meaning of the lines 
is the same as in Fig.~\ref{fig:WA102_ABS_noodd}
but here we added the \mbox{$\Ode$-$\Pom$ fusion} term
(see the red long-dashed line).
The results shown in panels (a) and (b) 
correspond to the approach~II and
the $\Pom \Ode \phi$ parameters in
(\ref{parameters_ode_a}) and (\ref{parameters_ode_b}),
respectively.
The results shown in panel (c) correspond to the approach I and (\ref{parameters_ode_c}).
The coherent sum of all contributions is shown by the black solid lines.
The lower line is for the parameter set A of photoproduction 
(\ref{photoproduction_setA})
and the upper line is for set B (\ref{photoproduction_setB}).
The absorption effects are included here.}
\end{figure}

\begin{figure}[!ht]
\includegraphics[width=0.46\textwidth]{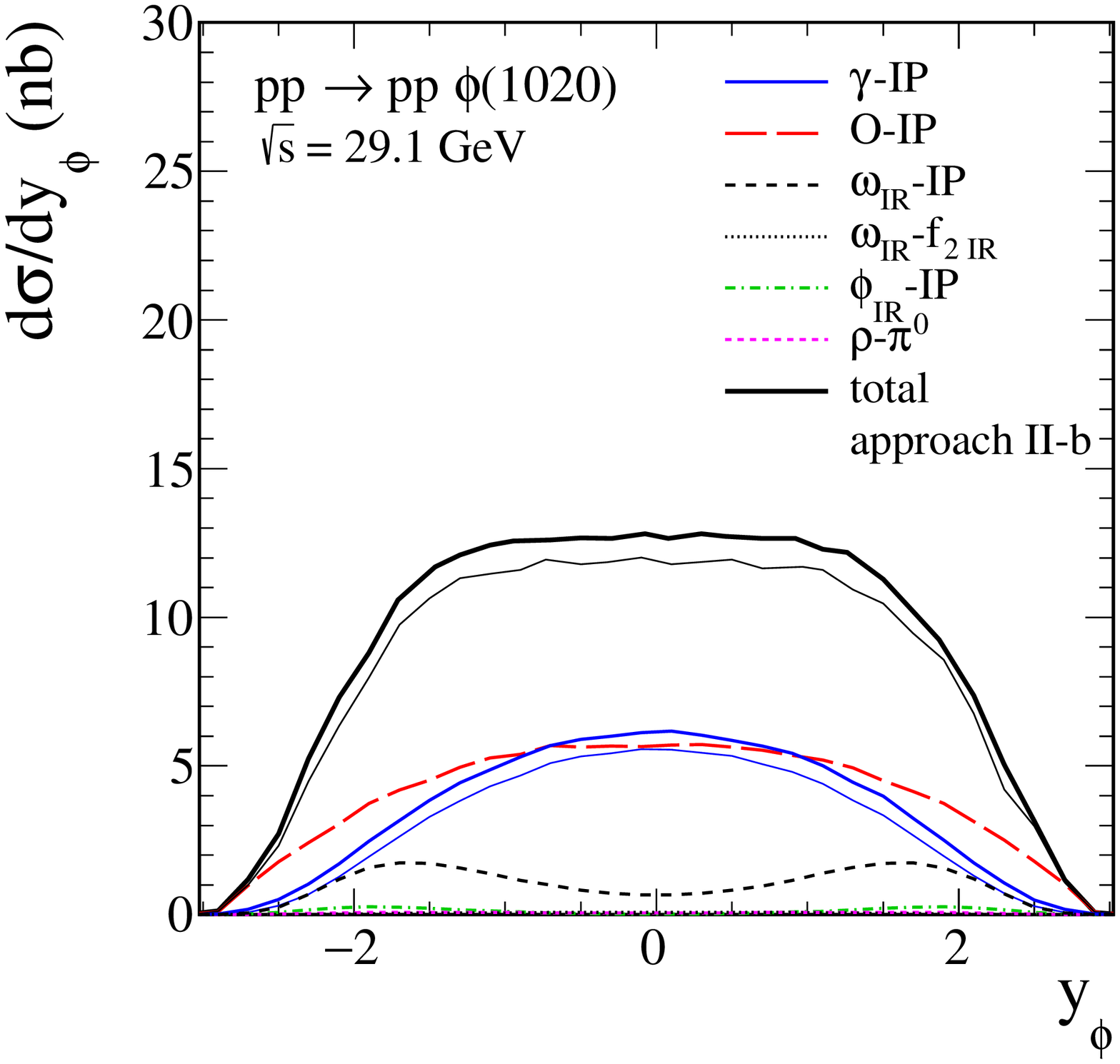}
\includegraphics[width=0.46\textwidth]{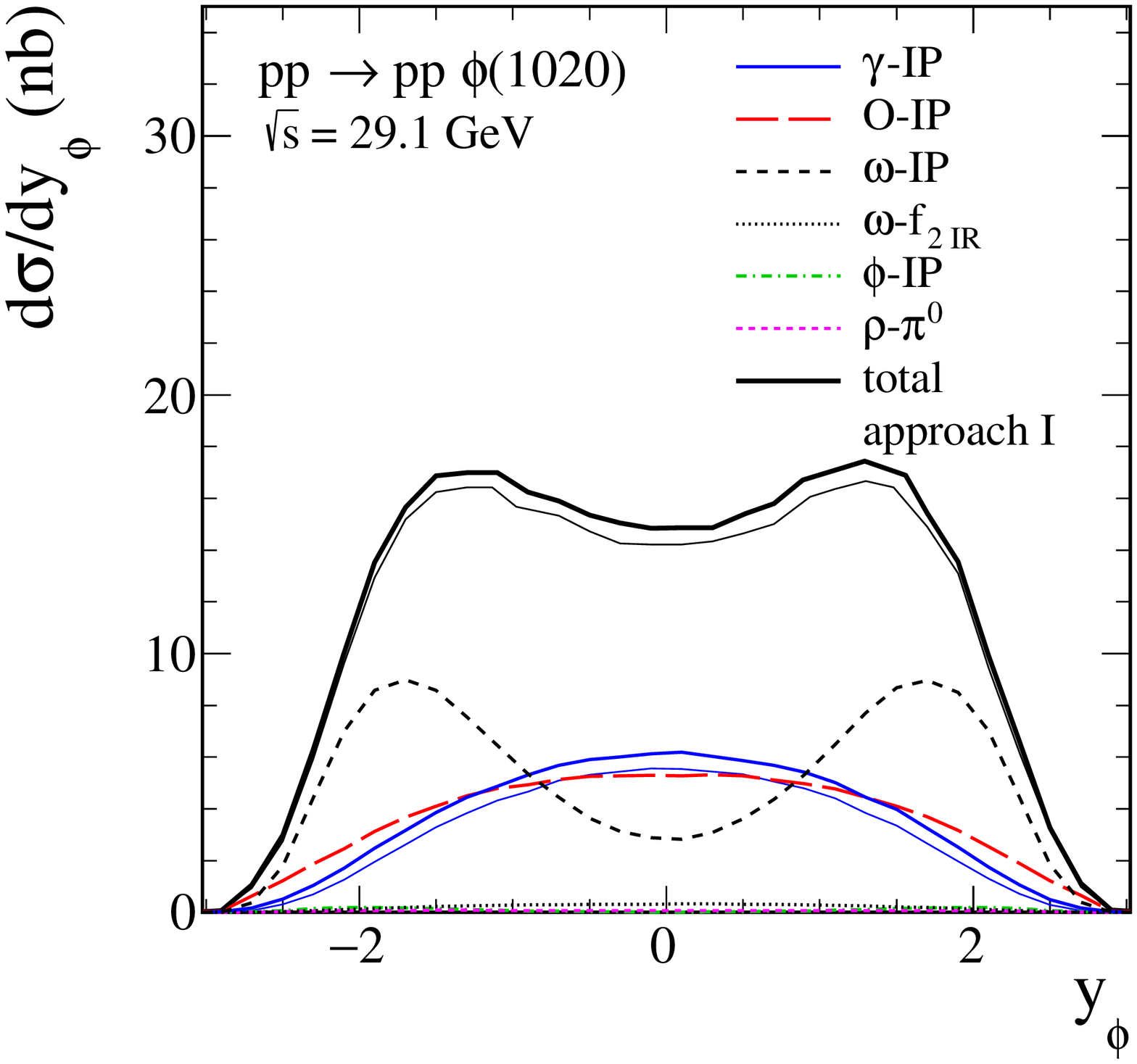}\\
\includegraphics[width=0.46\textwidth]{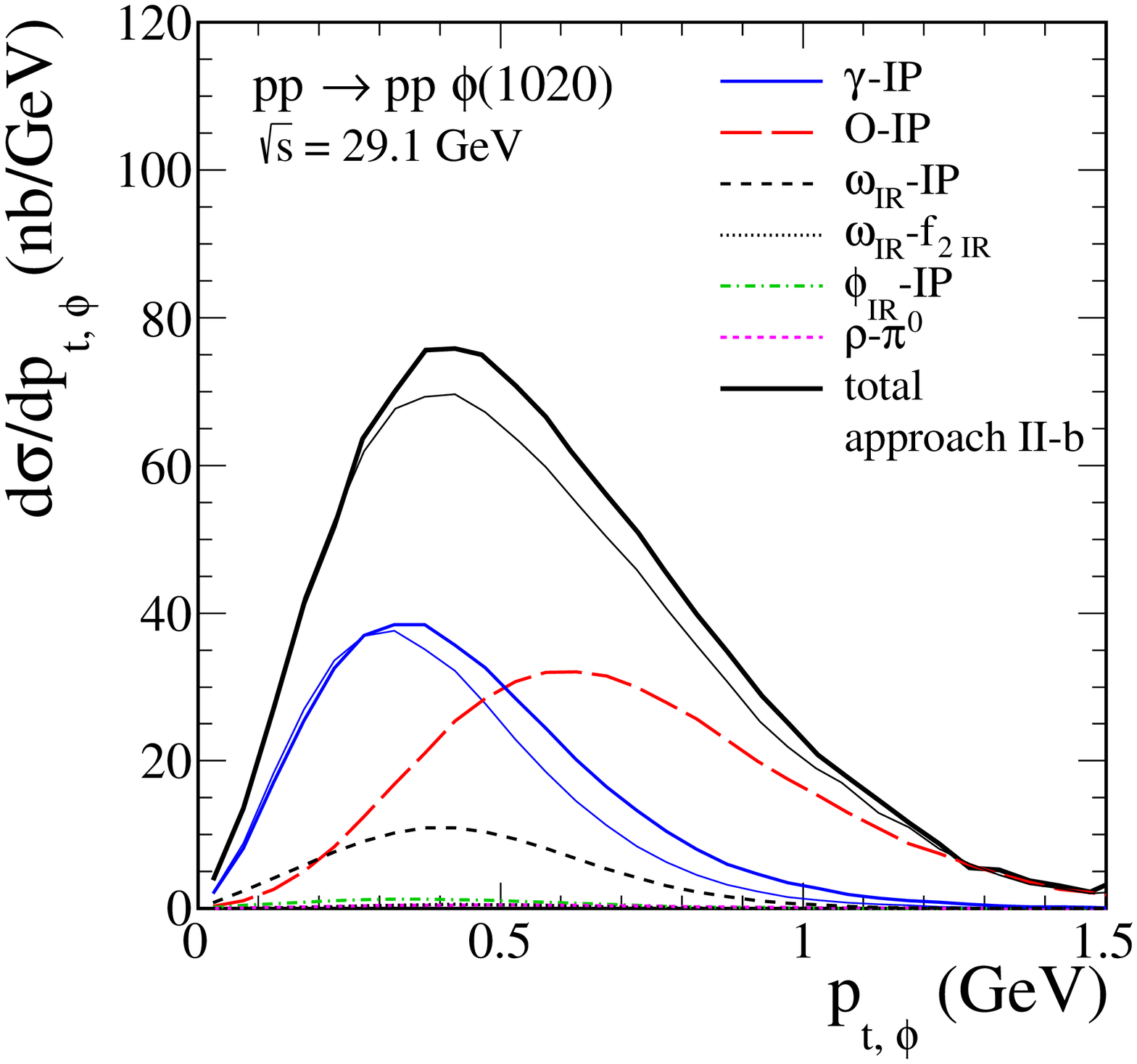}
\includegraphics[width=0.46\textwidth]{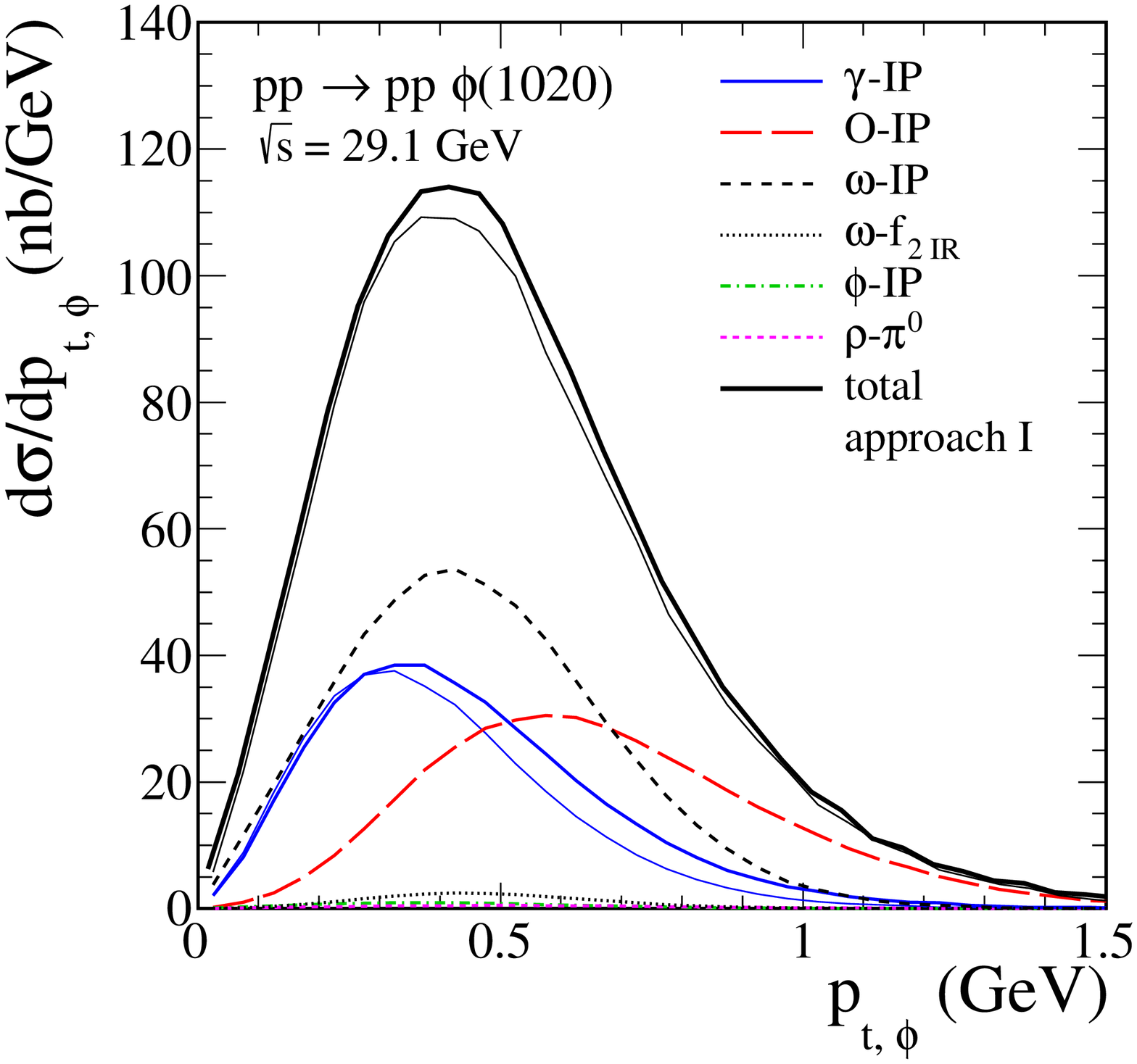}
\caption{\label{fig:WA102_ABS_rapidity}
Distributions in rapidity of the $\phi$ meson (top panels)
and in transverse momentum of the $\phi$ meson (bottom panels)
for the $pp \to pp \phi$ reaction at $\sqrt{s} = 29.1$~GeV.
The meaning of the lines is the same as in Fig.~\ref{fig:WA102_ABS_phi12}.}
\end{figure}

In \cite{Kirk:2000ws} experimental values for the cross sections
in three $\mathrm{dP_{t}}$ intervals and for
the ratio of $\phi$ production at small 
$\mathrm{dP_{t}}$ to large $\mathrm{dP_{t}}$ are given.
We show our corresponding results in Table~\ref{tab:ratio_dPt}
for the two approaches, I and II,
with appropriate $\Pom \Ode \phi$ coupling constants 
(\ref{parameters_ode_a}), (\ref{parameters_ode_b}), (\ref{parameters_ode_c}).
Here we take the parameter set~B (\ref{photoproduction_setB}) 
for the $\gamma$-$\Pom$ fusion contributions. 

Now we discuss our results concerning the WA102 data.
As already mentioned we find that the $\gamma$-$\Pom$ fusion processes alone
cannot describe the WA102 data for the $\phi_{pp}$ distribution.
This holds even if we scale down the experimental data by about 30~\%
corresponding to the quoted error on the total cross section in (\ref{WA102_xsection}).
Thus, we need other contributions, subleading ones or maybe
odderon-pomeron fusion. From the subleading ones we find that
the $\gamma$-$\pi^{0}$ and $\gamma$-$\eta$ contributions are very small;
see Fig.~\ref{fig:WA102_photo}.
Also the $\rho$-$\pi^{0}$-fusion contribution turns out to be very small.
According to our results, the important subleading contributions are
$\omega$-$\Pom$, $\omega$-$f_{2 \Reg}$ and $\phi$-$\Pom$ fusion.
We have treated them with two methods of reggeization, I and II.
The reggeized vector-meson approach~I, see (\ref{reggeization_2}), (\ref{reggeization_aux}),
almost certainly overestimates these contributions.
The reggeization means that we replace the vector-meson exchange 
by a coherent sum of exchanges with spin $1 + 3 + 5 + ...$.
The higher the spin the higher the mass of the exchanged particle.
In (\ref{reggeization_2}) this increase of mass is not taken
into account leading to the overestimate.
Also, the distribution in $\phi_{pp}$ in this approach~I
is too flat and does not fit the data; 
see the $\omega$-$\Pom$ contribution in the left bottom panel 
in Fig.~\ref{fig:WA102_ABS_noodd}.
The approach~II, on the other hand,
assumes reggeon exchanges, $\omega_{\Reg}$ and $\phi_{\Reg}$.
This approach maybe underestimates the contributions if
$s_{1}$ or $s_{2}$ are small, but should be very reasonable
for large $s_{1}$ or $s_{2}$.
But note that in our reaction the threshold for $s_{1}$ and $s_{2}$ 
is already quite large $s_{\rm thr} \approx 4$~GeV$^{2}$; see (\ref{sthr}).
We see clearly from Fig.~\ref{fig:WA102_ABS_noodd} that
in this approach the sum of the $\gamma$-$\Pom$, $\gamma$-$f_{2 \Reg}$,
$\omega_{\Reg}$-$\Pom$, $\omega_{\Reg}$-$f_{2 \Reg}$, 
$\phi_{\Reg}$-$\Pom$ and $\rho$-$\pi^{0}$ contributions
\footnote{For clarity: here we took into account
the $\Pom$ and $f_{2 \Reg}$ exchanges as a result of $\omega$-$\phi$ mixing; 
see the diagram~(b) of Fig.~\ref{fig:gamp_phip_diagrams}.
We neglect the $\phi_{\Reg}$-$f_{2 \Reg}$-fusion contribution
and the $f_{2 \Reg}$-exchange term 
from the diagram~(a) of Fig.~\ref{fig:gamp_phip_diagrams}
and the $a_{2 \Reg}$-exchange term from the diagram~(b) there.},
added coherently,
cannot explain the $\phi_{pp}$ data.
This gives a hint that the missing contribution could be the odderon-pomeron fusion.
And, indeed, with suitable odderon parameters we arrive at a decent
description of the $\phi_{pp}$ and the $\rm{dP_{t}}$ data
from WA102; see Fig.~\ref{fig:WA102_ABS_phi12}
and Table~\ref{tab:ratio_dPt}, respectively.
However, we have to remember that the $\phi_{pp}$ distributions
have a large normalisation uncertainty due to the relatively
large error on $\sigma_{\rm exp}$ (\ref{WA102_xsection}).
Therefore, we emphasise that our fits to the WA102 data
on single $\phi$ CEP only give a hint that
this reaction could be very interesting 
for a search of odderon effects.
It would be nice if we could fix the odderon
contribution to $\phi$ CEP at the WA102 energy more
quantitatively. But we must leave this to the experimentalists
who know in detail the statistical and systematic errors
of the data, including the error correlations.
Also the theoretical uncertainties of the subleading contributions
are relatively large at the WA102 energy.
These latter uncertainties should, however, 
be much smaller at LHC energies.
From Fig.~\ref{fig:WA102_ABS_rapidity} we see that 
the odderon-pomeron contribution dominates at larger
$|{\rm y}_\phi|$ and $p_{t, \phi}$ compared to the photon-pomeron contribution.
As we shall see this also holds at LHC energies and should help in searches
for odderon effects there.

\begin{table}
\caption{Results of central $\phi$ production as a function of $\mathrm{dP_{t}}$
expressed as a percentage of its total contribution 
at the WA102 collision energy $\sqrt{s}=29.1$~GeV.
In the last column the ratios of 
$\sigma(\mathrm{dP_{t}} \leqslant \,0.2~\mathrm{GeV})/
 \sigma(\mathrm{dP_{t}} \geqslant \,0.5~\mathrm{GeV})$ are given.
The experimental numbers are from Table~2 of \cite{Kirk:2000ws}.
The theoretical numbers correspond to the total results
including all terms contributing; 
see the upper black lines in 
the right panels of Figs.~\ref{fig:WA102_ABS_noodd} and \ref{fig:WA102_ABS_phi12}.}
\label{tab:ratio_dPt}
\begin{tabular}{|l|c|c|c|c|}
\hline
& 
$\mathrm{dP_{t}} \leqslant 0.2$~GeV & 
$0.2 \leqslant \mathrm{dP_{t}} \leqslant 0.5$~GeV & 
$\mathrm{dP_{t}} \geqslant 0.5$~GeV & Ratio\\
\hline
experiment & $8 \pm 3$ & $47 \pm 3$ & $45 \pm 4$ & $0.18 \pm 0.07$\\
\hline
approach II, no odderon & 22.0 & 46.9 & 31.1 & 0.71 \\
approach I,  no odderon & 19.5 & 48.0 & 32.5 & 0.60 \\
\hline
approach II-a  & 17.4 & 42.2 & 40.4 & 0.43 \\
approach II-b  & 13.3 & 37.0 & 49.7 & 0.27 \\
approach I     & 14.7 & 41.1 & 44.2 & 0.33 \\
\hline
\end{tabular}
\end{table}




\subsection{Predictions for the LHC experiments}
\label{predictions_LHC}

\subsubsection{The $pp \to pp K^{+}K^{-}$ reaction}
\label{pp_ppKK}

In this subsection we wish to show our predictions for the LHC experiments.
We start with the presentation of the differential distributions 
for the $pp \to pp (\phi \to K^{+}K^{-})$ reaction (\ref{2to4_reaction_KK_via_phi})
which we integrate in the $\phi$ resonance region (\ref{M43_phi_region}).
First we show, for orientation purposes,
results for the $\gamma \Pom$- and the $\Ode \Pom$-fusion contributions separately
(see the diagrams shown in Figs.~\ref{fig:diagrams_phi_photon}
and~\ref{fig:diagrams_phi_odderon}, respectively).
For the final results we shall, of course, add 
these contributions coherently
and calculate absorption corrections at the amplitude level.
We have checked that in the kinematic regimes
discussed in the following the subleading contributions 
(see Appendix~\ref{sec:subleading}) can be safely neglected.

In Figs.~\ref{fig:ATLASALFA_born_0}--\ref{fig:ATLASALFA_2} 
we show the results for $\sqrt{s} = 13$~TeV, 
and $|\eta_{K}| < 2.5$, $p_{t, K} > 0.1$~GeV
and sometimes with extra cuts on the leading protons of 
0.17~GeV~$< |p_{y,1}|, |p_{y,2}|<$~0.50~GeV
as will be the proton momentum window for the ALFA detectors 
placed on both sides of the ATLAS detector.
The choice of such cuts is based on the analysis 
initiated by the ATLAS Collaboration;
see \cite{Sikora_poster}.
For comparison, we will also show our predictions
for the ATLAS-ALFA experiment
for $p_{t, K} > 0.2$~GeV; 
see Figs.~\ref{fig:ATLASALFA_1_ptK0.2}--\ref{fig:ATLASALFA_1_ptK0.2_setBtestA}
and Table~\ref{tab:table2} below.

Figure~\ref{fig:ATLASALFA_born_0} shows the Born-level distributions 
in $|t_{1}|$ (top panels) and in transverse momentum 
$p_{t, 1} = |\bpta|$ of the proton $p\,(p_{1})$ (bottom panels).
In the left panels the photoproduction contributions are plotted
while in the right panels we show the results for the odderon contributions.
The results for the parameter set~B (\ref{photoproduction_setB}) for the photoproduction term
and for the parameters quoted in (\ref{parameters_ode}), (\ref{parameters_ode_lambda}), (\ref{parameters_ode_b}) 
for the $\Ode$-$\Pom$ fusion are presented.
We show results for two diagrams separately 
and for their coherent sum (denoted by ``total'').
The interference effects between the two diagrams are clearly visible,
especially for the $\Ode$-$\Pom$-fusion mechanism.
A different behaviour is seen at small $|t_{1}|$
for the $\gamma \Pom$ and the $\Ode \Pom$ components.
Due to the photon exchange the protons are scattered only at small angles
and the $\gamma \Pom$ distribution has a singularity for $|t_{1}| \to 0$.
Of course, $t_{1} = 0$ cannot be reached here from kinematics.
In contrast, the $\Ode \Pom$ distribution shows a dip for $|t_{1}| \to 0$.
The explanation of this type of behaviour
is given in Appendix~C of \cite{Bolz:2014mya}.
In the bottom panels we show the $p_{t}$ distributions for proton $p\,(p_{1})$.
Here these differences are also clearly visible.
\begin{figure}[!ht]
\includegraphics[width=0.45\textwidth]{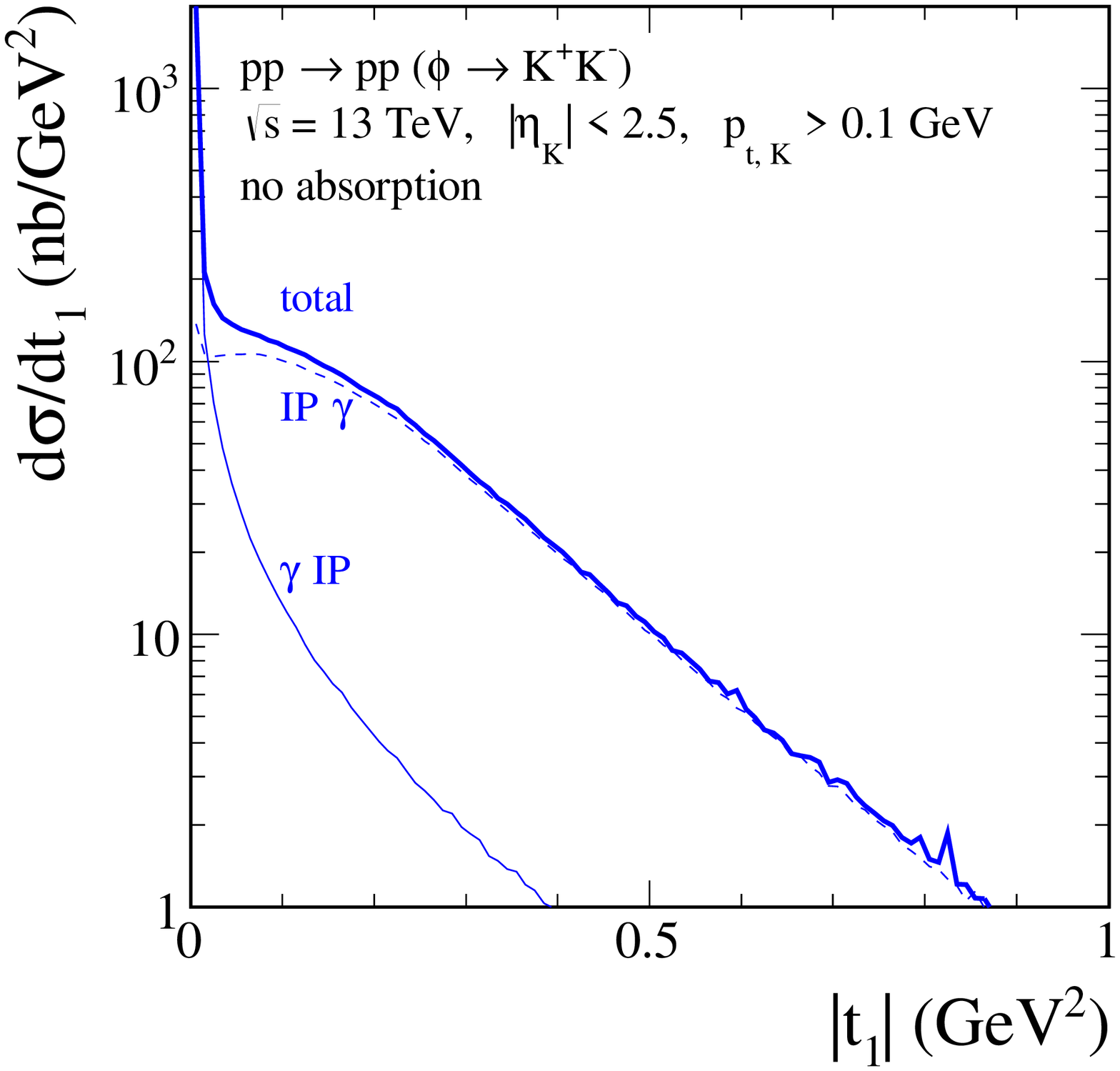}
\includegraphics[width=0.45\textwidth]{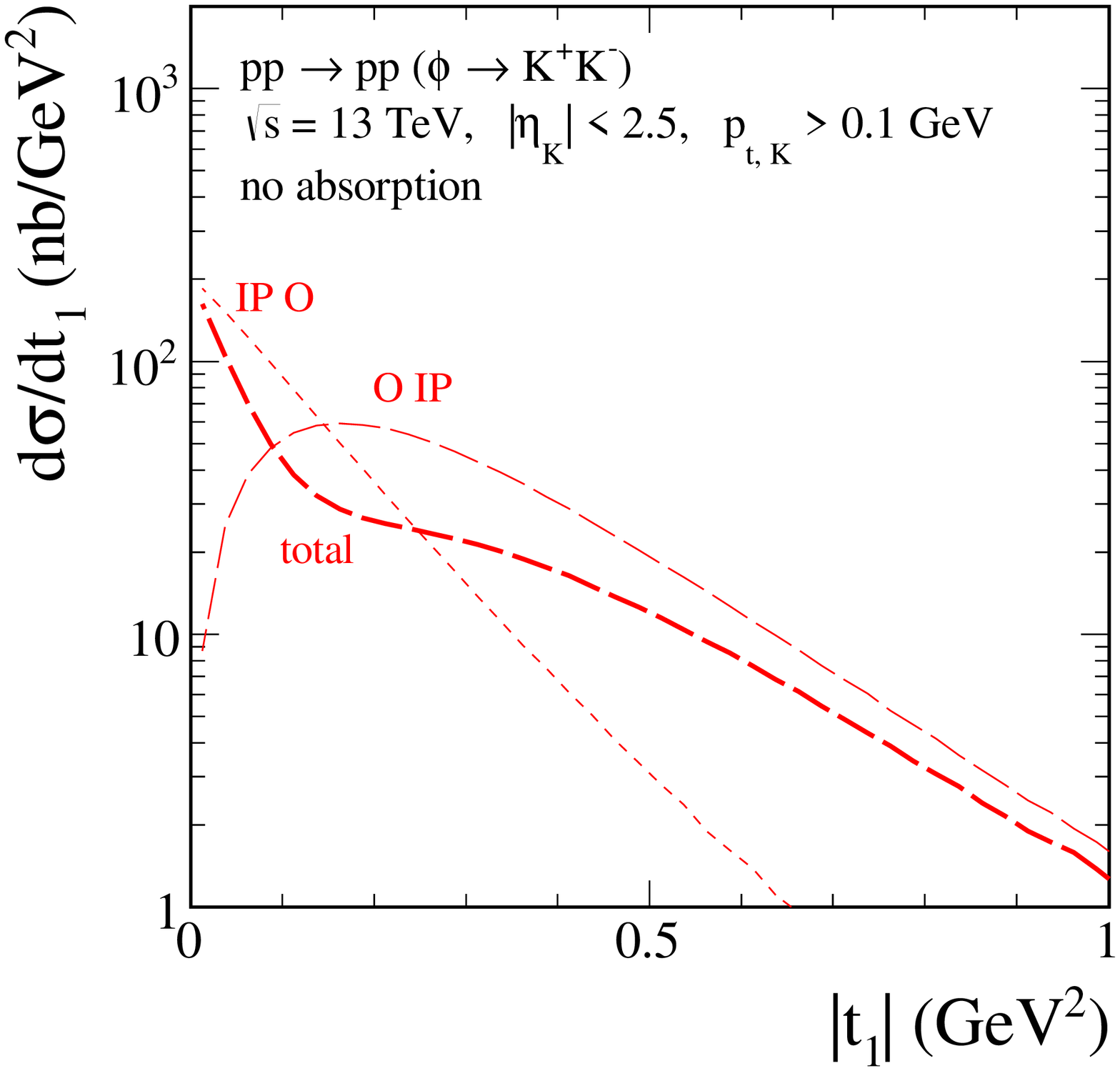}\\
\includegraphics[width=0.45\textwidth]{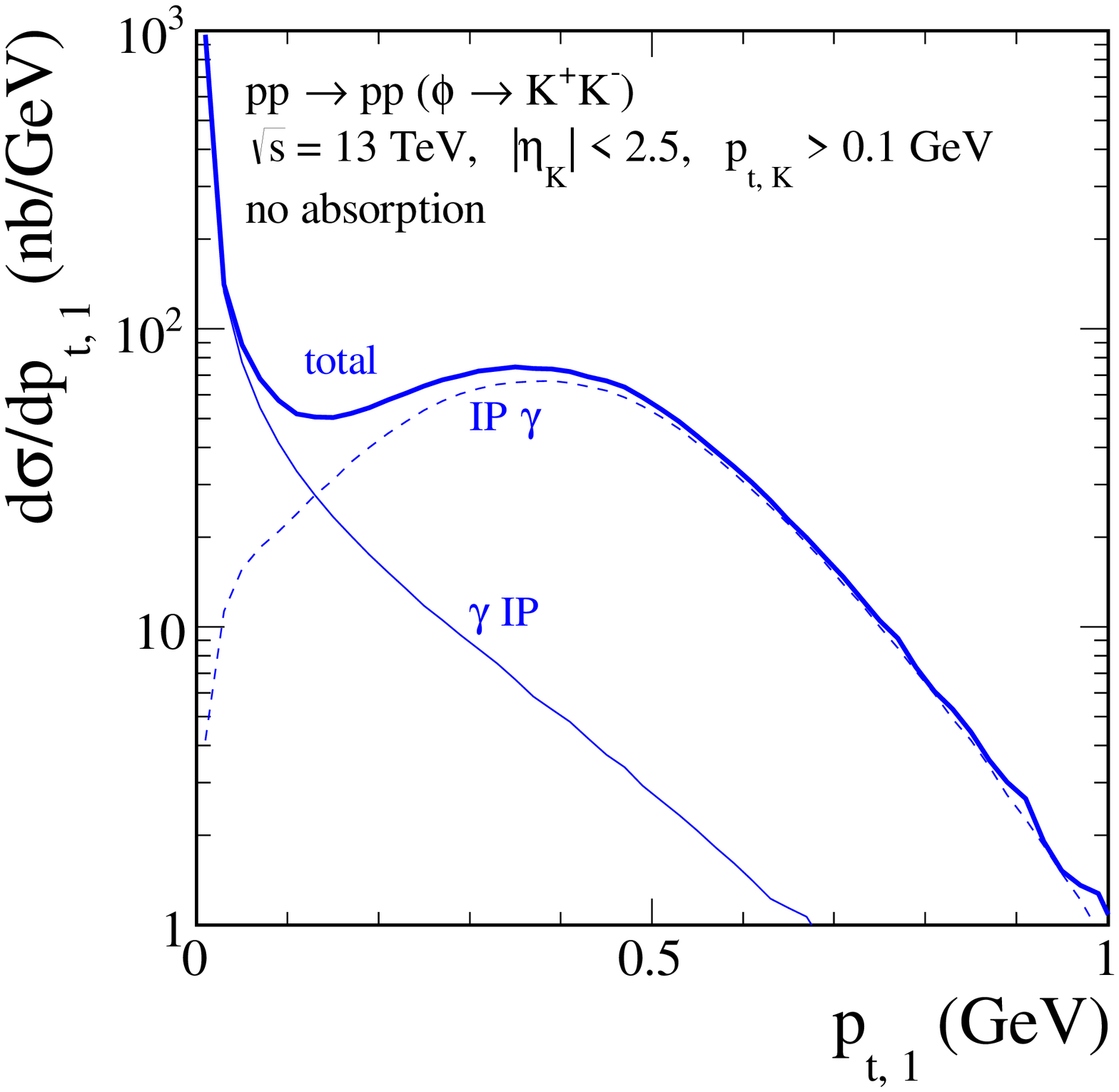}
\includegraphics[width=0.45\textwidth]{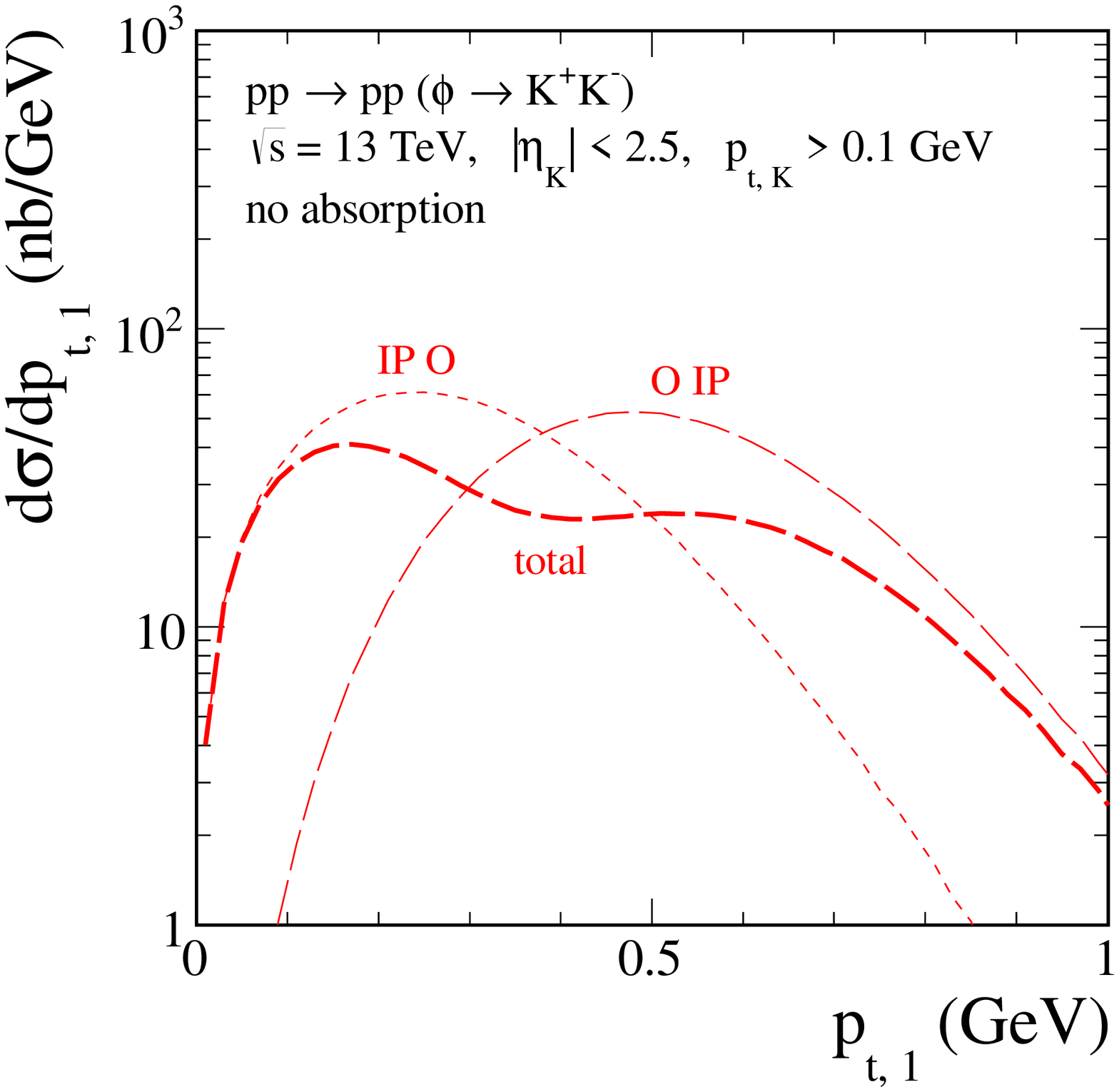}
\caption{\label{fig:ATLASALFA_born_0}
The distributions in four-momentum transfer squared $|t_{1}|$ (top panels) and 
in transverse momentum $p_{t, 1}$ of the proton $p\,(p_{1})$ (bottom panels)
for the $pp \to pp (\phi \to K^{+}K^{-})$ reaction at $\sqrt{s} = 13$~TeV
and for $|\eta_{K}| < 2.5$, $p_{t, K} > 0.1$~GeV.
Absorption effects are not included here.
In the left panels we show the results for the photoproduction mechanism
obtained with the parameter set~B (\ref{photoproduction_setB}).
The results for the $\gamma \Pom$- and $\Pom \gamma$-fusion contributions are presented.
Their coherent sum is shown by the blue solid thick line.
In the right panels we present the results for the odderon-pomeron-fusion mechanism
obtained with the parameters quoted in (\ref{parameters_ode}), 
(\ref{parameters_ode_lambda}), and (\ref{parameters_ode_b}).
Again, we show the $\Ode \Pom$- and $\Pom \Ode$-fusion contributions separately 
and their coherent sum (red long-dashed thick line).}
\end{figure}

In Fig.~\ref{fig:ATLASALFA_born_coup} we show results
for the hadronic diffractive contribution
for the two types of couplings in the $\Pom \Ode \phi$ vertex (\ref{A15})
separately and when both couplings are taken into account.
The distributions in $\phi_{pp}$,
the relative azimuthal angle between the outgoing protons, 
in $\rm{y_{diff}} = \rm{y}_{3} - \rm{y}_{4}$,
the rapidity distance between the two centrally produced kaons,
and in $\phi_{K^{+},\,{\rm CS}}$ and $\cos\theta_{K^{+},\,{\rm CS}}$ 
where the azimuthal and polar angles of the $K^{+}$ meson 
are defined in the Collins-Soper (CS) frame,
see Appendix~\ref{sec:appendixC}, are presented.
We can see that the complete result indicates 
a large interference effect of the $a_{\Pom \Ode \phi}$ 
and $b_{\Pom \Ode \phi}$ coupling contributions in the amplitudes. 
Note, in particular,
that both the $a$ and the $b$ term separately give a 
$\cos\theta_{K^{+},\,{\rm CS}}$ distribution with a maximum
at $\cos\theta_{K^{+},\,{\rm CS}} = 0$.
On the contrary, their coherent sum has a minimum there.
\begin{figure}[!ht]
\includegraphics[width=0.45\textwidth]{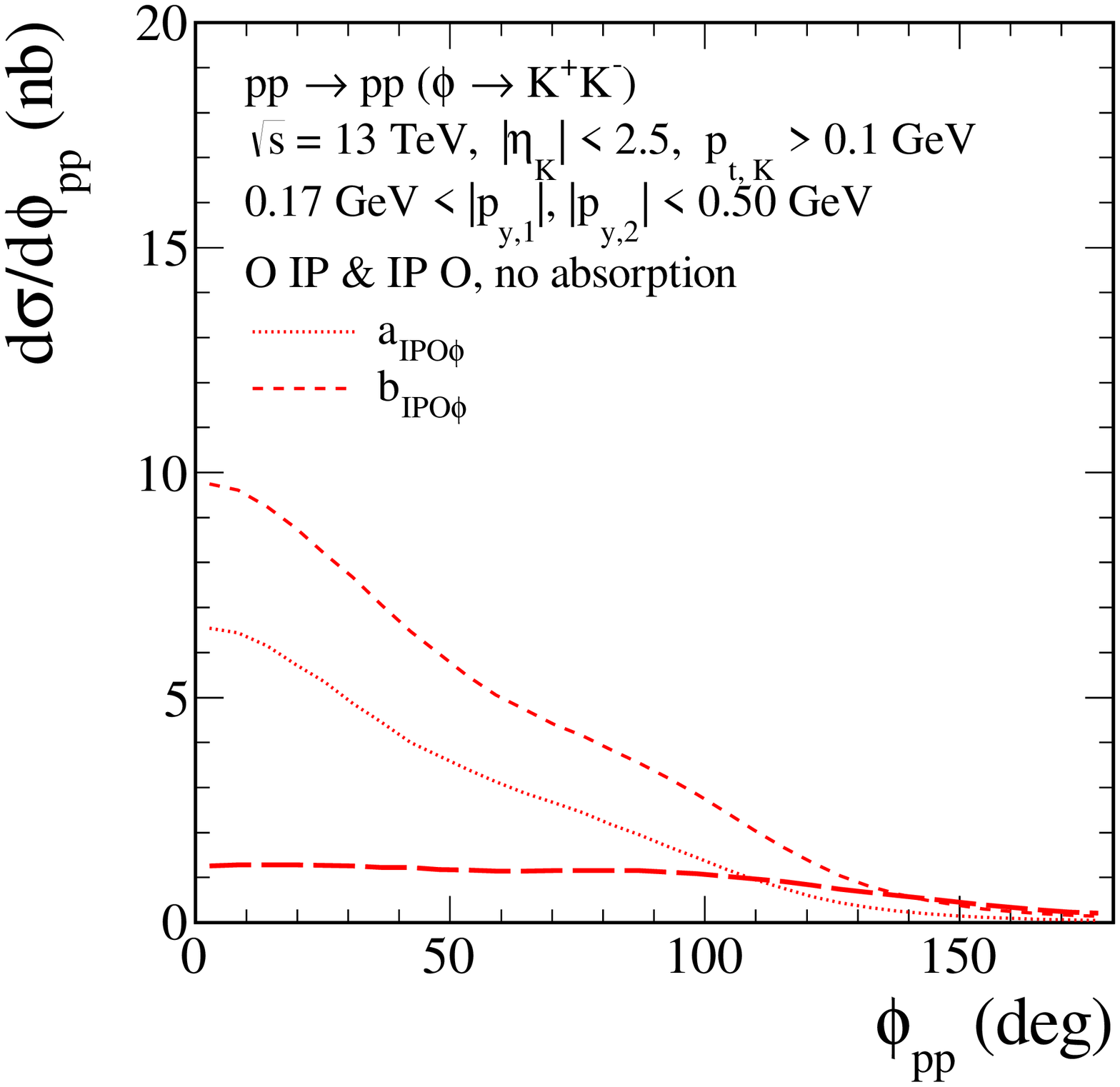}
\includegraphics[width=0.45\textwidth]{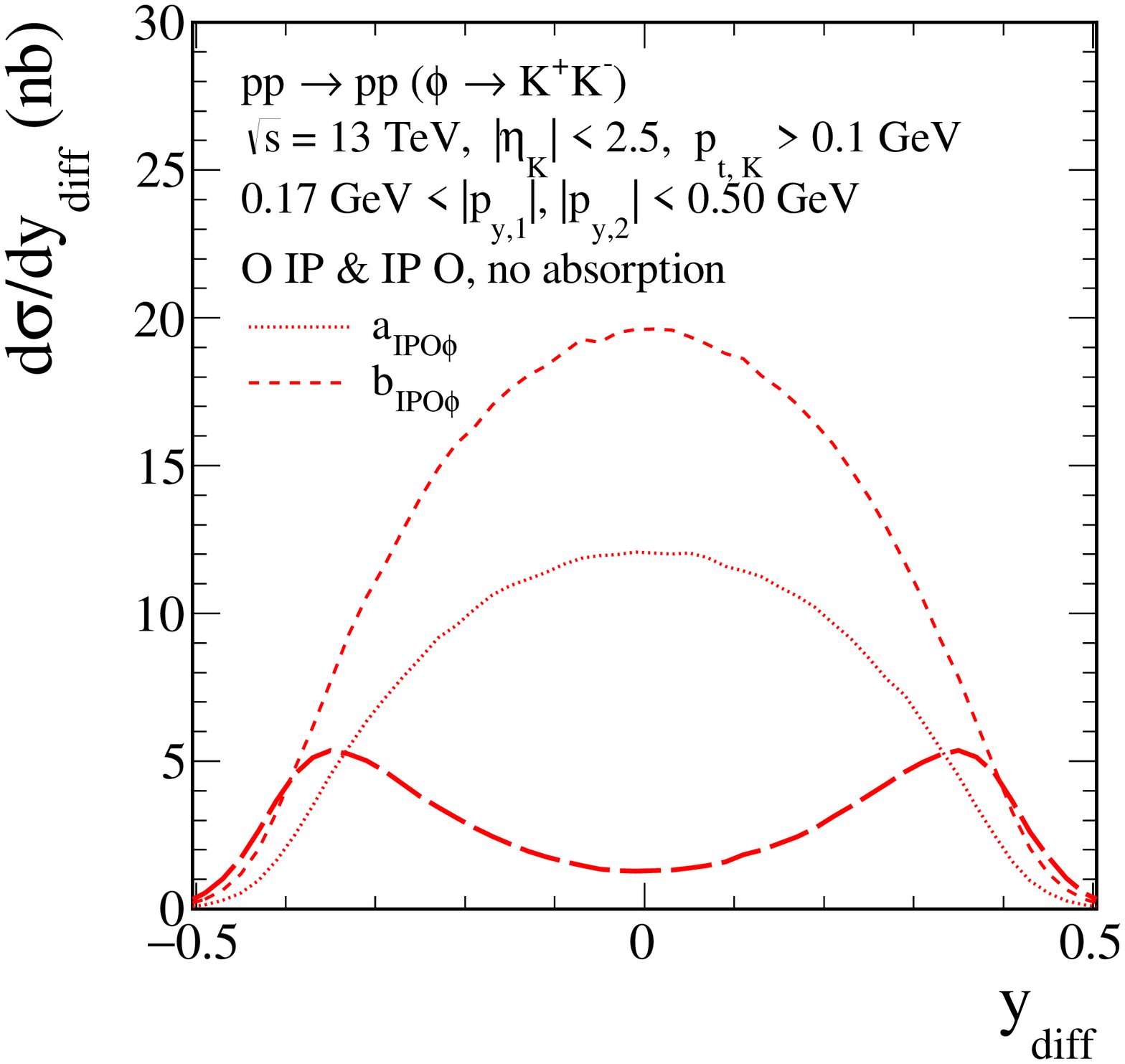}
\includegraphics[width=0.45\textwidth]{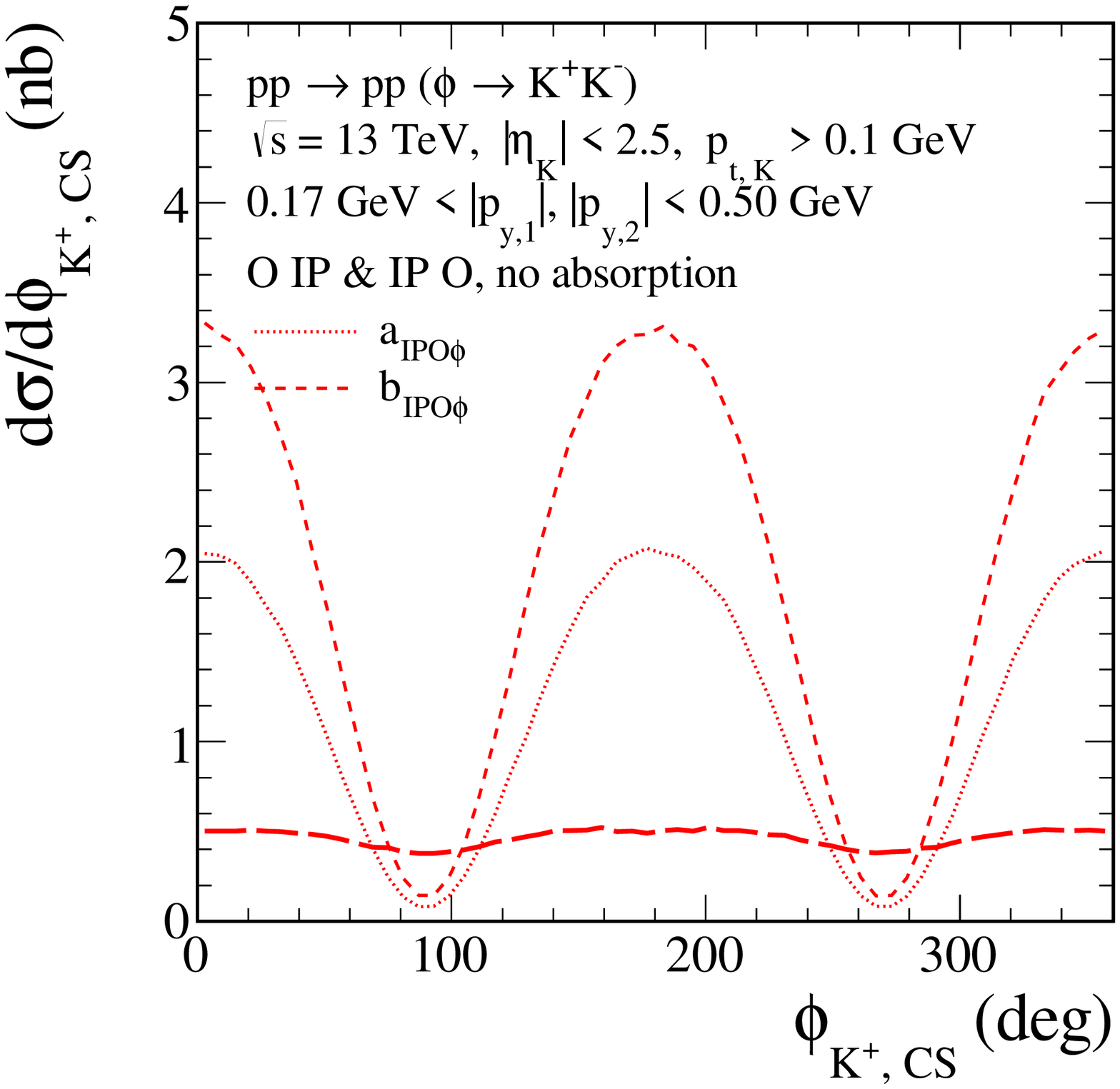}
\includegraphics[width=0.45\textwidth]{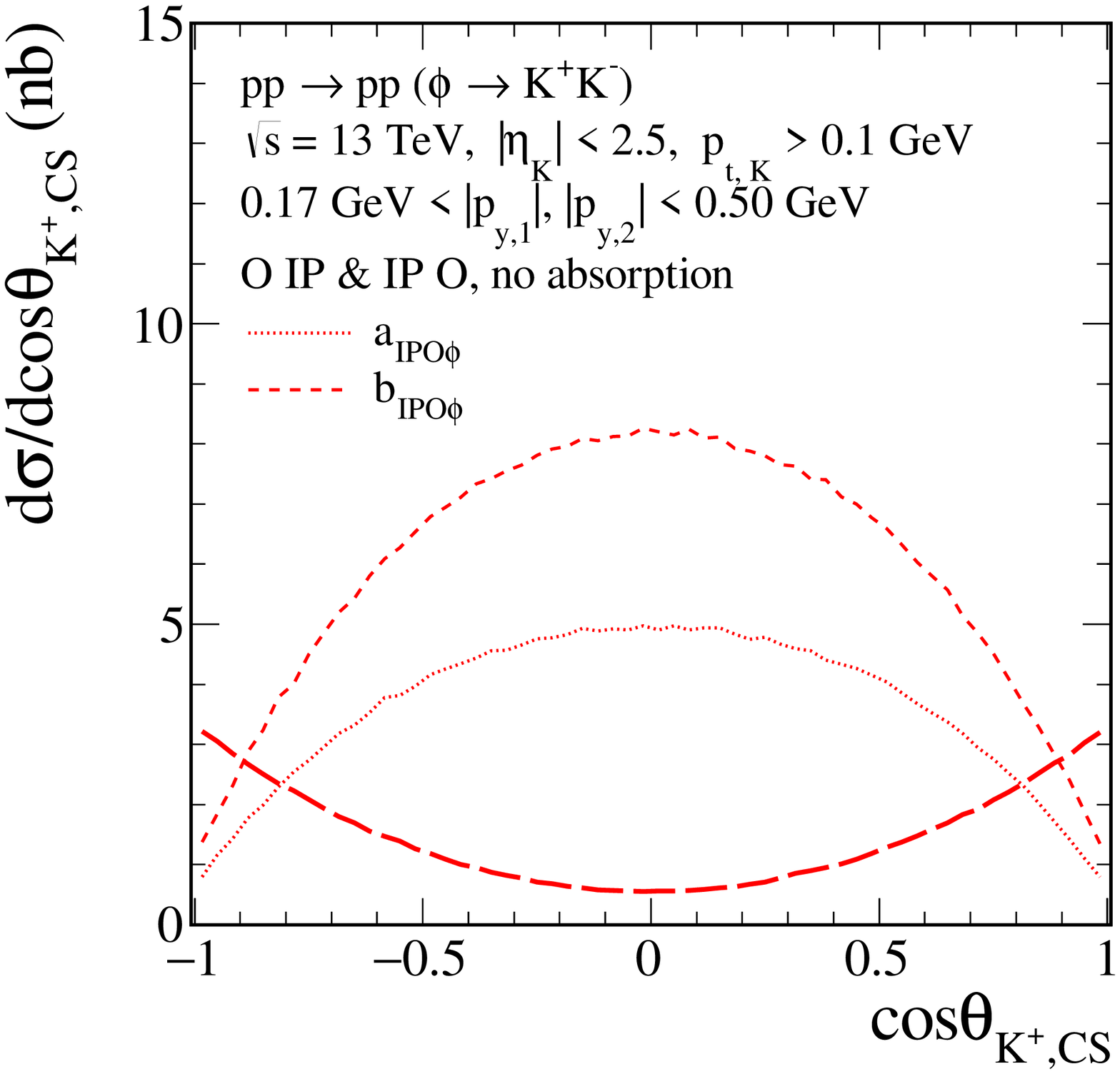}
\caption{\label{fig:ATLASALFA_born_coup}
The differential cross sections
for $\sqrt{s} = 13$~TeV and for the ATLAS-ALFA cuts 
($|\eta_{K}| < 2.5$, $p_{t, K} > 0.1$~GeV,
0.17~GeV~$< |p_{y,1}|, |p_{y,2}|<$~0.50~GeV).
We present the results for the hadronic diffractive contribution
neglecting absorption effects.
The thick long-dashed line represents the complete result with both
$a_{\Pom \Ode \phi}$ and $b_{\Pom \Ode \phi}$ couplings (\ref{parameters_ode_b}) 
included in the amplitude; see the $\Pom \Ode \phi$ vertex (\ref{A15}).
The contributions for the two type of couplings, 
$a$ and $b$ from (\ref{parameters_ode_b}), 
are shown separately:
the dotted line corresponds to the calculation only
with $a_{\Pom \Ode \phi}$,
and the short-dashed line corresponds to the calculation
only with $b_{\Pom \Ode \phi}$.}
\end{figure}

Figure~\ref{fig:ATLASALFA_born_1} shows 
the differential cross sections 
$d\sigma/d\phi_{pp}$ (see the top panels)
and $d\sigma/d\rm{y_{diff}}$ (see the bottom panels)
without (the left panels) and with (the right panels) limitations 
on the leading protons.
The blue lines correspond 
to the photoproduction contributions
while the red lines to the hadronic diffractive contributions.
The thin lines represent the results for one of the two diagrams separately 
($\gamma \Pom$ or $\Pom \gamma$ 
as well as $\Ode \Pom$ or $\Pom \Ode$)
and the thick lines represent their coherent sum 
($\gamma \Pom$ plus $\Pom \gamma$, 
$\Ode \Pom$ plus $\Pom \Ode$).
The reader is asked to note a reversed interference behaviour for
the photon-pomeron and odderon-pomeron mechanisms. 
The influence of kinematic cuts on the leading protons 
is also shown.
We see that due to the cuts on the leading protons
(0.17~GeV~$< |p_{y,1}|, |p_{y,2}|<$~0.50~GeV)
the photoproduction term is strongly suppressed.
The odderon-pomeron contribution dominates at larger
$|\rm{y_{diff}}|$ compared to the photon-pomeron contribution.
\begin{figure}[!ht]
\includegraphics[width=0.45\textwidth]{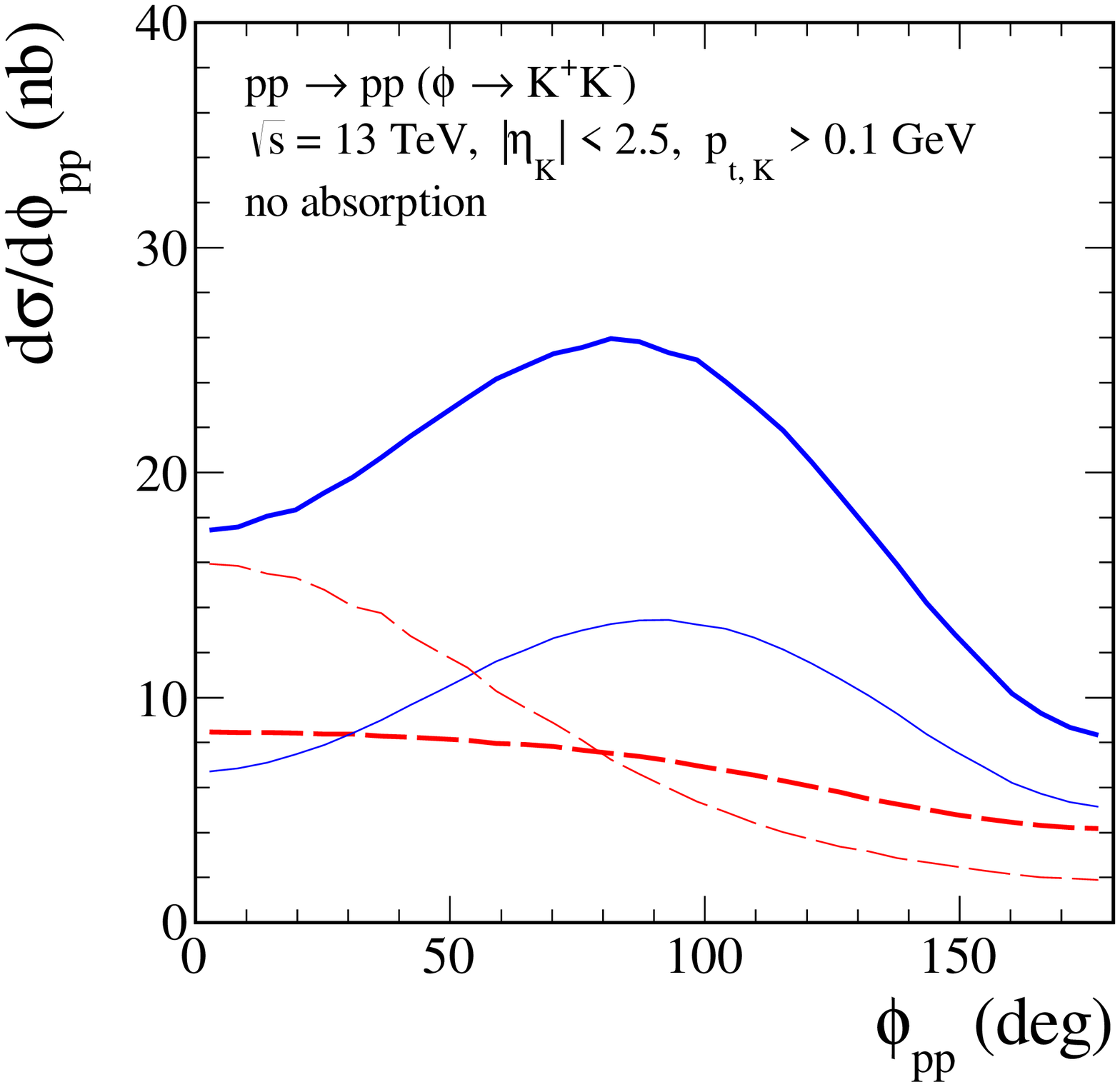}
\includegraphics[width=0.45\textwidth]{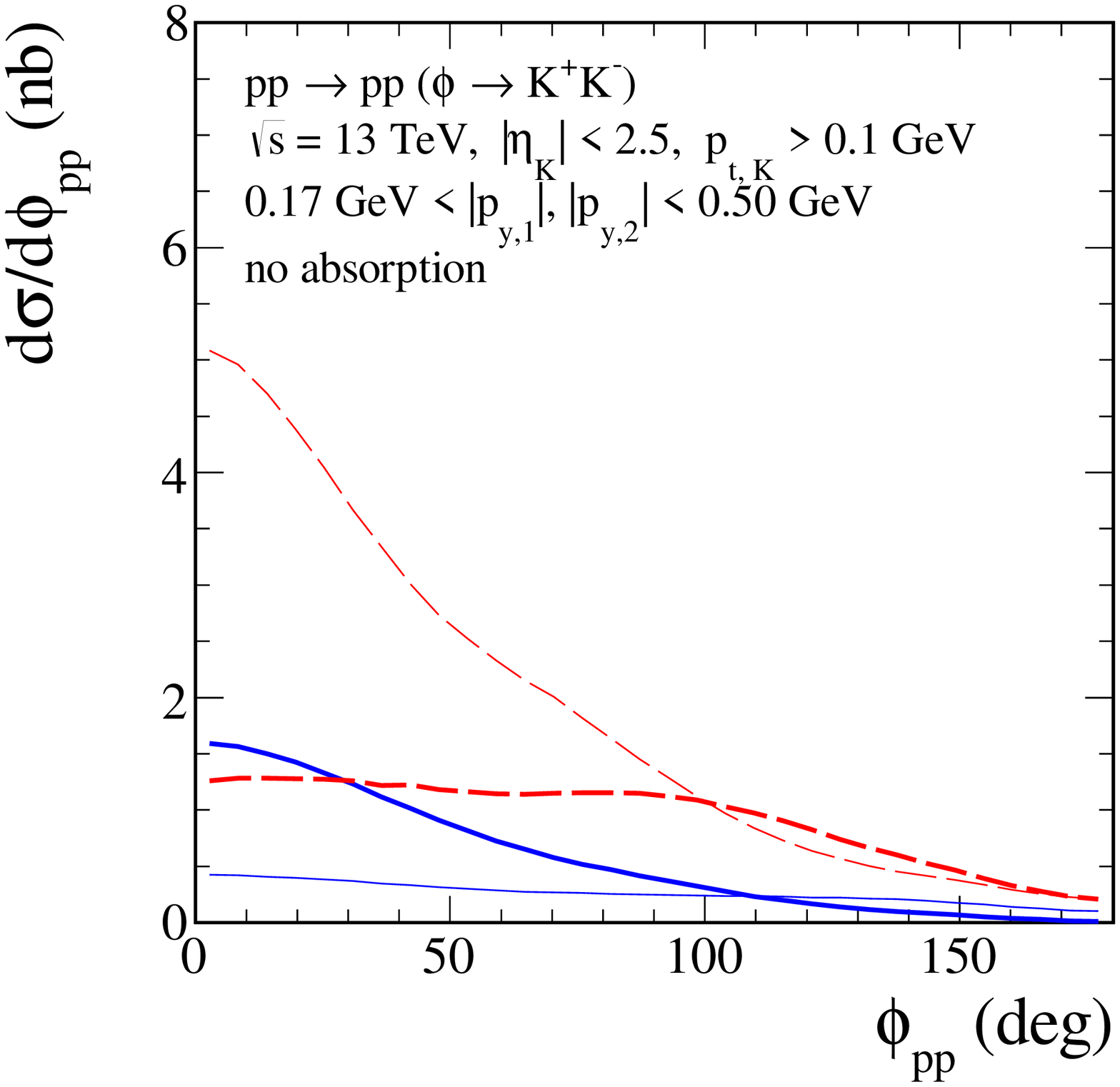}\\
\includegraphics[width=0.45\textwidth]{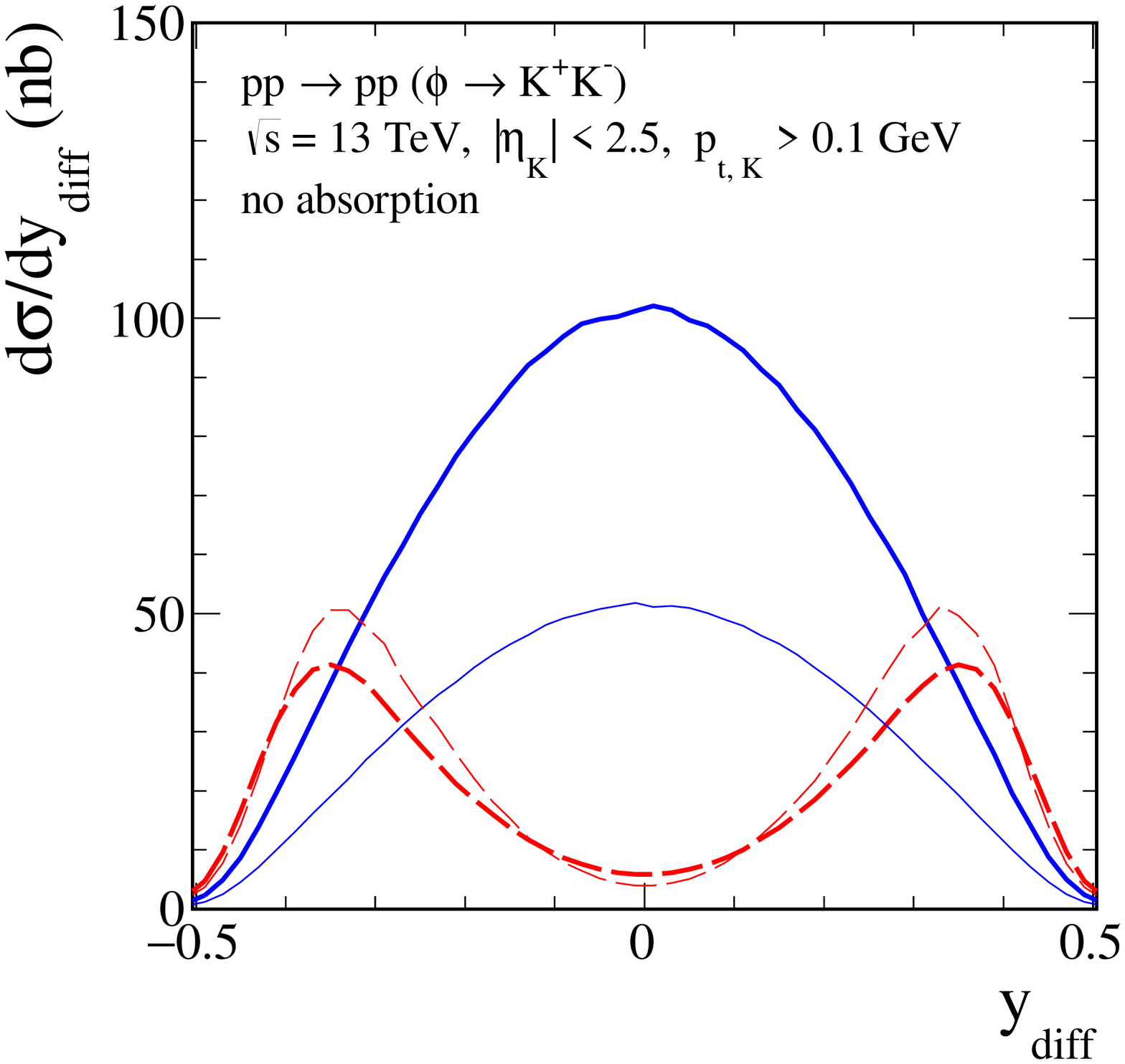}
\includegraphics[width=0.45\textwidth]{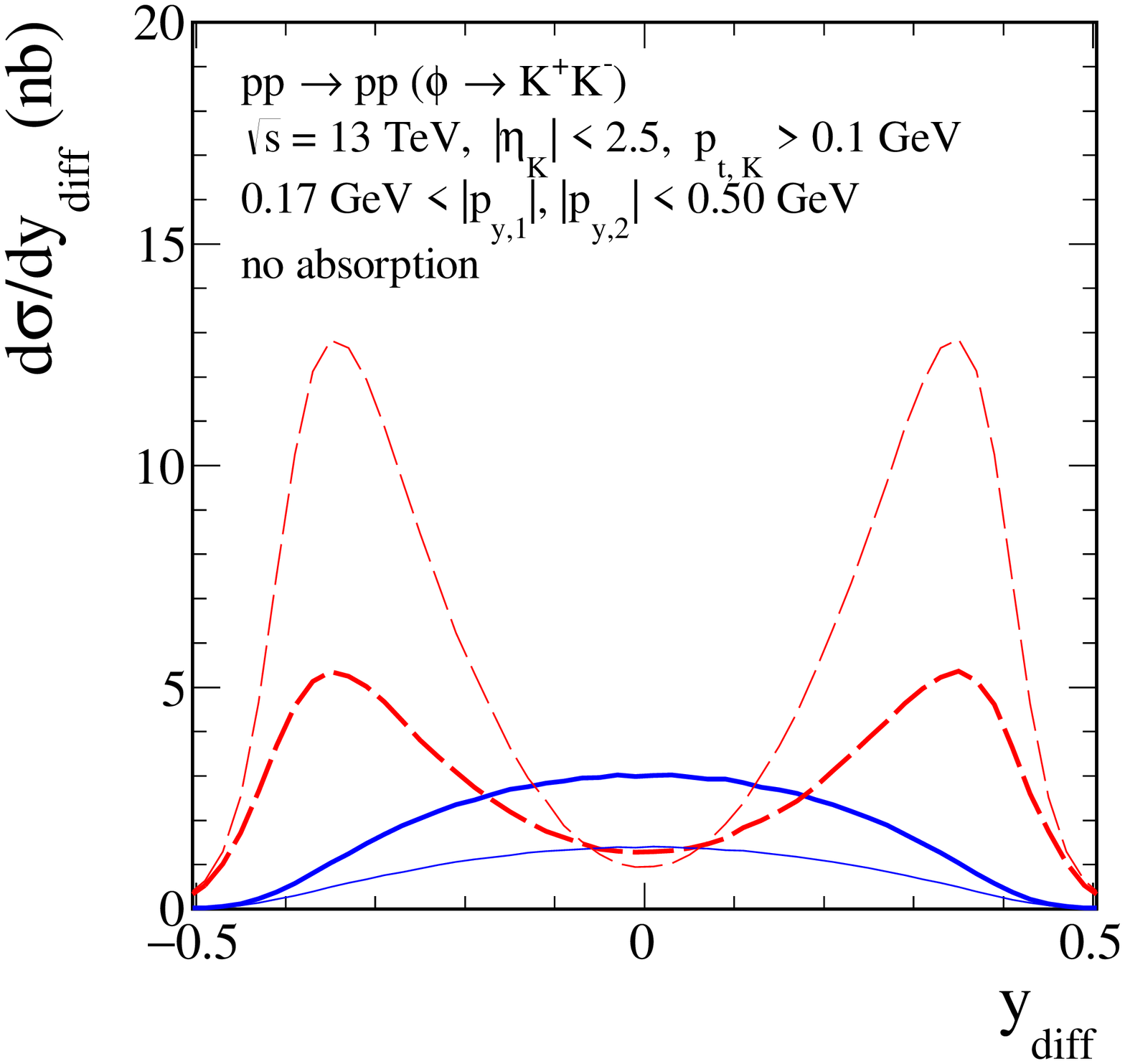}
\caption{\label{fig:ATLASALFA_born_1}
The distributions in azimuthal angle $\phi_{pp}$ 
between the transverse momentum vectors 
$\bpta$, $\bptb$ of the outgoing protons (top panels)
and in rapidity difference between kaons $\mathrm{y_{diff}}$ (bottom panels).
The calculations were performed for $\sqrt{s} = 13$~TeV
and for the ATLAS-ALFA experimental cuts
$|\eta_{K}| < 2.5$, $p_{t, K} > 0.1$~GeV (left panels),
and with extra cuts on the leading protons 
of 0.17~GeV~$< |p_{y,1}|, |p_{y,2}|<$~0.50~GeV (right panels).
The blue thick solid line corresponds to 
the coherent sum of the two diagrams 
($\gamma \Pom$ and $\Pom \gamma$).
The red thick dashed line corresponds 
to the coherent sum of the $\Ode \Pom$ and $\Pom \Ode$ contributions.
The thin lines correspond to the results for one of the two diagrams separately
(the second contribution is the same).
For the $\gamma$-$\Pom$-fusion contribution we take the parameter set~B (\ref{photoproduction_setB}).
For the $\Ode$-$\Pom$-fusion contribution
we take the parameters
quoted in (\ref{parameters_ode}), (\ref{parameters_ode_lambda}), 
and (\ref{parameters_ode_b}).}
\end{figure}

In Fig.~\ref{fig:ATLASALFA_born_2} we show the kaon angular distributions
in the $K^+ K^-$ rest system using the Collins-Soper (CS) frame;
see Appendix~\ref{sec:appendixC}.
The Collins-Soper frame which we use here is defined 
as in our recent paper on extracting the $\Pom \Pom f_2(1270)$ couplings 
in the $pp \to pp \pi^{+} \pi^{-}$ reaction \cite{Lebiedowicz:2019por}
with $K^{+}$ and $K^{-}$ in the place of $\pi^{+}$ and $\pi^{-}$, respectively.
For the $pp \to pp (\phi \to K^{+} K^{-})$ reaction
we can observe interesting structures in the $\phi_{K^{+},\,{\rm CS}}$ (top panel) 
and in the $\cos\theta_{K^{+},\,{\rm CS}}$ (bottom panel) distributions.
The distributions in $\phi_{K^{+},\,{\rm CS}}$ for the hadronic diffractive contribution
($\Ode \Pom$ plus $\Pom \Ode$) are relatively flat.
The photoproduction term, in contrast, shows pronounced maxima and minima
which are due to the interference 
of the $\gamma \Pom$ and $\Pom \gamma$ terms.
The cuts on leading protons considerably change 
the shape of the $\phi_{K^{+},\,{\rm CS}}$ distributions
for the photon-exchange contribution.
The angular distribution 
$d\sigma/d\cos\theta_{K^{+},\,{\rm CS}}$ looks promising 
for a search of odderon effects as it is very different 
for the $\gamma$-$\Pom$- and the $\Ode$-$\Pom$-fusion processes. 
\begin{figure}[!ht]
\includegraphics[width=0.45\textwidth]{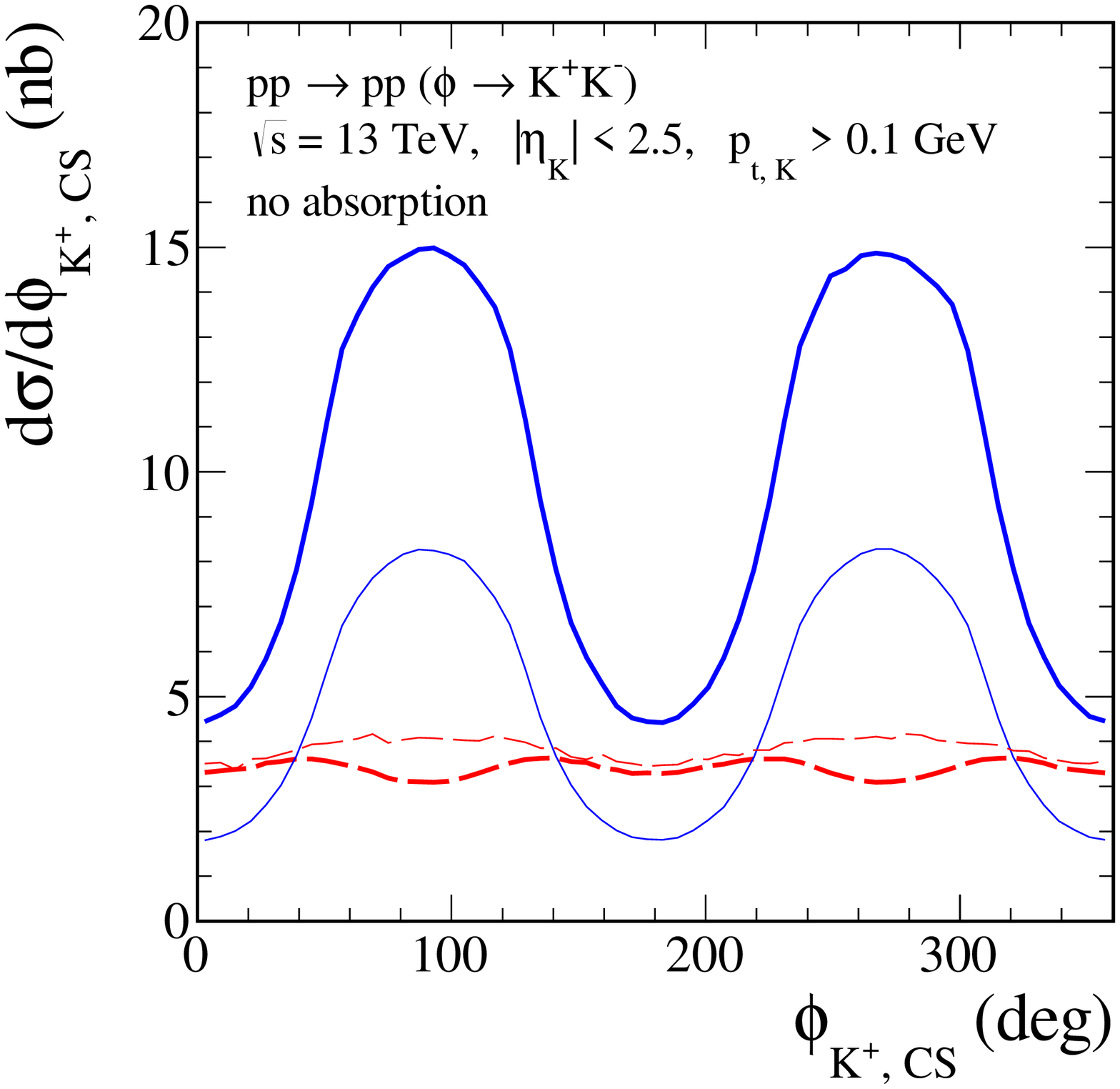}
\includegraphics[width=0.45\textwidth]{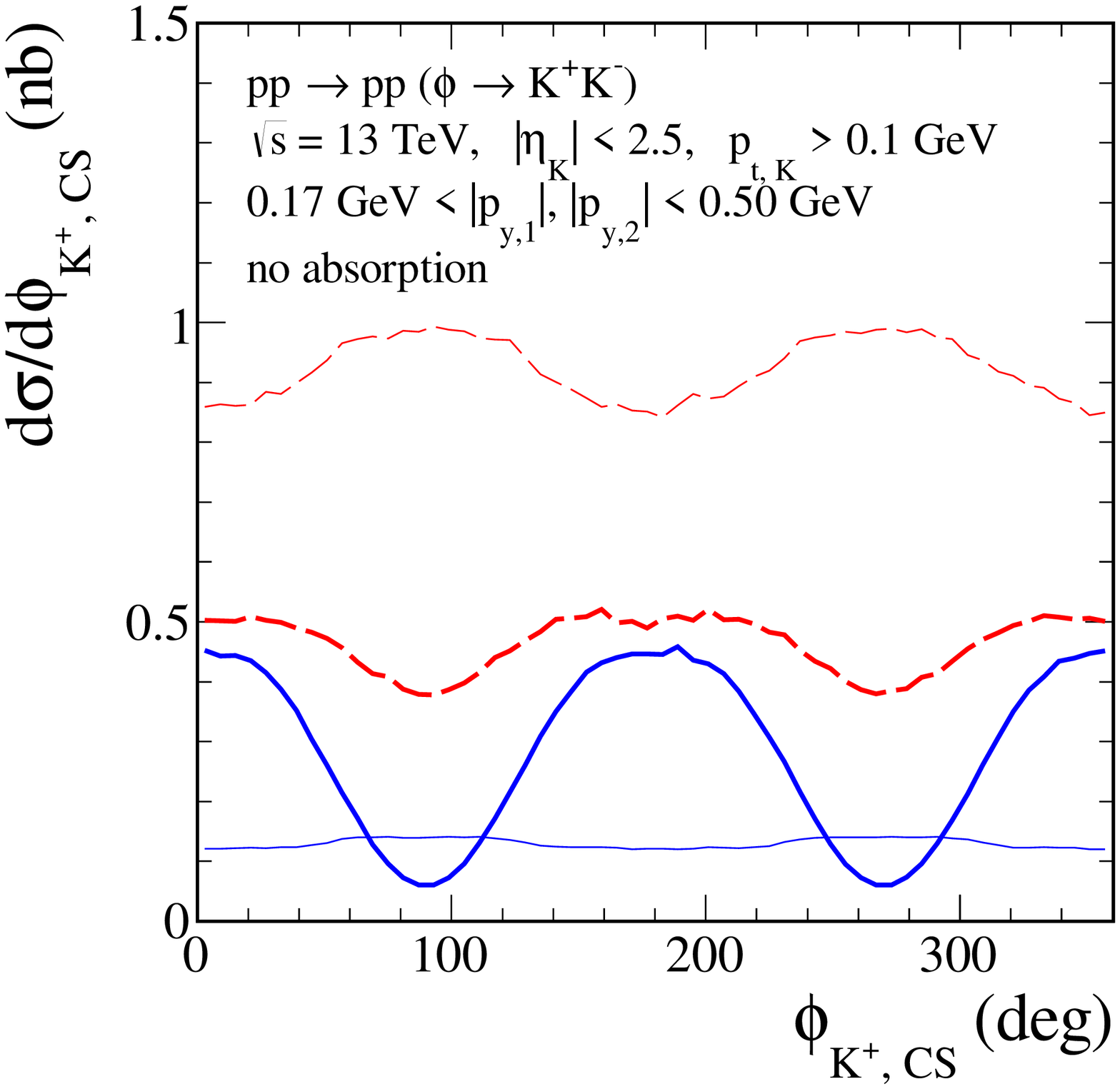}\\
\includegraphics[width=0.45\textwidth]{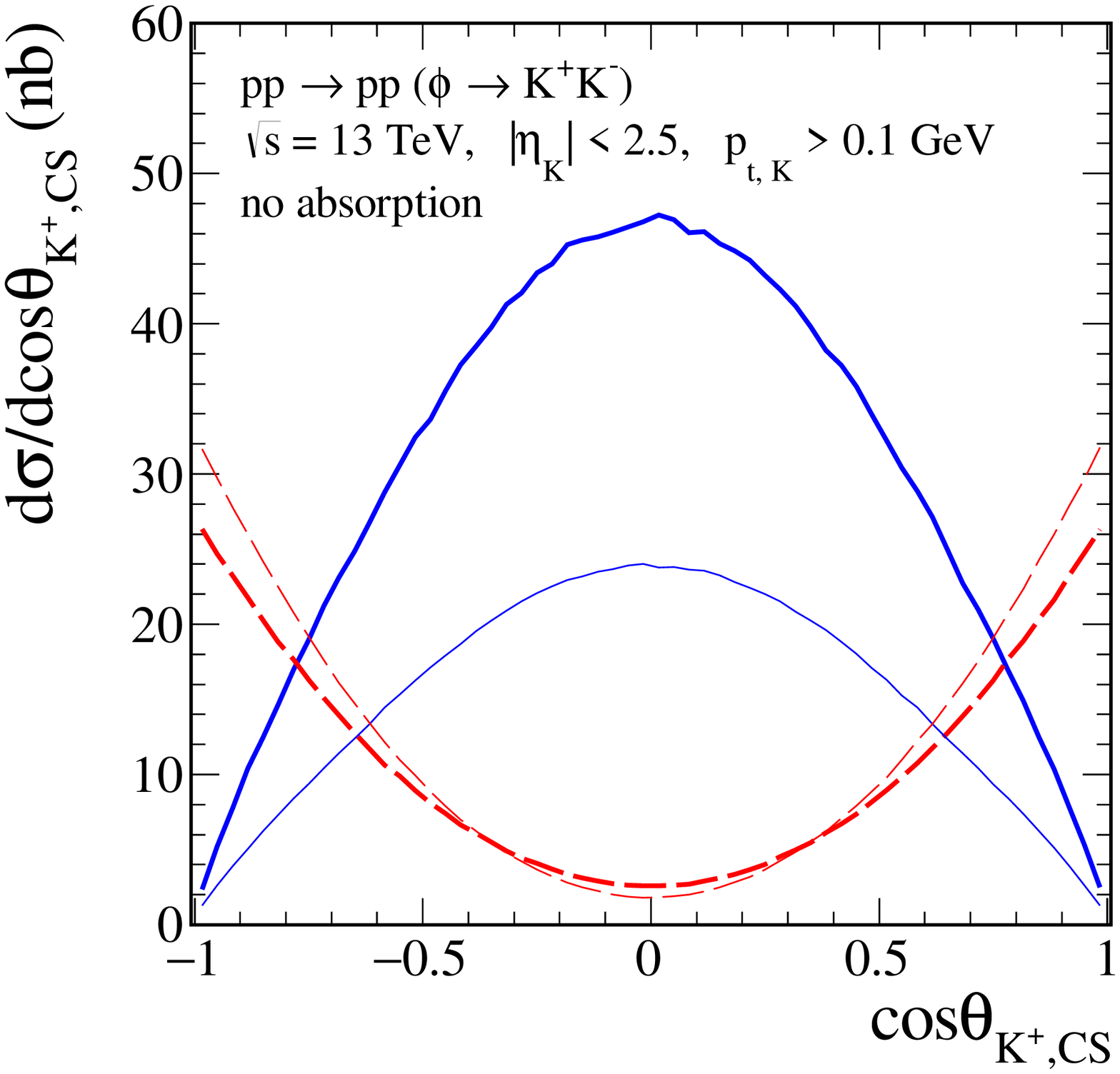}
\includegraphics[width=0.45\textwidth]{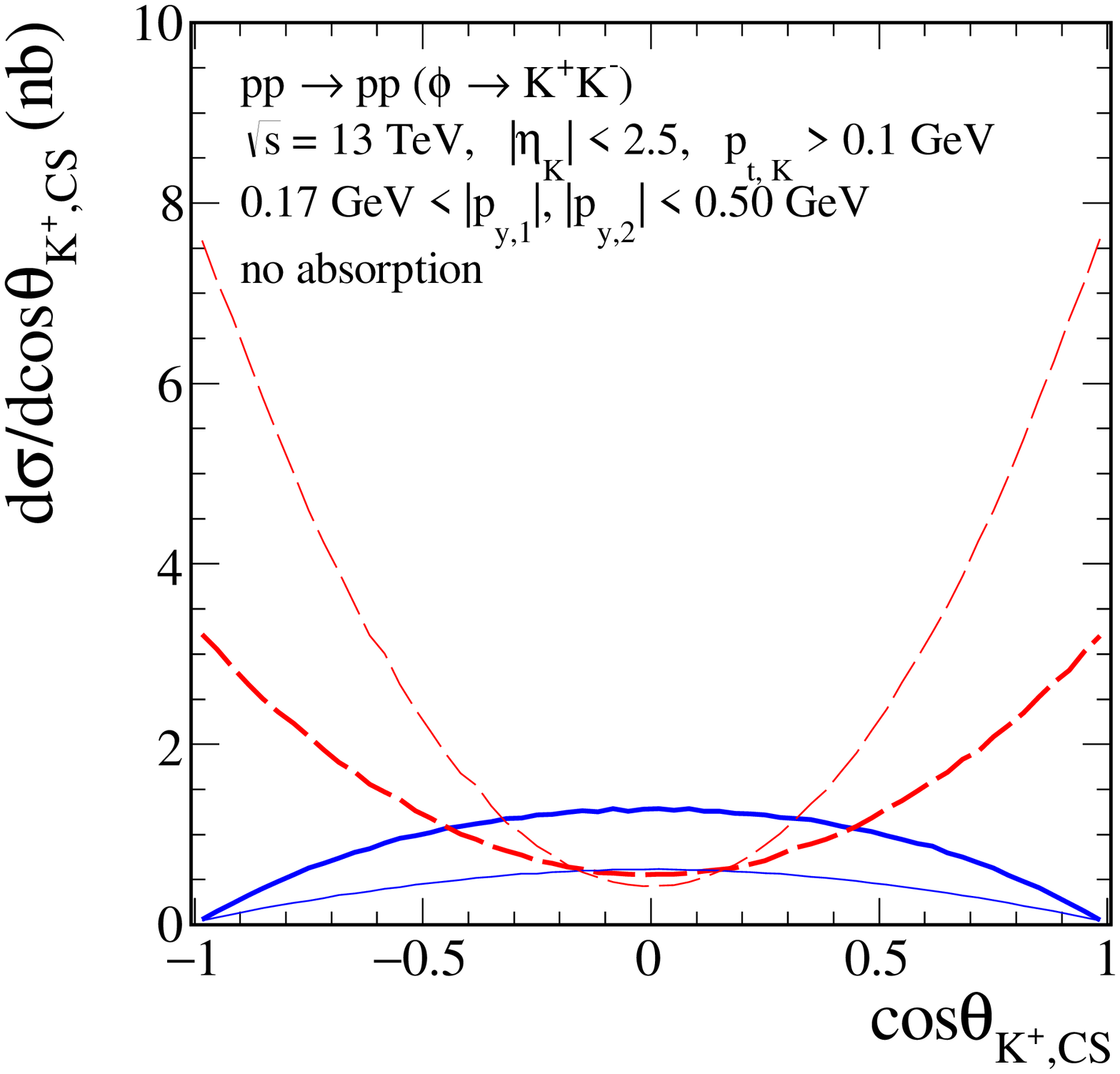}
\caption{\label{fig:ATLASALFA_born_2}
The distributions in $\phi_{K^{+}, {\rm CS}}$ (top panel) 
and in $\cos\theta_{K^{+},\,{\rm CS}}$ (bottom panel) 
for $\sqrt{s} = 13$~TeV, $|\eta_{K}| < 2.5$, $p_{t, K} > 0.1$~GeV (left panels),
and with extra cuts on the leading protons 
of 0.17~GeV~$< |p_{y,1}|, |p_{y,2}|<$~0.50~GeV (right panels).
The meaning of the lines is the same as in Fig.~\ref{fig:ATLASALFA_born_1}.}
\end{figure}

In Fig.~\ref{fig:ATLASALFA_absorption} we compare
results without (the thin lines) and with (the thick lines) absorption effects.
The absorption effects have been included in our analysis
within the one-channel-eikonal approach.
For the ATLAS-ALFA kinematics the absorption effects lead to a large
damping of the cross sections both for the hadronic diffractive 
and for the photoproduction mechanisms. 
We find a suppression factor of the cross section of
$\langle S^{2} \rangle \simeq 0.3$;
see Table~\ref{tab:table2}.
A similar value of suppression was found in \cite{Ryutin:2019khx}
(see Fig.~14 there)
for the exclusive $pp \to pp \pi^{+}\pi^{-}$ reaction 
for the diffractive continuum process at the LHC energy.
From Fig.~\ref{fig:ATLASALFA_absorption} 
we see that the absorption effects also modify the shape of the distributions.
\begin{figure}[!ht]
\includegraphics[width=0.42\textwidth]{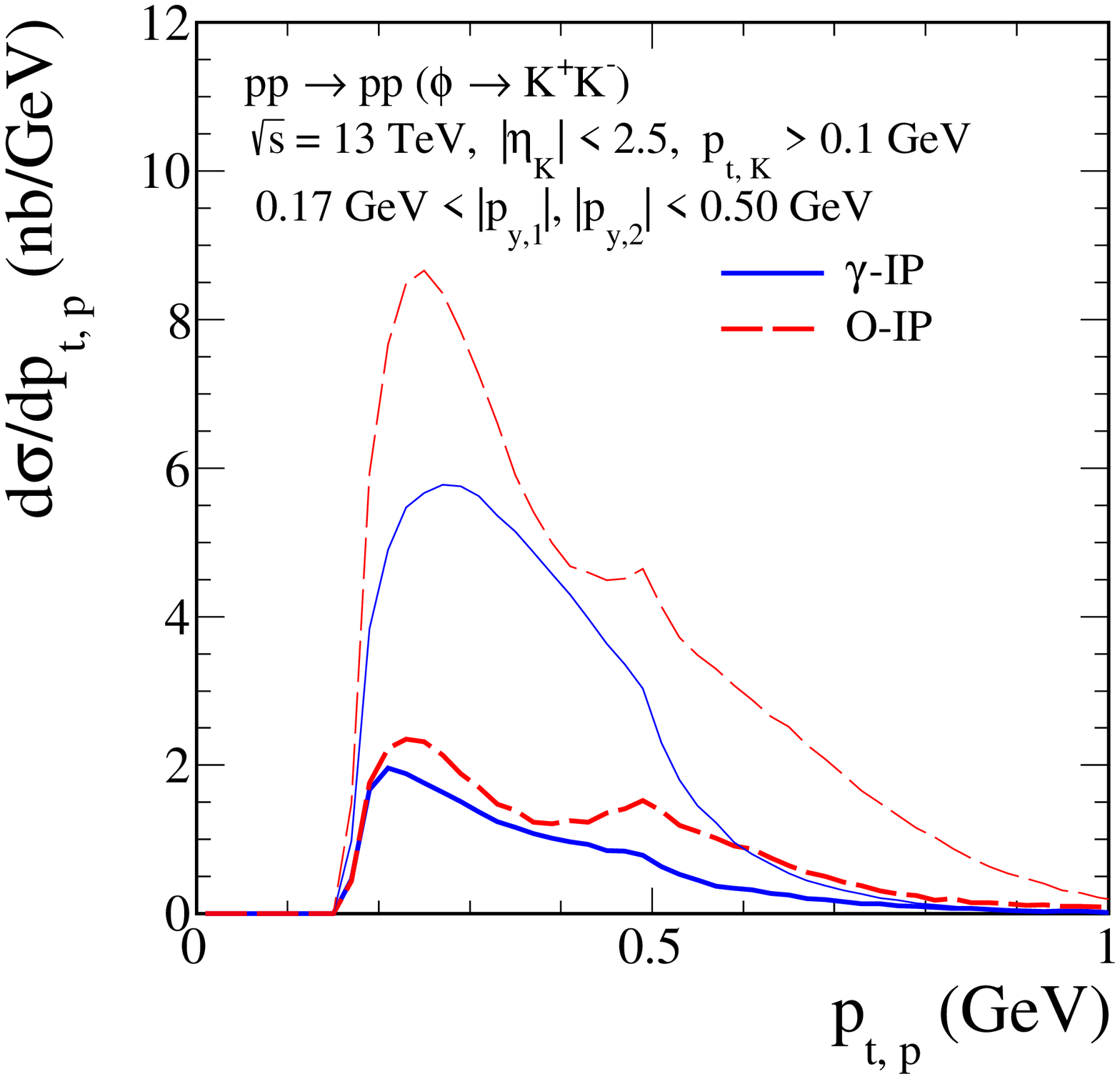}
\includegraphics[width=0.42\textwidth]{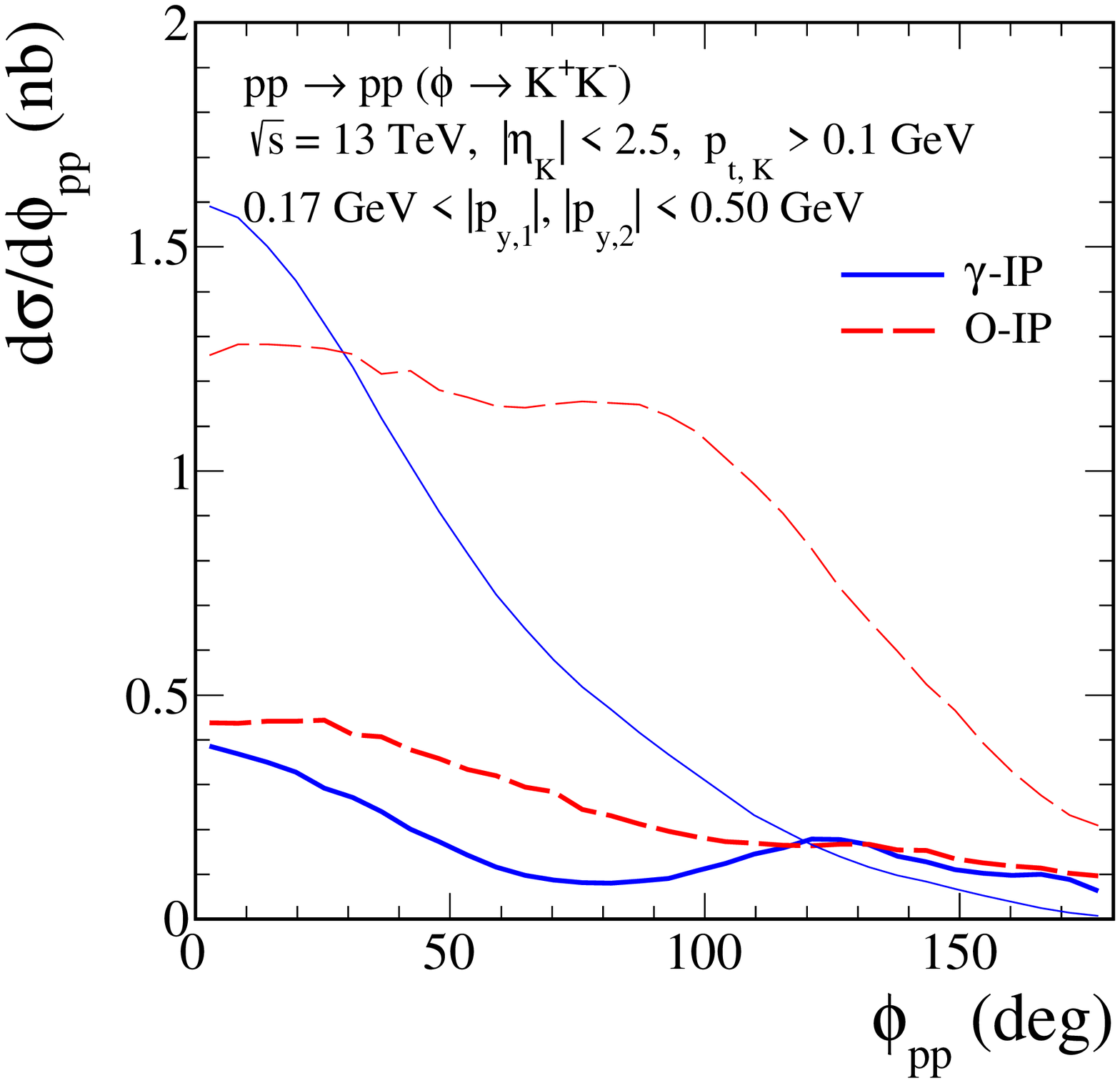}
\includegraphics[width=0.42\textwidth]{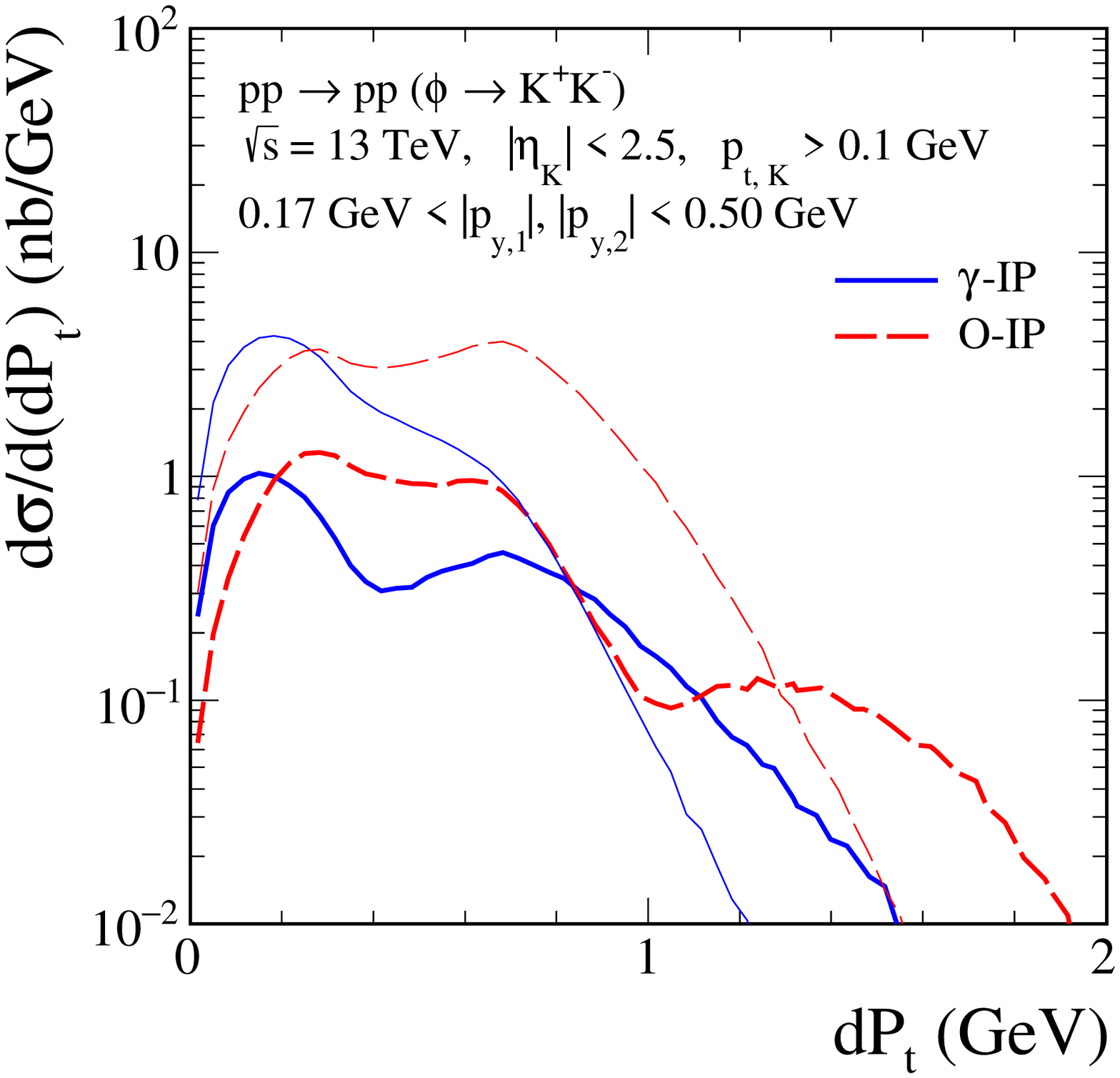}
\includegraphics[width=0.42\textwidth]{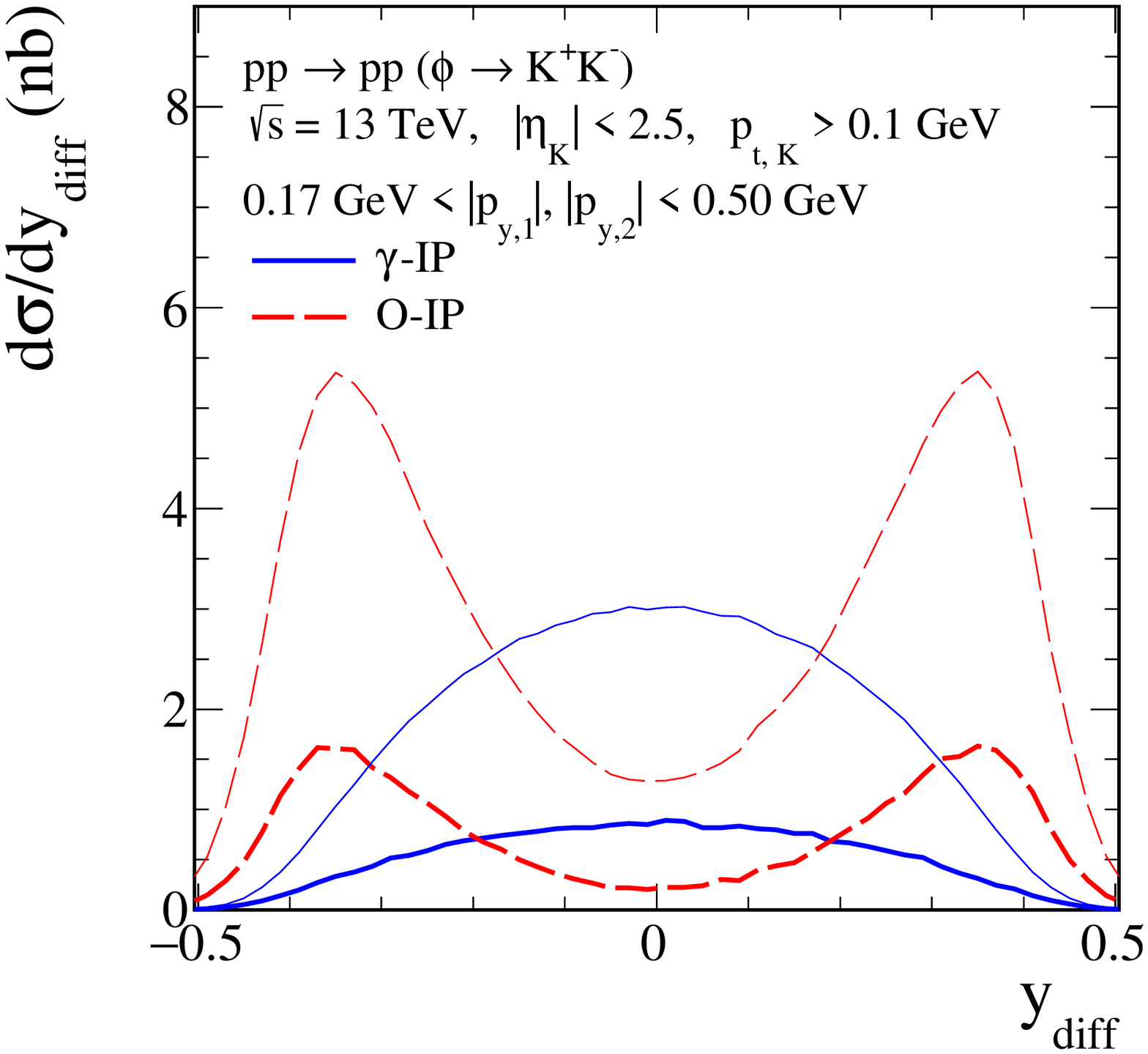}
\includegraphics[width=0.42\textwidth]{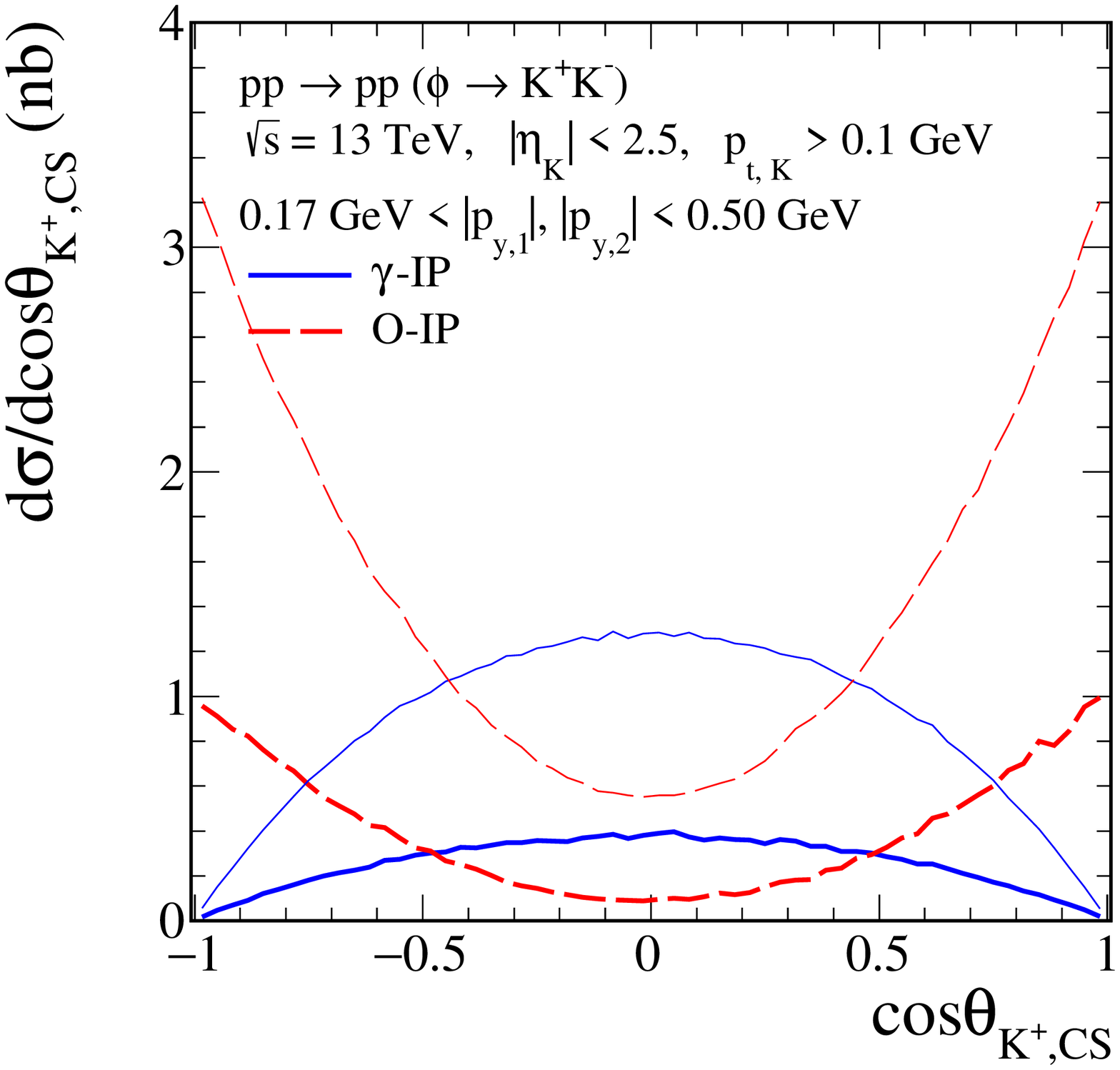}
\includegraphics[width=0.42\textwidth]{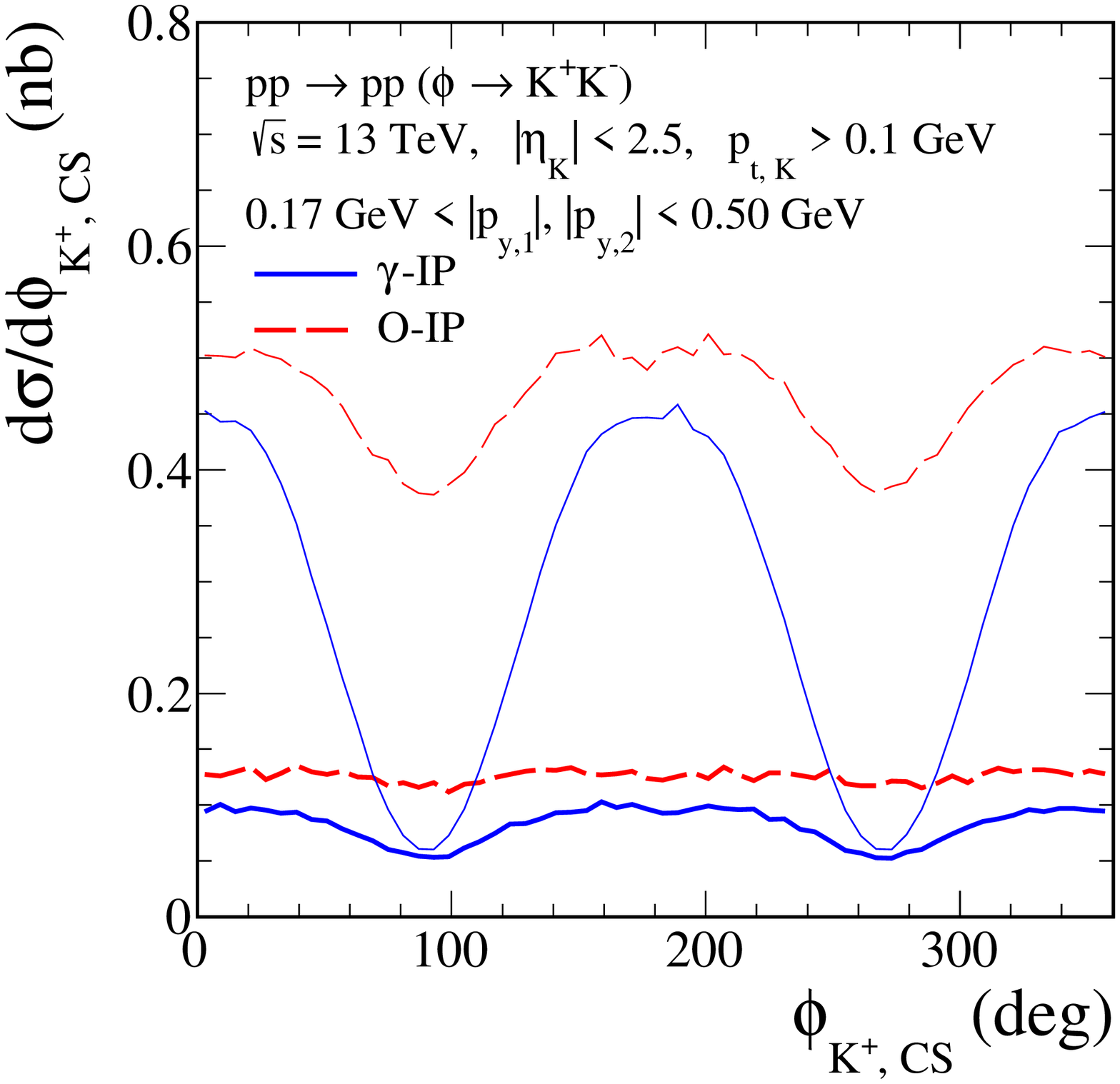}
\caption{\label{fig:ATLASALFA_absorption}
The differential cross sections for $\sqrt{s} = 13$~TeV
and the ATLAS-ALFA cuts without (the thin lines) 
and with (the thick lines) absorption effects.
For the $\gamma$-$\Pom$-fusion contribution 
we take the parameter set~B (\ref{photoproduction_setB}).
For the $\Ode$-$\Pom$-fusion contribution
we take the parameters
quoted in (\ref{parameters_ode}), (\ref{parameters_ode_lambda}), 
and (\ref{parameters_ode_b}).}
\end{figure}

From the $\cos\theta_{K^{+},\,{\rm CS}}$ distributions 
shown in Figs.~\ref{fig:ATLASALFA_born_2} and \ref{fig:ATLASALFA_absorption} 
we can conclude that from the $\gamma$-$\Pom$ fusion 
the $\phi$ meson gets preferentially a transverse polarisation
giving a distribution proportional to $\sin^{2}\theta_{K^{+},\,{\rm CS}}$.
For the $\Ode$-$\Pom$ fusion, on the other hand, we find
that the $\phi$ meson gets preferentially a longitudinal
polarisation with a distribution proportional 
to $\cos^{2}\theta_{K^{+},\,{\rm CS}}$.
This different behaviour can be understood using again the 
considerations of Appendix~C of \cite{Bolz:2014mya}.
The $\gamma$-$\Pom$ contribution is largest for very small~$|t|$,
see Fig.~\ref{fig:ATLASALFA_born_0}, where the virtual photon
has essentially only transverse polarisation
which it will transmit to the $\phi$. 
The $\Ode$-$\Pom$ fusion, on the other hand, gives a very
small contribution for very small~$|t|$.
For larger $|t|$, however, where the odderon contributes most,
the longitudinal cross section has a ``large'' factor $|t|$
relative to the transverse term.
(This is quite analogous to what happens in DIS for
the standard cross sections of the absorption of the
virtual photon on the proton, $\sigma_{T}$ and $\sigma_{L}$.
For $Q^{2} \to 0$ $\sigma_{T}$ goes to a constant,
$\sigma_{L}$ is proportional to $Q^{2}$;
see for instance \cite{Britzger:2019lvc}).

Up to now we have shown results including the ATLAS-ALFA experimental cuts
for a concrete set of parameters,
set~B (\ref{photoproduction_setB}) for the photoproduction term
and (\ref{parameters_ode_b}) for the $\Pom \Ode \phi$ coupling parameters.
In Fig. ~\ref{fig:ATLASALFA_unc} we show results for different parameter sets,
as discussed in Sec.~\ref{sec:comparison_WA102},
for the $\gamma$-$\Pom$- and $\Ode$-$\Pom$-fusion processes.
The upper blue solid line is for the parameter set~B
of photoproduction (\ref{photoproduction_setB})
and the lower blue solid line is for set~A (\ref{photoproduction_setA}).
The red long-dashed line corresponds to the odderon parameters 
quoted in (\ref{parameters_ode}), (\ref{parameters_ode_lambda}), and 
the $\Pom \Ode \phi$ coupling parameters~(b)~(\ref{parameters_ode_b}),
the red dash-dotted line is for 
the choice of $\Pom \Ode \phi$ coupling parameters~(a)~(\ref{parameters_ode_a}),
and the red dotted line is for (\ref{parameters_ode_c}).
\begin{figure}[!ht]
\includegraphics[width=0.45\textwidth]{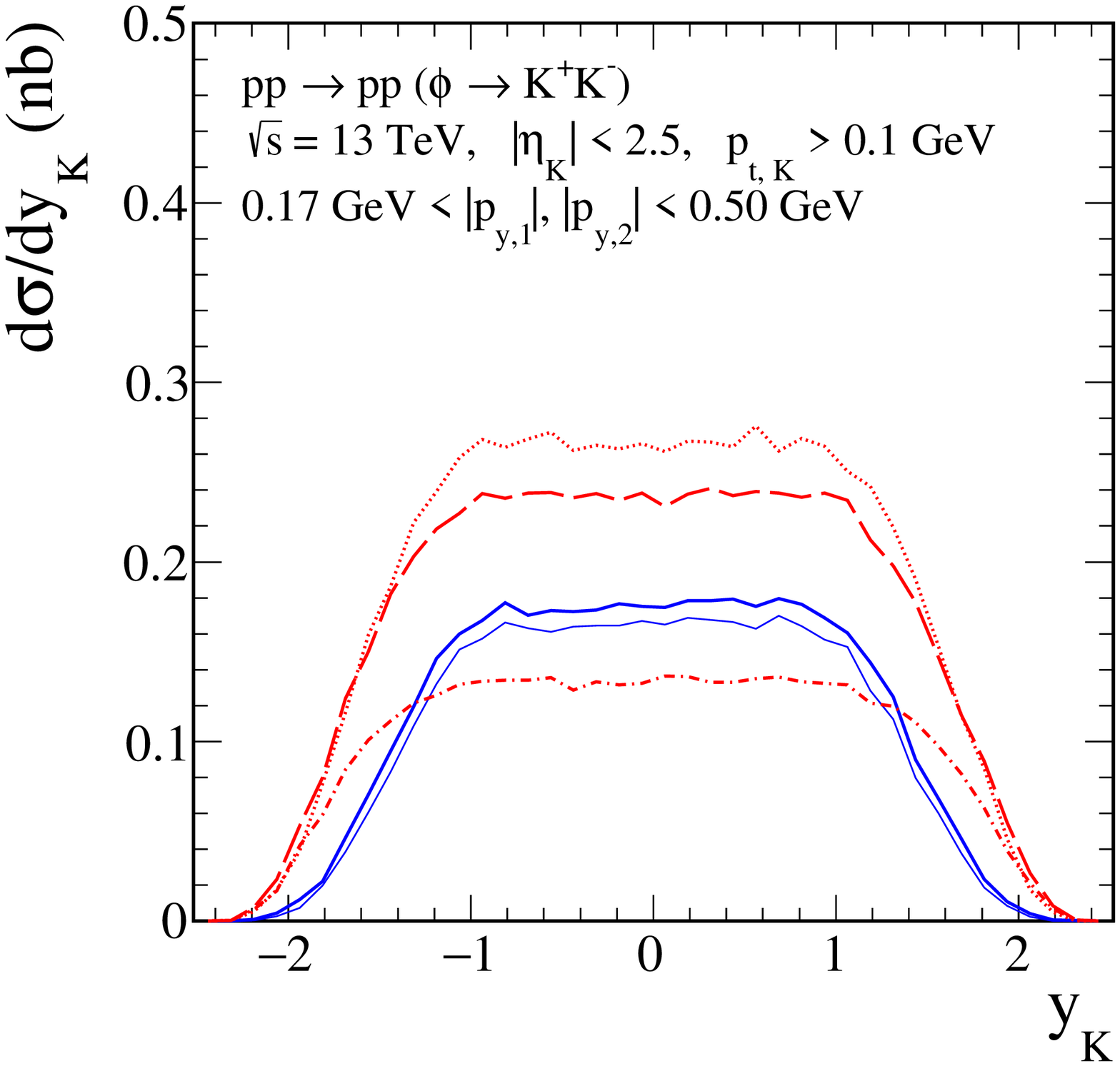}
\includegraphics[width=0.45\textwidth]{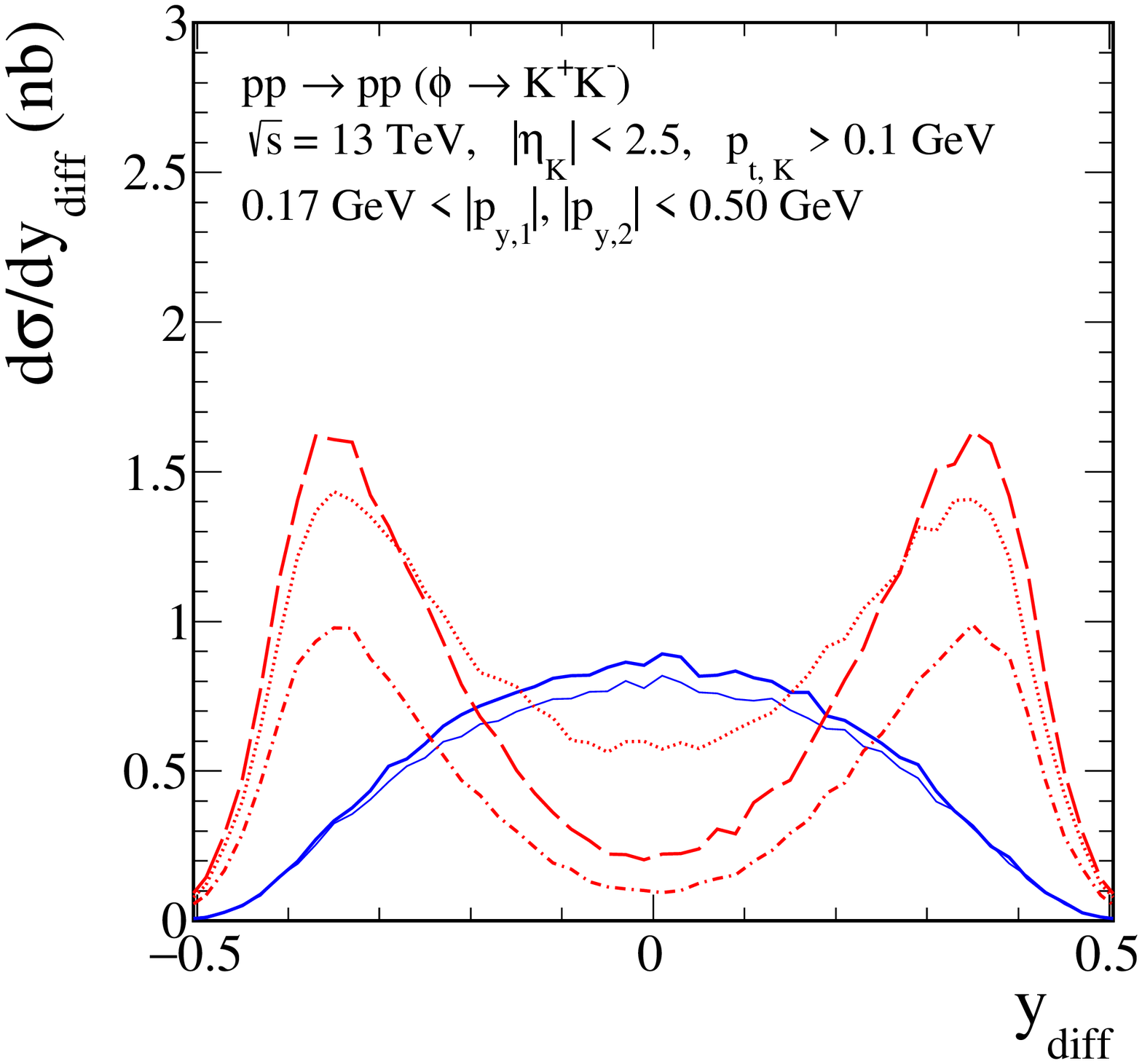}
\caption{\label{fig:ATLASALFA_unc}
Results for the ATLAS-ALFA experiment at $\sqrt{s} = 13$~TeV.
The lower blue solid line represents
the result for the parameter set~A of photoproduction (\ref{photoproduction_setA}) and the upper line is for
set~B (\ref{photoproduction_setB}).
The red long-dashed line represents the odderon-pomeron fusion
with the parameters quoted in (\ref{parameters_ode}), (\ref{parameters_ode_lambda}), and 
the $\Pom \Ode \phi$ coupling parameters (\ref{parameters_ode_b}).
the red dash-dotted line is for the choice (\ref{parameters_ode_a})
of the $\Pom \Ode \phi$ coupling parameters, 
and the red dotted line is for (\ref{parameters_ode_c}).
The absorption effects are included here.}
\end{figure}

In Figs.~\ref{fig:ATLASALFA_1}, 
\ref{fig:ATLASALFA_1_ptK0.2} and \ref{fig:ATLASALFA_2} 
we show distributions in several variables
for the ATLAS-ALFA experimental cuts,
$\sqrt{s} = 13$~TeV, $|\eta_{K}| < 2.5$, 
0.17~GeV~$< |p_{y,1}|, |p_{y,2}|<$~0.50~GeV,
$p_{t, K} > 0.1$~GeV and $p_{t, K} > 0.2$~GeV.
The absorption effects are included in the calculations.
We show results for the $\gamma$-$\Pom$- and $\Ode$-$\Pom$-fusion 
contributions separately (see the blue and red lines, respectively)
and when both terms are added coherently at the amplitude level (the black lines).
We take for the $\gamma$-$\Pom$- and $\Ode$-$\Pom$-fusion contributions
the coupling parameters (\ref{photoproduction_setB}) and (\ref{parameters_ode_b}), respectively.
In Fig.~\ref{fig:ATLASALFA_1_ptK0.2_setBtestA} we show
the results for (\ref{parameters_ode_a})
$a_{\Pom \Ode \phi}= -0.8 \; {\rm GeV}^{-3}$
and $b_{\Pom \Ode \phi}= 1.0 \; {\rm GeV}^{-1}$
[instead of $b_{\Pom \Ode \phi}= 1.6 \; {\rm GeV}^{-1}$
(\ref{parameters_ode_b})].
We can see that the complete result 
indicates a large interference effect of $\gamma$-$\Pom$-
and $\Ode$-$\Pom$-fusion terms.
The odderon-pomeron contribution dominates clearly at larger
$|\rm{y_{diff}}|$,
$p_{t, K^{+}K^{-}}$, the transverse momentum of the $K^{+}K^{-}$ pair,
and $\cos\theta_{K^{+},\,{\rm CS}} = \pm 1$,
compared to the photon-pomeron contribution.
We encourage the experimentalists associated to the ATLAS-ALFA experiment 
to prepare such distributions, especially $d\sigma/d\rm{y_{diff}}$,
$d\sigma/d\cos\theta_{K^{+},\,{\rm CS}}$,
and $d\sigma/d \phi_{K^{+},\,{\rm CS}}$.
Observation of the pattern of maxima and minima 
would be interesting by itself as it is due to interference effects.
Note, in particular,
the different pattern of $\phi_{K^{+},\,{\rm CS}}$
distributions in Figs.~\ref{fig:ATLASALFA_2} and \ref{fig:ATLASALFA_1_ptK0.2_setBtestA}.
Within the same kinematic cuts
we can observe for $\phi_{K^{+},\,{\rm CS}} = 0, \pi, 2\pi$
destructive interference for (\ref{parameters_ode_b})
and constructive interference for (\ref{parameters_ode_a}).
The same is clearly seen also for $\cos\theta_{K^{+},\,{\rm CS}} = 0$.

It is worth adding that much smaller interference effects
are predicted when no cuts on the outgoing protons are required; 
see the results in Table~\ref{tab:table2}
and Figs.~\ref{fig:LHCb_KK_1}, \ref{fig:LHCb_KK_2} below.
When cuts on transverse momenta of the outgoing protons are imposed
then the $\gamma$-$\Pom$- and $\Ode$-$\Pom$-fusion contributions
become comparable and large interference effects are in principle possible.

We have checked numerically that for $\alpha_{\Ode}(0) = 1.0$,
instead of $\alpha_{\Ode}(0) = 1.05$ [see (\ref{A14_aux})],
we get a bit smaller cross section for the $\Ode$-$\Pom$-fusion contribution
but the shape of the differential distributions 
(e.g., $d \sigma/d \phi_{pp}$, $d \sigma/d t_{1,2}$) is not changed.
In our plots for the LHC energies we have taken mainly the odderon
coupling parameters from (\ref{parameters_ode_b}).
This is to be understood as an example.
For the parameters from (\ref{parameters_ode_a}) the odderon effects
at the LHC are typically smaller than those from (\ref{parameters_ode_b})
by a factor of roughly 2; \mbox{see Figs.~\ref{fig:ATLASALFA_unc},
\ref{fig:ATLASALFA_1_ptK0.2},
\ref{fig:ATLASALFA_1_ptK0.2_setBtestA}}.
Figures~\ref{fig:ATLASALFA_1_ptK0.2} and \ref{fig:ATLASALFA_1_ptK0.2_setBtestA} 
show distinct interference effects between
the $\gamma$-$\Pom$- and \mbox{$\Ode$-$\Pom$-fusion} contributions
which depend on the choice of the odderon coupling parameters.
In an experimental analysis of single $\phi$ CEP at the LHC
clearly the odderon parameters from (\ref{A14}) and (\ref{A15})
should be considered as fit parameters to be determined
from the comparison of our theoretical results with the data.

\begin{figure}[!ht]
\includegraphics[width=0.42\textwidth]{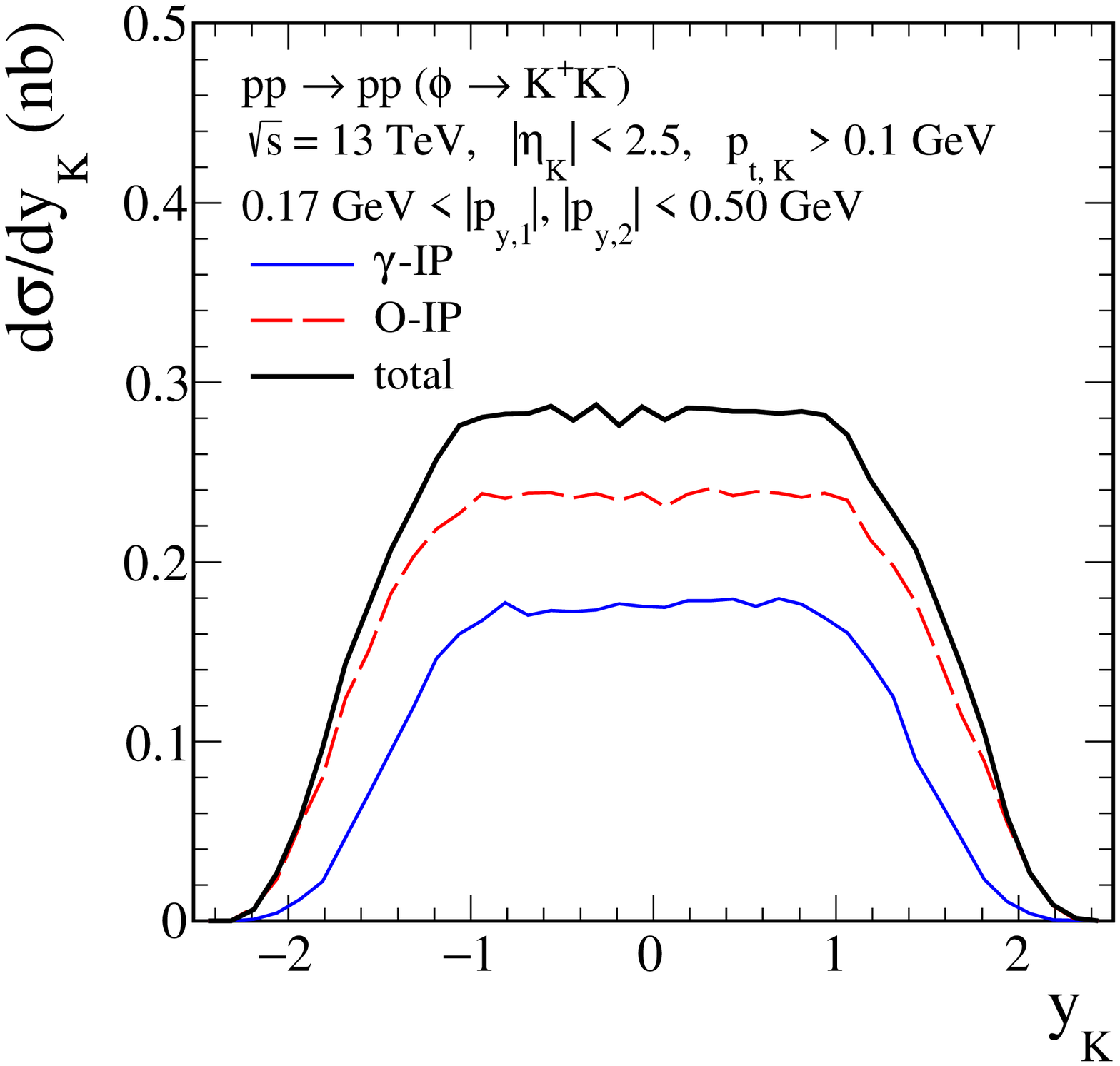}
\includegraphics[width=0.42\textwidth]{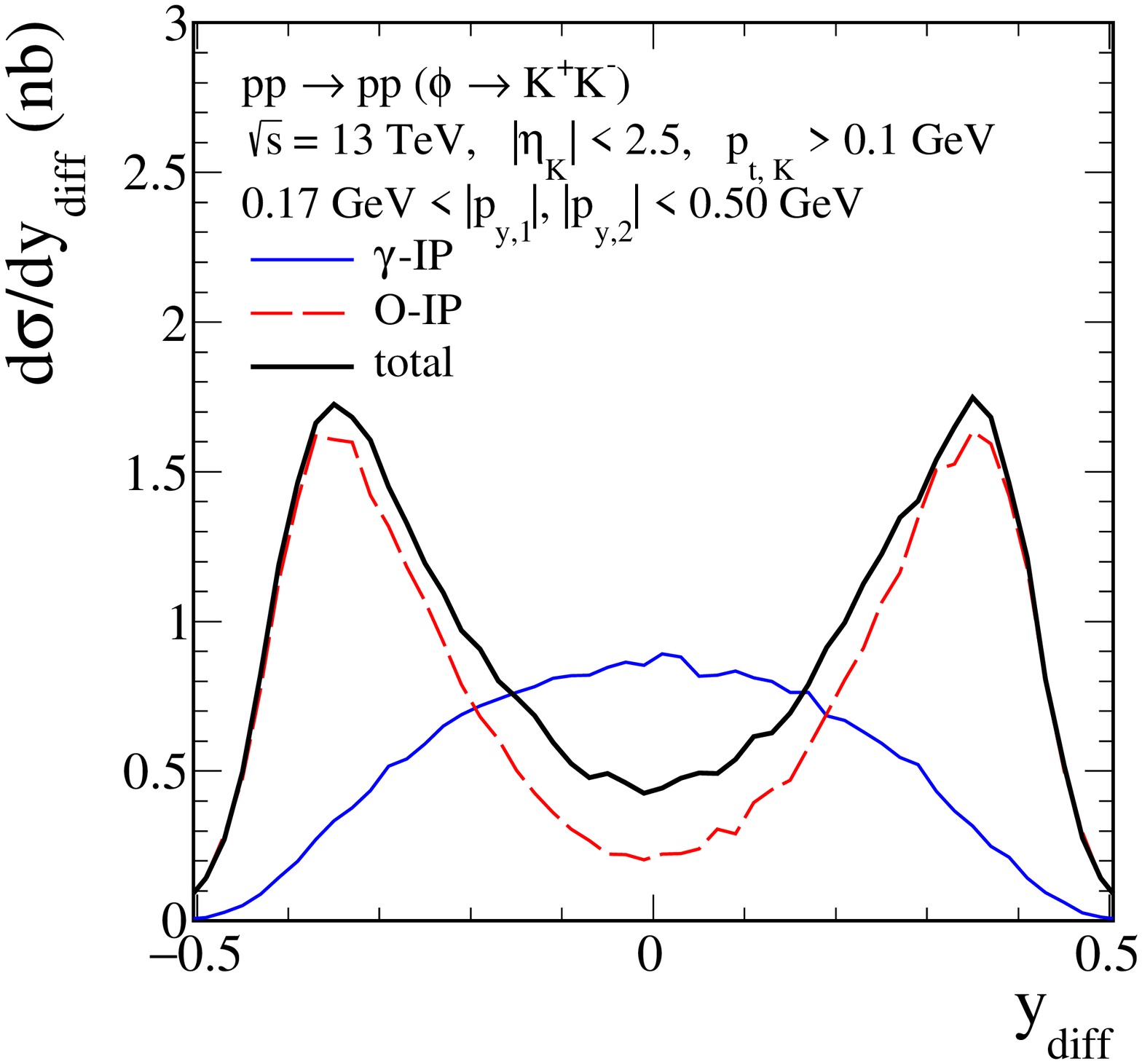}
\includegraphics[width=0.42\textwidth]{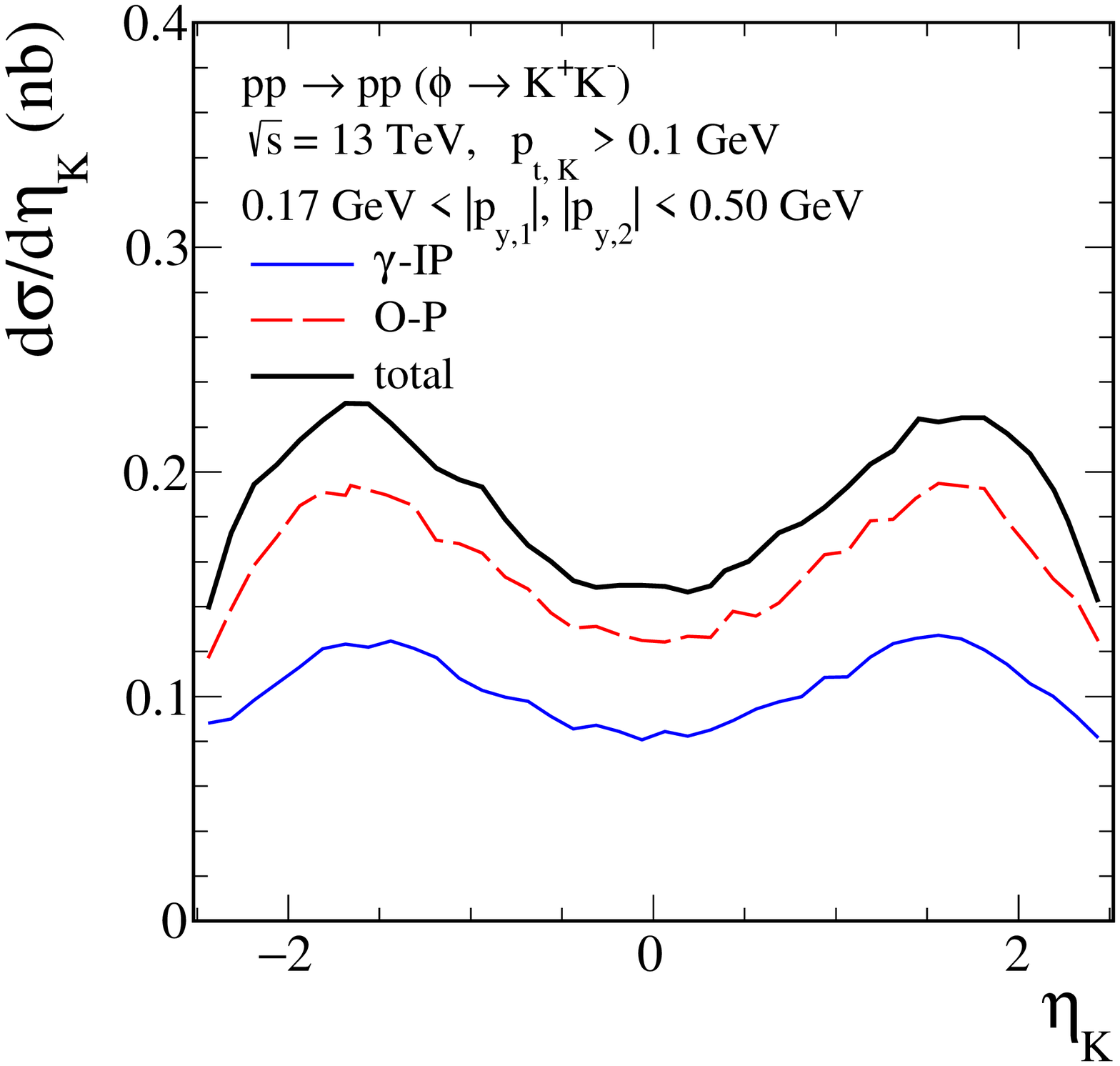}
\includegraphics[width=0.42\textwidth]{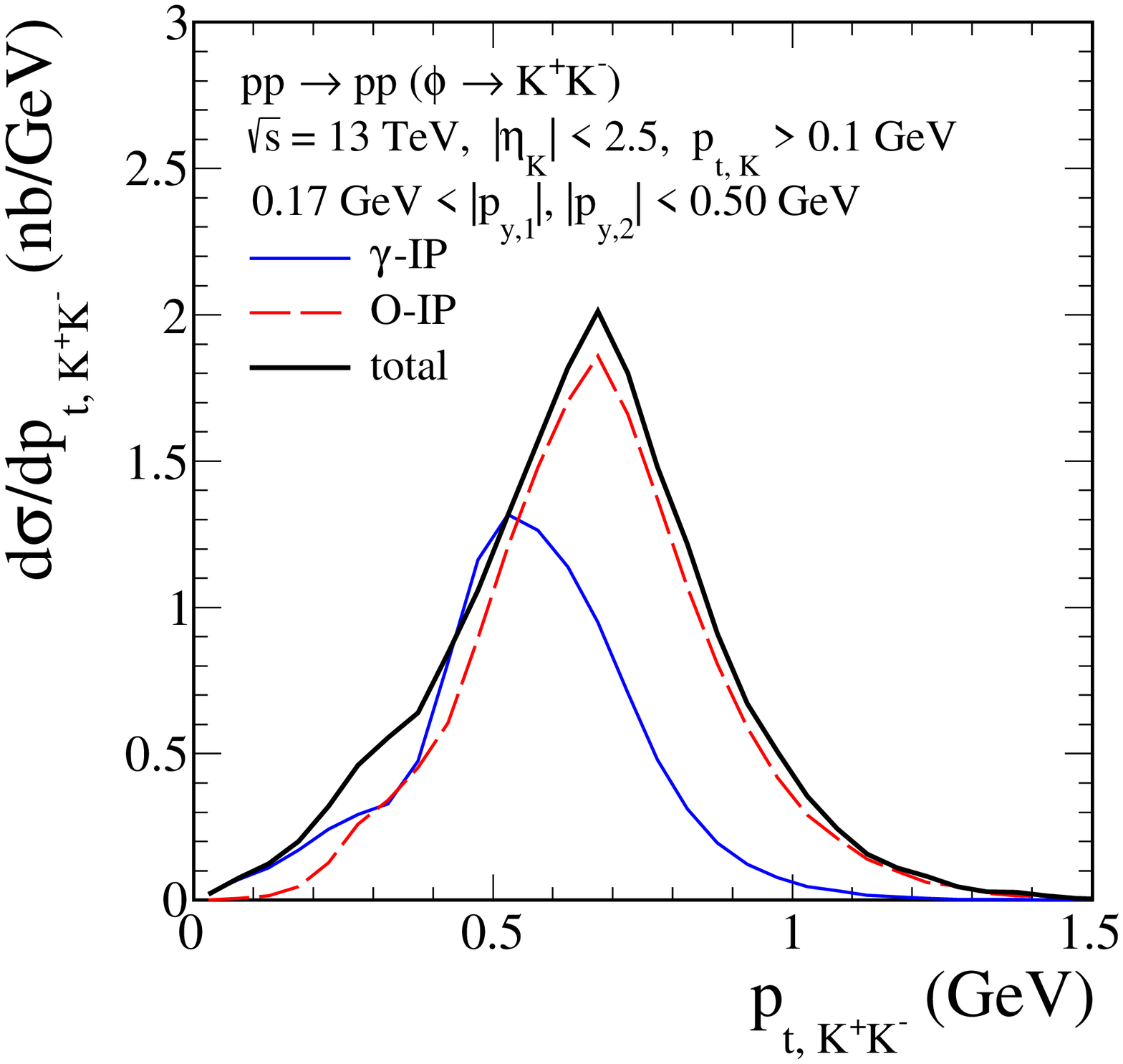}
\includegraphics[width=0.42\textwidth]{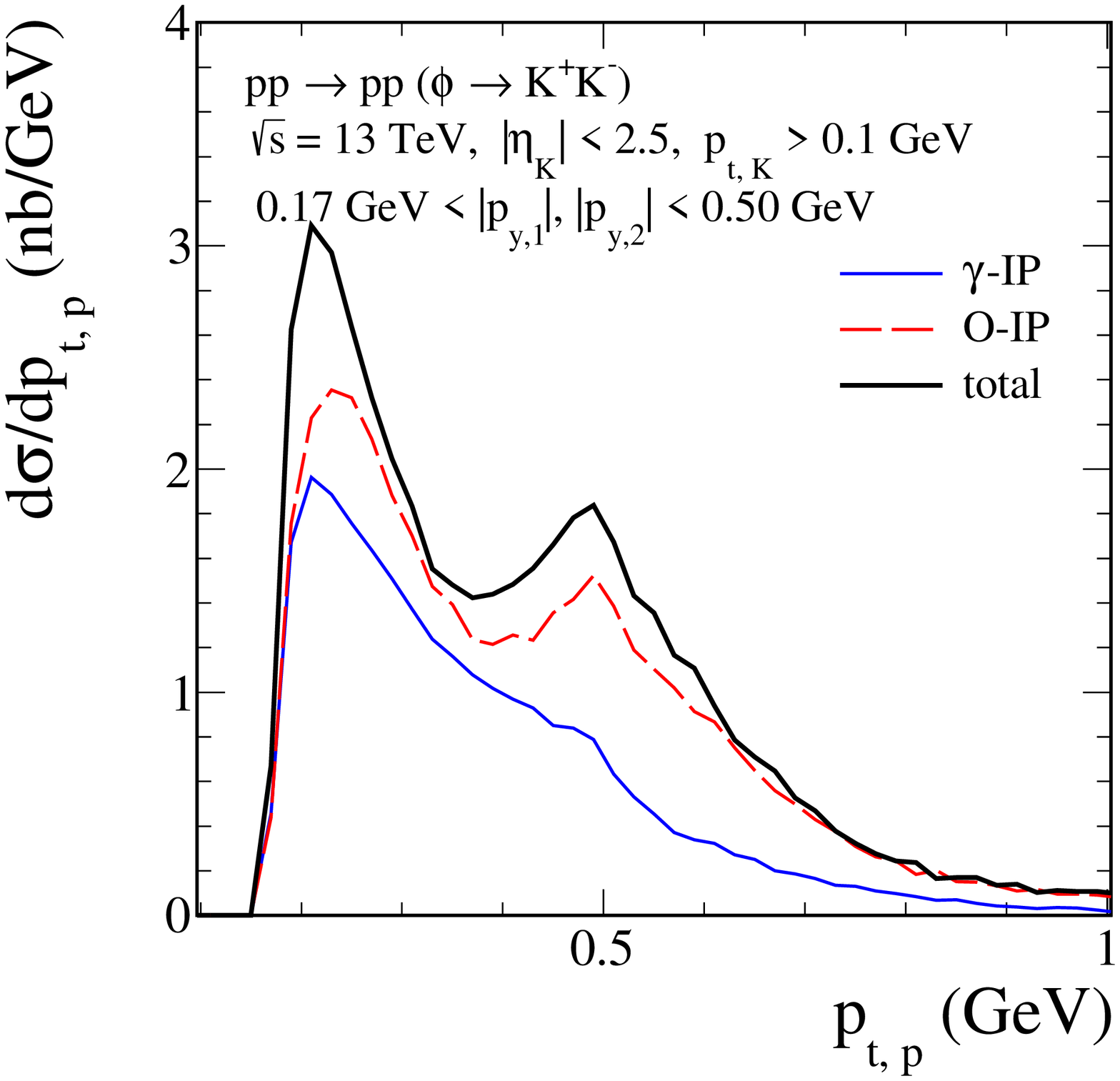}
\includegraphics[width=0.42\textwidth]{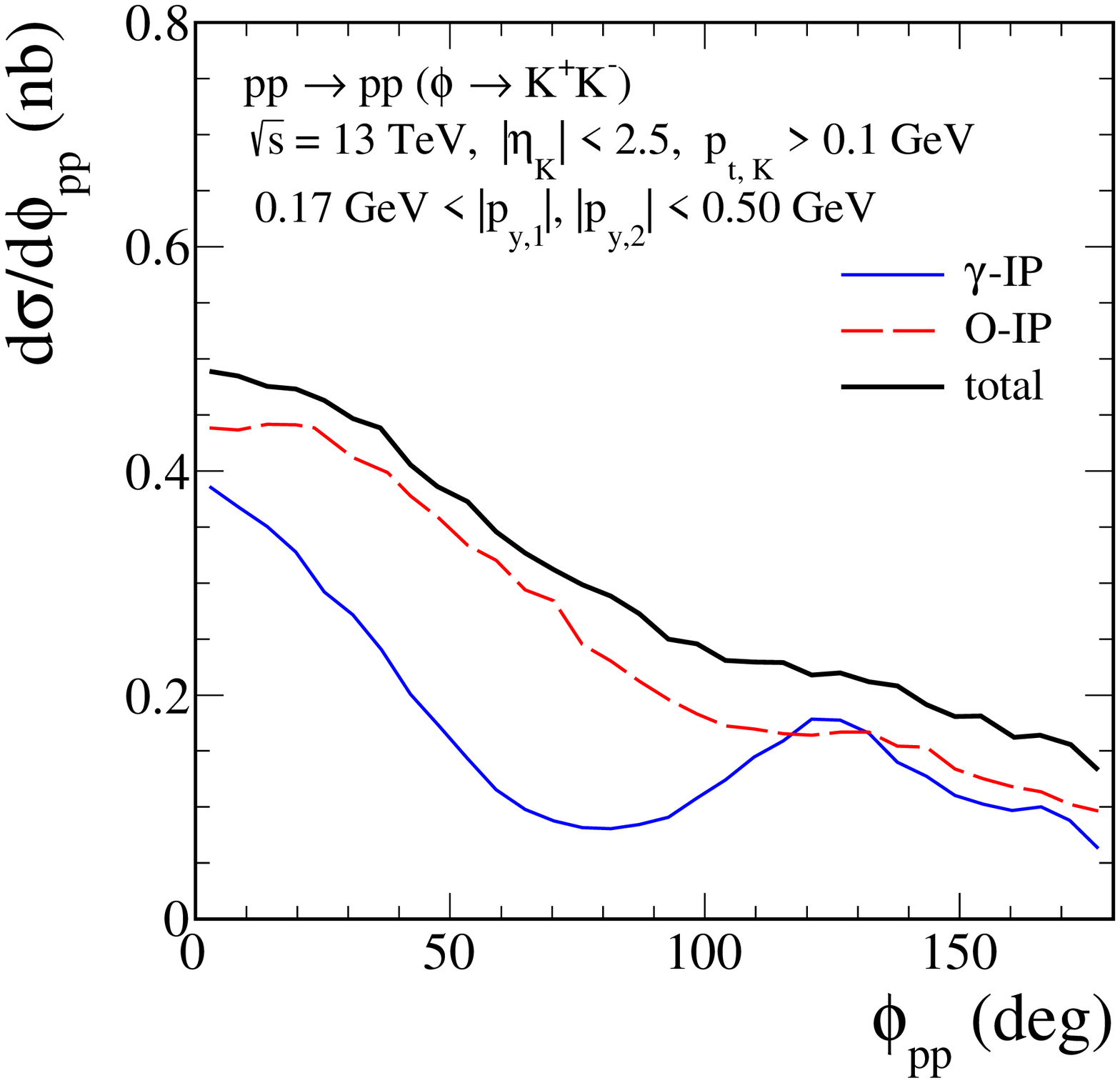}
\caption{\label{fig:ATLASALFA_1}
Selected predictions for the ATLAS-ALFA experiment at $\sqrt{s} = 13$~TeV.
The absorption effects are included here.
The blue solid line represents 
the result for the photoproduction mechanism for set~B (\ref{photoproduction_setB}) while
the red long-dashed line represents the odderon-pomeron fusion
with the parameters quoted in (\ref{parameters_ode}), (\ref{parameters_ode_lambda}), and 
the $\Pom \Ode \phi$ coupling parameters (\ref{parameters_ode_b}).
The coherent sum of the two fusion processes 
is shown by the black solid line.}
\end{figure}

\begin{figure}[!ht]
\includegraphics[width=0.42\textwidth]{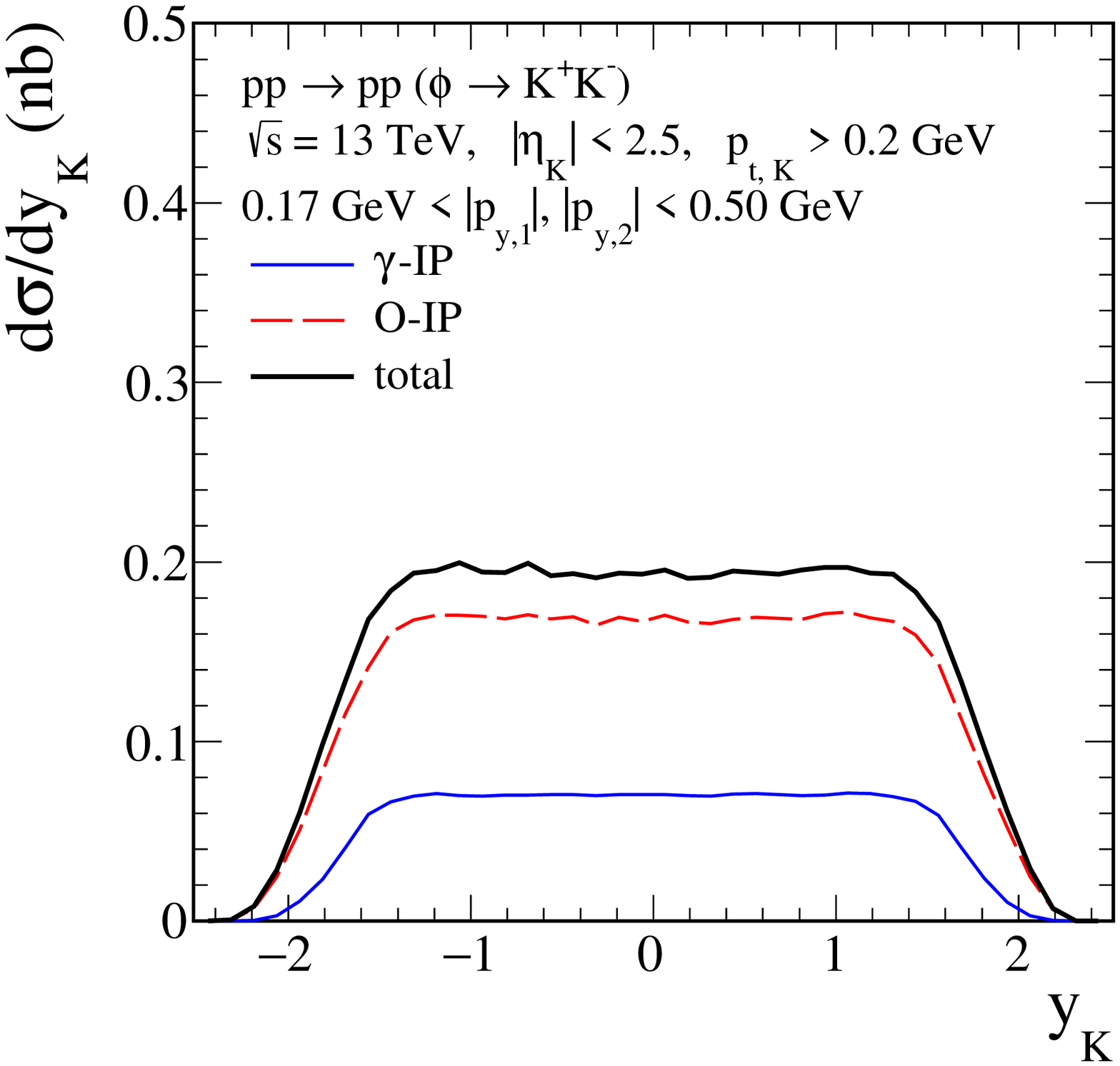}
\includegraphics[width=0.42\textwidth]{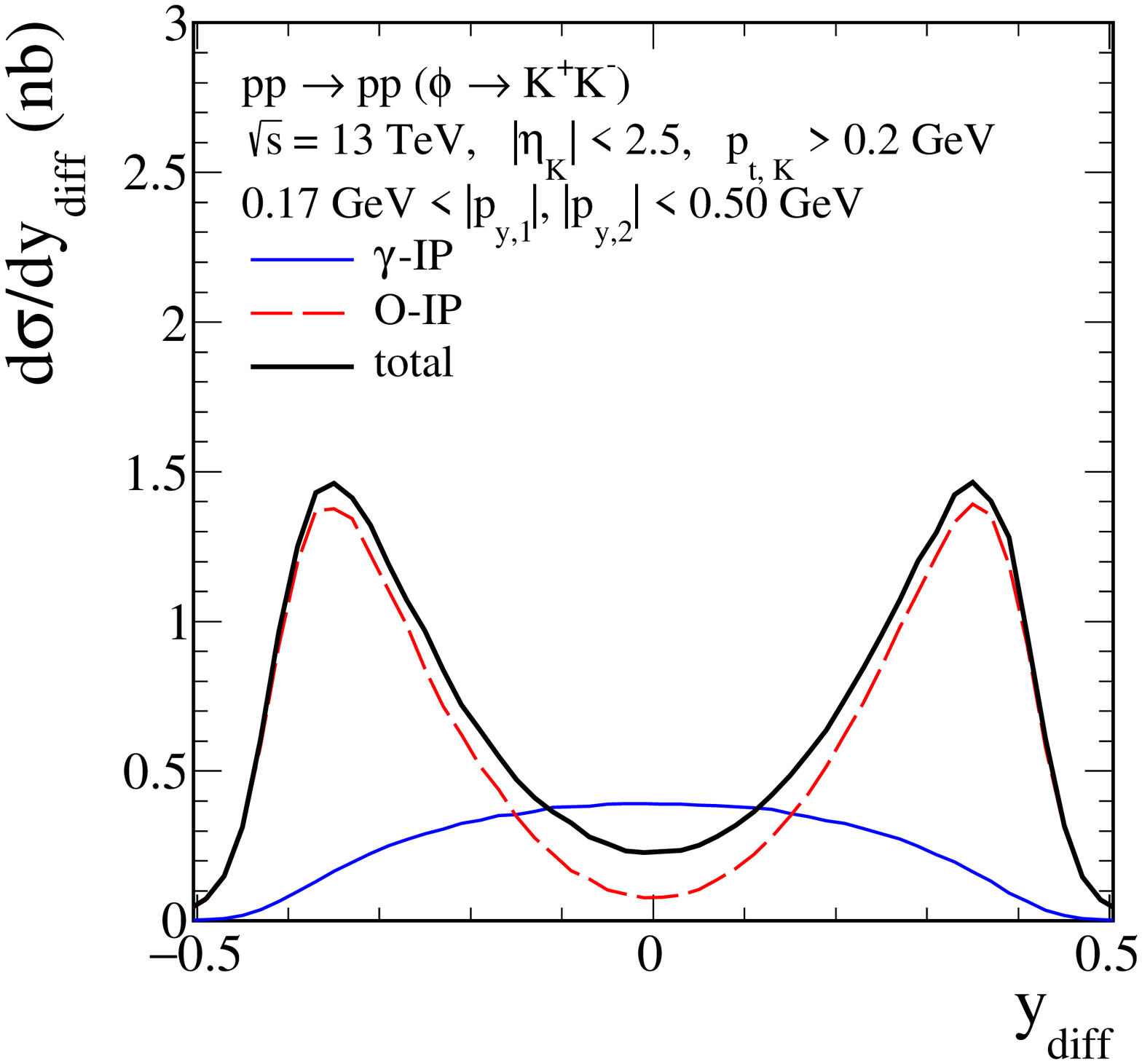}
\includegraphics[width=0.42\textwidth]{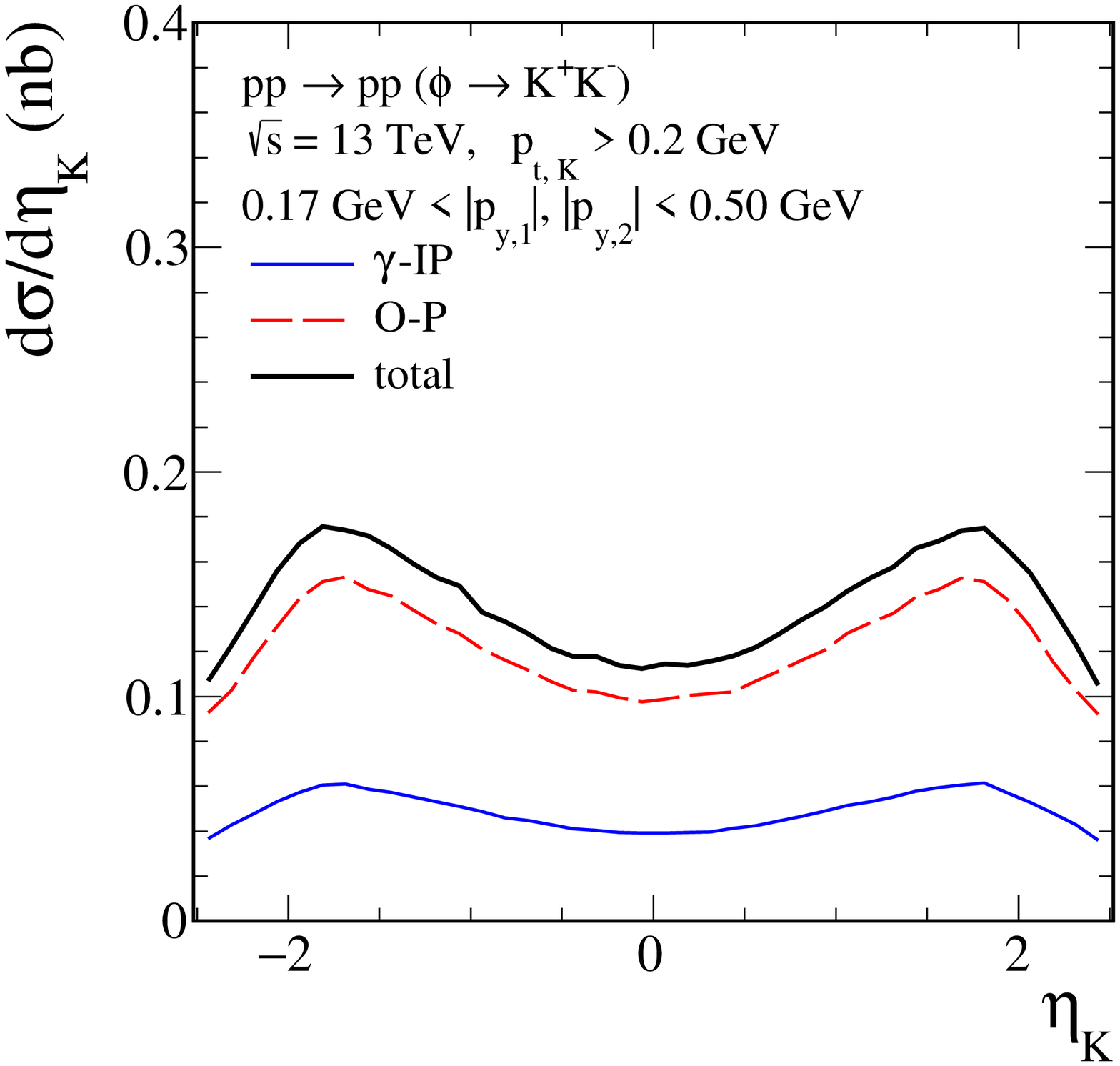}
\includegraphics[width=0.42\textwidth]{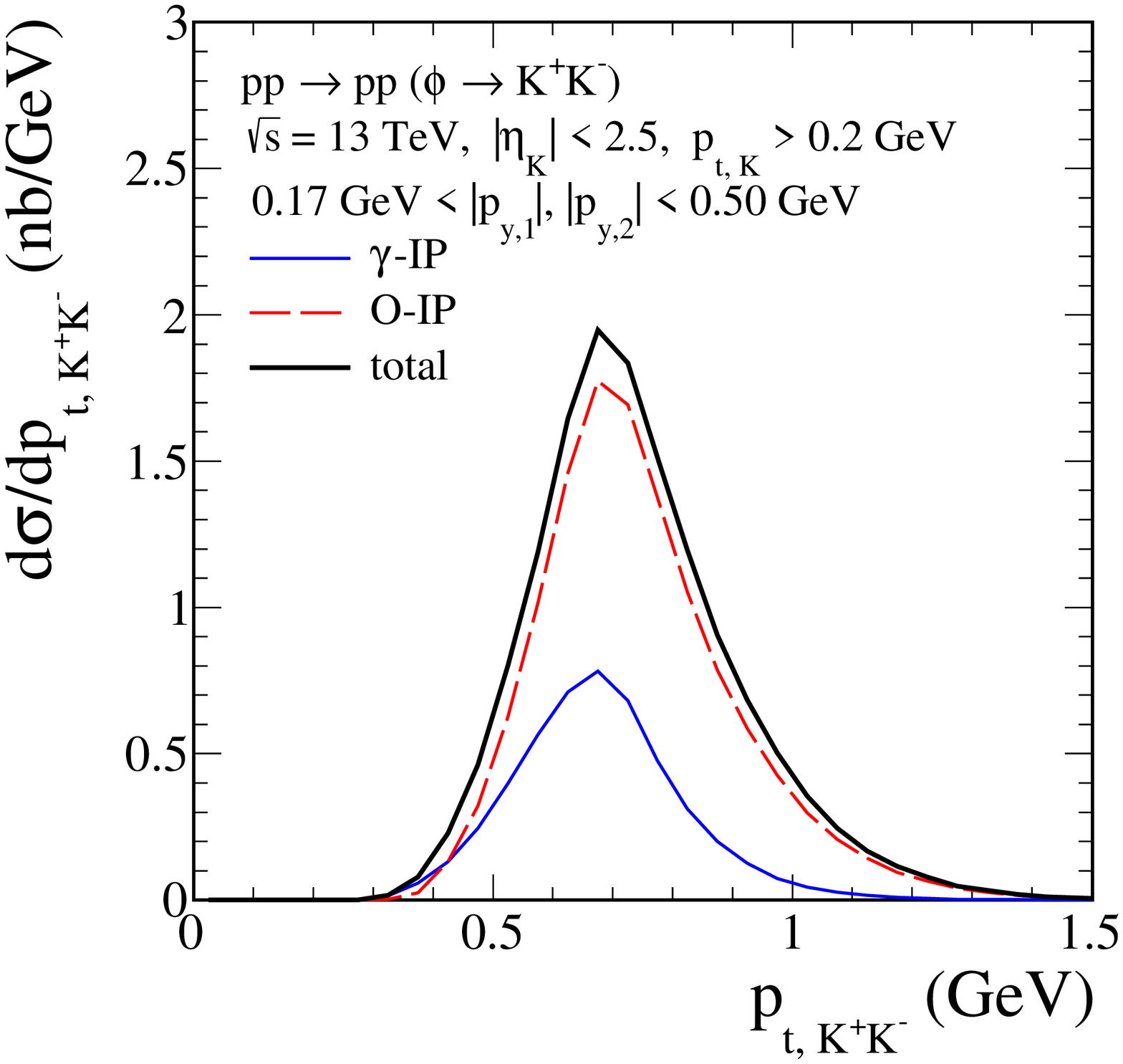}
\includegraphics[width=0.42\textwidth]{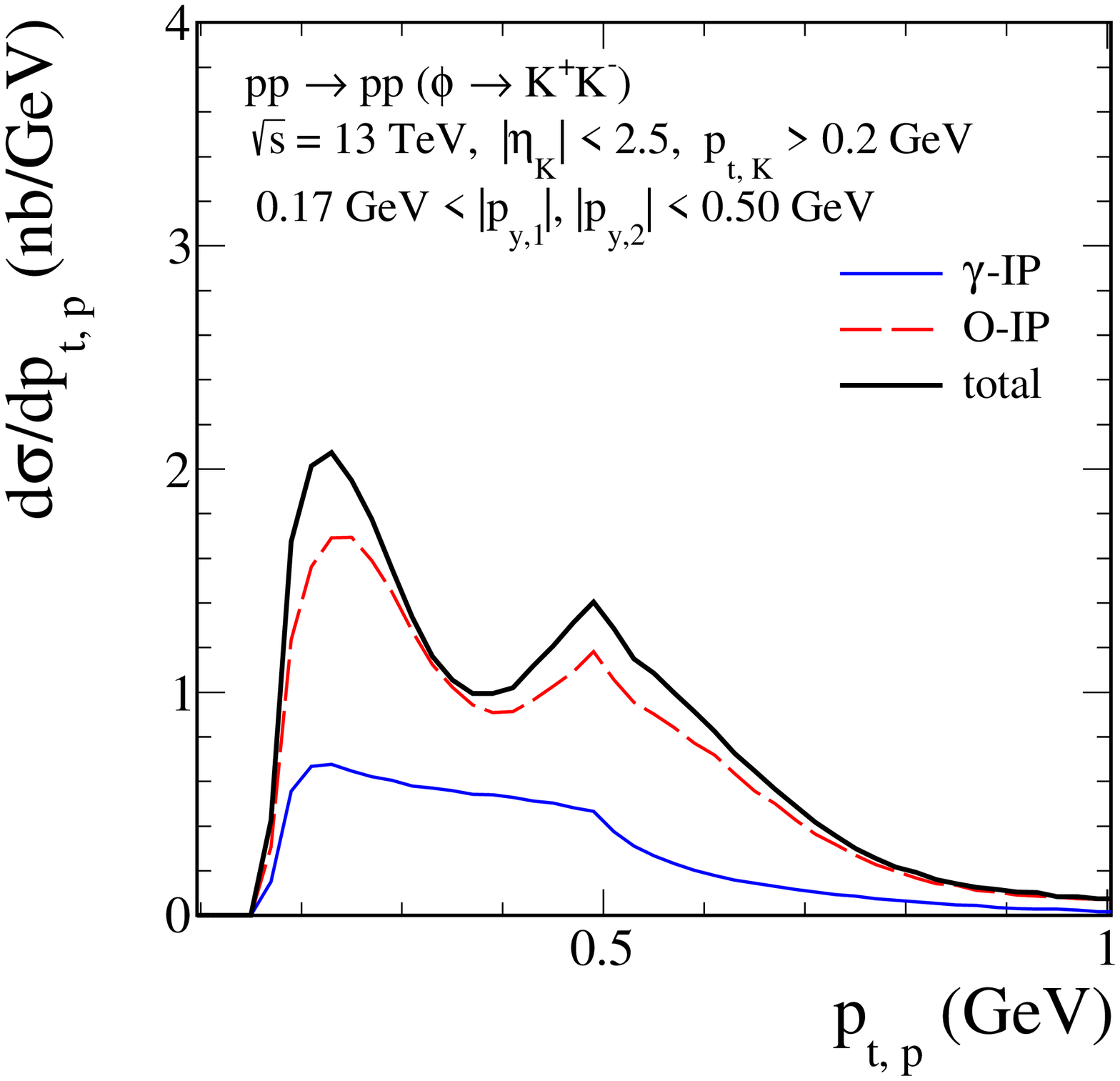}
\includegraphics[width=0.42\textwidth]{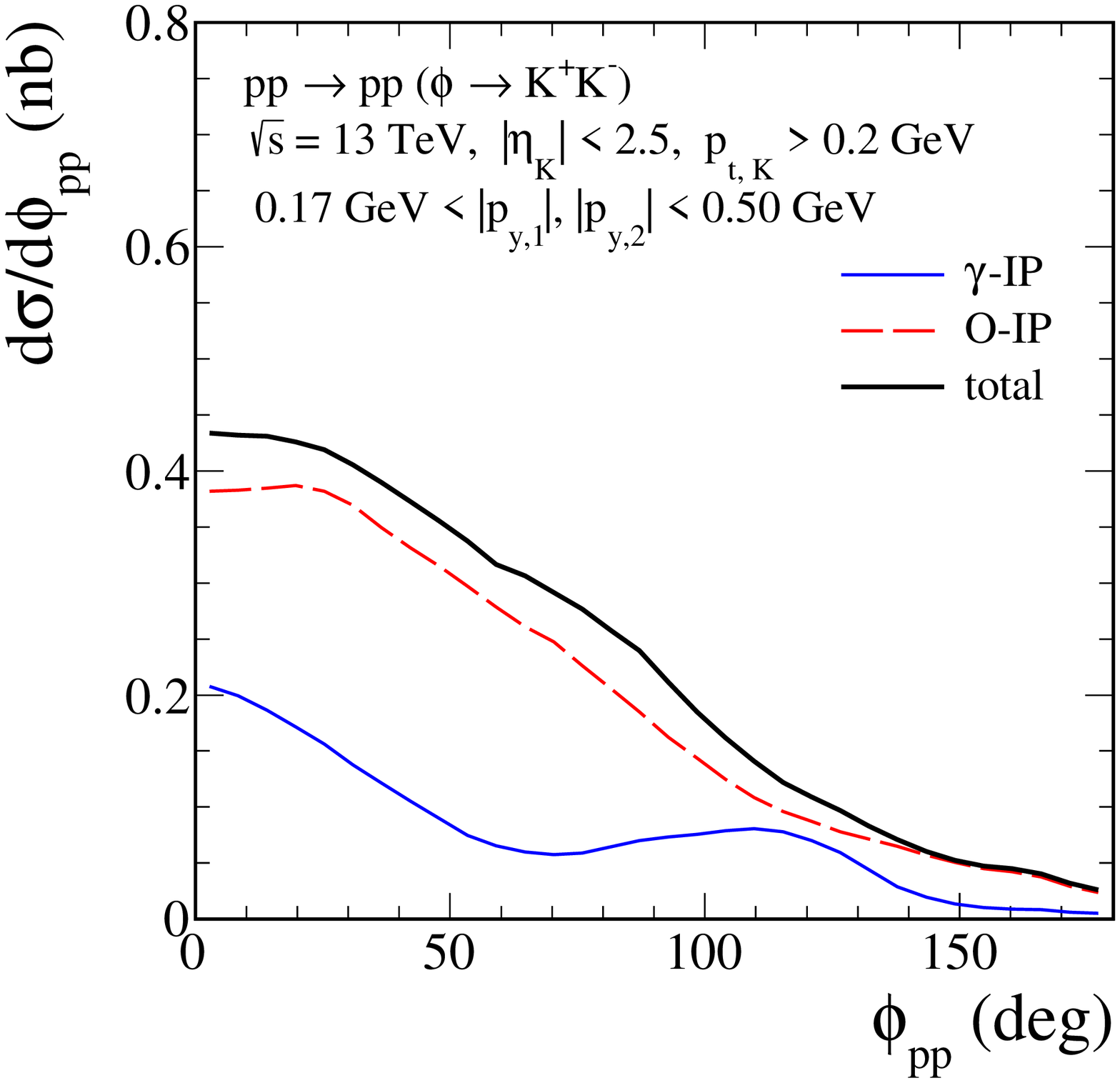}
\caption{\label{fig:ATLASALFA_1_ptK0.2}
The same as in Fig.~\ref{fig:ATLASALFA_1} but for $p_{t,K} > 0.2$~GeV.}
\end{figure}

\begin{figure}[!ht]
\includegraphics[width=0.42\textwidth]{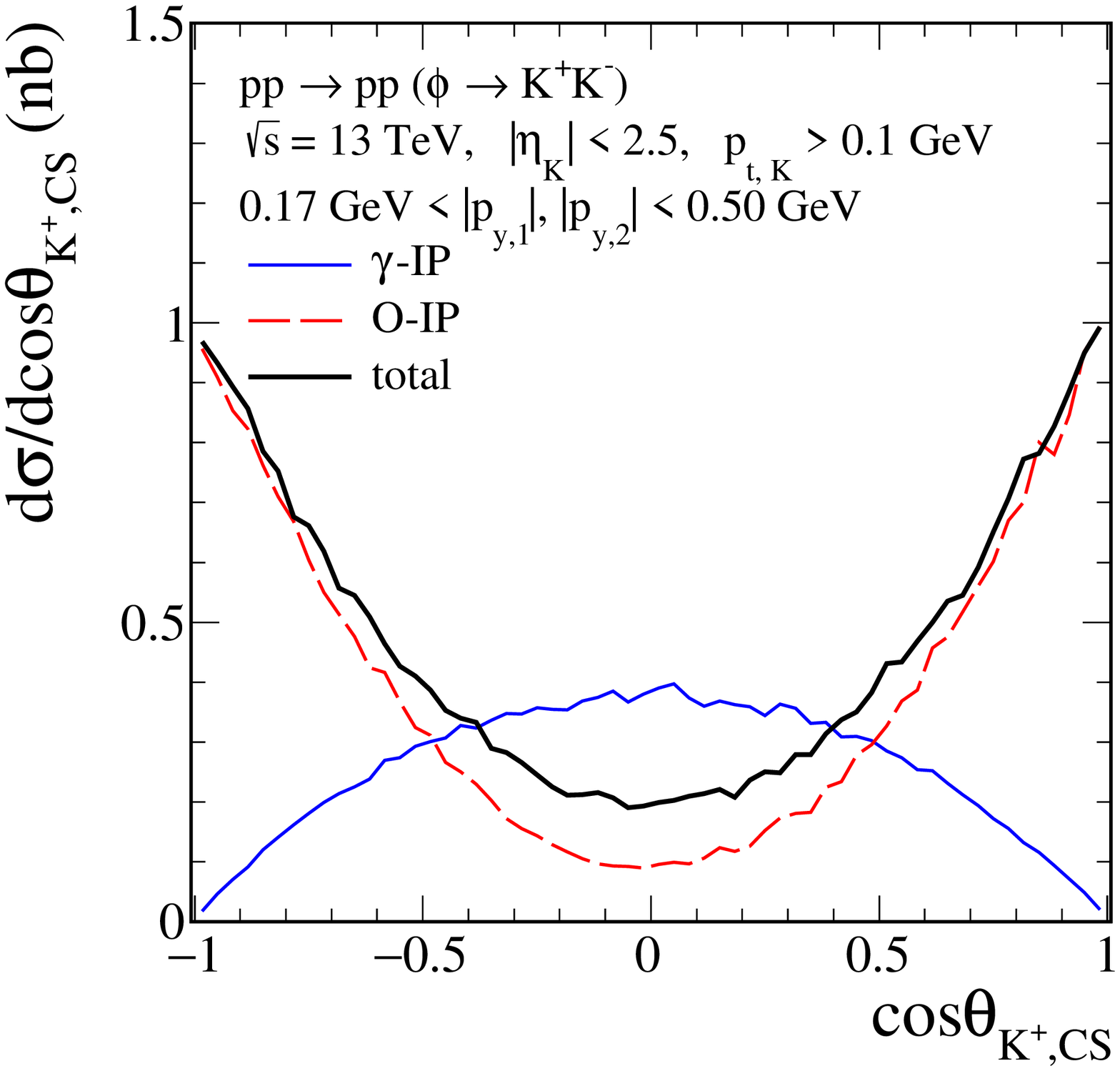}
\includegraphics[width=0.42\textwidth]{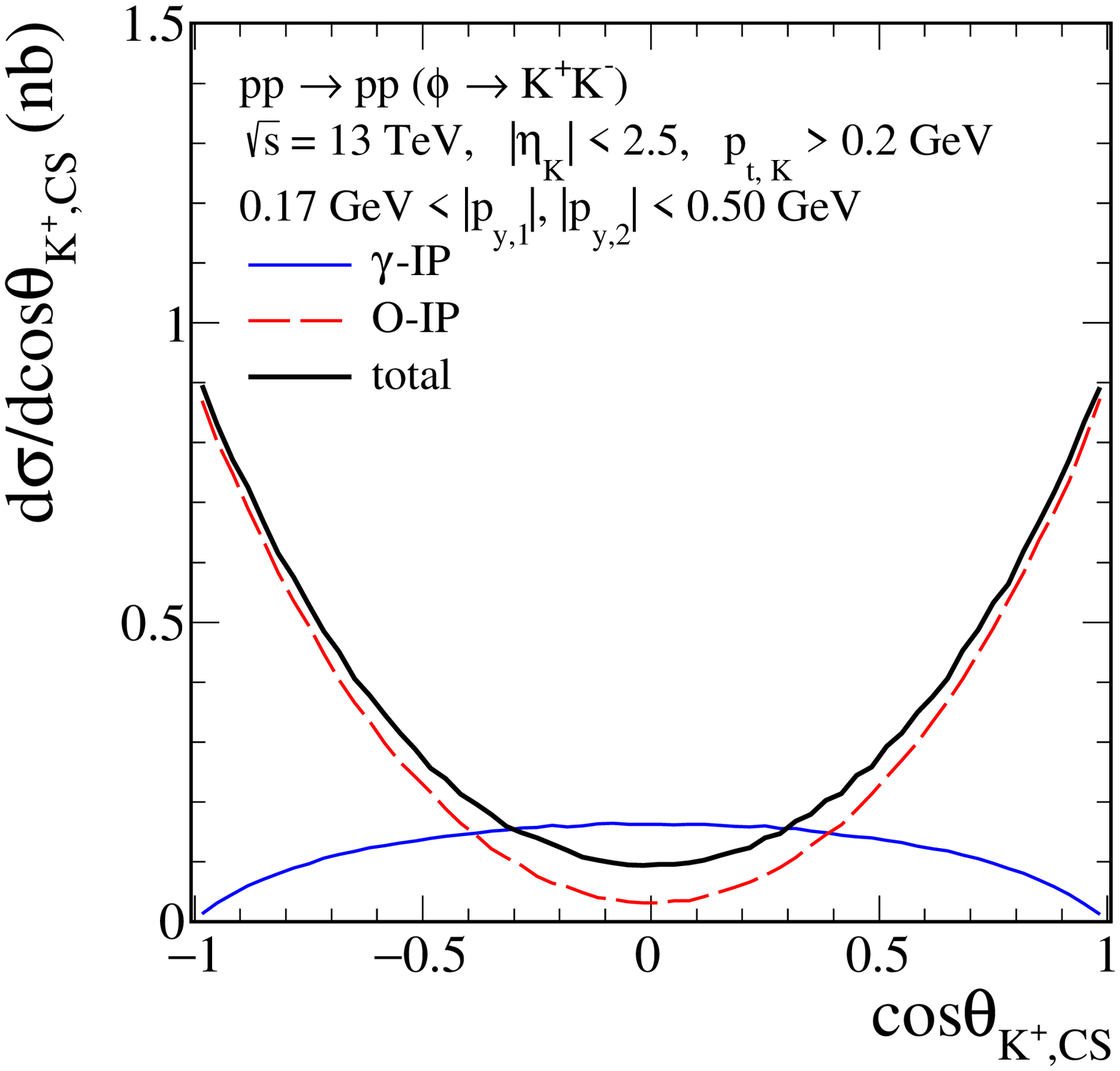}
\includegraphics[width=0.42\textwidth]{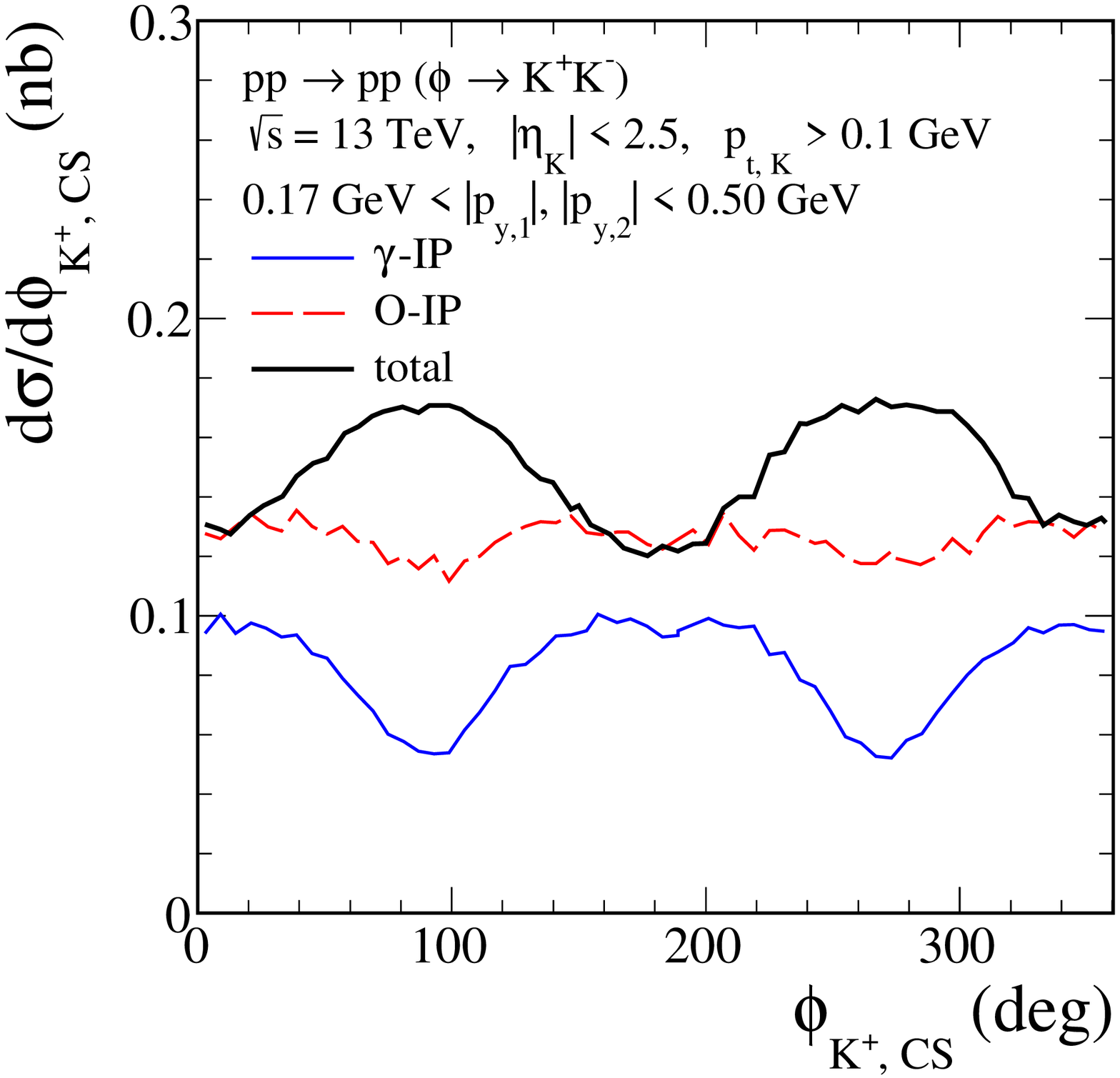}
\includegraphics[width=0.42\textwidth]{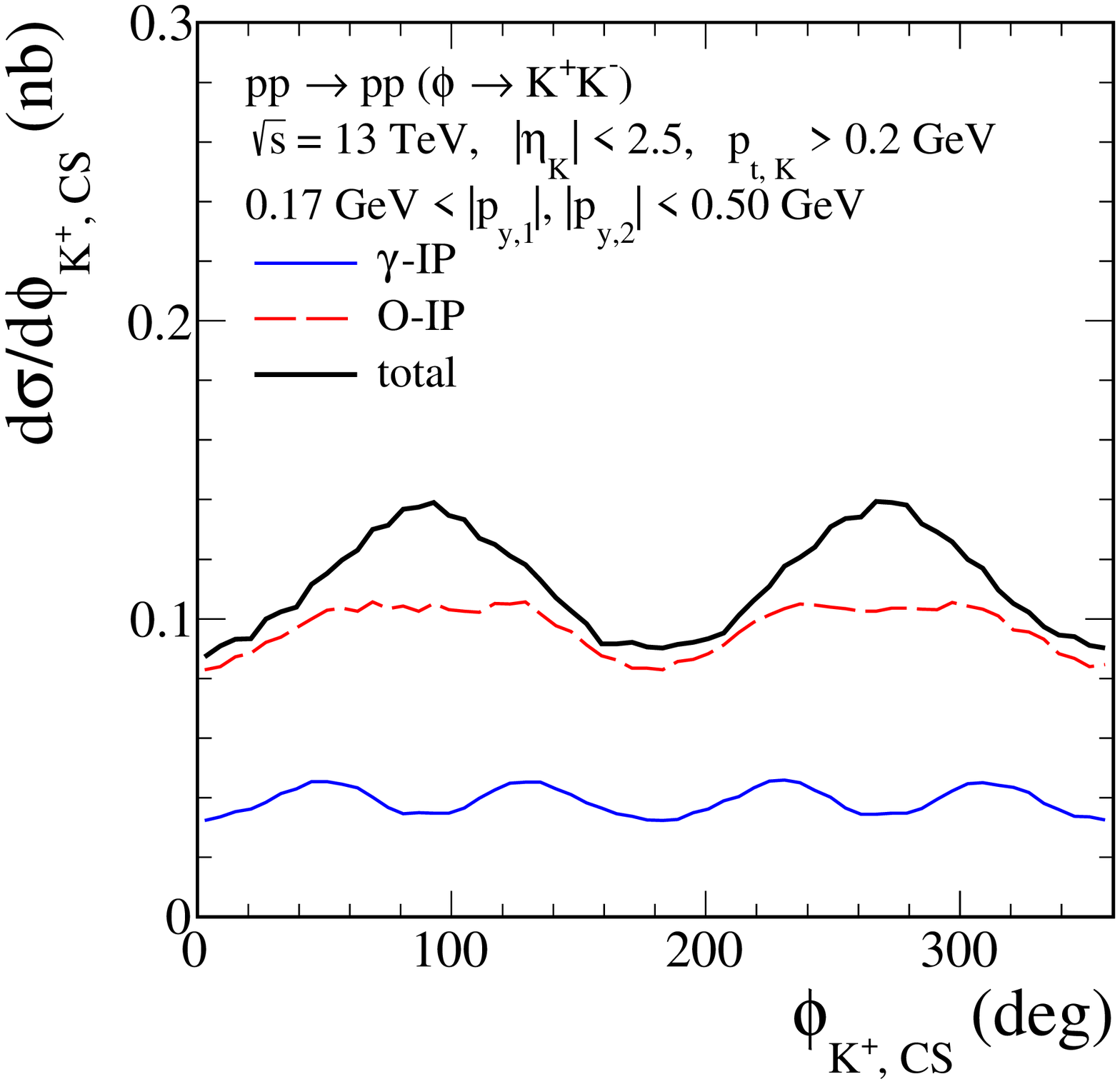}
\caption{\label{fig:ATLASALFA_2}
The distributions in $\cos\theta_{K^{+},\,{\rm CS}}$
(the top panels) and in $\phi_{K^{+},\,{\rm CS}}$ (the bottom panels).
The calculations were performed for $\sqrt{s} = 13$~TeV
and for the ATLAS-ALFA experimental cuts
\mbox{$|\eta_{K}| < 2.5$}, $p_{t, K} > 0.1$~GeV
(left panels)
or $p_{t, K} > 0.2$~GeV (right panels),
and with extra cuts on the leading protons of 0.17~GeV~$< |p_{y,1}|, |p_{y,2}|<$~0.50~GeV.
The meaning of the lines is the same as in Fig.~\ref{fig:ATLASALFA_1}.
The absorption effects are included here.}
\end{figure}

\begin{figure}[!ht]
\includegraphics[width=0.4\textwidth]{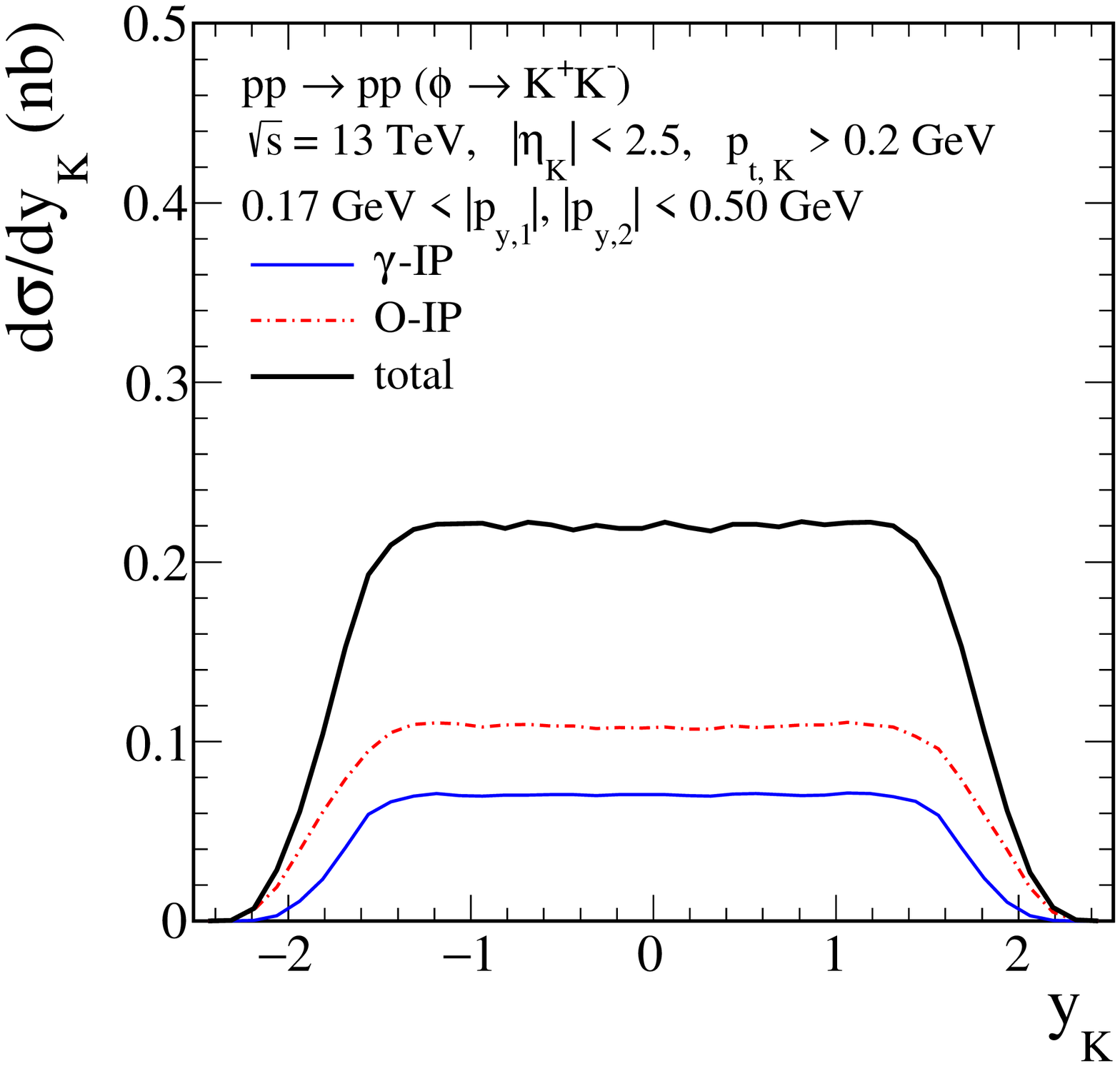}
\includegraphics[width=0.4\textwidth]{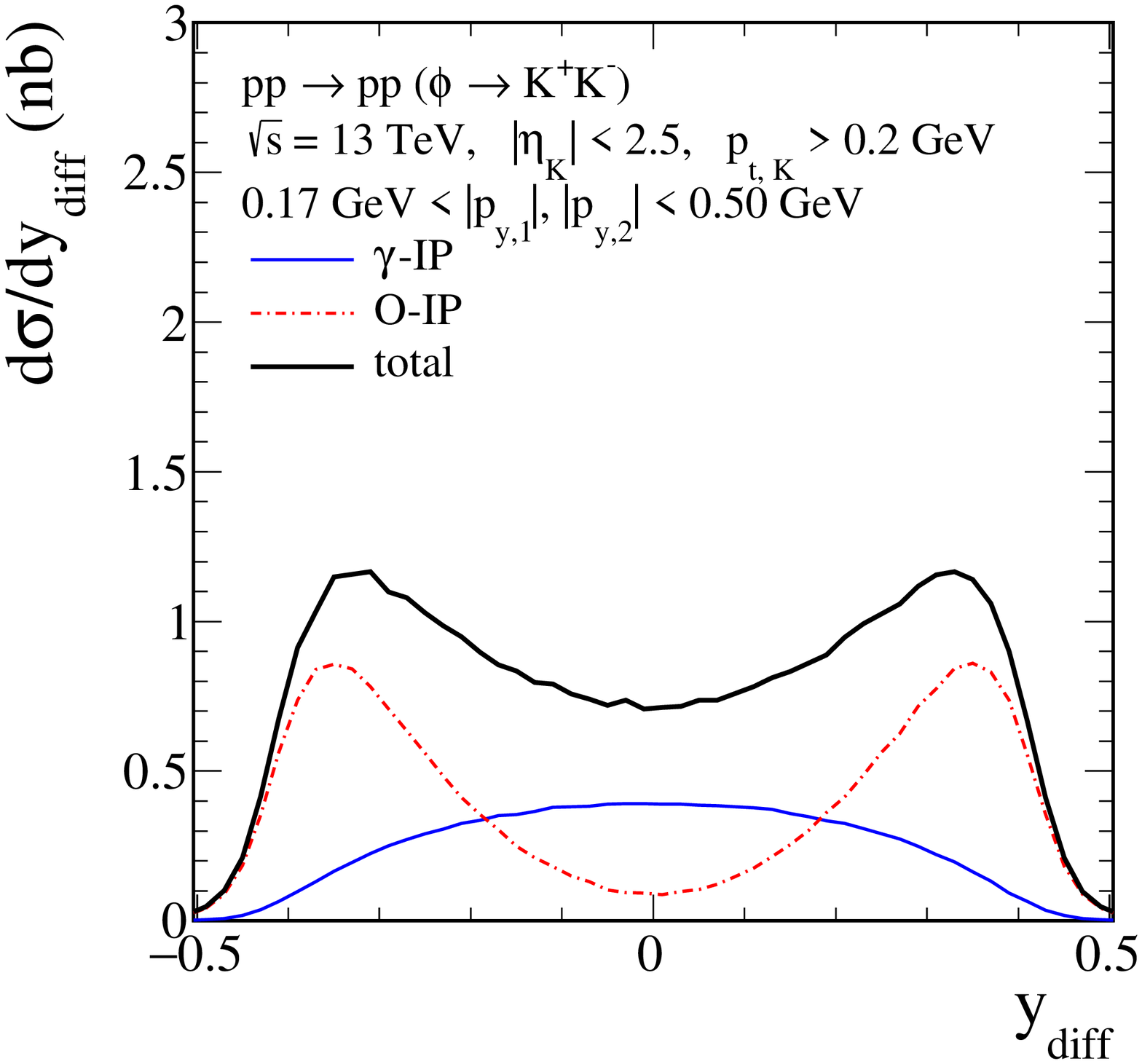}
\includegraphics[width=0.4\textwidth]{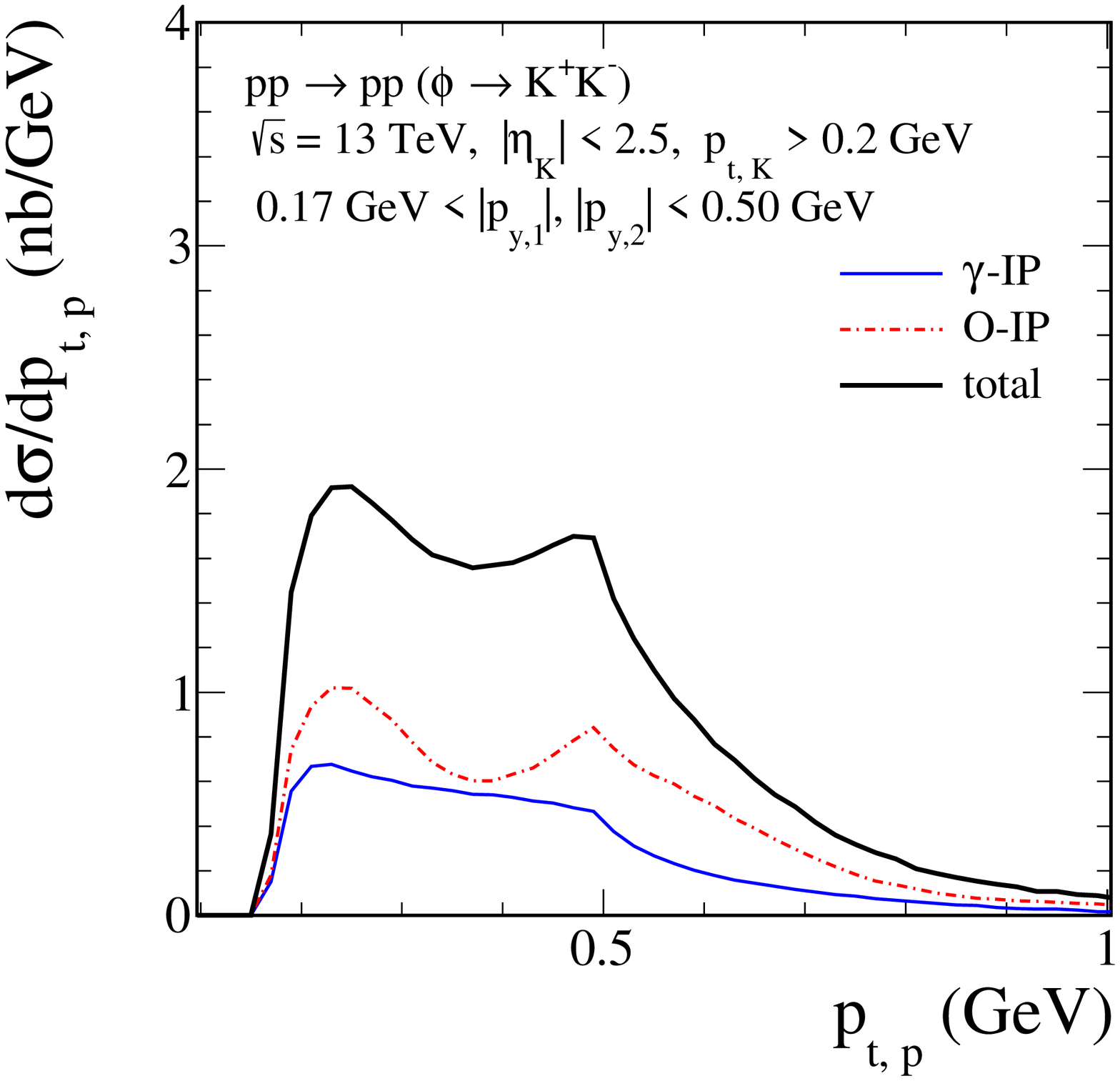}
\includegraphics[width=0.4\textwidth]{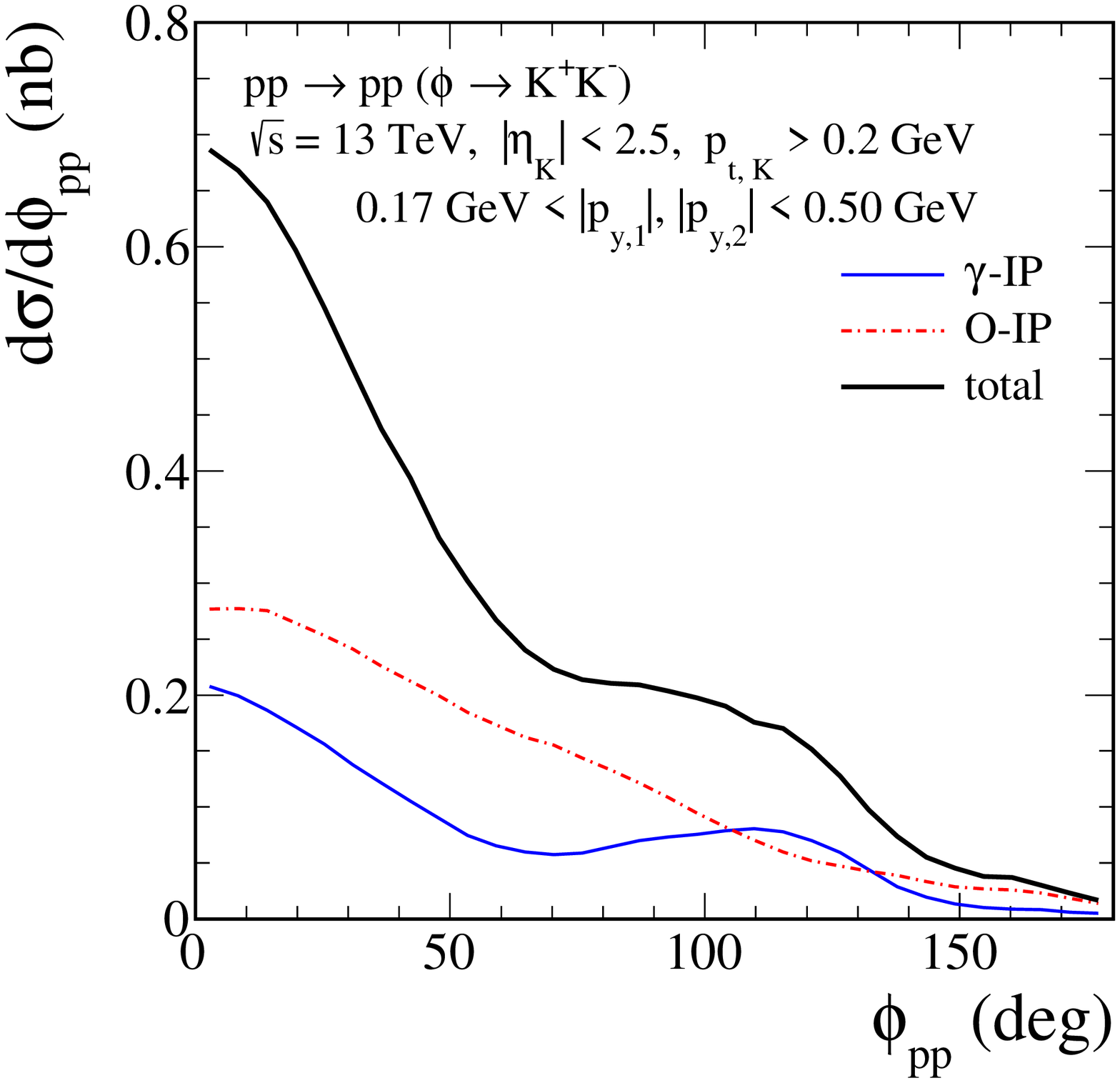}
\includegraphics[width=0.4\textwidth]{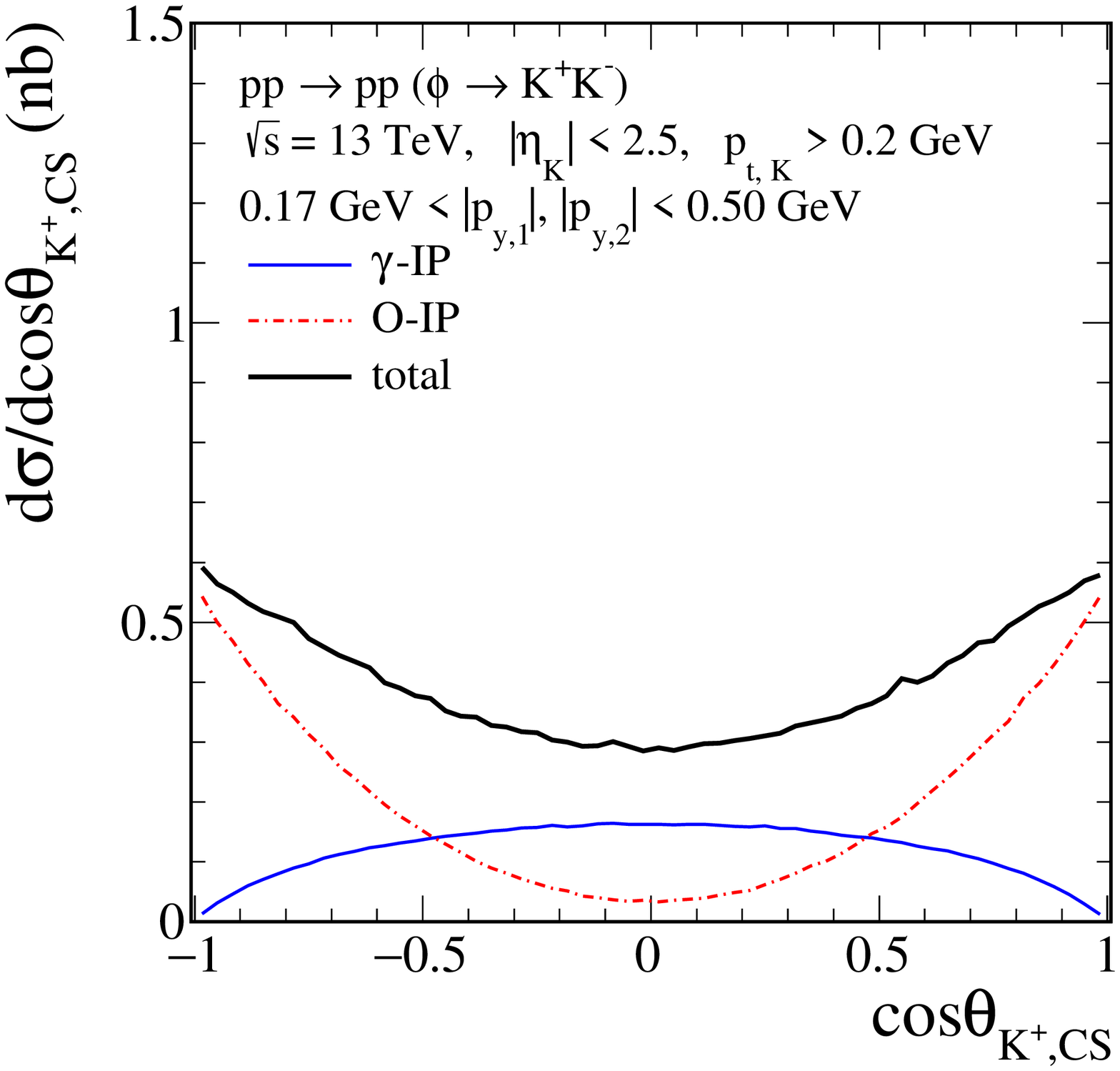}
\includegraphics[width=0.4\textwidth]{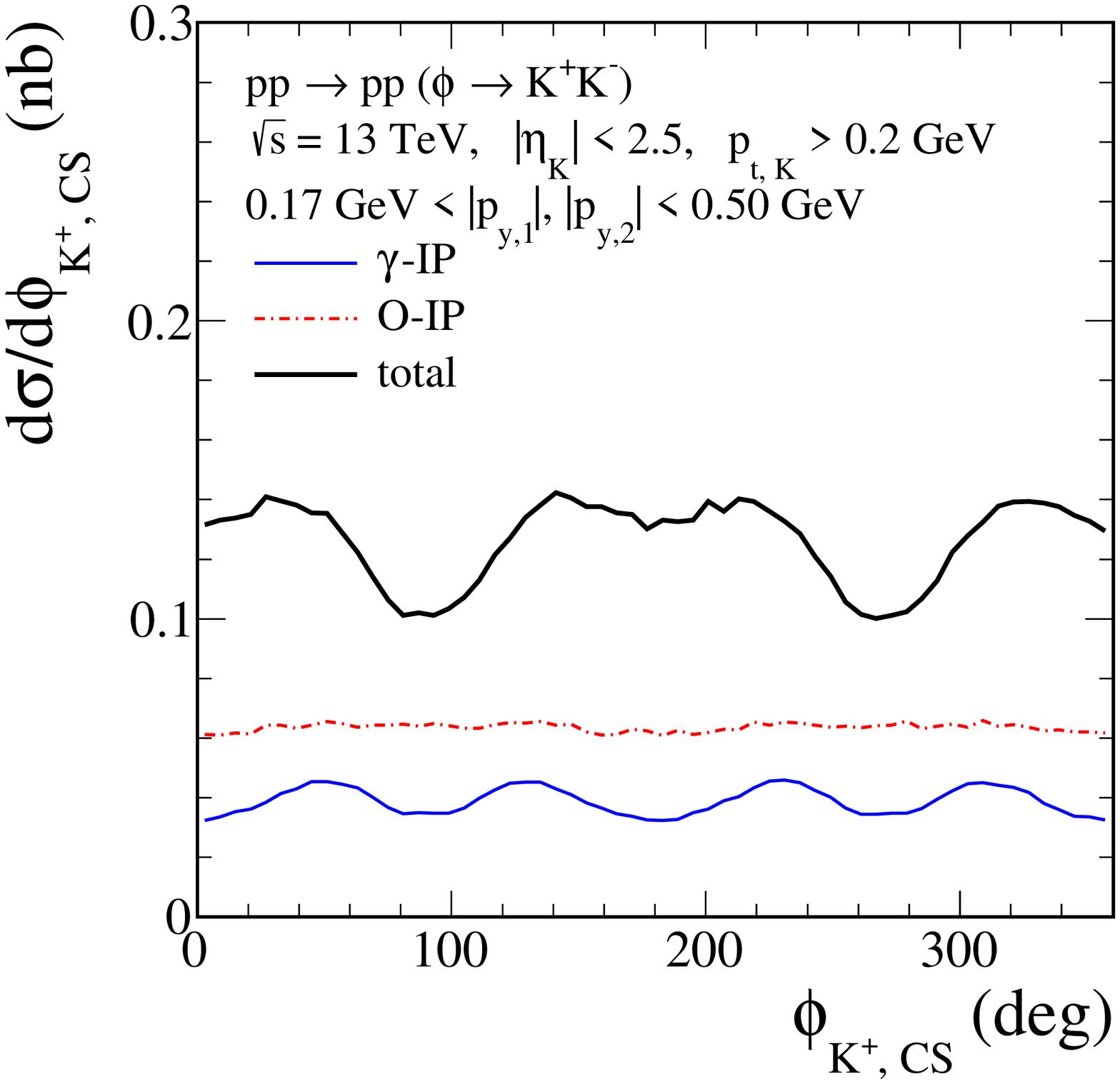}
\caption{\label{fig:ATLASALFA_1_ptK0.2_setBtestA}
The differential cross sections 
for the $pp \to pp (\phi \to K^{+}K^{-})$ reaction
calculated for $\sqrt{s} = 13$~TeV
and for the ATLAS-ALFA experimental cuts
$|\eta_{K}| < 2.5$, $p_{t, K} > 0.2$~GeV,
0.17~GeV~$< |p_{y,1}|, |p_{y,2}|<$~0.50~GeV.
The meaning of the lines is the same as in Fig.~\ref{fig:ATLASALFA_1}
but here we have taken the smaller value of the $b_{\Pom \Ode \phi}$ coupling parameter; see (\ref{parameters_ode_a}).
The absorption effects are included here.}
\end{figure}

Now we shall discuss results for the LHCb experimental conditions.
In Fig.~\ref{fig:LHCb_KK_2D_pt3pt4} we show the two-dimensional distributions 
in ($p_{t,K^{+}}$, $p_{t,K^{-}}$) for 
$\sqrt{s} = 13$~TeV, $2.0 < \eta_{K} < 4.5$, and $p_{t, K} > 0.1$~GeV.
In the left panel we show the result
for $\gamma$-$\Pom$ fusion
obtained with the parameter set~B (\ref{photoproduction_setB}).
In the right panel we show the result for $\Ode$-$\Pom$ fusion
for the parameters
quoted in (\ref{parameters_ode}), (\ref{parameters_ode_lambda}), and (\ref{parameters_ode_b}).
We can see that the $\gamma$-$\Pom$-fusion contribution
is larger at smaller $p_{t, K}$ than the $\Ode$-$\Pom$-fusion contribution.
Therefore, a low-$p_{t, K}$ cut on transverse momenta 
of the kaons can be helpful to reduce 
the $\gamma$-$\Pom$-fusion contribution;
compare the left and right panels
in Figs.~\ref{fig:LHCb_KK_1} and \ref{fig:LHCb_KK_2} below.
\begin{figure}[!ht]
\includegraphics[width=0.45\textwidth]{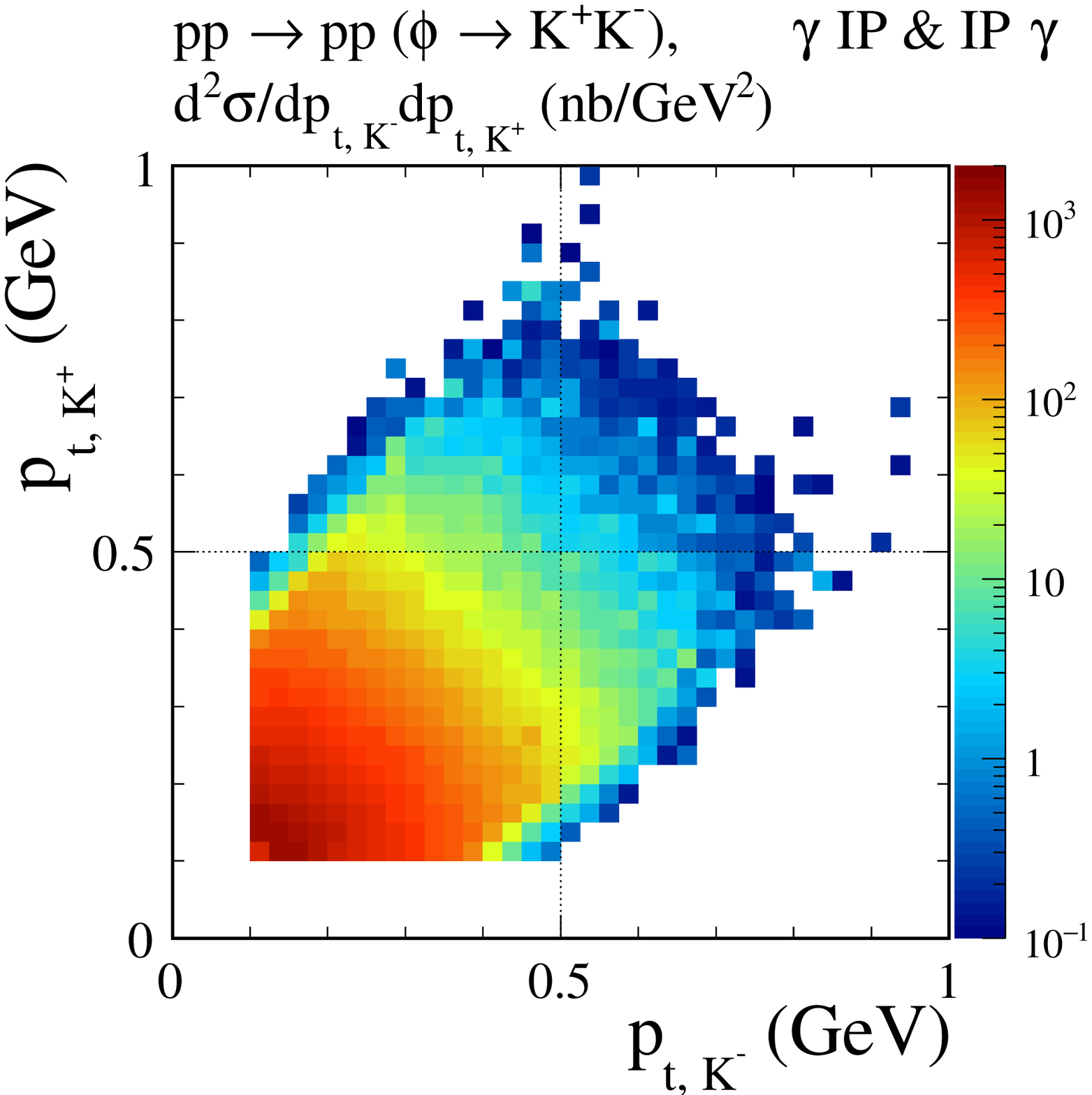}
\includegraphics[width=0.45\textwidth]{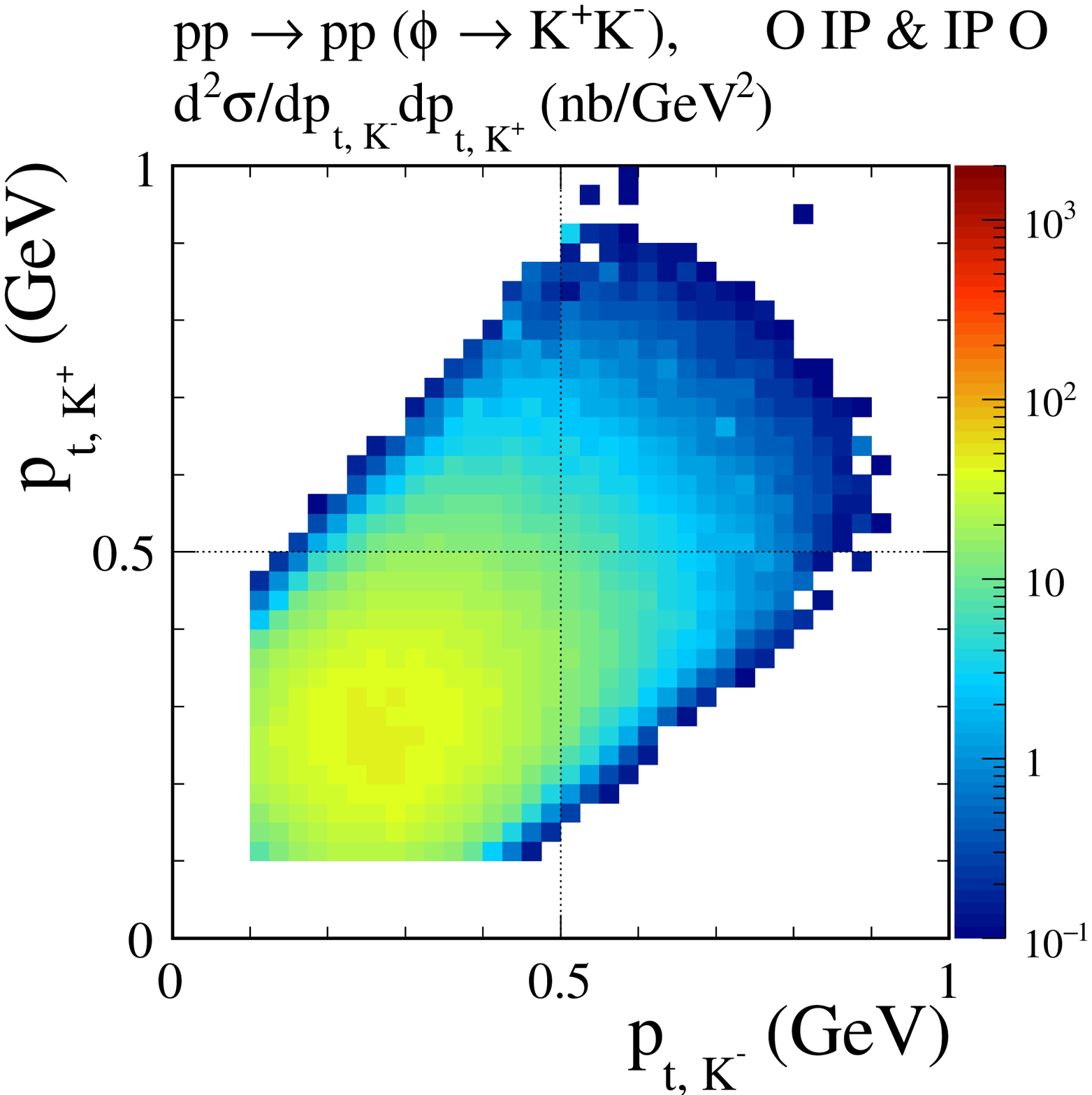}
\caption{\label{fig:LHCb_KK_2D_pt3pt4}
\small
The two-dimensional distributions in ($p_{t,K^{+}}$, $p_{t,K^{-}}$) 
for the $pp \to pp (\phi \to K^{+}K^{-})$ reaction 
via $\gamma$-$\Pom$-fusion (left panel) and 
via $\Ode$-$\Pom$-fusion (right panel) processes.
The calculations were done for 
$\sqrt{s} = 13$~TeV and with cuts on $2.0 < \eta_{K} < 4.5$ and 
$p_{t, K} > 0.1$~GeV.
Here we show the result for $\gamma$-$\Pom$ fusion
obtained with the parameter set~B (\ref{photoproduction_setB})
while the result for $\Ode$-$\Pom$ fusion
was obtained with the parameters
quoted in (\ref{parameters_ode}), (\ref{parameters_ode_lambda}), and (\ref{parameters_ode_b}).
The absorption effects are included here.}
\end{figure}

In Figs.~\ref{fig:LHCb_KK_1} and \ref{fig:LHCb_KK_2}
we show several distributions for $\gamma$-$\Pom$- 
and $\Ode$-$\Pom$-fusion contributions
and their coherent sum
for the LHCb experimental conditions,
$\sqrt{s} = 13$~TeV, $2.0 < \eta_{K} < 4.5$,
$p_{t, K} > 0.3$~GeV (left panels) or $p_{t, K} > 0.5$~GeV (right panels).
The absorption effects were included in the calculations.
For larger kaon transverse momenta 
(or transverse momentum of the $K^+ K^-$ pair)
the odderon-exchange contribution, using our parameters for the odderon,
is bigger than the photon-exchange one.

As in the previous (ATLAS-ALFA) case the angular distributions 
in the $K^+ K^-$ Collins-Soper rest system seem interesting. 
In Fig.~\ref{fig:LHCb_KK_2D_CS} we show the two-dimensional distributions 
in ($\phi_{K^{+},\,{\rm CS}}$, $\cos\theta_{K^{+},\,{\rm CS}}$)
for $2.0 < \eta_{K} < 4.5$ and $p_{t, K} > 0.3$~GeV.
We see here again that the \mbox{$\gamma$-$\Pom$ fusion} leads
predominantly to transverse polarisation of the $\phi$ meson.
The distribution for the $\Ode$-$\Pom$ fusion
(the right panel of Fig.~\ref{fig:LHCb_KK_2D_CS}) shows clearly a strong
longitudinal $\phi$-meson component but, 
due to the marked $\phi_{K^{+},\,{\rm CS}}$ dependence,
also transverse $\phi$ components must be present.
\begin{figure}[!ht]
\includegraphics[width=0.42\textwidth]{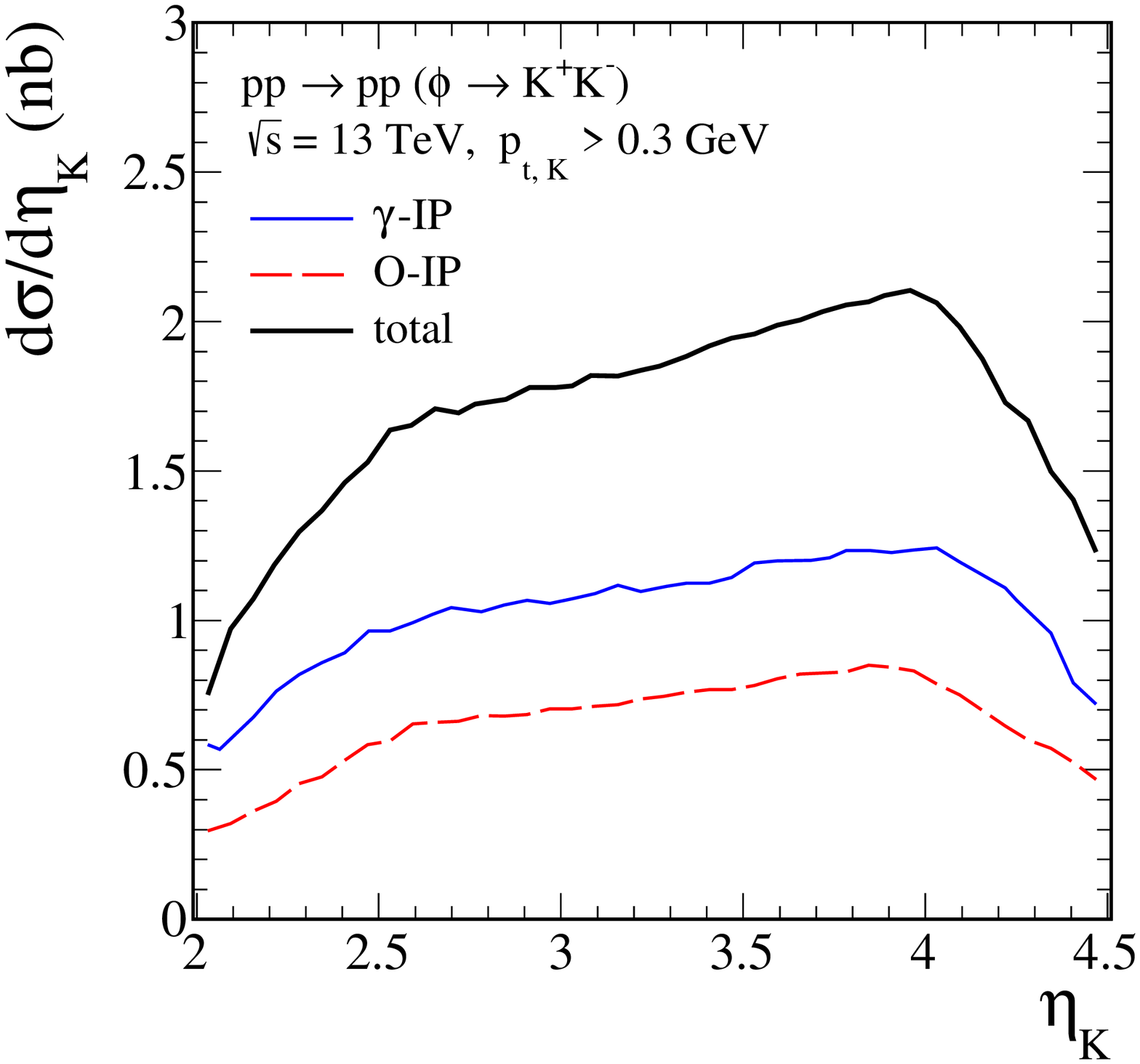}
\includegraphics[width=0.42\textwidth]{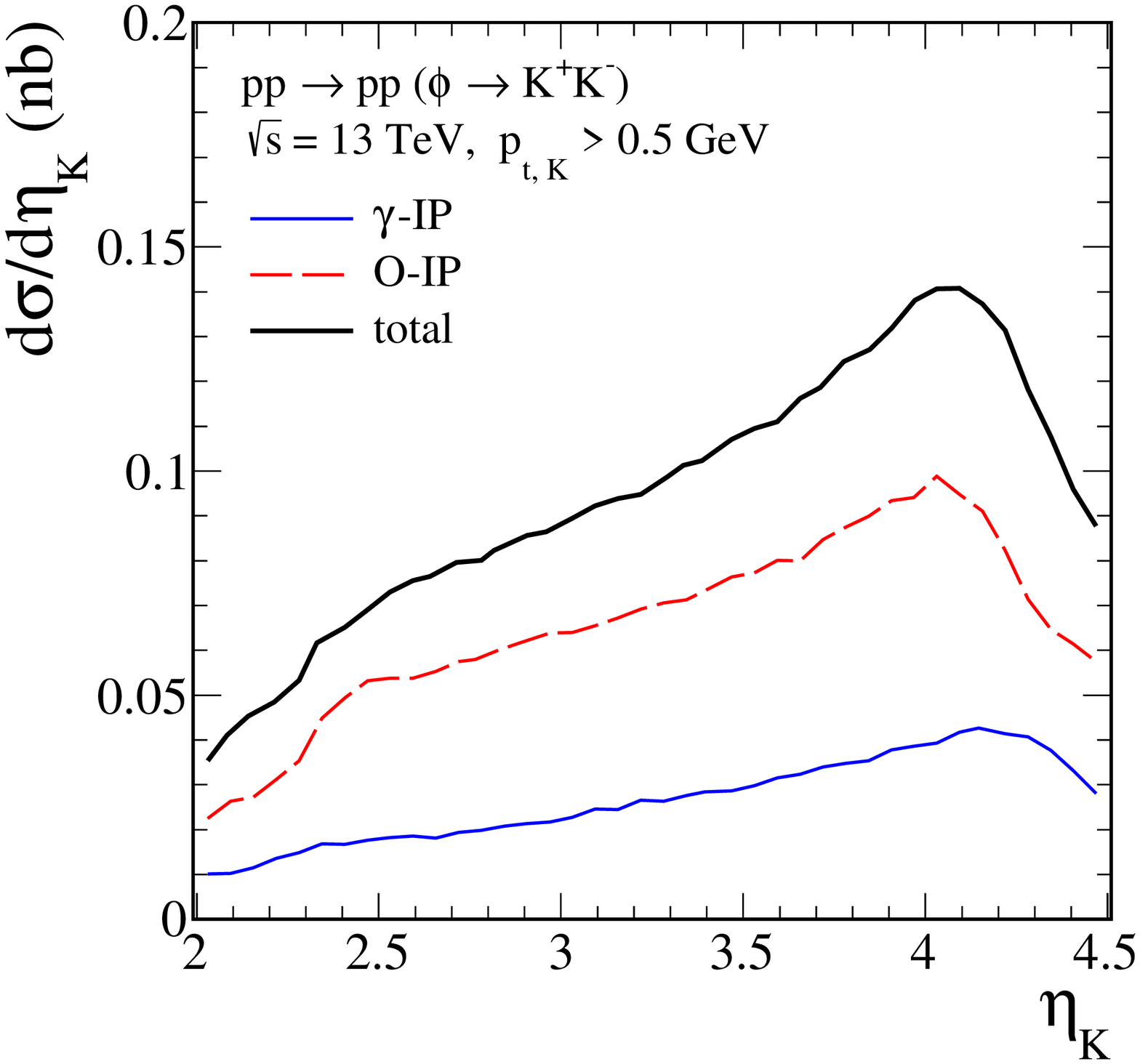}
\includegraphics[width=0.42\textwidth]{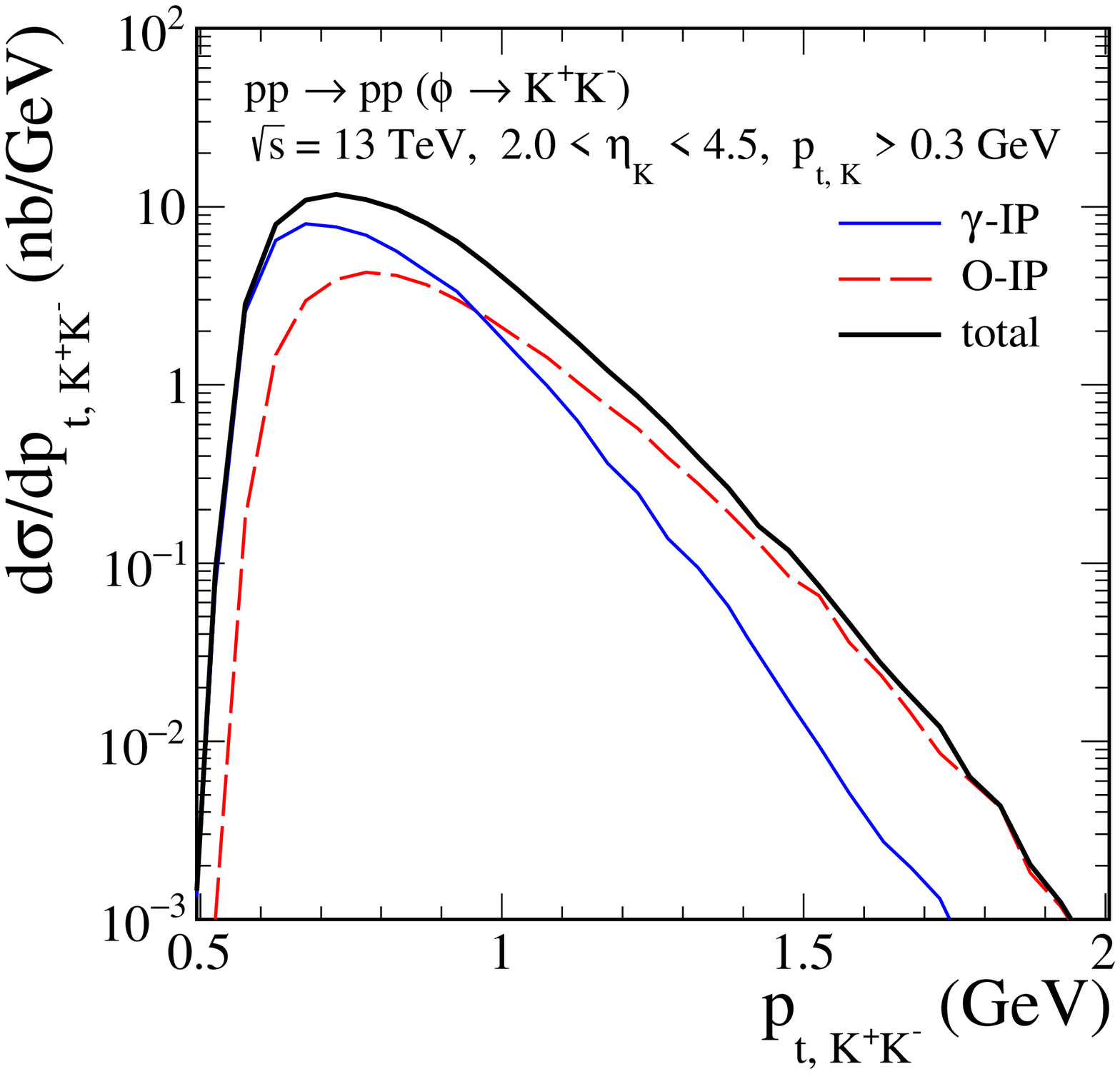}
\includegraphics[width=0.42\textwidth]{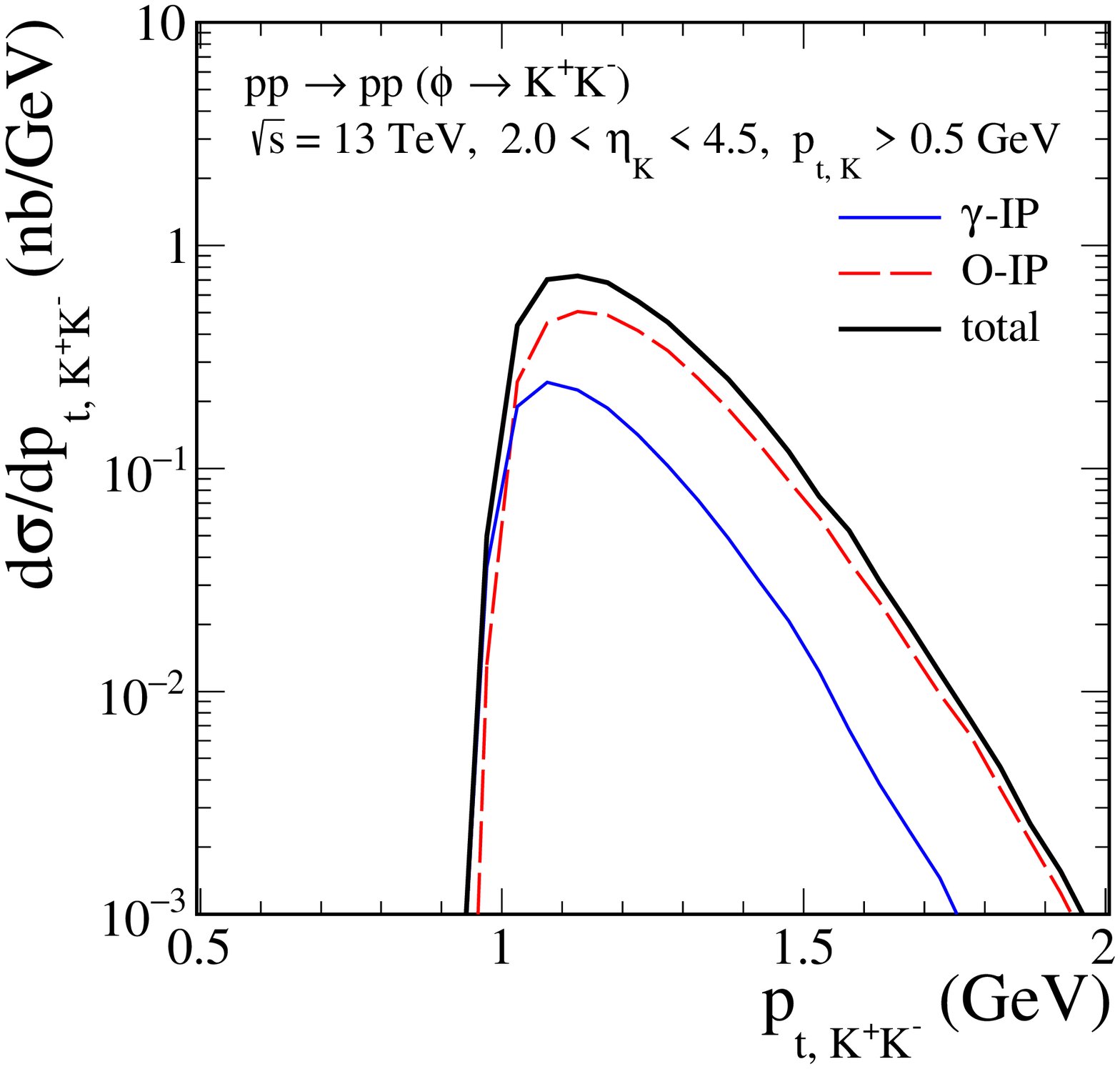}
\includegraphics[width=0.42\textwidth]{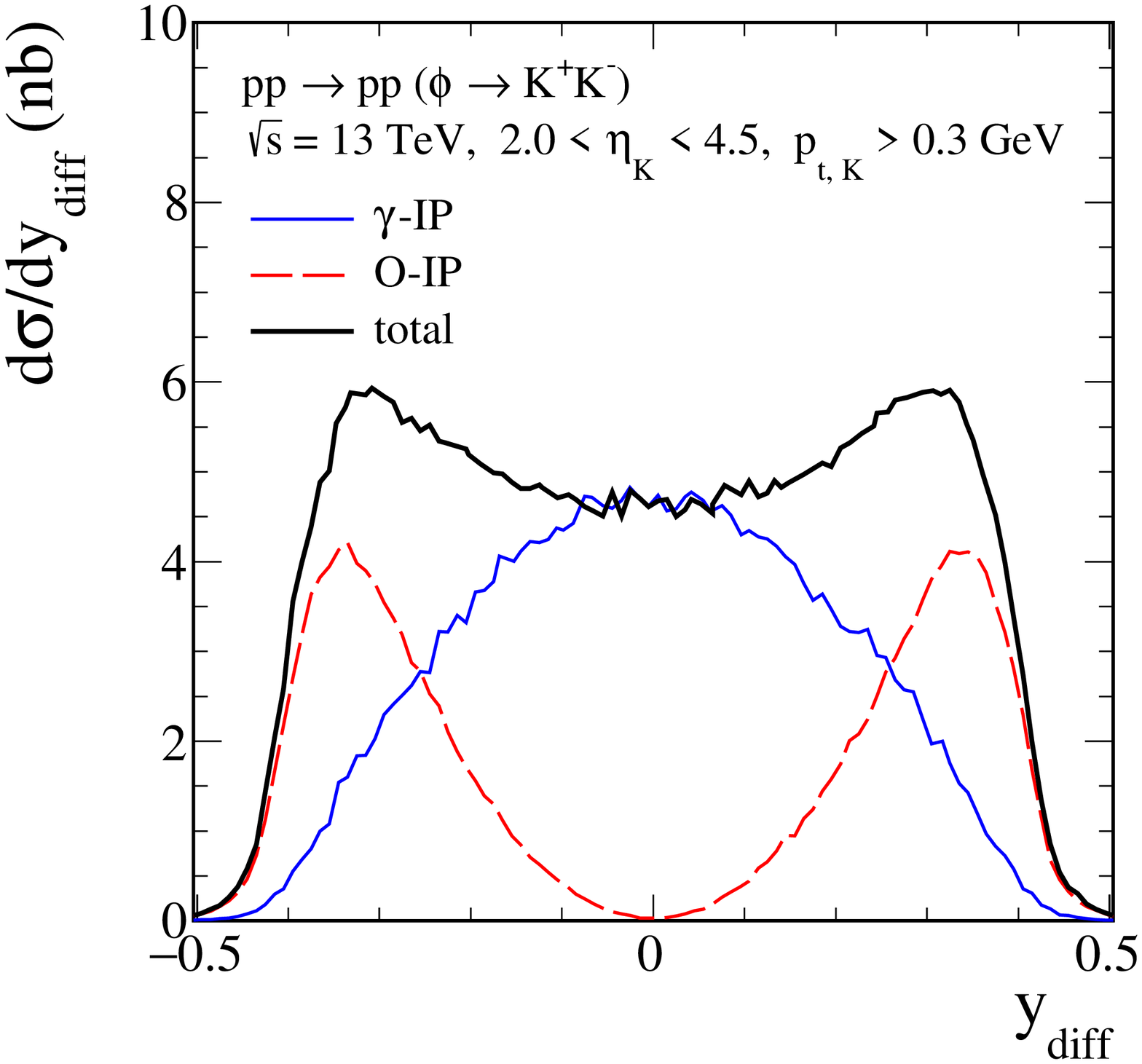}
\includegraphics[width=0.42\textwidth]{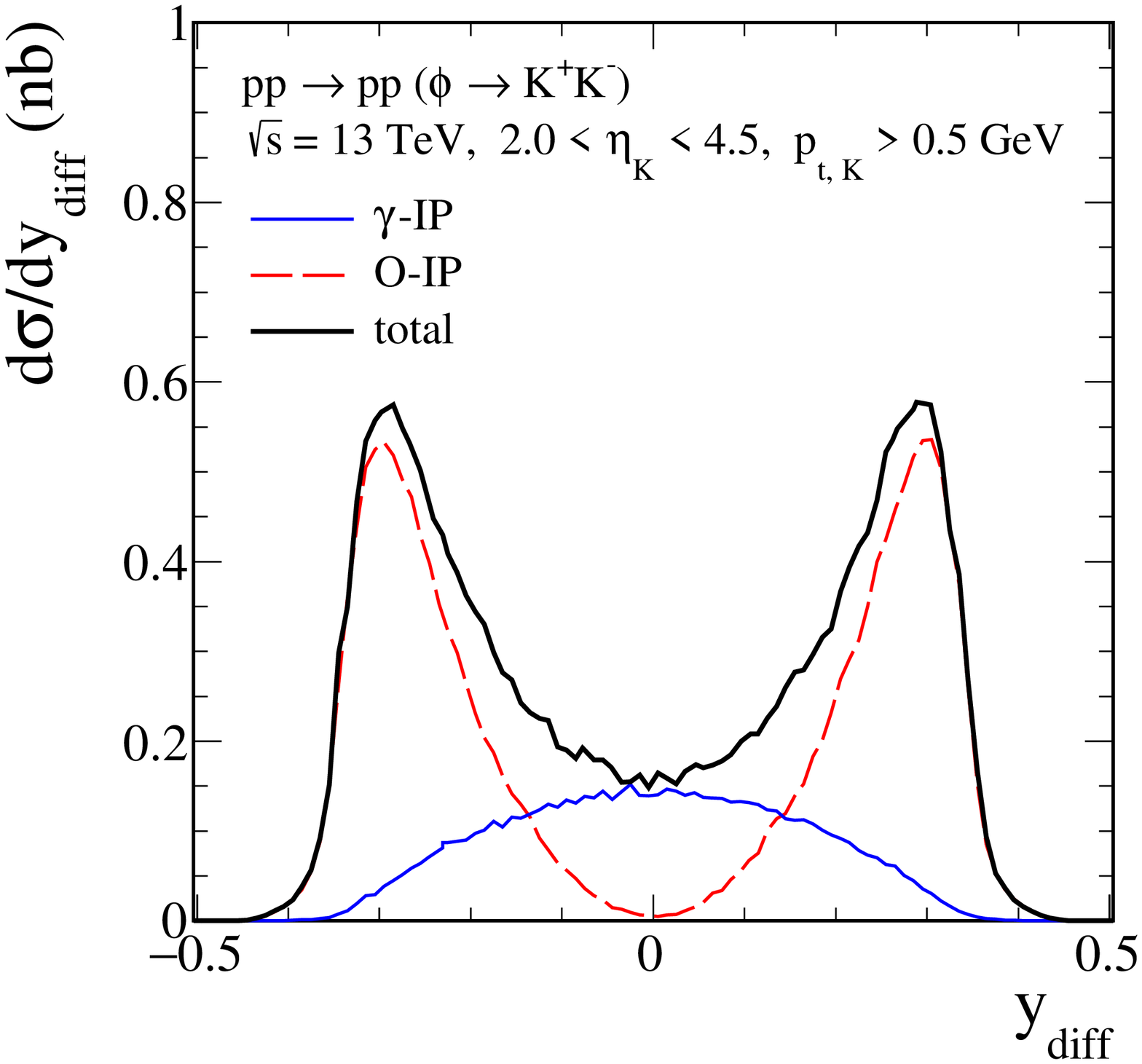}
\caption{\label{fig:LHCb_KK_1}
The differential cross sections for the $pp \to pp (\phi \to K^{+}K^{-})$ reaction.
Calculations were done for $\sqrt{s} = 13$~TeV, $2.0 < \eta_{K} < 4.5$, 
and $p_{t, K} > 0.3$~GeV (left panels) or $p_{t, K} > 0.5$~GeV (right panels). 
The meaning of the lines is the same as in Fig.~\ref{fig:ATLASALFA_1}.
Results for the photoproduction (blue solid lines) 
and the $\Ode$-$\Pom$-fusion (red lines) 
contributions are shown separately.
The black solid line corresponds 
to the coherent sum of the $\gamma$-$\Pom$-
and $\Ode$-$\Pom$-fusion processes with the coupling parameters
(\ref{photoproduction_setB}) and (\ref{parameters_ode_b}), respectively.
The absorption effects are included here.}
\end{figure}
\begin{figure}[!ht]
\includegraphics[width=0.45\textwidth]{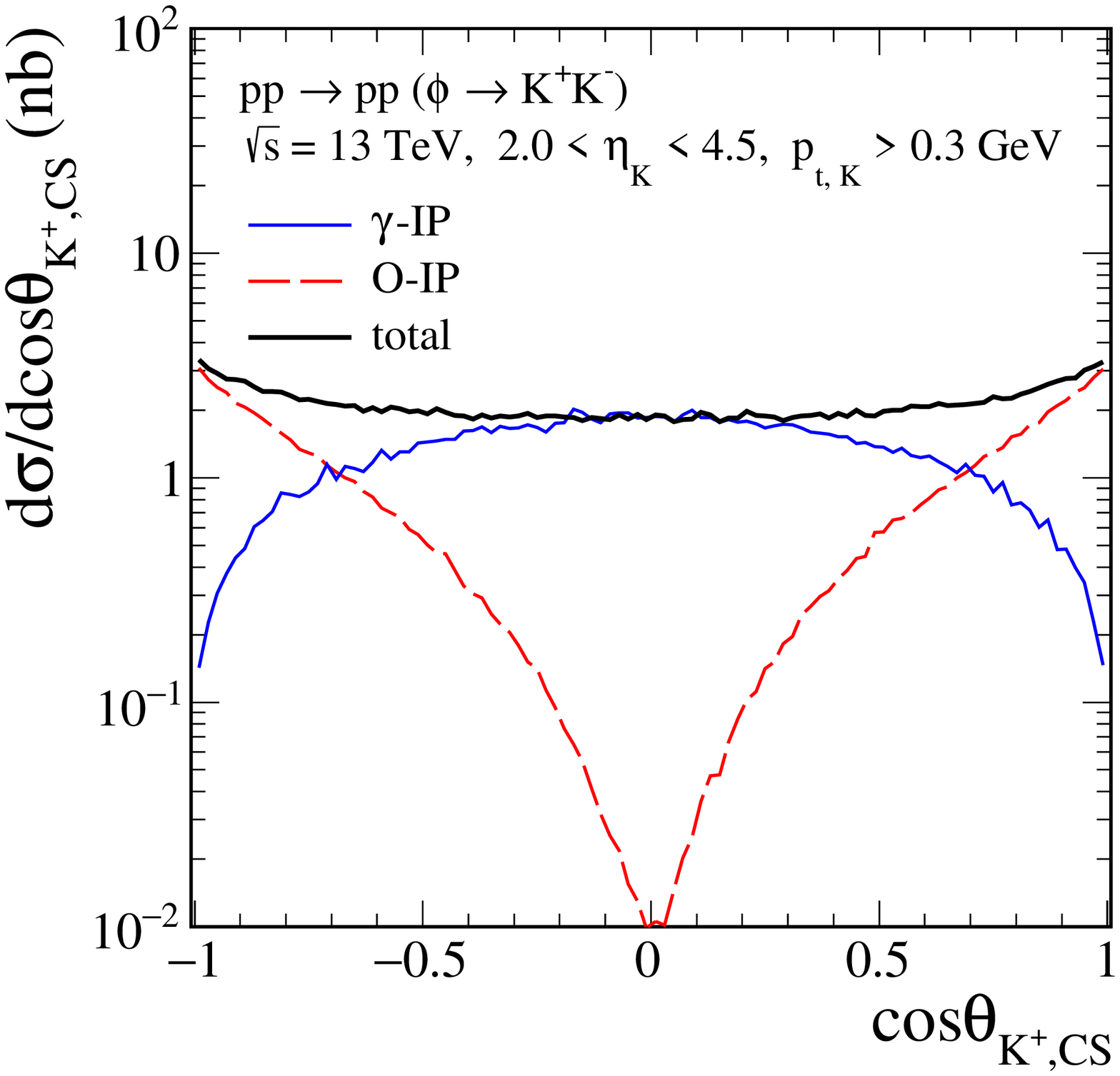}
\includegraphics[width=0.45\textwidth]{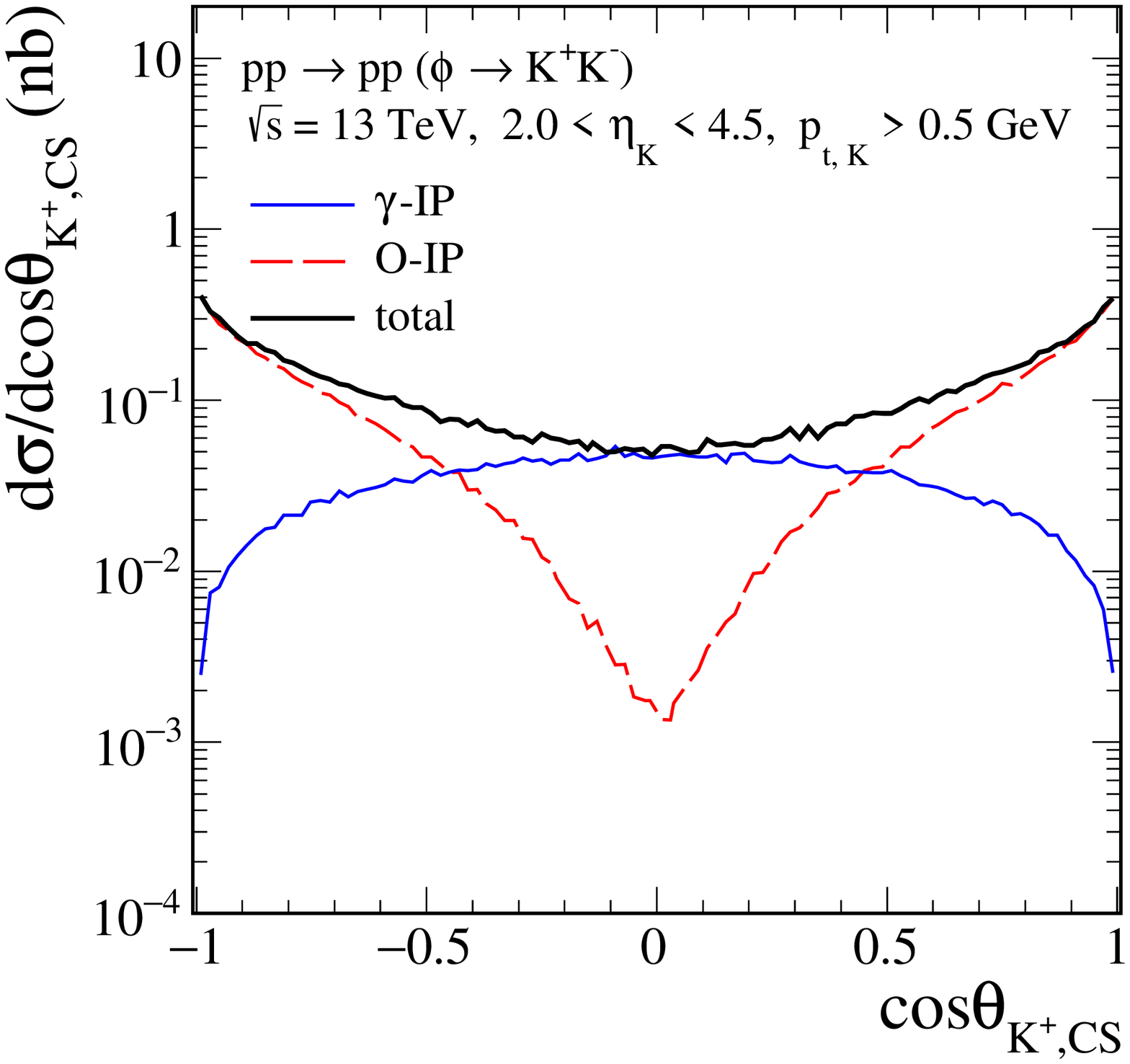}\\
\includegraphics[width=0.45\textwidth]{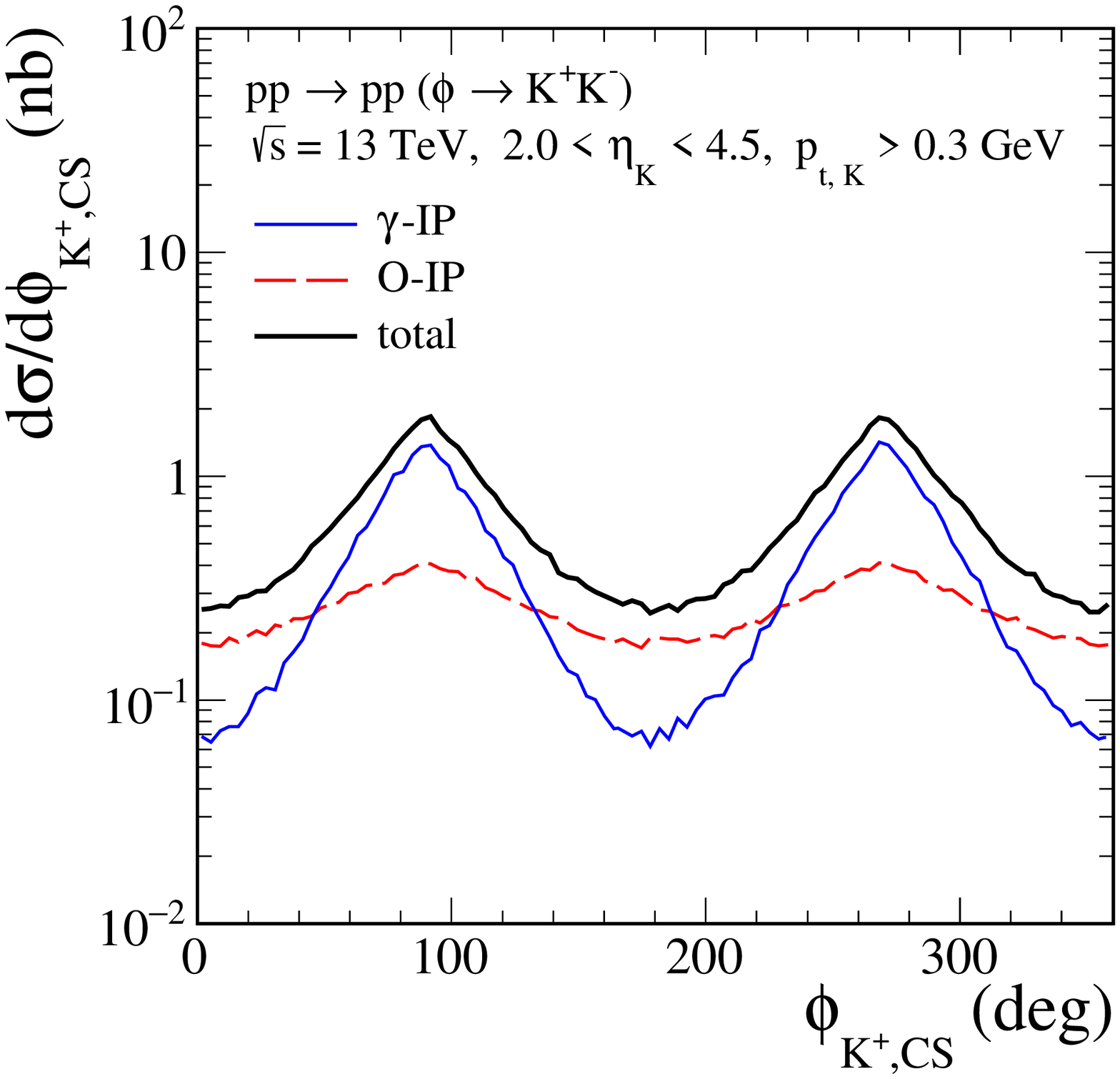}
\includegraphics[width=0.45\textwidth]{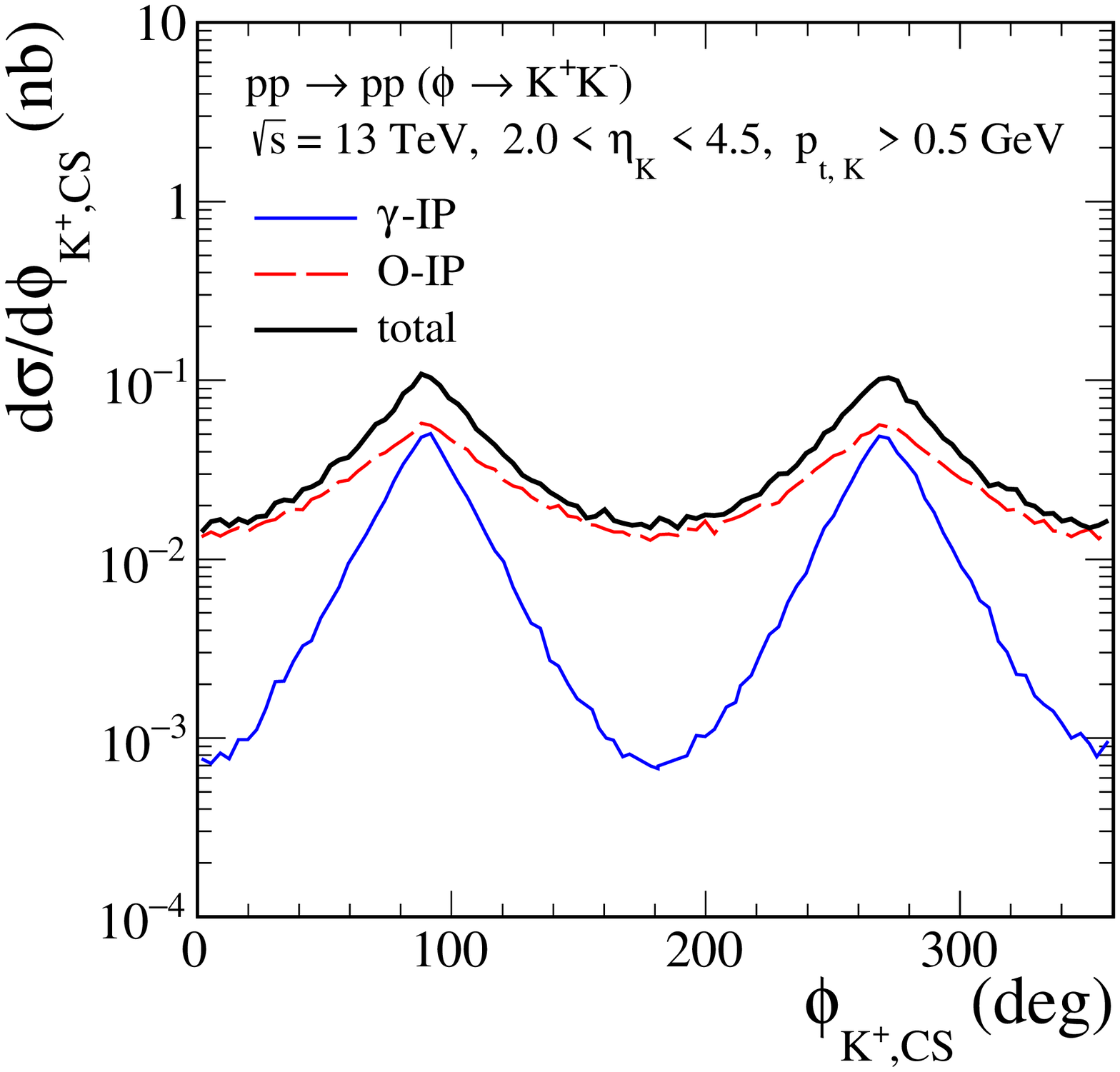}
\caption{\label{fig:LHCb_KK_2}
The distributions in $\cos\theta_{K^{+},\,{\rm CS}}$ and $\phi_{K^{+},\,{\rm CS}}$
for the same experimental cuts as in Fig.~\ref{fig:LHCb_KK_1}.
Also the meaning of the lines is as in Fig.~\ref{fig:LHCb_KK_1}.}
\end{figure}
\begin{figure}[!ht]
\includegraphics[width=0.45\textwidth]{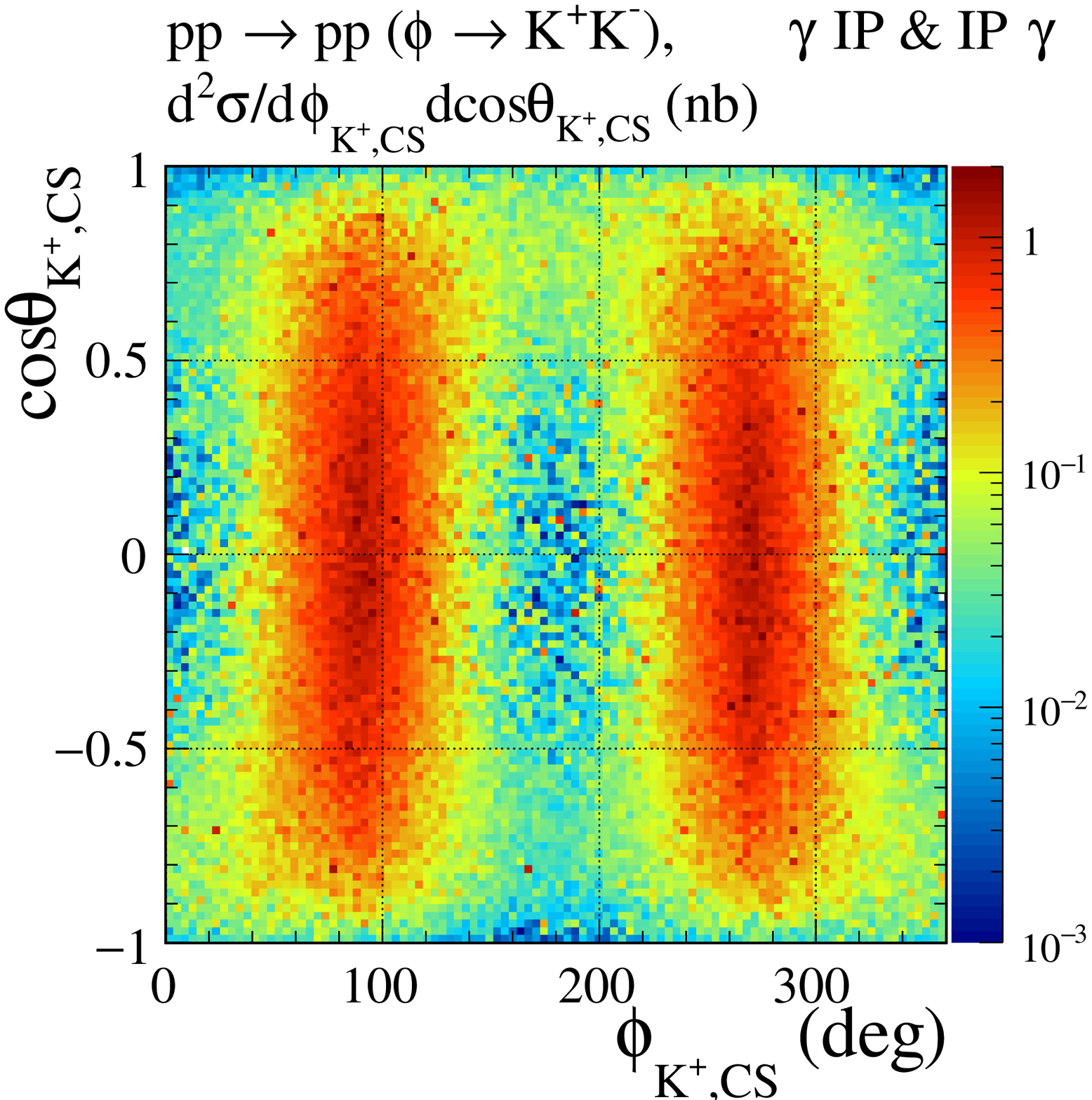}
\includegraphics[width=0.45\textwidth]{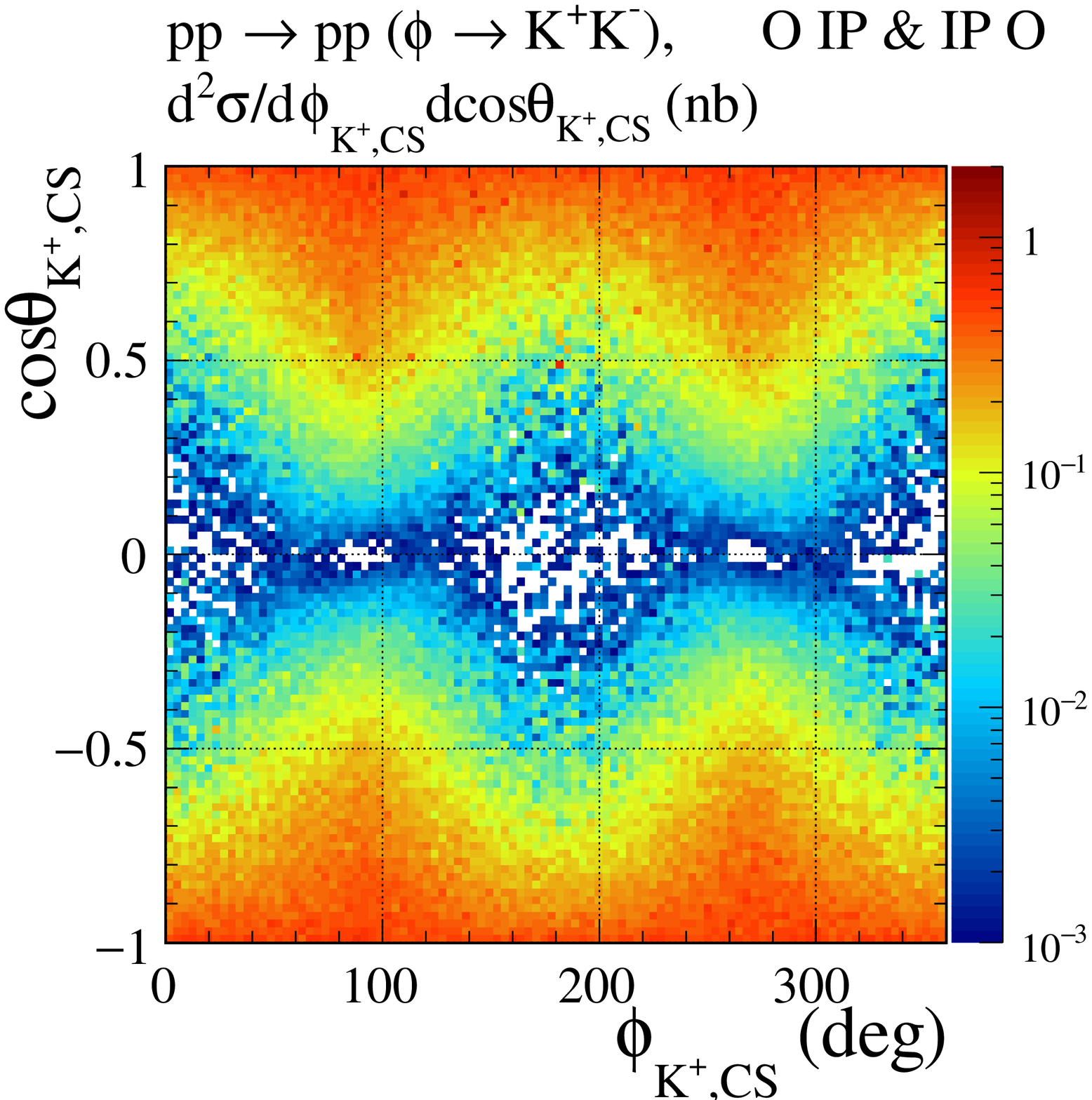}
\caption{\label{fig:LHCb_KK_2D_CS}
\small
The two-dimensional distributions in ($\phi_{K^{+},\,{\rm CS}}$, $\cos\theta_{K^{+},\,{\rm CS}}$) 
for the $pp \to pp (\phi \to K^{+}K^{-})$ reaction 
via $\gamma$-$\Pom$ fusion (left panel) and 
via $\Ode$-$\Pom$ fusion (right panel).
The calculations were done for 
$\sqrt{s} = 13$~TeV and with the cuts $2.0 < \eta_{K} < 4.5$ and $p_{t, K} > 0.3$~GeV.
We show the result for $\gamma$-$\Pom$ fusion
obtained with the parameter set~B (\ref{photoproduction_setB})
while the result for $\Ode$-$\Pom$ fusion
was obtained with the parameters quoted in (\ref{parameters_ode}),
(\ref{parameters_ode_lambda}), and (\ref{parameters_ode_b}).
The absorption effects are included here.}
\end{figure}

\subsubsection{The $pp \to pp \mu^{+}\mu^{-}$ reaction}
\label{pp_ppLL}

The $\phi$ meson can also be observed in the $\mu^{+}\mu^{-}$ channel.
In this subsection we wish to show our predictions
for the $pp \to pp \mu^{+}\mu^{-}$ reaction for the LHCb experiment
at $\sqrt{s} = 13$~TeV for the $2.0 < \eta_{\mu} < 4.5$ pseudorapidity range.
Here we require no detection of the leading protons.

In Fig.~\ref{fig:LHCb_mumu_0} we present the $\mu^{+}\mu^{-}$ invariant mass
distributions in the $\phi(1020)$ resonance region.
We show the contributions from the $\gamma$-$\Pom$- 
and $\Ode$-$\Pom$-fusion processes
and the continuum $\gamma \gamma \to \mu^{+}\mu^{-}$ term.
The dimuon-continuum process ($\gamma \gamma \to \mu^{+}\mu^{-}$)
was discussed, e.g., in \cite{Lebiedowicz:2018muq}
in the context of the ATLAS measurement \cite{Aaboud:2017oiq}.
In our analysis here we are looking at the dimuon invariant mass region 
$M_{\mu^{+}\mu^{-}} \in (1.01, 1.03)$~GeV.
\begin{figure}[!ht]
\includegraphics[width=0.45\textwidth]{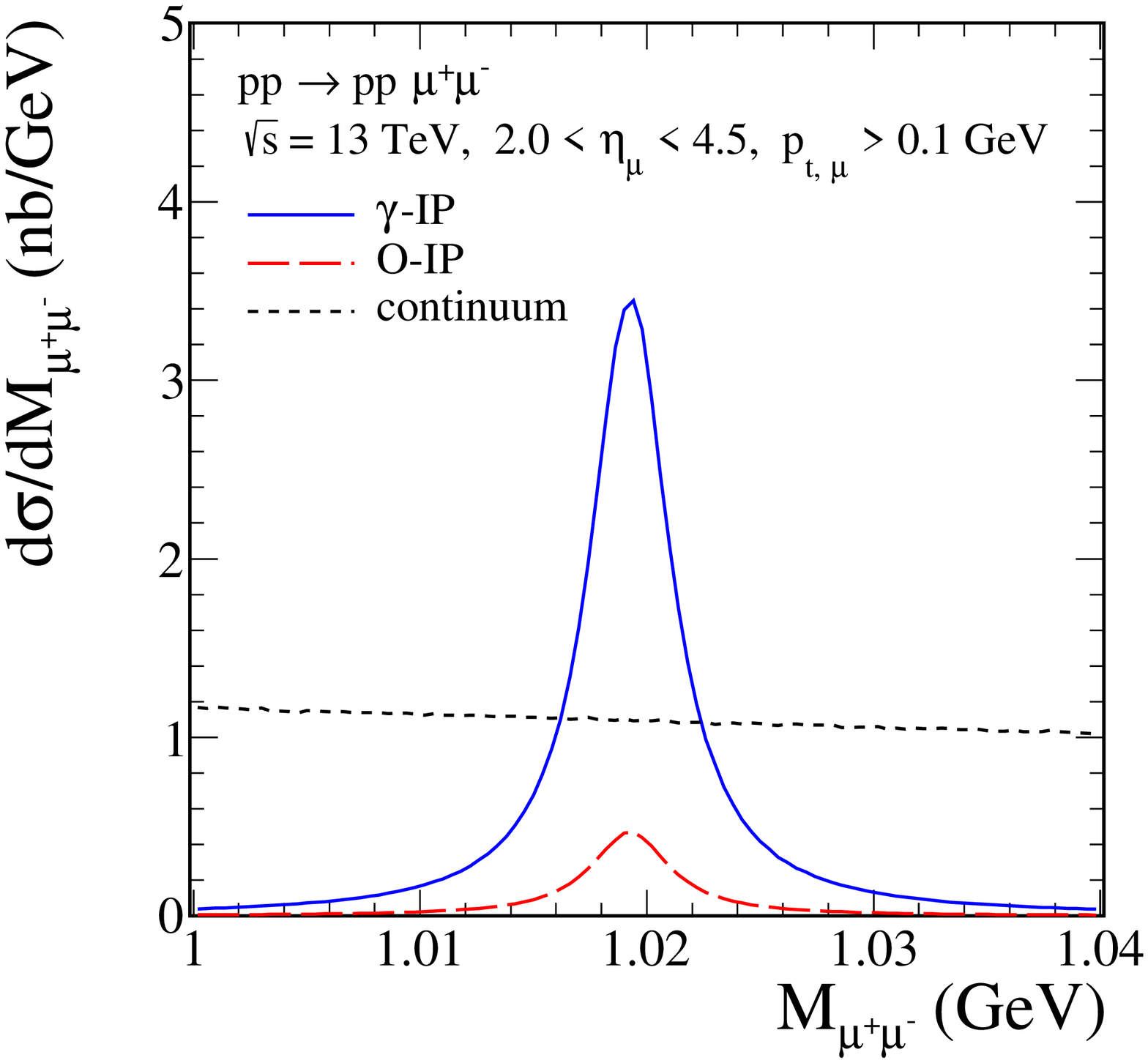}
\caption{\label{fig:LHCb_mumu_0}
The distributions in $\mu^{+}\mu^{-}$ invariant mass
for the exclusive $pp \to pp \mu^{+}\mu^{-}$ reaction
including the $\phi$-meson production via
the $\gamma$-$\Pom$- and the $\Ode$-$\Pom$-fusion processes
and the nonresonant $\gamma \gamma \to \mu^{+}\mu^{-}$ continuum term.
The calculations were done for $\sqrt{s} = 13$~TeV, $2.0 < \eta_{\mu} < 4.5$,
and $p_{t, \mu} > 0.1$~GeV.
Here we show the result for $\gamma$-$\Pom$ fusion (the blue solid line)
obtained with the parameter set~B (\ref{photoproduction_setB}).
The result for $\Ode$-$\Pom$ fusion (the red long-dashed line) 
was obtained with the parameters
quoted in (\ref{parameters_ode}), (\ref{parameters_ode_lambda}), and (\ref{parameters_ode_b}).
The black short-dashed line corresponds to the continuum contribution.
The absorption effects are included here.}
\end{figure}

Note, that in the continuum term, $\gamma \gamma \to \mu^{+}\mu^{-}$,
the $\mu^{+}\mu^{-}$ are in a state of charge conjugation $C = +1$.
For $\phi \to \mu^{+}\mu^{-}$ we have a state of $C = -1$.
Thus, the interference of the continuum and the $\phi$-production reactions
will lead to $\mu^{+}$-$\mu^{-}$ asymmetries.
We have checked, however, that the interference in the $\mu^{+} \mu^{-}$ channel
is smaller than our numerical precision, definitely smaller than 2\%.

In Fig.~\ref{fig:LHCb_mumu_2D_pt3pt4} we show two-dimensional distributions 
in ($p_{t,\mu^{+}}$, $p_{t,\mu^{-}}$) for three different processes.
The result in the panel (a) corresponds to the continuum contribution
without the cut on $M_{\mu^{+}\mu^{-}}$.
Here the maximum of the cross section is placed along 
the $p_{t,\mu^{+}} = p_{t,\mu^{-}}$ line
which is due to the predominantly small transverse momenta of the photons 
in this photon-exchange process.
The results in the panels (b), (c), and (d) correspond to 
the continuum term, the $\gamma$-$\Pom$- and $\Ode$-$\Pom$-fusion processes, respectively,
including the limitation on $M_{\mu^{+}\mu^{-}}$.
\begin{figure}[!ht]
(a)\includegraphics[width=0.4\textwidth]{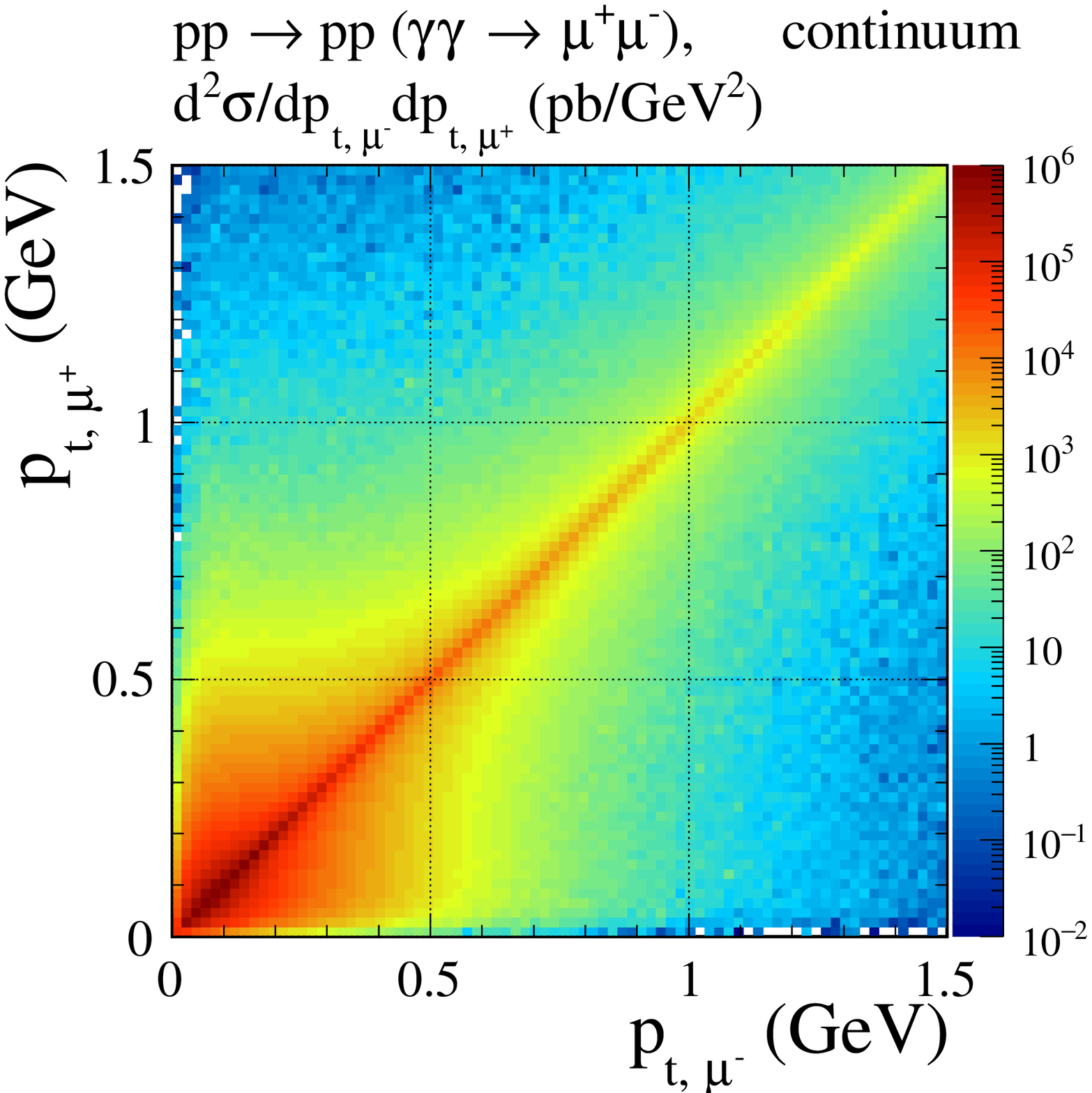}
(b)\includegraphics[width=0.4\textwidth]{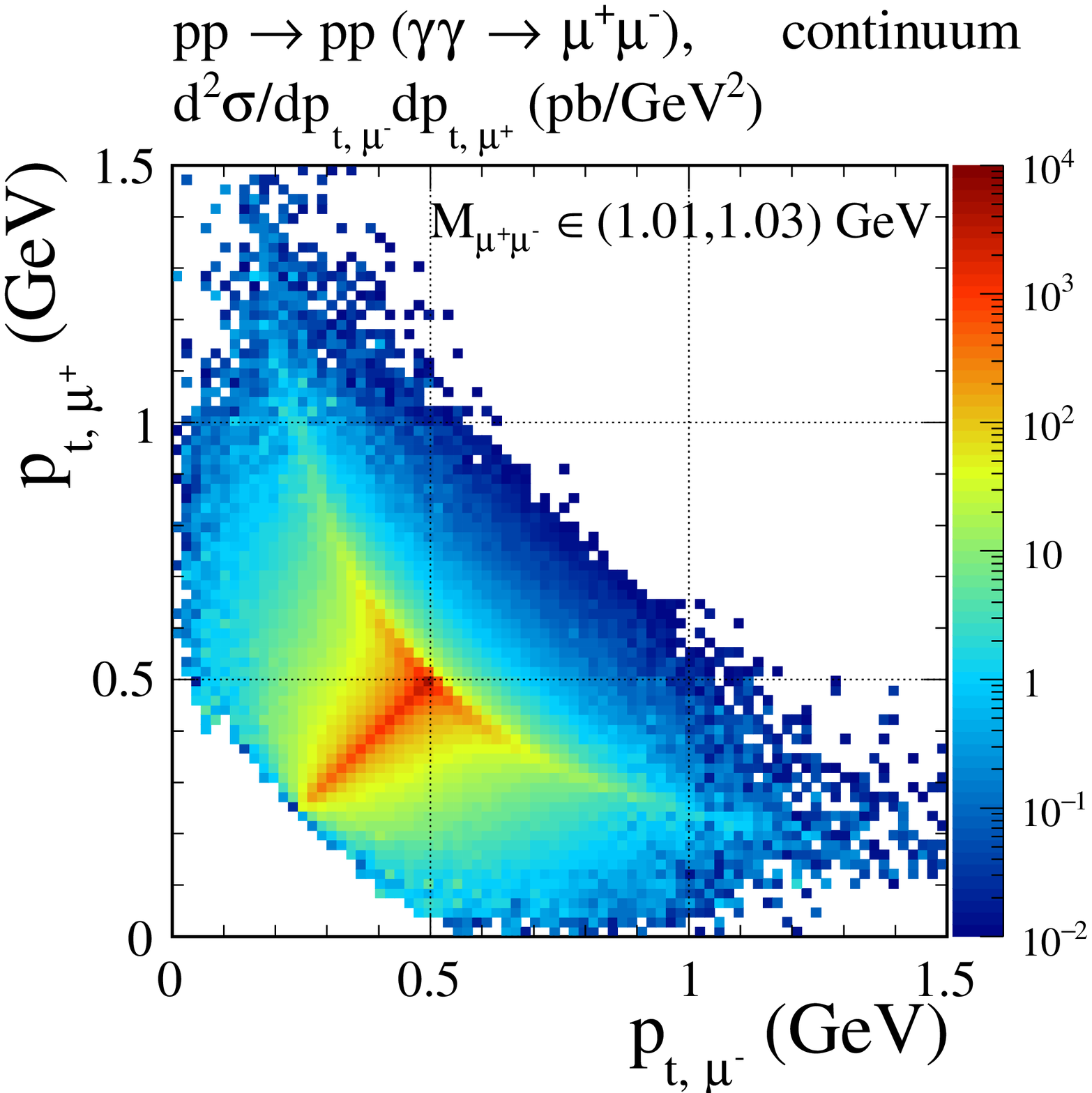}
(c)\includegraphics[width=0.4\textwidth]{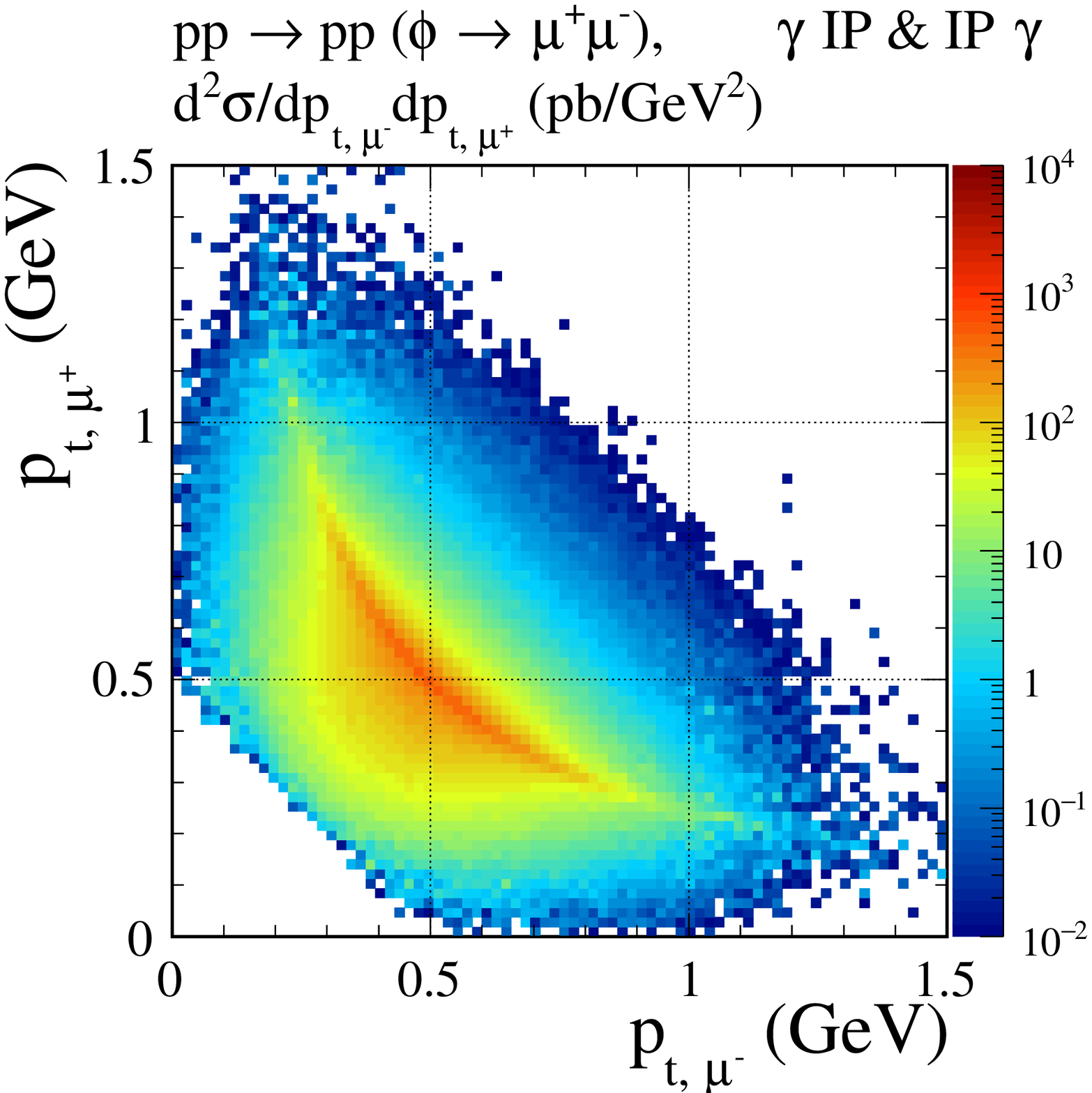}
(d)\includegraphics[width=0.4\textwidth]{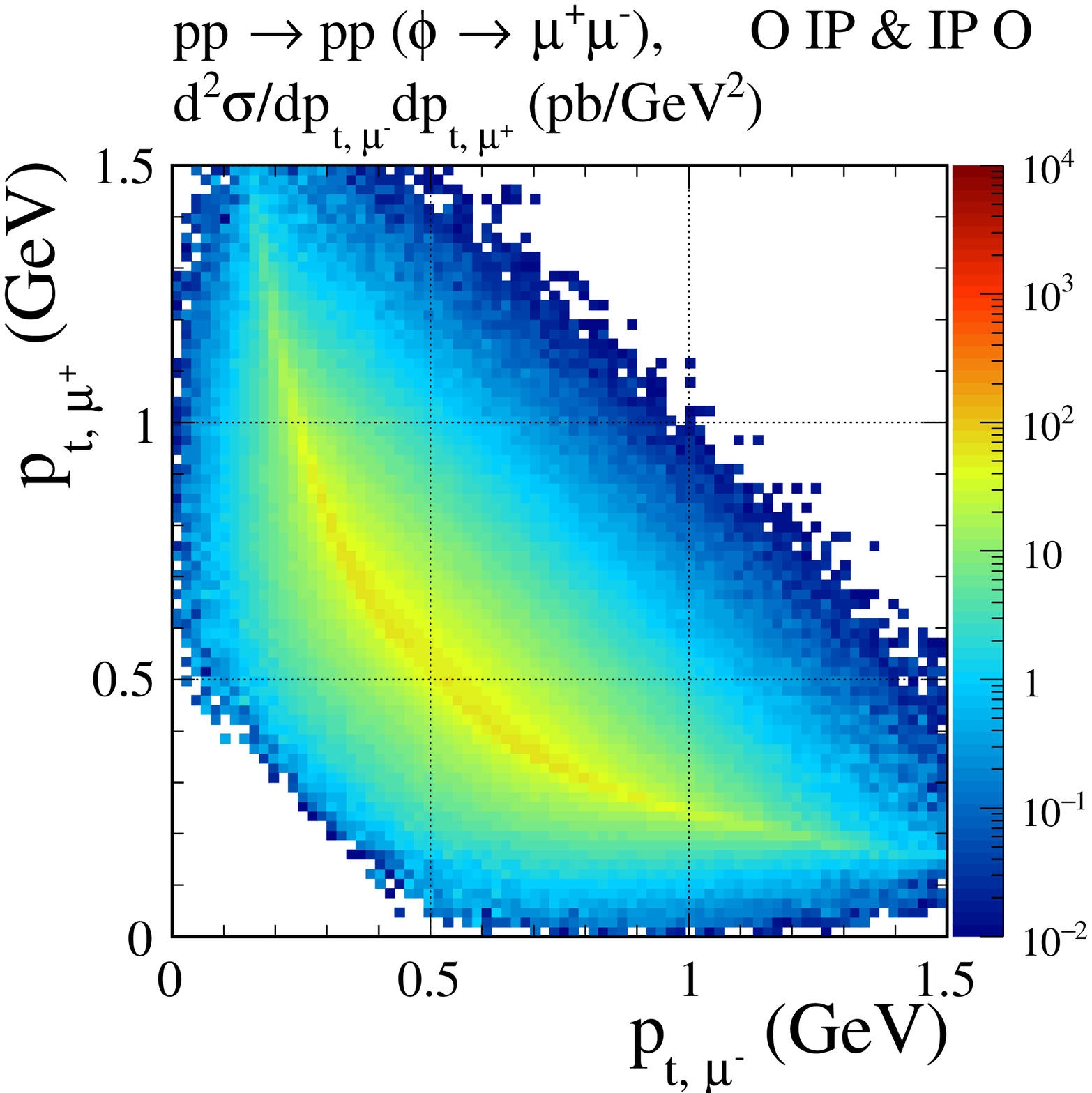}
\caption{\label{fig:LHCb_mumu_2D_pt3pt4}
\small
The two-dimensional distributions in ($p_{t,\mu^{+}}$, $p_{t,\mu^{-}}$)
for the $pp \to pp \mu^{+}\mu^{-}$ reaction.
The calculations were done for $\sqrt{s} = 13$~TeV and $2.0 < \eta_{\mu} < 4.5$.
The results in the panels (a) and (b) correspond to the $\mu^{+}\mu^{-}$ continuum 
without and with the cut on $M_{\mu^{+}\mu^{-}} \in (1.01, 1.03)$~GeV, respectively.
The results in the panels (c) and (d) correspond to
the $\phi$ production via $\gamma$-$\Pom$ fusion
and via $\Ode$-$\Pom$ fusion, respectively.
No absorption effects are included here.}
\end{figure}

In Figs.~\ref{fig:LHCb_mumu_1} and \ref{fig:LHCb_mumu_2},
we show the predictions for the $pp \to pp \mu^{+}\mu^{-}$ reaction
for typical experimental lower cuts on the transverse momentum of the muons,
$p_{t, \mu} > 0.1$~GeV and $p_{t, \mu} > 0.5$~GeV, respectively.
In contrast to dikaon production here there is 
for both the $\gamma$-$\Pom$- and the $\Ode$-$\Pom$-fusion contributions
a maximum at $\rm{y_{diff}} = 0$ (or $\cos\theta_{\mu^{+},{\rm CS}} = 0$).
In Fig.~\ref{fig:LHCb_mumu_1} the continuum contribution is large. 
Imposing a larger cut on the transverse momenta of the muons
reduces the continuum contribution which, however,
still remains sizeable at $\rm{y_{diff}} = 0$.
Such a cut reduces the statistics of the measurement;
see the results in Table~\ref{tab:table2}.
In Fig.~\ref{fig:LHCb_mumu_2} 
we show our predictions for different choices of parameters.
The $\mu^+ \mu^-$ channel seems to be less promising
in identifying the odderon exchange 
at least when only the $p_{t, \mu}$ cuts are imposed.
Eventually, the absolute normalization of the cross section and detailed
studies of shapes of distributions
should provide a clear answer whether one can observe 
the odderon-exchange mechanism here.
\begin{figure}[!ht]
\includegraphics[width=0.45\textwidth]{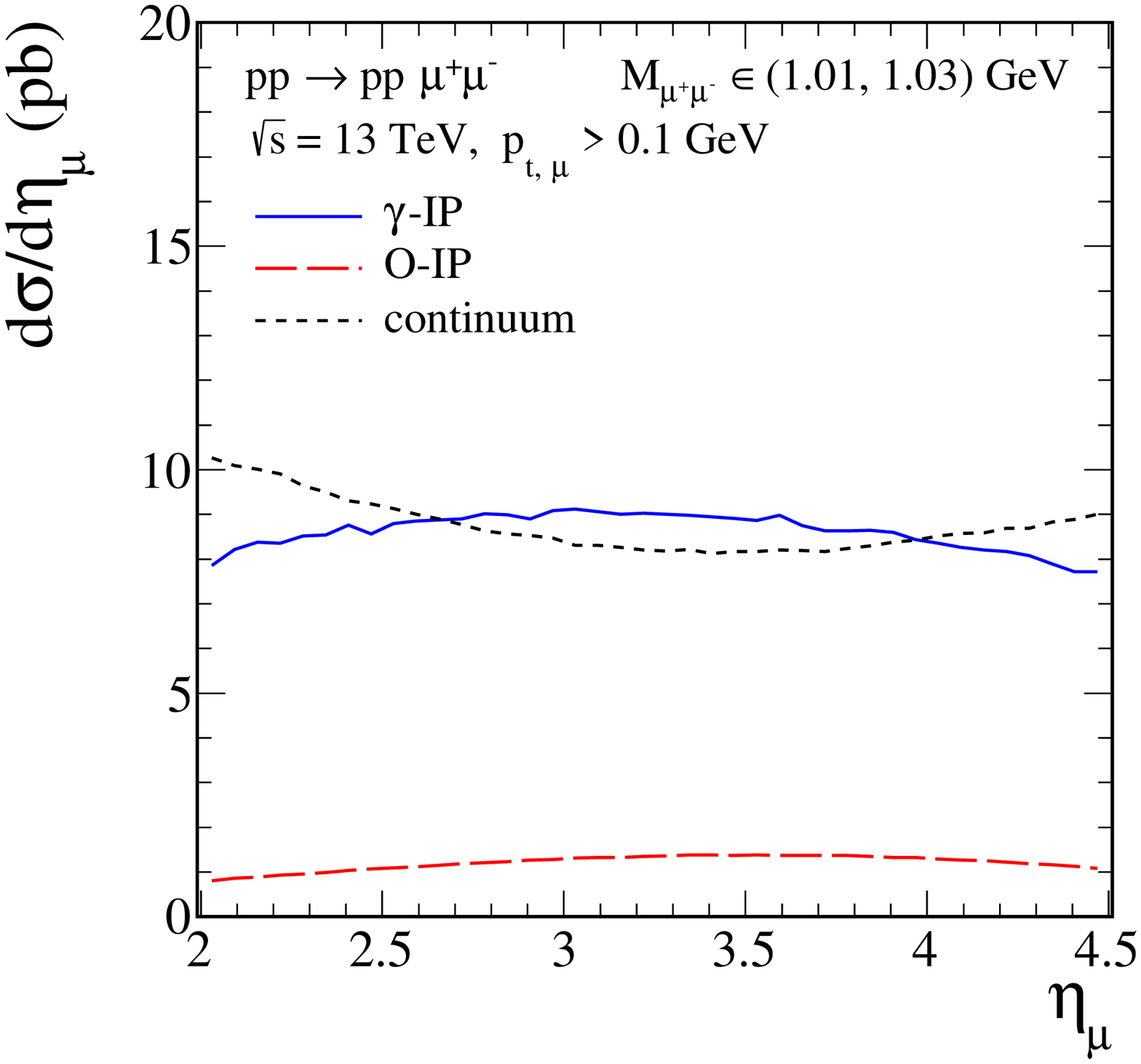}
\includegraphics[width=0.45\textwidth]{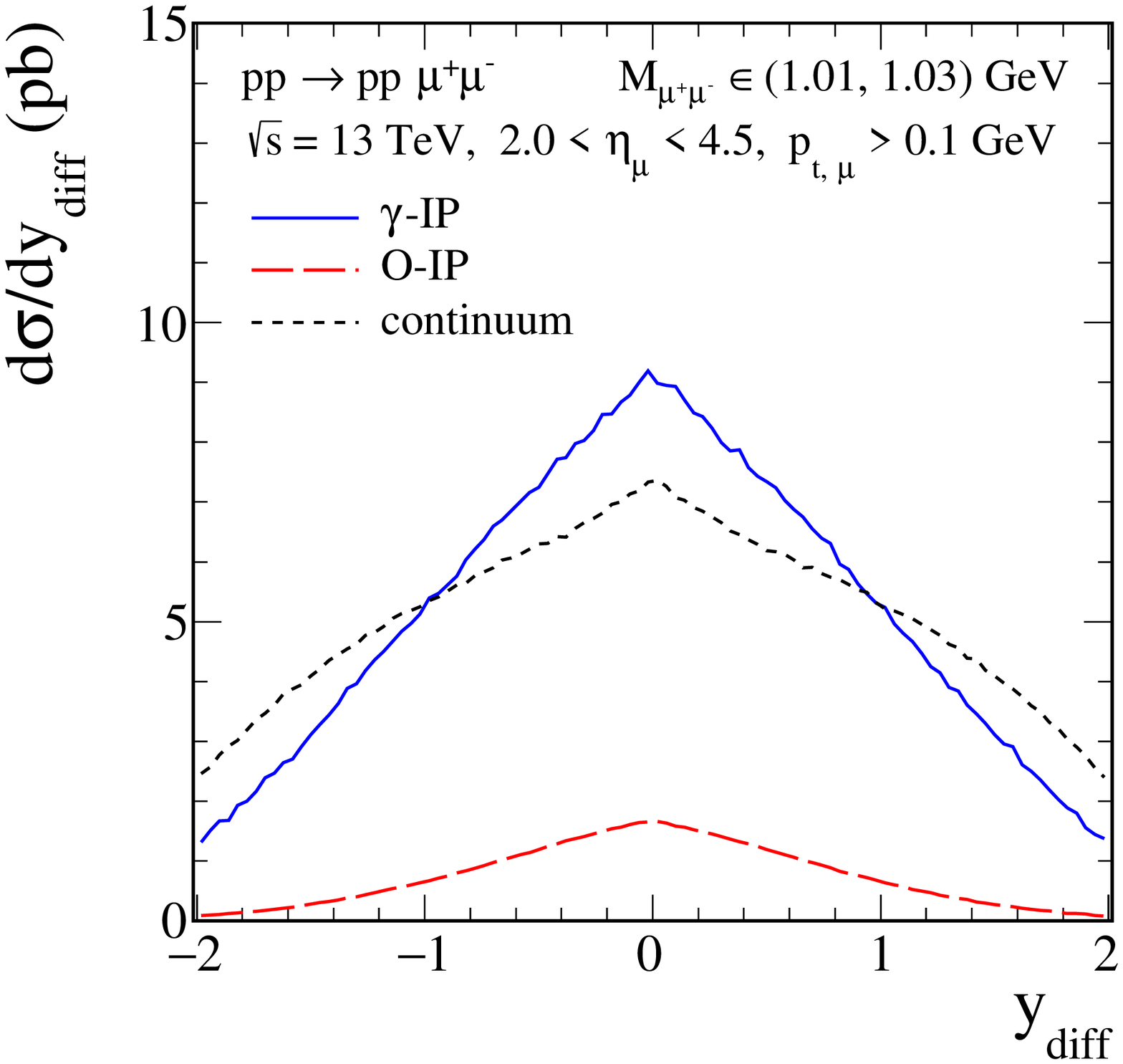}
\includegraphics[width=0.45\textwidth]{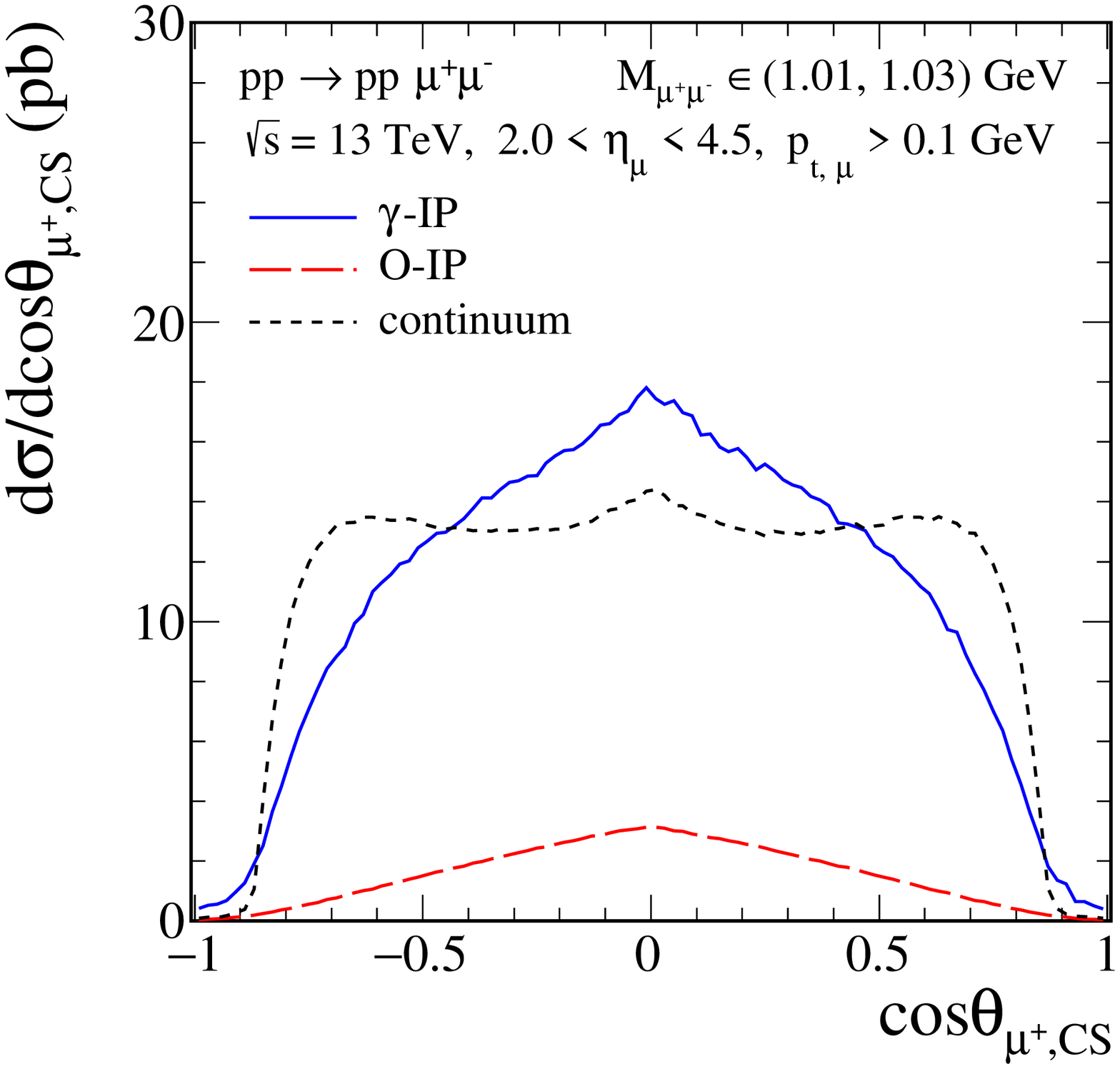}
\includegraphics[width=0.45\textwidth]{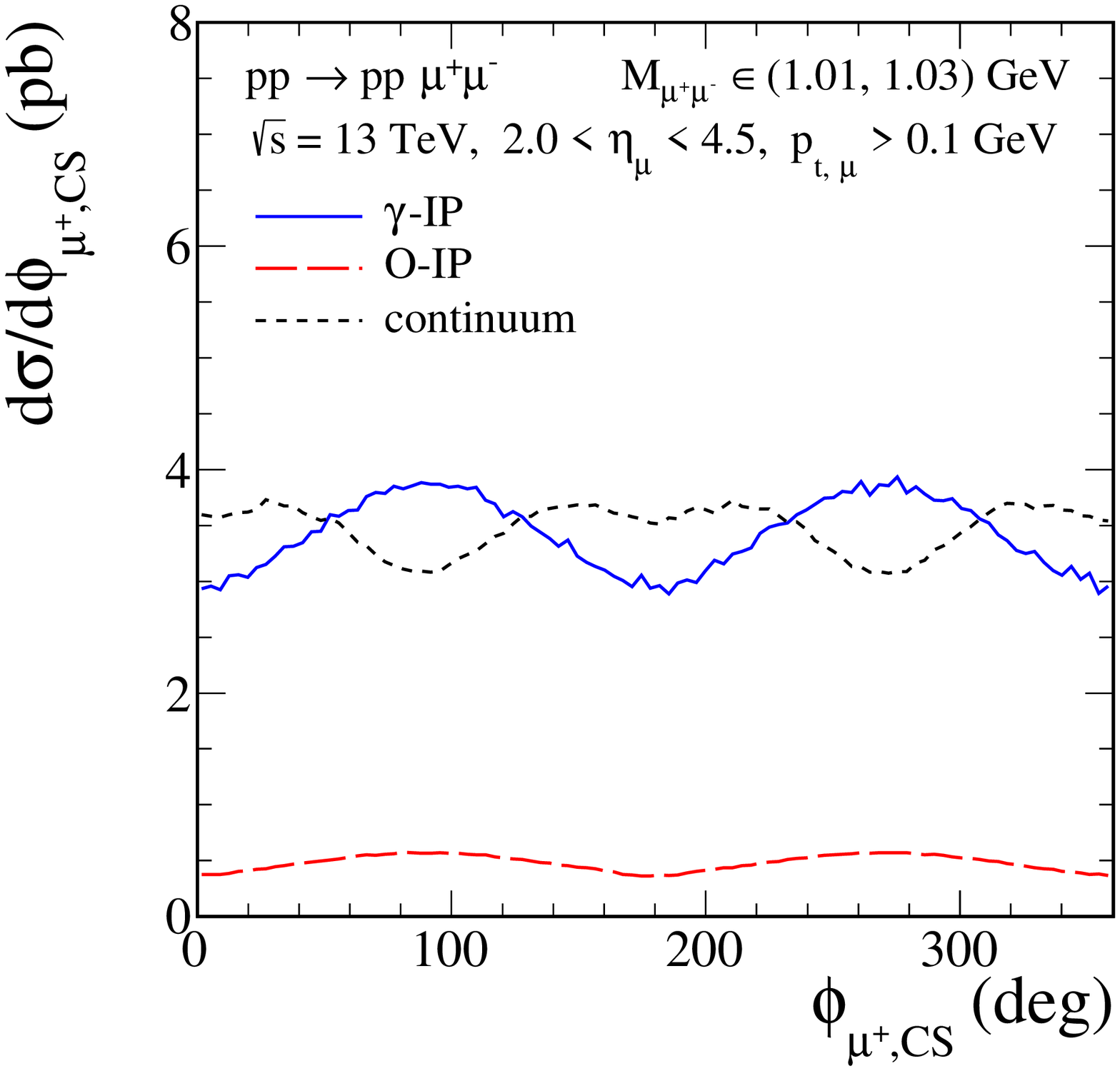}
\caption{\label{fig:LHCb_mumu_1}
The differential cross sections for the $pp \to pp \mu^{+}\mu^{-}$ reaction
in the dimuon invariant mass region $M_{\mu^{+}\mu^{-}} \in (1.01, 1.03)$~GeV.
Calculations were done for $\sqrt{s} = 13$~TeV, $2.0 < \eta_{\mu} < 4.5$, 
and $p_{t, \mu} > 0.1$~GeV.
The meaning of the lines is the same as in Fig.~\ref{fig:LHCb_mumu_0}.
We take the $\gamma$-$\Pom$- and $\Ode$-$\Pom$-fusion contributions
for the coupling parameters (\ref{photoproduction_setB}) 
and (\ref{parameters_ode_b}), respectively. 
The absorption effects are included here.}
\end{figure}
\begin{figure}[!ht]
\includegraphics[width=0.45\textwidth]{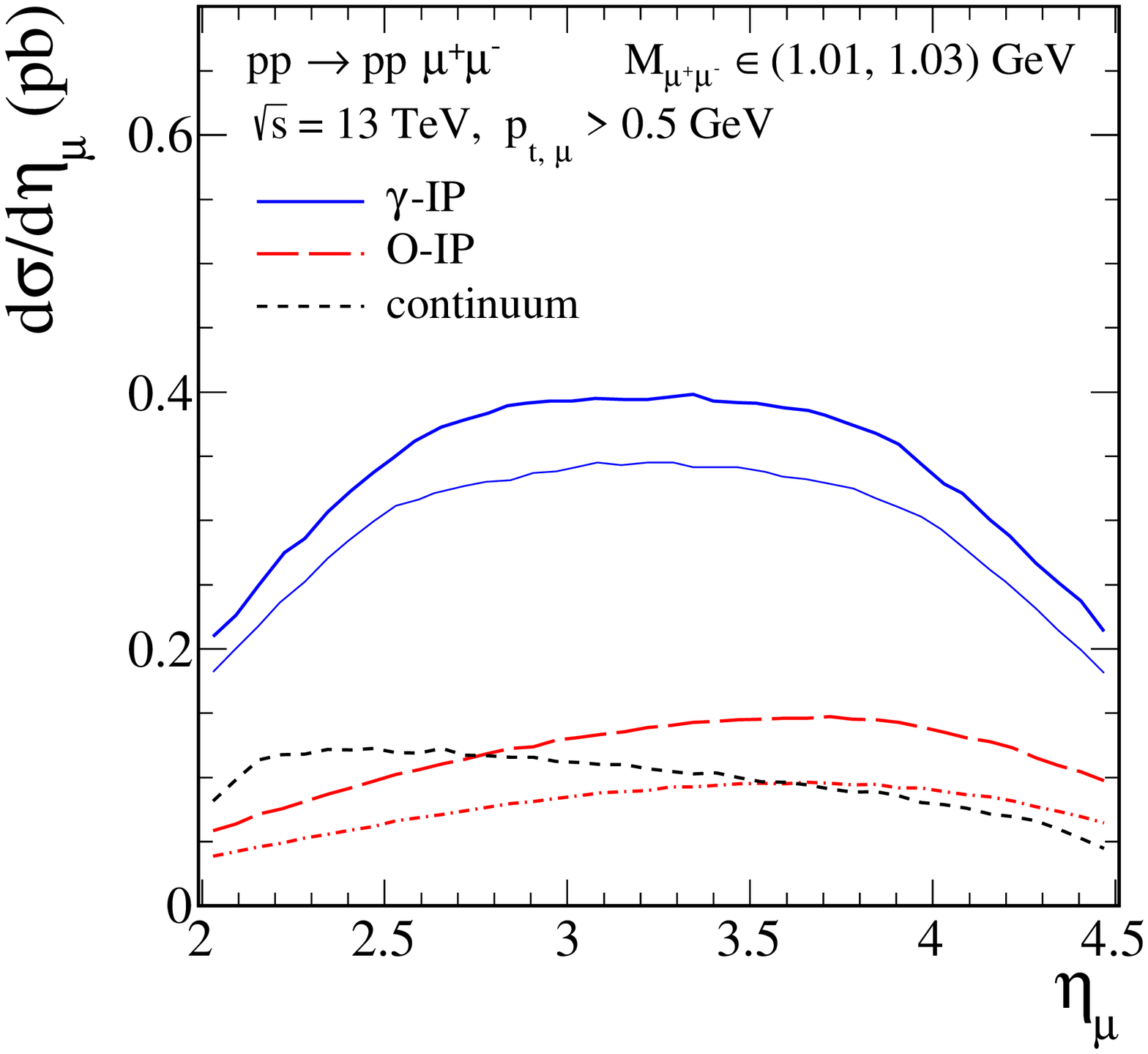}
\includegraphics[width=0.45\textwidth]{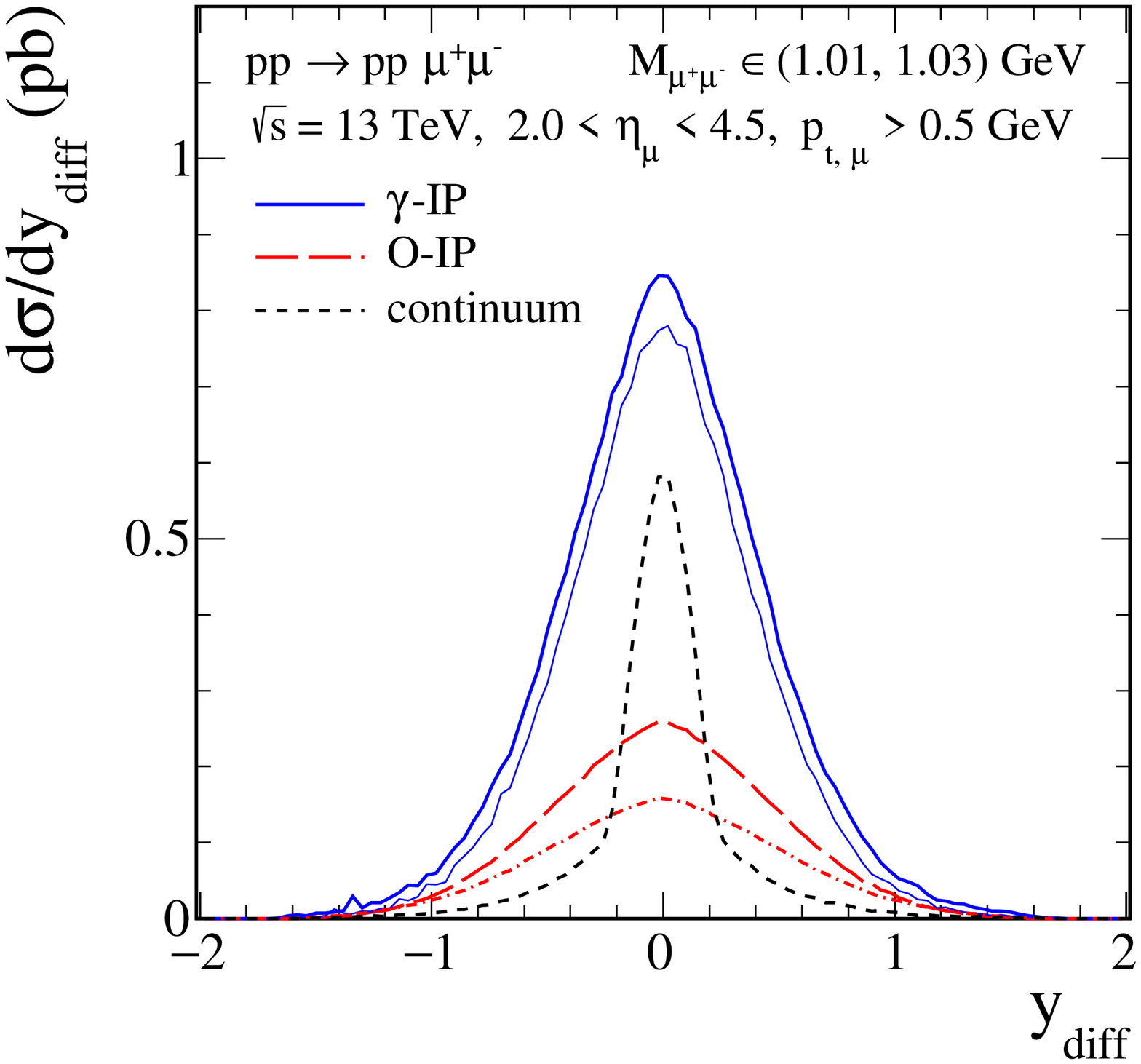}
\includegraphics[width=0.45\textwidth]{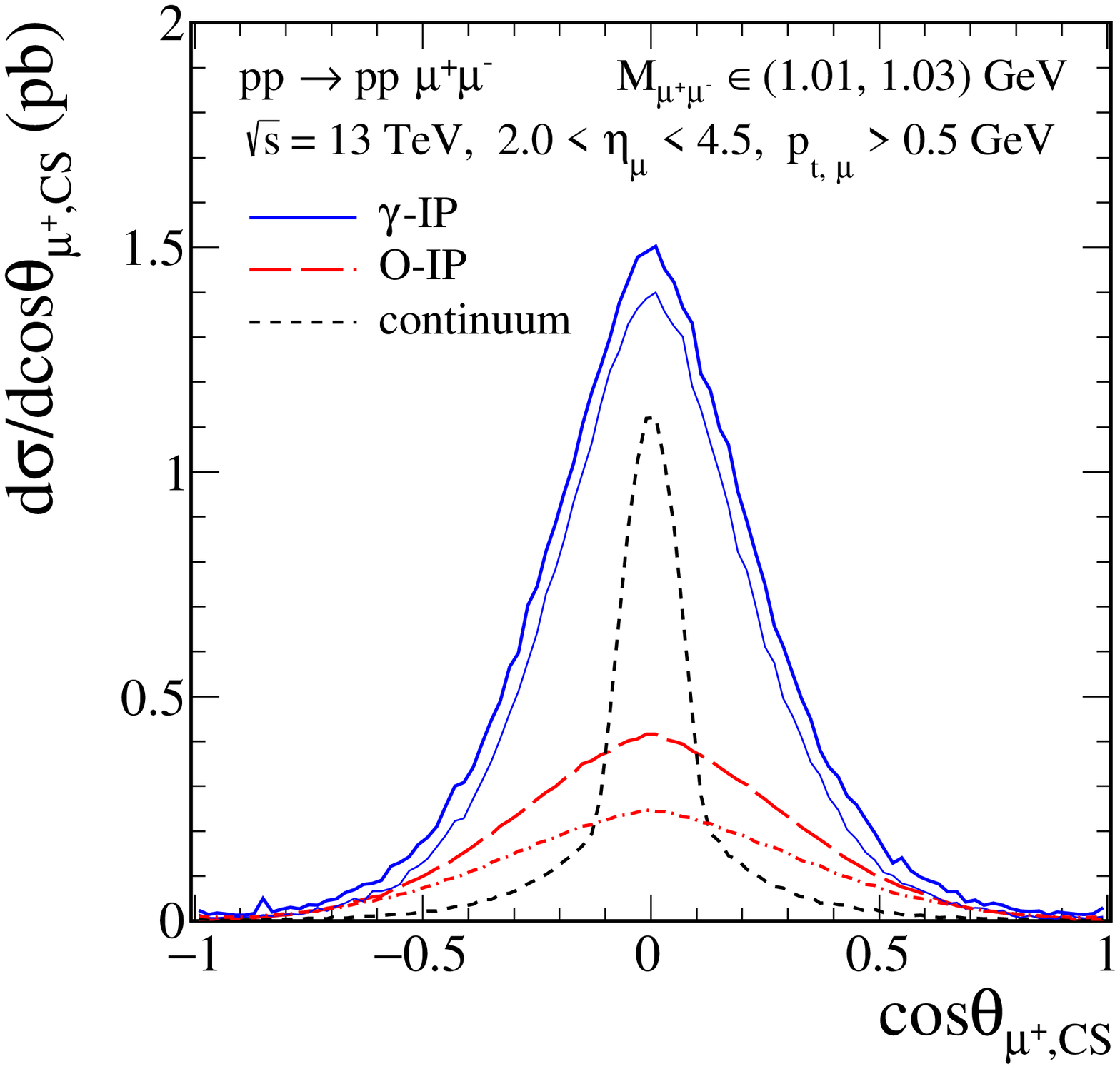}
\includegraphics[width=0.45\textwidth]{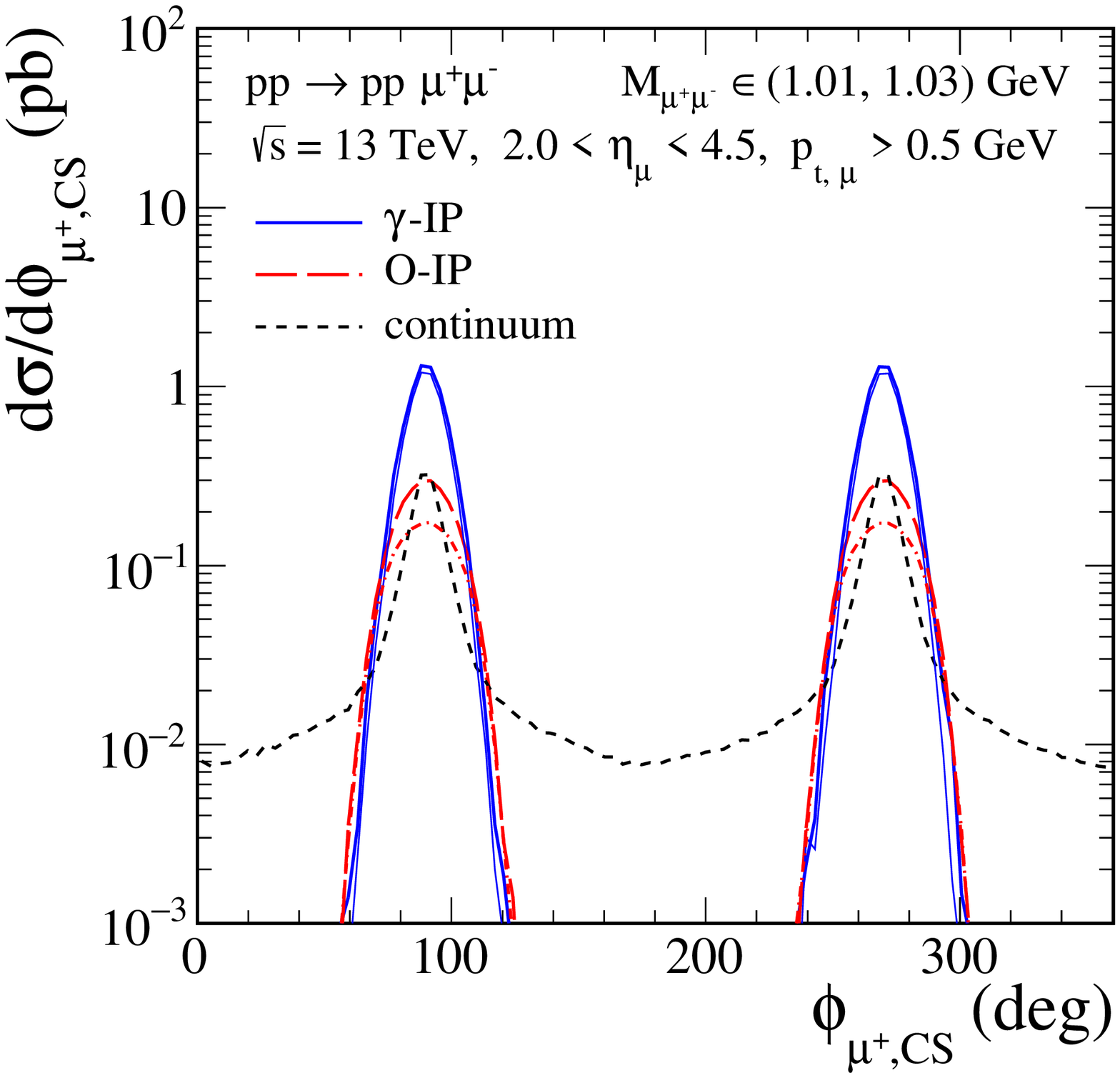}
\caption{\label{fig:LHCb_mumu_2}
The same as in Fig.~\ref{fig:LHCb_mumu_1} 
but for $p_{t, \mu} > 0.5$~GeV.
The upper blue solid line 
is for the $\gamma$-$\Pom$-fusion contribution 
for the parameter set~B (\ref{photoproduction_setB})
while the lower blue solid line is for set~A (\ref{photoproduction_setA}).
The red long-dashed line corresponds to 
the $\Ode$-$\Pom$-fusion contribution
with the parameters quoted in (\ref{parameters_ode}), (\ref{parameters_ode_lambda}), 
and (\ref{parameters_ode_b}),
the red dash-dotted line 
is for another choice of the $\Pom \Ode \phi$ coupling 
parameter (\ref{parameters_ode_a}). 
The black short-dashed line corresponds to the continuum contribution.
The absorption effects are included here.}
\end{figure}

In Fig.~\ref{fig:LHCb_mumu_3} we present the distributions 
in transverse momentum of the $\mu^{+}\mu^{-}$ pair.
We can see that the low-$p_{t,\mu^{+}\mu^{-}}$ cut can be helpful
to reduce the continuum ($\gamma \gamma \to \mu^{+}\mu^{-}$) 
and photon-pomeron-fusion contributions.
\begin{figure}[!ht]
\includegraphics[width=0.45\textwidth]{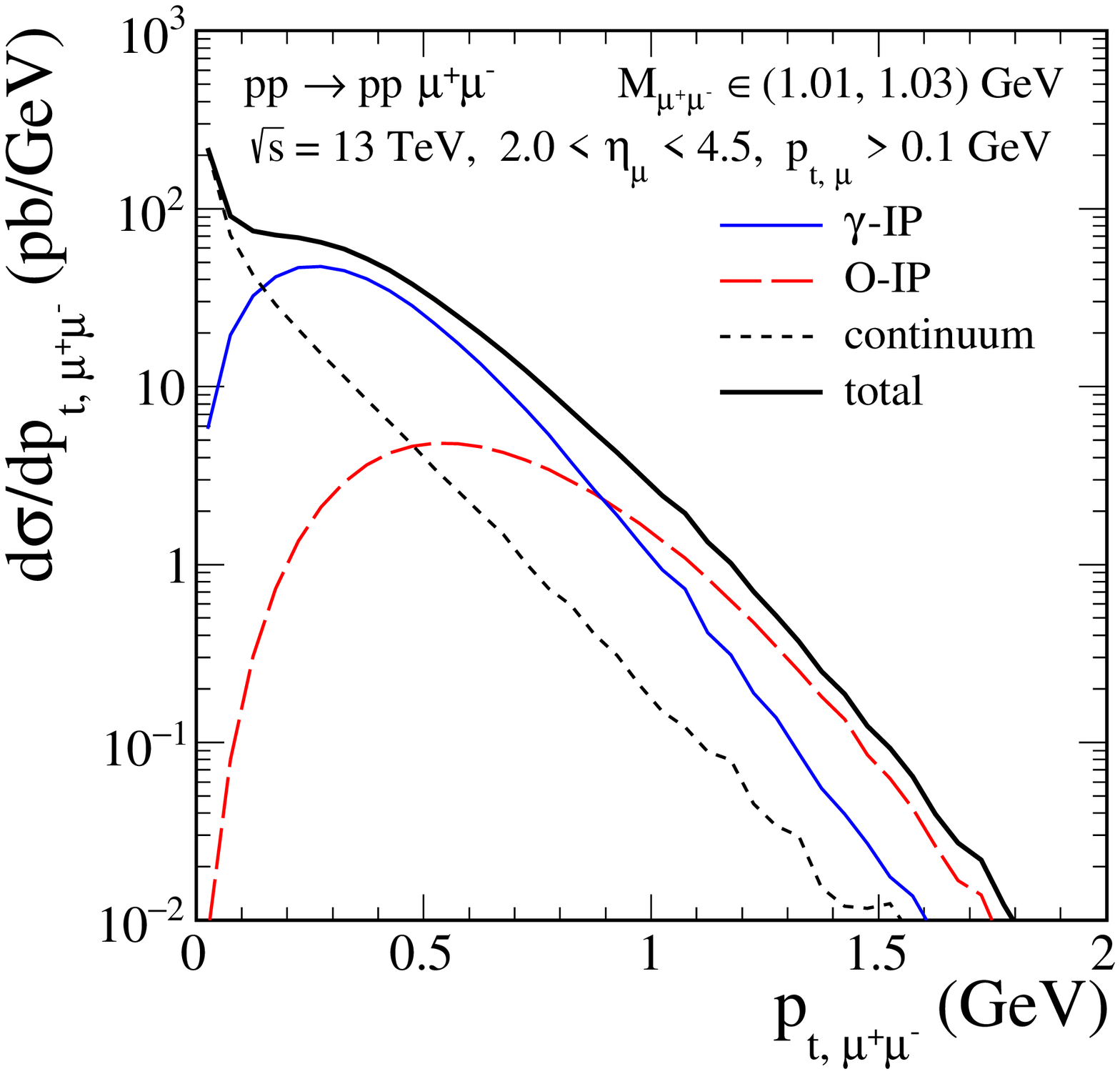}
\includegraphics[width=0.45\textwidth]{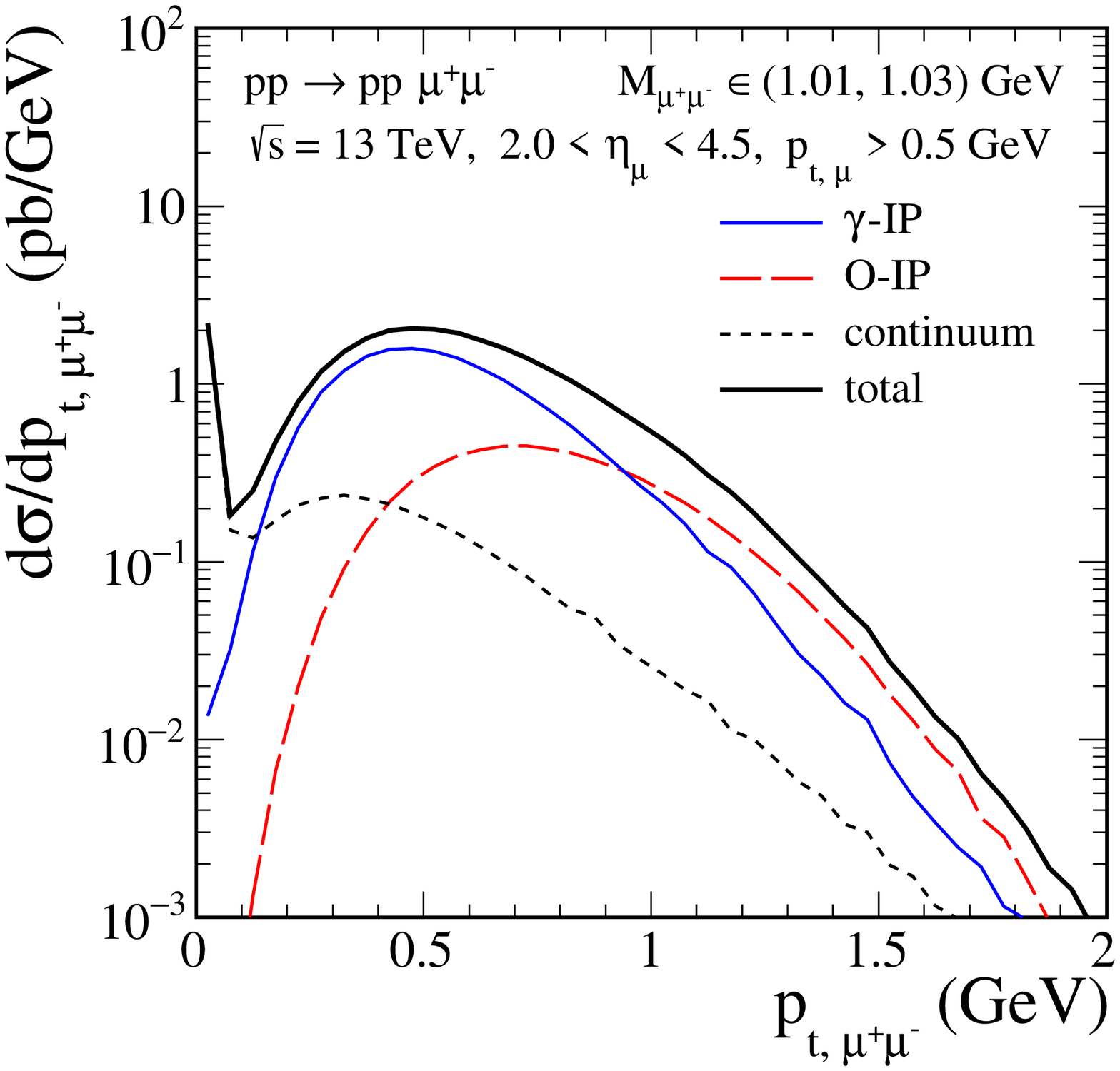}
\caption{\label{fig:LHCb_mumu_3}
The distributions in transverse momentum of the $\mu^{+}\mu^{-}$ pair
for the $pp \to pp \mu^{+}\mu^{-}$ reaction
in the dimuon invariant mass region $M_{\mu^{+}\mu^{-}} \in (1.01, 1.03)$~GeV.
Calculations were done for $\sqrt{s} = 13$~TeV, $2.0 < \eta_{\mu} < 4.5$
and for $p_{t, \mu} > 0.1$~GeV (left panel)
and for $p_{t, \mu} > 0.5$~GeV (right panel).
The meaning of the lines is the same as in Fig.~\ref{fig:LHCb_mumu_0} 
but here we added the coherent sum of all contributions
shown by the black solid line.
Here we take the $\gamma$-$\Pom$- and $\Ode$-$\Pom$-fusion contributions
for the coupling parameters (\ref{photoproduction_setB}) 
and (\ref{parameters_ode_b}), respectively.
The absorption effects are included here.}
\end{figure}

In Fig.~\ref{fig:LHCb_mumu_4} we show the results
when imposing in addition a cut $p_{t, \mu^{+}\mu^{-}} > 0.8$~GeV.
The $\gamma \gamma \to \mu^{+}\mu^{-}$ contribution is now very small.
We can see from the $\rm{y_{diff}}$ distribution that the photon-pomeron term gives
a broader distribution than the odderon-pomeron term. 
At $\rm{y_{diff}} = 0$ the odderon-exchange term 
is now bigger than the photoproduction terms.
\begin{figure}[!ht]
\includegraphics[width=0.45\textwidth]{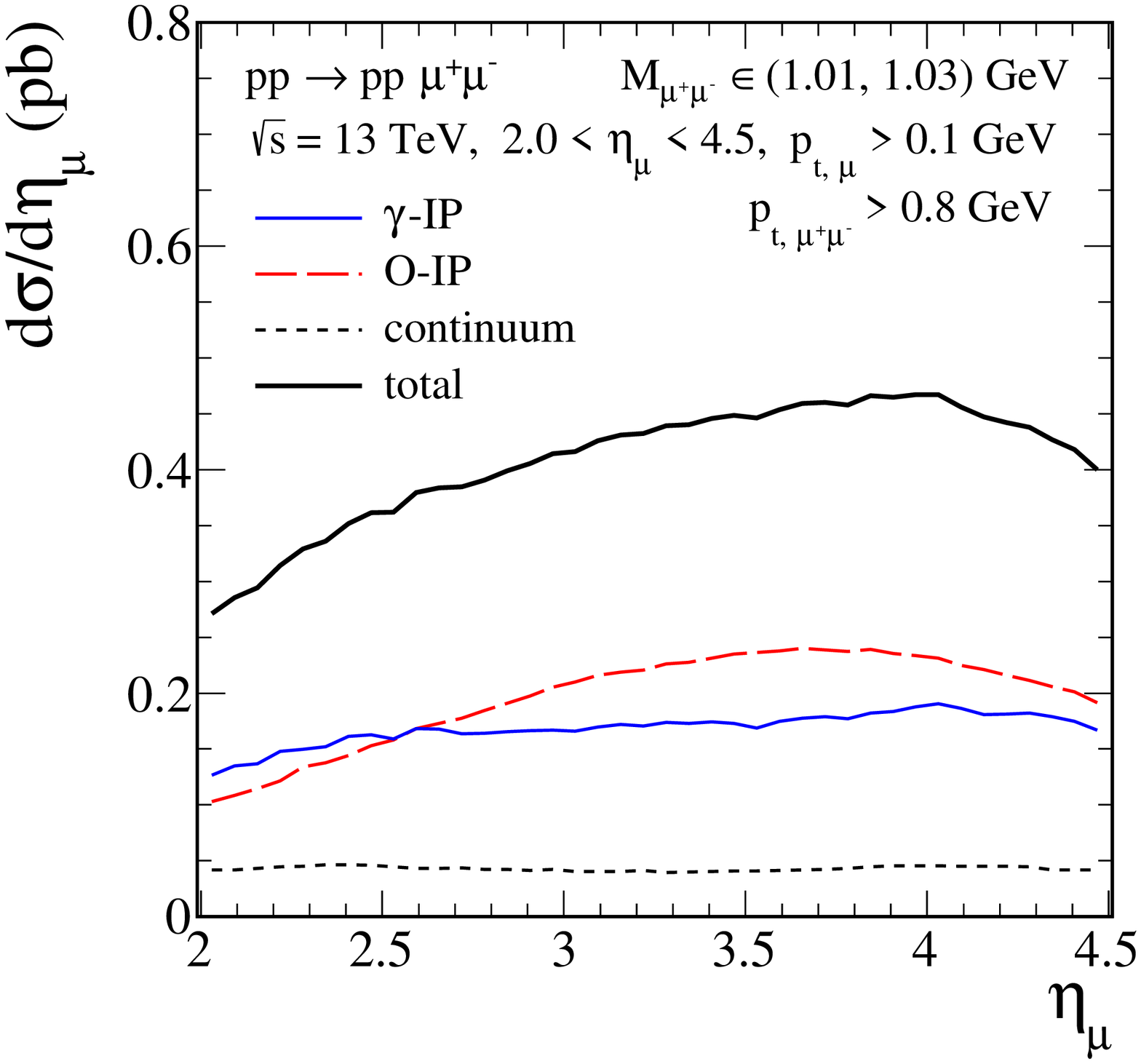}
\includegraphics[width=0.45\textwidth]{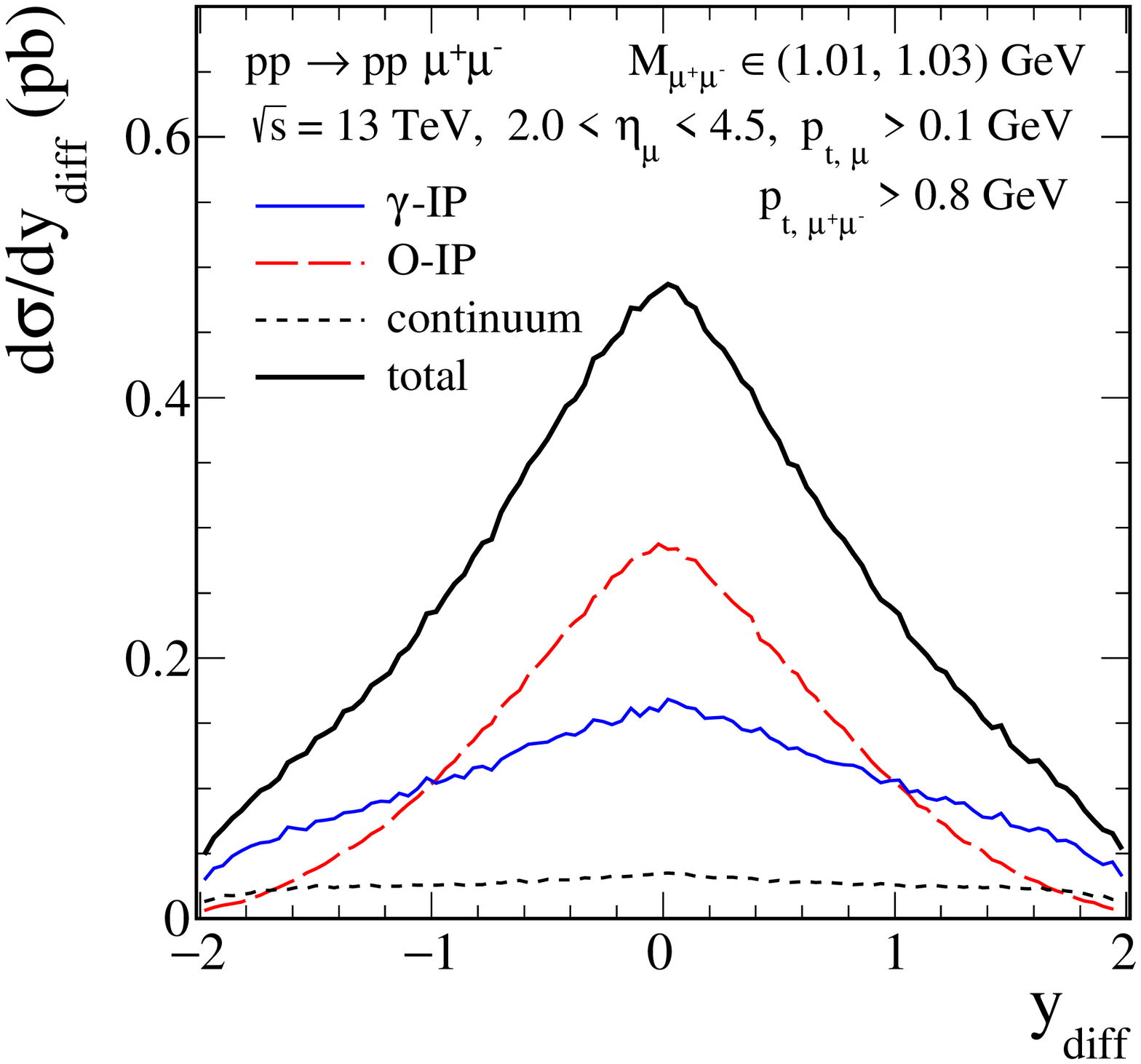}
\includegraphics[width=0.45\textwidth]{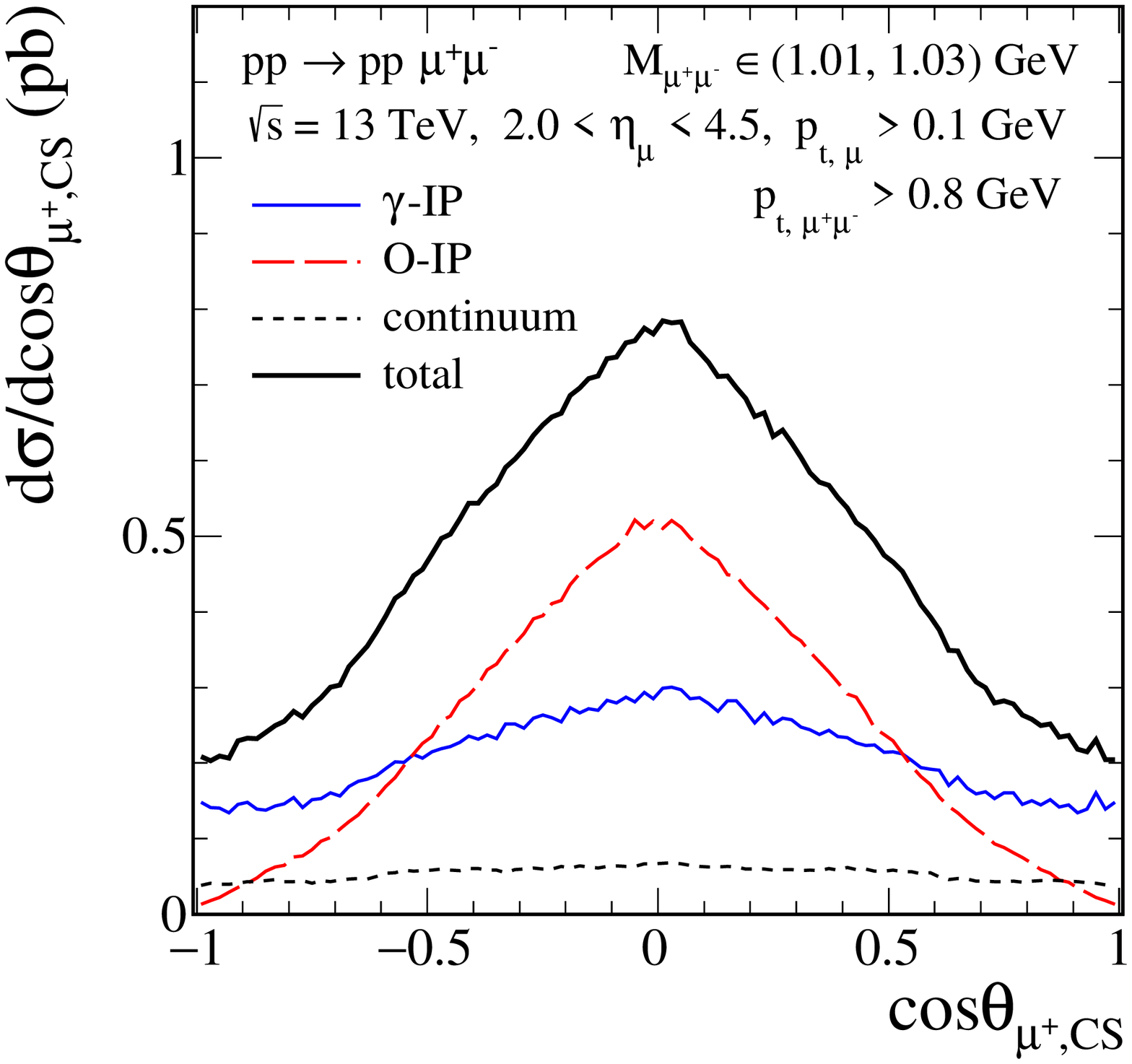}
\includegraphics[width=0.45\textwidth]{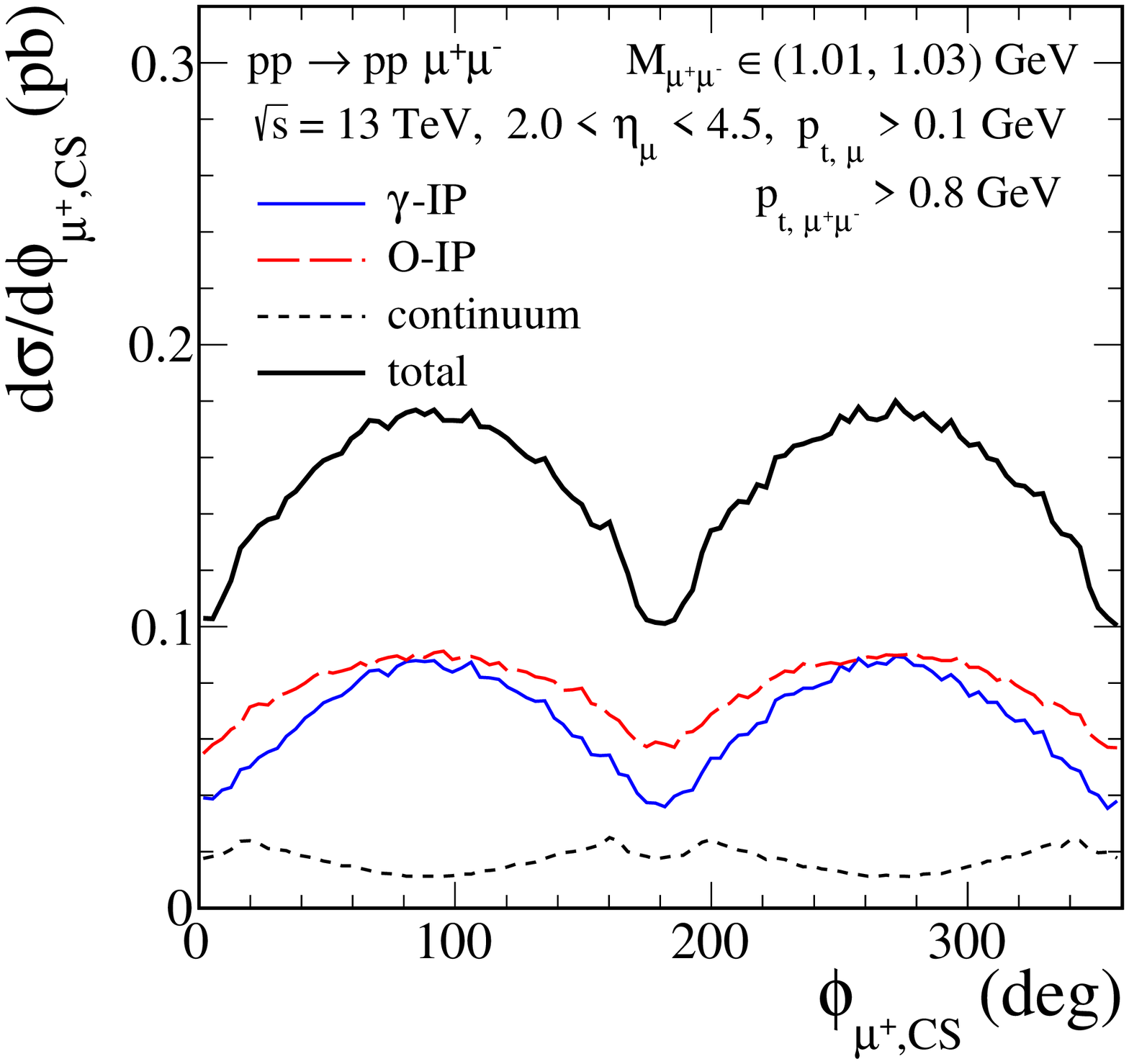}
\caption{\label{fig:LHCb_mumu_4}
The differential cross sections for the $pp \to pp \mu^{+}\mu^{-}$ reaction
in the dimuon invariant mass region $M_{\mu^{+}\mu^{-}} \in (1.01, 1.03)$~GeV.
Calculations were done for $\sqrt{s} = 13$~TeV, $2.0 < \eta_{\mu} < 4.5$, 
$p_{t, \mu} > 0.1$~GeV, and $p_{t, \mu^{+}\mu^{-}} > 0.8$~GeV.
The meaning of the lines is the same as 
in Fig.~\ref{fig:LHCb_mumu_3}.
The absorption effects are included here.}
\end{figure}

In Table~\ref{tab:table2} we have collected integrated cross sections 
in nb for $\sqrt{s} = 13$~TeV and with different experimental cuts
for the exclusive $pp \to ppK^{+}K^{-}$ and $pp \to pp \mu^{+}\mu^{-}$ reactions
including the $\gamma$-$\Pom$- and $\Ode$-$\Pom$-fusion processes separately.
We also show the results 
for the coherent sum of the $\gamma$-$\Pom$- and $\Ode$-$\Pom$-fusion
processes including absorption corrections.
Here we take for the $\gamma$-$\Pom$- and $\Ode$-$\Pom$-fusion contributions
the coupling parameters (\ref{photoproduction_setB}) and (\ref{parameters_ode_b}), respectively.
The ratios of full and Born cross sections 
$\langle S^{2} \rangle$ (the gap survival factors) are also presented.
We obtain $\langle S^{2} \rangle \simeq 0.2 - 0.3$ 
for the purely diffractive $\Ode$-$\Pom$ contribution.
For the $\gamma$-$\Pom$ contribution we find that 
$\langle S^{2} \rangle$
strongly depends on the cuts on the leading protons.
\begin{table}[]
\centering
\caption{The integrated cross sections in nb 
for the central exclusive production of single $\phi$ mesons in proton-proton collisions
with the subsequent decays $\phi \to K^{+}K^{-}$ or $\phi \to \mu^{+}\mu^{-}$.
The results have been calculated for $\sqrt{s}=13$~TeV
in the dikaon/dimuon invariant mass region \mbox{$M_{34} \in (1.01,1.03)$~GeV}
and for some typical experimental cuts.
We show results for the $\gamma$-$\Pom$-
and $\Ode$-$\Pom$-fusion contributions separately 
and for their coherent sum (``total'').
The ratios of full and Born cross sections 
$\langle S^{2} \rangle$ (the gap survival factors) 
are shown in the last column.}
\label{tab:table2}
\begin{tabular}{|l|c|c|c|c|}
\hline
Cuts & Contributions & $\sigma^{(\rm{Born})}$ (nb) & $\sigma^{(\rm{full})}$ (nb)  & $\langle S^{2} \rangle$ \\ 
\hline
$|\eta_{K}| < 2.5$, $p_{t, K} > 0.1$~GeV
&$\gamma$-$\Pom$ & 60.07 & 55.09 & 0.9 \\ 
&$\Ode$-$\Pom$ & 21.40 & \;\,6.44 & 0.3 \\ 
&total   & & 58.58 & \\ 
\hline
$|\eta_{K}| < 2.5$, $p_{t, K} > 0.1$~GeV, 
&$\gamma$-$\Pom$ & \;\,1.77 & \;\,0.52 & 0.3 \\ 
0.17~GeV $< |p_{y,1}|, |p_{y,2}| < 0.5$~GeV
&$\Ode$-$\Pom$  & \;\,2.91 & \;\,0.79 & 0.3 \\ 
&total & & \;\,0.93 & \\ 
\hline
$|\eta_{K}| < 2.5$, $p_{t, K} > 0.2$~GeV,
&$\gamma$-$\Pom$ & \;\,1.07 & \;\,0.24 & 0.2 \\ 
0.17~GeV $< |p_{y,1}|, |p_{y,2}| < 0.5$~GeV
&$\Ode$-$\Pom$   & \;\,2.10 & \;\,0.61 & 0.3 \\ 
&total    & & \;\,0.70 & \\ 
\hline
$|\eta_{K}| < 2.5$, $p_{t, K} > 0.5$~GeV,
&$\gamma$-$\Pom$ & \;\,$6.74 \times 10^{-3}$ & \;\,$0.76 \times 10^{-3}$ & 0.1 \\ 
0.17~GeV $< |p_{y,1}|, |p_{y,2}| < 0.5$~GeV
&$\Ode$-$\Pom$  & $87.94 \times 10^{-3}$ & $18.97 \times 10^{-3}$ & 0.2 \\ 
&total   & & $20.47 \times 10^{-3}$ & \\ 
\hline
$2.0 < \eta_{K} < 4.5$, $p_{t, K} > 0.1$~GeV
&$\gamma$-$\Pom$ & 43.18 & 40.07 & 0.9 \\ 
&$\Ode$-$\Pom$  & 16.73 & \;\,4.70 & 0.3 \\
&total   & & 43.28 & \\  
\hline
$2.0 < \eta_{K} < 4.5$, $p_{t, K} > 0.3$~GeV
&$\gamma$-$\Pom$ & \;\,3.09 & \;\,2.57 & 0.8 \\ 
&$\Ode$-$\Pom$ & \;\,6.57 & \;\,1.64 & 0.3 \\ 
&total   & & \;\,4.24 & \\ 
\hline
$2.0 < \eta_{K} < 4.5$, $p_{t, K} > 0.5$~GeV
&$\gamma$-$\Pom$ & $\;\,0.93 \times 10^{-1}$ & $\;\,0.66 \times 10^{-1}$ & 0.7 \\ 
&$\Ode$-$\Pom$ & \;\,0.88 & \;\,0.16 & 0.2 \\ 
&total   & & \;\,0.24 & \\ 
\hline\hline
$2.0 < \eta_{\mu} < 4.5$, $p_{t, \mu} > 0.1$~GeV
&$\gamma$-$\Pom$ & $23.93 \times 10^{-3}$ & $20.96 \times 10^{-3}$ & 0.9 \\ 
&$\Ode$-$\Pom$ & $10.06 \times 10^{-3}$ & $\;\,3.02 \times 10^{-3}$ & 0.3 \\ 
&total   & & $21.64 \times 10^{-3}$ & \\  
\hline
$2.0 < \eta_{\mu} < 4.5$, $p_{t, \mu} > 0.5$~GeV
&$\gamma$-$\Pom$& $\;\,1.21 \times 10^{-3}$ & $\;\,0.85 \times 10^{-3}$ & 0.7 \\ 
&$\Ode$-$\Pom$ & $\;\,1.49 \times 10^{-3}$ & $\;\,0.45 \times 10^{-3}$ & 0.2 \\ 
&total   & & $\;\,1.07 \times 10^{-3}$ & \\  
\hline
$2.0 < \eta_{\mu} < 4.5$, $p_{t, \mu} > 0.1$~GeV, 
&$\gamma$-$\Pom$ & $\;\,0.70 \times 10^{-3}$ & $\;\,0.41 \times 10^{-3}$ & 0.6 \\ 
$p_{t, \mu^{+}\mu^{-}} > 0.8$~GeV
&$\Ode$-$\Pom$& $\;\,2.46 \times 10^{-3}$ & $\;\,0.51 \times 10^{-3}$ & 0.2 \\ 
&total& & $\;\,0.91 \times 10^{-3}$ & \\ 
\hline
\end{tabular}
\end{table}

We close this section with a brief comment on the absorptive corrections 
in the nonperturbative (soft) diffractive and in pQCD processes.

The survival factor for the soft exclusive process
$pp \to pp \pi^{+}\pi^{-}$
via the pomeron-pomeron fusion for $\sqrt{s} = 7$~TeV
was calculated also in \cite{Ryutin:2019khx}.
From Fig.~14 of \cite{Ryutin:2019khx} 
we see that the survival factor
(only the $pp$ rescattering corrections)
is about $\langle S^2 \rangle = 0.2$.

In the perturbative case there is an additional factor for 
the gluon-gluon fusion vertex.
This factor suppresses the emission of virtual ``soft'' gluons
that could fill rapidity gaps (Sudakov-like suppression).
For ``hard'' pQCD processes at the LHC energies the expected
$\langle S^2 \rangle$ value is about 0.03 (or smaller);
see, e.g., \cite{Petrov:2007kn, Ryskin:2009tk, Harland-Lang:2013dia}.
Besides the effect of eikonal screening, 
there is some suppression caused by the rescatterings of the protons 
with the intermediate partons (inside the unintegrated gluon distribution).
This effect, 
neglected in the present calculations, 
is described by the so-called
enhanced reggeon diagrams and usually denoted as $S^{2}_{\rm enh}$.
The precise size of this effect is uncertain,
but due to the relatively large transverse momentum 
(and so smaller absorptive cross section) of the intermediate partons, 
it is only expected to reduce the corresponding CEP cross section 
by a factor of at most a ``few'', that is a much weaker suppression 
than in the case of $\langle S^2 \rangle$, the eikonal survival factor;
see, e.g., \cite{Ryskin:2009tk, Harland-Lang:2013dia}.

A similar method of calculation of the soft survival factor, 
$\langle S^2 \rangle$, as in our paper,
was used in the {\tt GRANITTI} Monte Carlo event generator \cite{Mieskolainen:2019jpv}.
For instance, for central exclusive $\pi^{+}\pi^{-}$ production
(via pomeron-pomeron fusion), 
denoted in Table~1 of \cite{Mieskolainen:2019jpv}
by $\pi^{+}\pi^{-}_{EL}$, 
the author gets $\langle S^2 \rangle \simeq 0.2$
at the LHC energies.
Note, that a much smaller $\langle S^2 \rangle = 0.06$ is obtained
in \cite{Mieskolainen:2019jpv} for a pQCD process,
production of a gluon pair $gg$ at $\sqrt{s} = 13$~TeV,
using the pQCD based Durham model.

Finally, we note that for the $\gamma \gamma$-fusion
processes the values of $\langle S^2 \rangle$ also 
depend on kinematic regions considered; see, e.g., 
\cite{Lebiedowicz:2018muq}.

\section{Conclusions}
\label{sec:conclusions}

In the present paper we have discussed the possibility to search
for odderon exchange in the $p p \to p p \phi$ reaction
with the $\phi$ meson observed in the $K^+ K^-$ or $\mu^{+}\mu^{-}$ channels. 
There are two basic processes: 
the relatively well known (at the Born level) photon-pomeron fusion
and the rather elusive odderon-pomeron fusion.
In our previous analysis on two $\phi$-meson production 
in proton-proton collisions \cite{Lebiedowicz:2019jru} 
we tried to tentatively (optimistically) fix the parameters of 
the pomeron-odderon-$\phi$ vertex to describe the relatively large
$\phi \phi$ invariant mass distribution measured 
by the WA102 Collaboration \cite{Barberis:1998bq}.
The calculation for the $p p \to p p \phi$ process requires in addition
knowledge of the rather poorly known coupling of the odderon to the proton.
The latter can be fixed, in principle, by a careful study of elastic
proton-proton scattering. The present estimates suggest
$\beta_{\Ode pp} \simeq 0.1\, \beta_{\Pom NN}$ [see Eq.~(\ref{beta_ONN})]. 
In the present study we therefore fixed the odderon coupling to the proton 
at this reasonable value and tried to make predictions
for central exclusive $\phi$-meson production. 
Our results also depend on the assumptions made 
for the Regge trajectory of the odderon, Eqs.~(\ref{A14}) and (\ref{A14_aux}).
In this context the photon-pomeron fusion is a background for 
the odderon-pomeron fusion. The parameters of photoproduction
were fixed to describe the HERA $\phi$-meson photoproduction data; 
see Appendices~\ref{sec:appendixA} and \ref{sec:appendixB}.
There, we pay special attention 
to the importance of the $\phi$-$\omega$ mixing effect 
in the description of the $\gamma p \to \phi p$ and $\gamma p \to \omega p$ reactions.
We would like to invite experimentalists to perform further studies
of these reactions both with still unanalysed HERA data
and data from ultraperipheral $Ap$ collisions.
This should include $\omega$ and $\phi$ polarisation studies
in order to get precise values for the relevant coupling parameters defined in Appendices~\ref{sec:appendixA} and \ref{sec:appendixB}.
To fix the parameters of the pomeron-odderon-$\phi$ vertex
(coupling constants and cutoff parameters)
we have considered several subleading contributions
and compared our theoretical predictions for the $pp \to pp \phi$ reaction
with the WA102 experimental data from \cite{Kirk:2000ws}.

Having fixed the parameters of the model we have 
made estimates of the integrated cross sections as well as shown
several differential distributions for $pp \to pp \phi$
at the WA102 energy $\sqrt{s} = 29.1$~GeV.
In addition we have discussed in detail exclusive production
of single $\phi$ mesons at the LHC, both in the $K^+ K^-$ and $\mu^+ \mu^-$
observation channels, for two possible distinct types of measurements:
(a) at midrapidity and without or with forward measurement of protons 
(relevant for ATLAS-ALFA or CMS-TOTEM),
(b) at forward rapidities and without measurement of protons (relevant for LHCb).
In contrast to low energies, where several processes may compete,
at the large LHC energies the odderon-exchange contribution competes only
with the photoproduction mechanism.
We have considered different dedicated observables.
Some of them seem to be promising.
The distributions in $\rm{y_{diff}}$ (rapidity difference between kaons)
and the angular distributions of kaons in the Collins-Soper frame seem
particularly interesting for the $K^+ K^-$ final state.
These angular distributions give information on the polarisation
state of the produced $\phi$ meson.
It is a main result of our paper that,
according to our odderon model, the polarisation of the $\phi$ and,
as a consequence, the angular distribution of the kaons in the Collins-Soper frame are very different for the $\gamma$-$\Pom$-
and $\Ode$-$\Pom$-fusion processes.
This should be a big asset for an odderon search.
Increasing the value of the cut on the transverse momenta of kaons improves the signal 
(pomeron-odderon fusion) to the background (photon-pomeron fusion) ratio.
Of course, in this way the rates are reduced; see Table~\ref{tab:table2}.
In general, the $\mu^+ \mu^-$ channel seems to be less promising
in identifying the odderon exchange. 
In this case detailed studies of shapes 
of $d\sigma/d\rm{y_{diff}}$ or/and $d\sigma/d\cos\theta_{\mu^{+},{\rm CS}}$ 
would be very useful in understanding the general situation.
To observe a sizeable deviation from photoproduction
a $p_{t,\mu^+ \mu^-} > 0.8$~GeV cut on the transverse momentum 
of the $\mu^+ \mu^-$ pair seems necessary. 
Such a cut reduces then the statistics of the measurement considerably.
A combined analysis of both the $K^+ K^-$ and the $\mu^+ \mu^-$ channels
should be the ultimate goal in searches for odderon exchange.
We are looking forward to first experimental 
results on single $\phi$ CEP at the LHC.

In summary, we have presented results for single $\phi$ CEP
both at the Born level as well as including
absorption effects in the eikonal approximation. 
We have argued that the WA102 experimental results at c.m. energy
$\sqrt{s} = 29.1$~GeV leave room for a possible odderon-exchange contribution there.
Then we have turned to LHC energies where single $\phi$ CEP can be studied by experiments
such as: ATLAS-ALFA, CMS-TOTEM, ALICE, and LHCb.
Using our results it should be possible to see experimentally 
if odderon effects as calculated are present,
if our odderon parameters have to be changed,
or if it is only possible to derive limits on the odderon parameters.
We are looking forward also to relevant data 
from the lower energy COMPASS experiment.
At high energies the deviations from the $\gamma$-$\Pom$-fusion 
contribution can be treated
as a signal of odderon exchange. 
In our opinion several distributions should be studied 
to draw a definite conclusion on the odderon exchange.
So far the odderon exchange was not unambiguously identified in any reaction. 
In the present paper we have shown that for the odderon search the study of
central exclusive production of single $\phi$ mesons is a valuable
addition and alternative to the study of elastic proton-proton scattering or
production of two $\phi$ mesons in the $p p \to p p \phi \phi$ reaction 
discussed by us very recently; see \cite{Lebiedowicz:2019jru}.
But the results of our paper are not limited to
the odderon search. We give in the Appendices~\ref{sec:appendixA} and \ref{sec:appendixB} also all the necessary formulas for the analyses
of $\omega$ and $\phi$ photoproduction
in the framework of our tensor-pomeron model.
We hope that experimentalists will perform such analysis
using both data from HERA and from ultraperipheral $Ap$
collisions at the LHC. Such results will then be very useful
to make refined predictions for $\phi$ CEP 
via the $\gamma$-$\Pom$ fusion. This process is not only
a background for an odderon search but also interesting by itself.

\appendix

\section{Off-diagonal diffractive $\omega \to \phi$ transition}
\label{sec:appendixA}

In the naive quark model the nucleon has no $s\bar{s}$ content,
whereas the $\phi$ meson is a pure $s\bar{s}$ state (ideal mixing of the vector mesons).
Thus, the coupling of the $\phi$ meson to the nucleon is expected to be very weak.
In practice there is a slight deviation from ideal mixing of the vector mesons,
which means that the $\phi$ meson has a small $u \bar{u} + d \bar{d}$ component.
Therefore, one should worry about diffractive off-diagonal
$\omega \to \phi$ transitions ($\omega$ strongly couples to the nucleon). 
We should consider the diagrams shown below in Fig.~\ref{fig:RP}.
How to treat the off-diagonal diffractive transitions due to pomeron exchange?

The physical states $\omega$ and $\phi$ are usually written in terms 
of flavour eigenstates $\omega_1$ and $\omega_8$ 
and the so-called mixing angle $\theta_V$
[see (B1) of \cite{Titov:1999eu}]
\begin{eqnarray}
\omega &=& \omega_8 \cos \theta_V + \omega_1 \sin \theta_V \,, \nonumber \\
-\phi   &=& -\omega_8 \sin \theta_V + \omega_1 \cos \theta_V \,,
\label{mixing}
\end{eqnarray}
where 
$\omega_1 = \frac{1}{\sqrt{3}}\left(u \bar{u} + d \bar{d} + s \bar{s} \right)$,
$\omega_8 = \frac{1}{\sqrt{6}}\left(u \bar{u} + d \bar{d} -2 s \bar{s} \right)$.
The mixing angle can be written as:
\begin{equation}
\theta_V = \theta_{V,i} - \Delta \theta_V \; .
\label{mixing_angle}
\end{equation}
The first component corresponds to the so-called ideal mixing angle
and the second one quantifies the deviation from the ideal mixing.
For the ideal mixing angle $\theta_{V,i}$ we have :
\begin{eqnarray}
&&\sin \theta_{V,i} = \sqrt{\frac{2}{3}}\,, \quad \cos \theta_{V,i} = \frac{1}{\sqrt{3}}\,, 
\quad \tan \theta_{V,i} = \sqrt{2}\,, \quad \theta_{V,i} = 54.74^{\circ}\,.
\label{mixing_aux}
\end{eqnarray}
Then it is easy to show, using (\ref{mixing_angle}) and (\ref{mixing_aux}), that:
\begin{eqnarray}
\sin \theta_V &=& \sqrt{\frac{2}{3}} \cos \Delta \theta_V 
               - \frac{1}{\sqrt{3}} \sin \Delta \theta_V \,, \nonumber \\
\cos \theta_V &=& \frac{1}{\sqrt{3}} \cos \Delta \theta_V
               + \sqrt{\frac{2}{3}} \sin \Delta \theta_V \,.
\label{auxilliary}
\end{eqnarray}

Inserting this in (\ref{mixing}) and defining
$\omega_0 = \frac{1}{\sqrt{2}} \left(u \bar u + d \bar d \right)$ 
and $\phi_0 = -s \bar s$ the mixing equation reads:
\begin{eqnarray}
\omega &=& \omega_{0} \cos(\Delta \theta_V) + \phi_{0} \sin(\Delta \theta_V) \,, \nonumber \\
\phi  &=& -\omega_{0} \sin(\Delta \theta_V) + \phi_{0} \cos(\Delta \theta_V) \,.
\label{mixing_V}
\end{eqnarray}
The reverse reads
\begin{eqnarray}
\omega_{0} &=& \omega \cos(\Delta \theta_V) - \phi \sin(\Delta \theta_V) \,, \nonumber \\
\phi_{0}  &=&  \omega \sin(\Delta \theta_V) + \phi \cos(\Delta \theta_V) \,.
\end{eqnarray}

It is well known that experimentally the angle $\Delta \theta_V$ is small.
Thus, the physical $\omega$ and $\phi$ 
are nearly equal to $\omega_{0}$ and $\phi_{0}$, respectively.

Now we consider the $\Pom \omega_{\Reg} \omega$, $\Pom \omega_{\Reg} \phi$,
$\Pom \phi_{\Reg} \omega$, and $\Pom \phi_{\Reg} \phi$ vertices
for which we assume a structure as in (\ref{vertex_pomphiphi}) 
with appropriate coupling constants $a$ and $b$.
In our case (CEP of $\phi$ meson in proton-proton collisions)
the $\omega_{\Reg}$ ($\omega$ reggeon) is, however, 
off-mass shell and we neglect the rather unknown mixing 
in this Regge-like state and include mixing in the on-shell $\phi$ only.
We shall argue, therefore, that 
in the $\Pom \omega_{\Reg} \omega$ and $\Pom \omega_{\Reg} \phi$ vertices 
only the $\omega_{0}$ will couple.
In this way we get for our coupling constants $a$ and $b$ 
\begin{eqnarray}
a_{\Pom \omega_{\Reg} \omega} &=& a_{\Pom \omega_{\Reg} \omega_0} \cos(\Delta \theta_V) \,, \nonumber \\
b_{\Pom \omega_{\Reg} \omega} &=& b_{\Pom \omega_{\Reg} \omega_0} \cos(\Delta \theta_V) \,; 
\label{transition_constants_ome}\\
a_{\Pom \omega_{\Reg} \phi} &=& -a_{\Pom \omega_{\Reg} \omega_0} \sin(\Delta \theta_V) \,, \nonumber \\
b_{\Pom \omega_{\Reg} \phi} &=& -b_{\Pom \omega_{\Reg} \omega_0} \sin(\Delta \theta_V) \,;
\label{transition_constants_phi}\\
\frac{a_{\Pom \omega_{\Reg} \phi}}{a_{\Pom \omega_{\Reg} \omega}} 
&=& -\tan(\Delta \theta_V) \,, \nonumber \\
\frac{b_{\Pom \omega_{\Reg} \phi}}{b_{\Pom \omega_{\Reg} \omega}} 
&=& -\tan(\Delta \theta_V) \,.
\label{transition_constants}
\end{eqnarray}

In an analogous way we shall assume that in the $\Pom \phi_{\Reg} \omega$
and $\Pom \phi_{\Reg} \phi$ vertices only the $\phi_{0}$ will couple.
This gives
\begin{eqnarray}
a_{\Pom \phi_{\Reg} \omega} &=& a_{\Pom \phi_{\Reg} \phi_0} \sin(\Delta \theta_V) \,, \nonumber \\
b_{\Pom \phi_{\Reg} \omega} &=& b_{\Pom \phi_{\Reg} \phi_0} \sin(\Delta \theta_V) \,; 
\label{transition_constants_ome2}\\
a_{\Pom \phi_{\Reg} \phi} &=& a_{\Pom \phi_{\Reg} \phi_0} \cos(\Delta \theta_V) \,, \nonumber \\
b_{\Pom \phi_{\Reg} \phi} &=& b_{\Pom \phi_{\Reg} \phi_0} \cos(\Delta \theta_V) \,;
\label{transition_constants_phi2}\\
\frac{a_{\Pom \phi_{\Reg} \omega}}{a_{\Pom \phi_{\Reg} \phi}} 
&=& \tan(\Delta \theta_V) \,, \nonumber \\
\frac{b_{\Pom \phi_{\Reg} \omega}}{b_{\Pom \phi_{\Reg} \phi}} 
&=& \tan(\Delta \theta_V) \,.
\label{transition_constants2}
\end{eqnarray}
In Sec.~\ref{sec:pp_ppKK} and in Appendix~\ref{sec:subleading}
we consider also 
the couplings of the pomeron to reggeized vector mesons and vector mesons.
In Appendix~\ref{sec:appendixB} below we need the couplings
of the pomeron to the off-shell vector mesons at $q^{2} = 0$
and the vector mesons.
We denote here, for clarity, these reggeized or off-shell mesons by $\widetilde{V}$.
In the following we shall assume that
\begin{eqnarray}
a_{\Pom \omega_{\Reg} \omega} &=& a_{\Pom \widetilde{\omega} \omega} 
= a_{\Pom \omega \omega}\,, \nonumber \\
a_{\Pom \omega_{\Reg} \phi} &=& a_{\Pom \widetilde{\omega} \phi} 
= -\tan(\Delta \theta_V) \,a_{\Pom \omega \omega}\,, \nonumber \\
b_{\Pom \omega_{\Reg} \omega} &=& b_{\Pom \widetilde{\omega} \omega} 
= b_{\Pom \omega \omega}\,, \nonumber \\
b_{\Pom \omega_{\Reg} \phi} &=& b_{\Pom \widetilde{\omega} \phi} 
= -\tan(\Delta \theta_V) \,b_{\Pom \omega \omega}\,;
\label{AA10a}\\
a_{\Pom \phi_{\Reg} \phi} &=& a_{\Pom \widetilde{\phi} \phi} 
= a_{\Pom \phi \phi}\,, \nonumber \\
a_{\Pom \phi_{\Reg} \omega} &=& a_{\Pom \widetilde{\phi} \omega} 
= \tan(\Delta \theta_V) \,a_{\Pom \phi \phi}\,, \nonumber \\
b_{\Pom \phi_{\Reg} \phi} &=& b_{\Pom \widetilde{\phi} \phi} 
= b_{\Pom \phi \phi}\,, \nonumber \\
b_{\Pom \phi_{\Reg} \omega} &=& b_{\Pom \widetilde{\phi} \omega} 
= \tan(\Delta \theta_V) \,b_{\Pom \phi \phi}\,.
\label{AA10b}
\end{eqnarray}
From (\ref{transition_constants_ome}) to (\ref{AA10b}) we obtain the coupling constants to be inserted in 
(\ref{amplitude_omeR_pomeron_aux}) and (\ref{amplitude_V_pomeron_aux}).

The deviation $\Delta \theta_V$ from the ideal mixing in (\ref{mixing_V}) 
can be estimated through the decay widths
of $\phi \to \pi^0 \gamma$ and $\omega \to \pi^0 \gamma$
($\pi^0$ is assumed not to have any $s \bar{s}$ component);
see Eq.~(B2) of \cite{Titov:1999eu}.
Using the most recent values from \cite{Tanabashi:2018oca} we have~\footnote{To calculate 
the coupling constants the expression (\ref{coupling_phigamPS}) was used; 
see (31) of \cite{Titov:1999eu}.}
\begin{eqnarray}
\frac{g_{\phi \gamma \pi^{0}}}{g_{\omega \gamma \pi^{0}}} = \frac{-0.137}{1.811} = -0.076 \,
\label{delta_theta_V}
\end{eqnarray}
and
$\Delta \theta_V = \arctan(0.076) = 4.35^{\circ}$.
In Refs.~\cite{Choi:1997iq,Kucukarslan:2006wk,Qian:2008px} a smaller value was found,
$\Delta \theta_V \simeq 3.7^{\circ}$.
In the following we shall use this latter value for $\Delta \theta_V$.

\section{Photoproduction of $\omega$ and $\phi$ mesons}
\label{sec:appendixB}

In order to estimate the coupling constants
$a_{\Pom \omega \omega}$ and $b_{\Pom \omega \omega}$ 
we consider the reaction $\gamma p \to \omega p$.
It is known, that in order to describe the intermediate $\gamma p$ energy region
we should include not only pomeron exchange 
but also subleading reggeon exchanges.
In Fig.~\ref{fig:gamp_omep_diagrams} we show the two diagrams
with diffractive exchanges which we shall take into account in our analysis.
\begin{figure}[!ht]
(a)\includegraphics[width=0.35\textwidth]{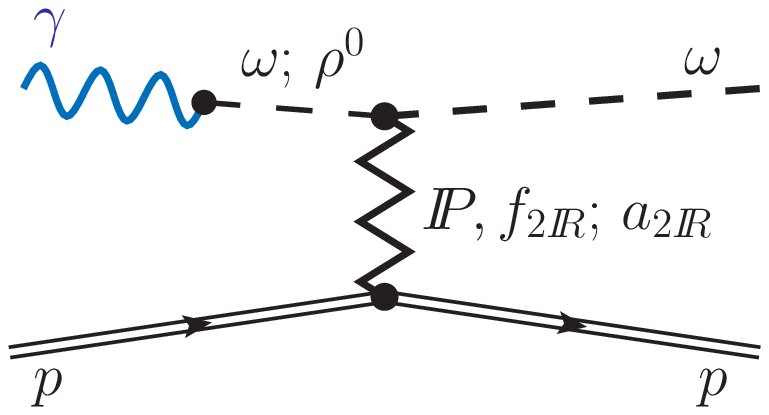}
(b)\includegraphics[width=0.35\textwidth]{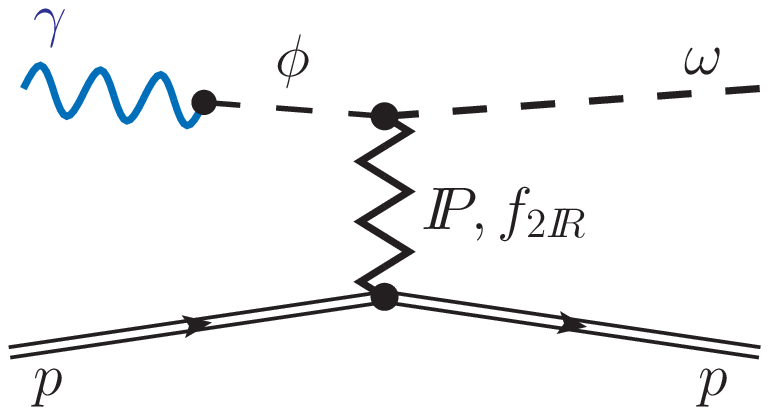}
  \caption{\label{fig:gamp_omep_diagrams}
  \small
Photoproduction of an $\omega$ meson (a) via pomeron and subleading reggeon exchanges,
and (b) as a result of $\phi$-$\omega$ mixing.}
\end{figure}
The diffractive amplitude for the $\gamma p \to \omega p$ reaction
represented by the diagram (a) of Fig.~\ref{fig:gamp_omep_diagrams}
can be treated analogously as for the $\gamma p \to \rho^{0} p$ reaction,
see Sec.~II and Eqs.~(2.9)--(2.11) of \cite{Lebiedowicz:2014bea},
but with the replacements:
$m_{\rho} \to m_{\omega}$, 
$\gamma_{\rho} \to \gamma_{\omega}$ (see (3.25) of \cite{Ewerz:2013kda}),
$a_{\Pom \rho \rho} \to a_{\Pom \omega \omega}$,
$b_{\Pom \rho \rho} \to b_{\Pom \omega \omega}$.
$a_{f_{2 \Reg} \rho \rho} \to a_{f_{2 \Reg} \omega \omega}$,
$b_{f_{2 \Reg} \rho \rho} \to b_{f_{2 \Reg} \omega \omega}$.
In our case ($\gamma p \to \omega p$) the $a_{2 \Reg}$-reggeon exchange
cannot be neglected due to the large value of the $\gamma$-$\rho^{0}$ coupling constant;
see (3.23)--(3.25) of \cite{Ewerz:2013kda}.
The propagators for $\Pom$, $f_{2 \Reg}$, and $a_{2 \Reg}$ will be taken
as in (3.10), and (3.12), respectively, of \cite{Ewerz:2013kda}.
The couplings of $\Pom$, $f_{2 \Reg}$, and $a_{2 \Reg}$ to the proton
will be taken according to (3.43), (3.49), and (3.51), respectively, of \cite{Ewerz:2013kda}.
Here, in analogy to $\gamma p \to \rho^{0} p$, we take
$\Lambda_{0}^{2} = 0.5$~GeV$^{2}$ in the form factor $F_{M}(t)$; 
see (2.11) of \cite{Lebiedowicz:2014bea} and (3.34) of \cite{Ewerz:2013kda}.
In Fig.~\ref{fig:gamp_omep_diagrams} the diagram (b) represents the $\phi$-$\omega$ mixing term
to the process $\gamma p \to \omega p$.
The procedure for determining the appropriate constants for this process is outlined below;
see Eqs.~(\ref{coupling_constant_pom_phi_ome_setA}), \ref{coupling_constant_pom_phi_ome_setB}).

In order to estimate the relevant coupling parameters we shall assume
that the $f_{2 \Reg} \omega \omega$ couplings are similar to the $f_{2 \Reg} \rho \rho$ ones.
Then we take the default values 
for the $f_{2 \Reg} \rho \rho$ and $a_{2 \Reg} \rho \omega$ couplings 
estimated from VMD in Sec.~7.2,
Eqs. (7.31), (7.32), (7.36), and (7.43), of \cite{Ewerz:2013kda}:
\begin{eqnarray}
&&a_{f_{2 \Reg} \omega \omega} \approx a_{f_{2 \Reg} \rho \rho} = 2.92\;{\rm GeV^{-3}} \,, \quad 
  b_{f_{2 \Reg} \omega \omega} \approx b_{f_{2 \Reg} \rho \rho} = 5.02\;{\rm GeV^{-1}}  \,,
\label{couplings_omega_reggeon_f2} \\
&&a_{a_{2 \Reg} \rho \omega} = 2.56\;{\rm GeV^{-3}} \,, \quad 
  b_{a_{2 \Reg} \rho \omega} = 4.68\;{\rm GeV^{-1}} \,.
\label{couplings_omega_reggeon_a2}
\end{eqnarray}
In (\ref{couplings_omega_reggeon_a2}) we assume that both coupling constants are positive.
To estimate the $\Pom \omega \omega$ coupling constants
we use the relation:
\begin{eqnarray}
2 m_{\omega}^{2} \, a_{\Pom \omega \omega} 
                  + b_{\Pom \omega \omega}
     = 4 \beta_{\Pom \pi \pi} = 7.04 \; \mathrm{GeV}^{-1} \,,
\label{omep_tot_opt_aux_pom}
\end{eqnarray}
in analogy to the corresponding one for the $\rho$ meson;
see (7.27) of \cite{Ewerz:2013kda} and (2.13) of \cite{Lebiedowicz:2014bea}.
Note that $a_{\Pom \omega \omega}$ must be positive in order to have a positive
$\omega p$ total cross section for all $\omega$ polarisations.
This follows from (7.21) of \cite{Ewerz:2013kda}
replacing there the $\rho$ by the $\omega$ meson.

In Fig.~\ref{fig:gamp_omep} we show the cross sections
for the $\gamma p \to \omega p$ reaction together with the experimental data.
From the comparison of our results to the experimental data,
taking first only the diagrams of Fig.~\ref{fig:gamp_omep_diagrams}~(a)
into account, we found
that even a small (and positive) value of the $a_{\Pom \omega \omega}$ coupling
leads to a reduction of the cross section. 
Therefore, for simplicity, we choose $a_{\Pom \omega \omega} = 0$ in (\ref{omep_tot_opt_aux_pom}).
The black solid line corresponds to the calculation
including only the terms shown in the diagram (a) of Fig.~\ref{fig:gamp_omep_diagrams}.
We used here the $\Pom \omega \omega$ coupling constants
\begin{eqnarray}
a_{\Pom \omega \omega} = 0 \,, \quad 
b_{\Pom \omega \omega} = 7.04\;{\rm GeV^{-1}} \,
\label{couplings_omega_pomeron}
\end{eqnarray}
and the parameters (\ref{couplings_omega_reggeon_f2}) and (\ref{couplings_omega_reggeon_a2})
for the reggeon exchanges.
We recall that for all exchanges participating in the diagram (a)
we take $\Lambda_{0}^{2} = 0.5$~GeV$^{2}$ in the form factor $F_{M}(t)$;
see (3.34) of \cite{Ewerz:2013kda}.

Now we include the off-diagonal terms from the diagram of Fig.~\ref{fig:gamp_omep_diagrams}~(b).
For estimating the coupling constants $a_{\Pom \widetilde{\phi} \omega}$ and
$b_{\Pom \widetilde{\phi} \omega}$ we use (\ref{AA10b})
and the determination of $a_{\Pom \phi \phi}$ and $b_{\Pom \phi \phi}$
from the discussion of the $\gamma p \to \phi p$ reaction below.
We get with the sets~A and B, respectively, with $\Delta \theta_V = 3.7^{\circ}$
\begin{eqnarray}
&&{\rm set\,A}: \;a_{\Pom \widetilde{\phi} \omega} = 0.05\;{\rm GeV^{-3}} \,, \quad 
                  b_{\Pom \widetilde{\phi} \omega} = 0.23\;{\rm GeV^{-1}} \,,\quad
\Lambda_{0,\,\Pom \widetilde{\phi} \omega}^{2} = 1.0\;{\rm GeV^{2}};
\label{coupling_constant_pom_phi_ome_setA} \\
&&{\rm set\,B}: \;a_{\Pom \widetilde{\phi} \omega} = 0.07\;{\rm GeV^{-3}} \,, \quad 
                  b_{\Pom \widetilde{\phi} \omega} = 0.19\;{\rm GeV^{-1}} \,,\quad
\Lambda_{0,\,\Pom \widetilde{\phi} \omega}^{2} = 4.0\;{\rm GeV^{2}}.
\label{coupling_constant_pom_phi_ome_setB}
\end{eqnarray}
In a similar way the coupling parameters for $f_{2 \Reg}$ exchange,
$a_{f_{2 \Reg} \widetilde{\phi} \omega}$ and
$b_{f_{2 \Reg} \widetilde{\phi} \omega}$, can be obtained.
However, the $f_{2 \Reg} \phi \phi$ couplings are expected to be very small.
In practice, we do not consider an $f_{2 \Reg}$-exchange contribution from
the diagram of Fig.~\ref{fig:gamp_phip_diagrams}~(a) below.
Here, we neglect also the $f_{2 \Reg}$ exchange from 
the diagram of Fig.~\ref{fig:gamp_omep_diagrams}~(b).

The blue solid line in Fig.~\ref{fig:gamp_omep} 
corresponds to the calculation 
including in addition to the processes from diagram (a) of Fig.~\ref{fig:gamp_omep_diagrams}
the $\phi$-$\omega$ mixing effect for the $\Pom$ exchange
[see diagram (b) of Fig.~\ref{fig:gamp_omep_diagrams}].
Our model calculation describes the total cross section 
fairly well~\footnote{A slight mismatch of our complete result 
with the ZEUS data may be due to the fact 
that the formula given by Eq.~(\ref{omep_tot_opt_aux_pom}),
assuming that at high energies the total cross section
for transversely polarised $\omega$ mesons equals 
the average of the $\pi^{\pm} p$ cross sections,
is an approximate relation.}
for energies ${\rm W}_{\gamma p} > 10$~GeV.
At low $\gamma p$ energies there are other processes contributing,
such as the $\pi^{0}$-meson exchange, and the $\omega$ bremsstrahlung;
see, e.g., \cite{Cisek:2011vt,Yu:2017vvp} for reviews and details concerning 
the exclusive $\omega$ production.
We nicely describe also the differential cross section $d\sigma/d|t|$.
We have checked that the complete results including
the $\phi$-$\omega$ mixing effect
with sets~A (\ref{coupling_constant_pom_phi_ome_setA})
and B (\ref{coupling_constant_pom_phi_ome_setB}) 
differ only marginally.
\begin{figure}[!ht]
\includegraphics[width=0.49\textwidth]{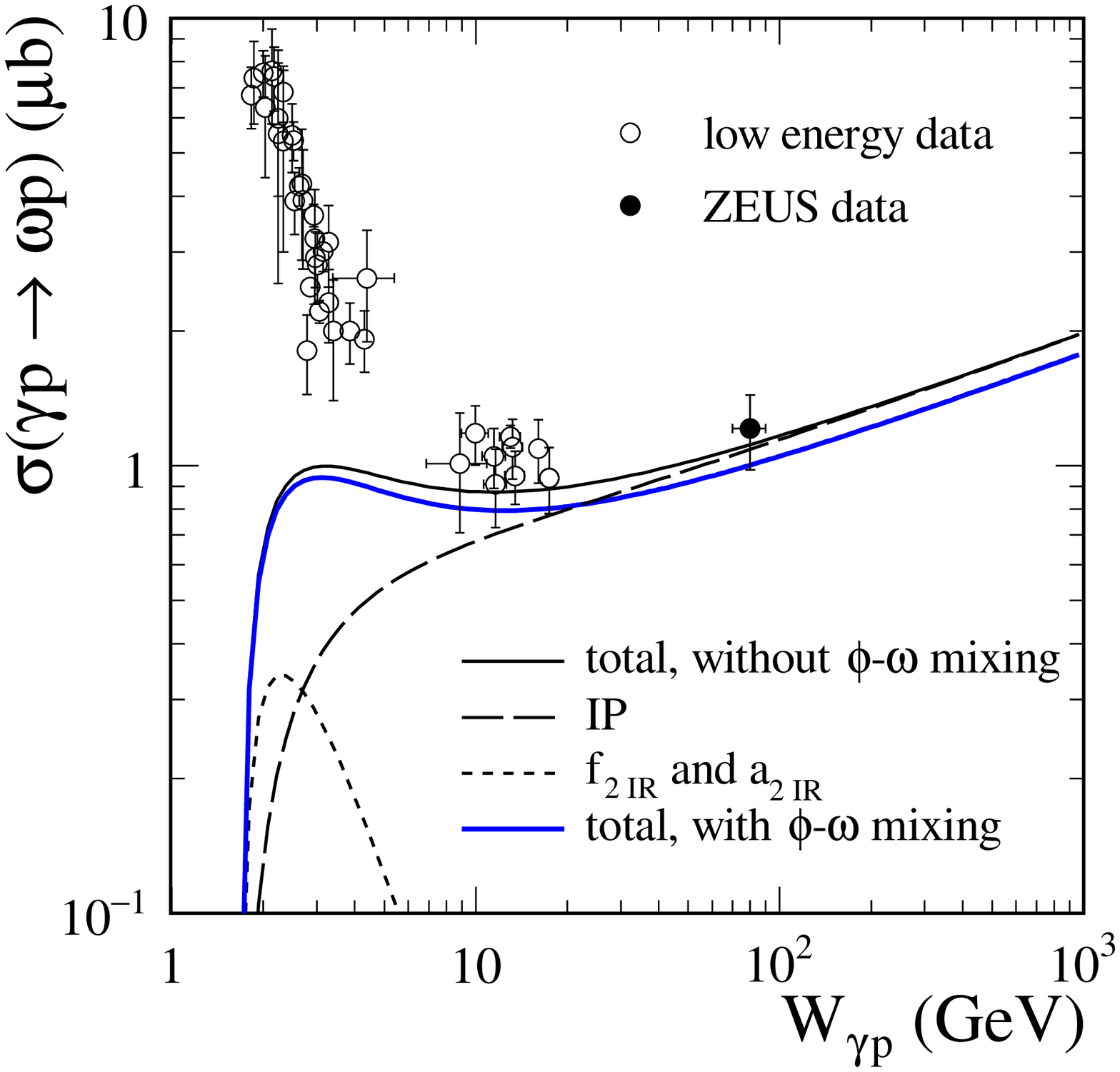}
\includegraphics[width=0.49\textwidth]{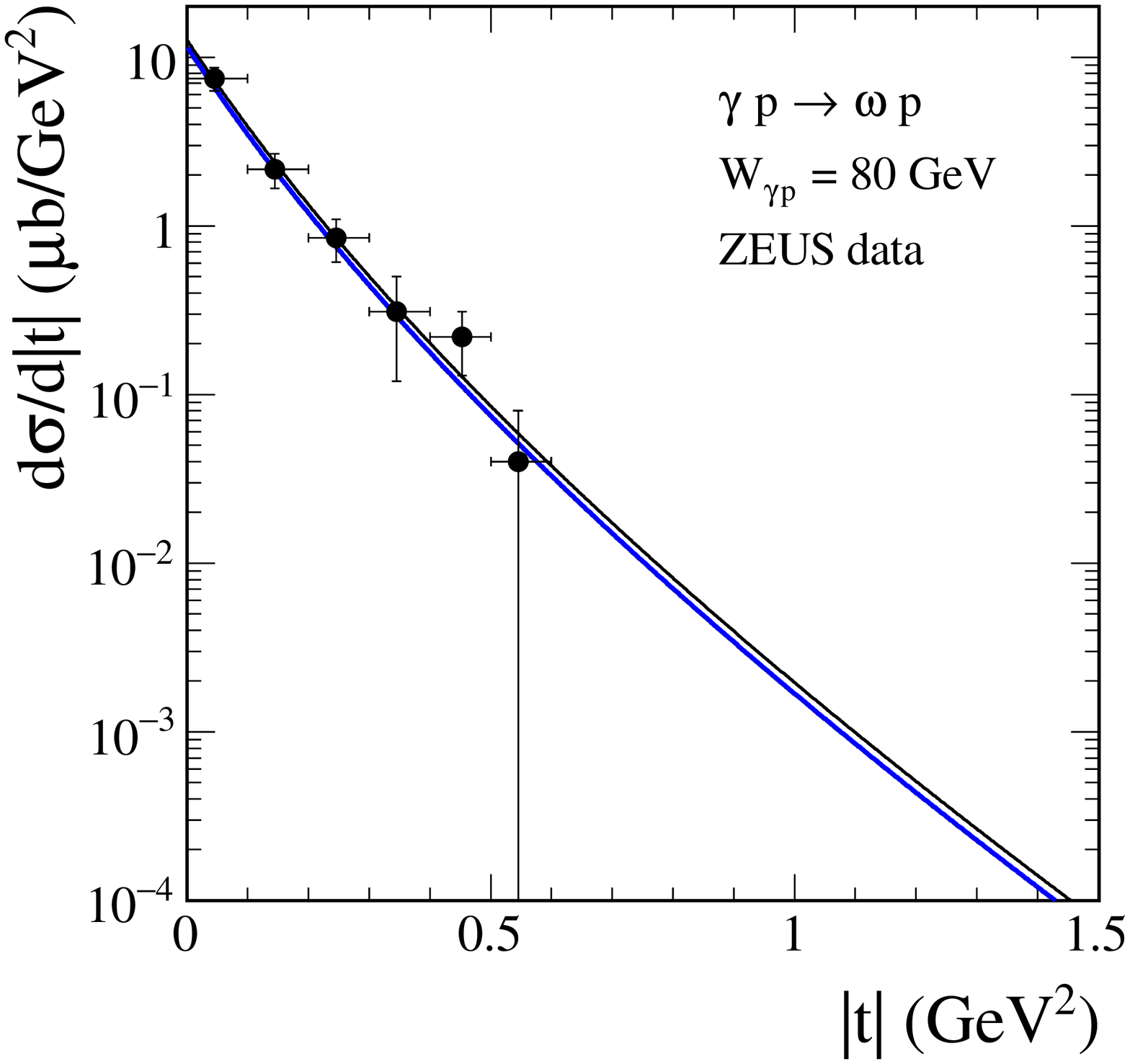}
\caption{\label{fig:gamp_omep}
\small
Left panel: The elastic $\omega$ photoproduction cross section
as a function of the center-of-mass energy ${\rm W}_{\gamma p}$.
Our results are compared with the ZEUS data \cite{Derrick:1996yt}
(at $\gamma p$ average c.m. energy $\langle {\rm W}_{\gamma p}\rangle = 80~{\rm GeV}$)
and with a compilation of low-energy experimental data (open circles;
see the caption of Fig.~2 of \cite{Cisek:2011vt} for more references).
The black solid line corresponds to results with 
both the pomeron and reggeon ($f_{2 \Reg}$, $a_{2 \Reg}$) exchanges.
The black long-dashed line corresponds to the pomeron exchange alone
while the black short-dashed line corresponds to the reggeon term.
In the calculation we used the parameters of the coupling constants 
given by (\ref{couplings_omega_reggeon_f2}), (\ref{couplings_omega_reggeon_a2}),
and (\ref{couplings_omega_pomeron}).
The blue solid line corresponds to the complete result
including the $\phi$-$\omega$ mixing effect (for the $\Pom$ exchange)
with the parameter set~A (\ref{coupling_constant_pom_phi_ome_setA}).
Right panel: The differential cross section 
for the $\gamma p \to \omega p$ reaction at ${\rm W}_{\gamma p} = 80~{\rm GeV}$.
Our complete results, 
without (the black line) and with (the blue line) the mixing effect, 
are compared to the ZEUS data \cite{Derrick:1996yt}.}
\end{figure}

\begin{figure}[!ht]
(a)\includegraphics[width=0.35\textwidth]{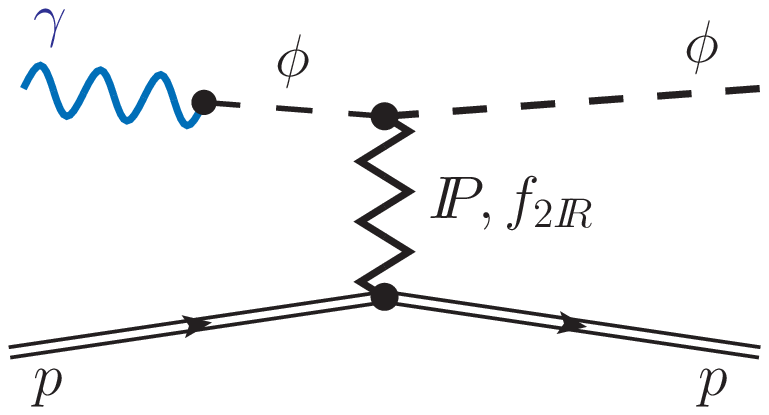}
(b)\includegraphics[width=0.35\textwidth]{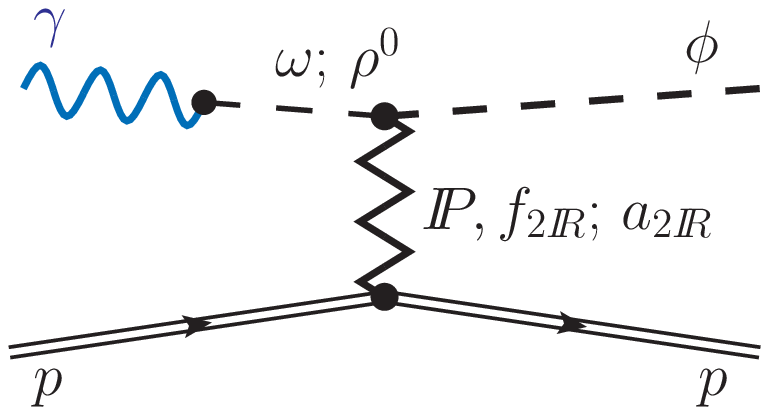}
  \caption{\label{fig:gamp_phip_diagrams}
  \small
Photoproduction of a $\phi$ meson 
(a) via pomeron and subleading $f_{2 \Reg}$ exchanges,
and (b) as a result of $\omega$-$\phi$ mixing.}
\end{figure}

Next, we discuss the $\gamma p \to \phi p$ reaction.
At high $\gamma p$ energies the pomeron exchange contribution,
shown by the diagram (a) of Fig.~\ref{fig:gamp_phip_diagrams}, is the dominant one;
see Sec.~IV~B of \cite{Lebiedowicz:2018eui}.
As was mentioned in Sec.~\ref{sec:intro},
in the low-energy region the corresponding production mechanism is not well established yet.
There the nondiffractive processes of the pseudoscalar 
$\pi^{0}$- and $\eta$-meson exchange are known to contribute
and are not negligible due to constructive $\eta$-$\pi^{0}$ interference; 
see, e.g., \cite{Titov:1999eu, Titov:2003bk}.
In addition, many other processes, e.g., 
direct $\phi$ meson radiation via the $s$- and $u$-channel proton exchanges
\cite{Titov:1999eu, Kim:2019kef},
$s \bar{s}$-cluster knockout \cite{Oh:2001bq},
$t$-channel $\sigma$-, $f_{2}(1270)$- and $f_{1}(1285)$-exchanges \cite{Kong:2016scm} were considered.
In \cite{Kong:2016scm} no vertex form factors were taken into account 
for the reggeized meson exchange contributions
and instead of the $f_{2}(1270)$-exchange 
there one should consider $f_{2}'$-exchange with appropriate parameters.
However, a peak in the differential cross sections 
$(d\sigma/dt)_{t = t_{\rm min}}$ at forward angles around $E_{\gamma} \sim 2$~GeV
(${\rm W}_{\gamma p} \sim 2.3$~GeV) observed by
the LEPS~\cite{Mibe:2005er, Mizutani:2017wpg} 
and CLAS~\cite{Dey:2014tfa} collaborations
cannot be explained by the processes mentioned above.
To explain the near-threshold bump structure
the authors of \cite{Kiswandhi:2010ub, Kiswandhi:2011cq, Kim:2019kef}
propose to include exchanges with the excitation of nucleon resonances.
In \cite{Ozaki:2009mj, Ryu:2012tw} 
another explanation, using the coupled-channel contributions 
with the $\Lambda(1520)$ resonance, was investigated.
In \cite{Ryu:2012tw} the hadronic box diagrams 
with the dominant $K \Lambda(1520)$ rescattering amplitude in the intermediate state
were treated only approximately in a coupled-channel formalism  
neglecting the real part of the transition amplitudes.

Implementation of the box diagrams in our four-body calculation 
is rather cumbersome. 
On the other hand, we expect that they do not play a crucial role
for the $p p \to p p \phi$ reaction at the high energies of interest to us here.

\begin{figure}[!ht]
\includegraphics[width=0.49\textwidth]{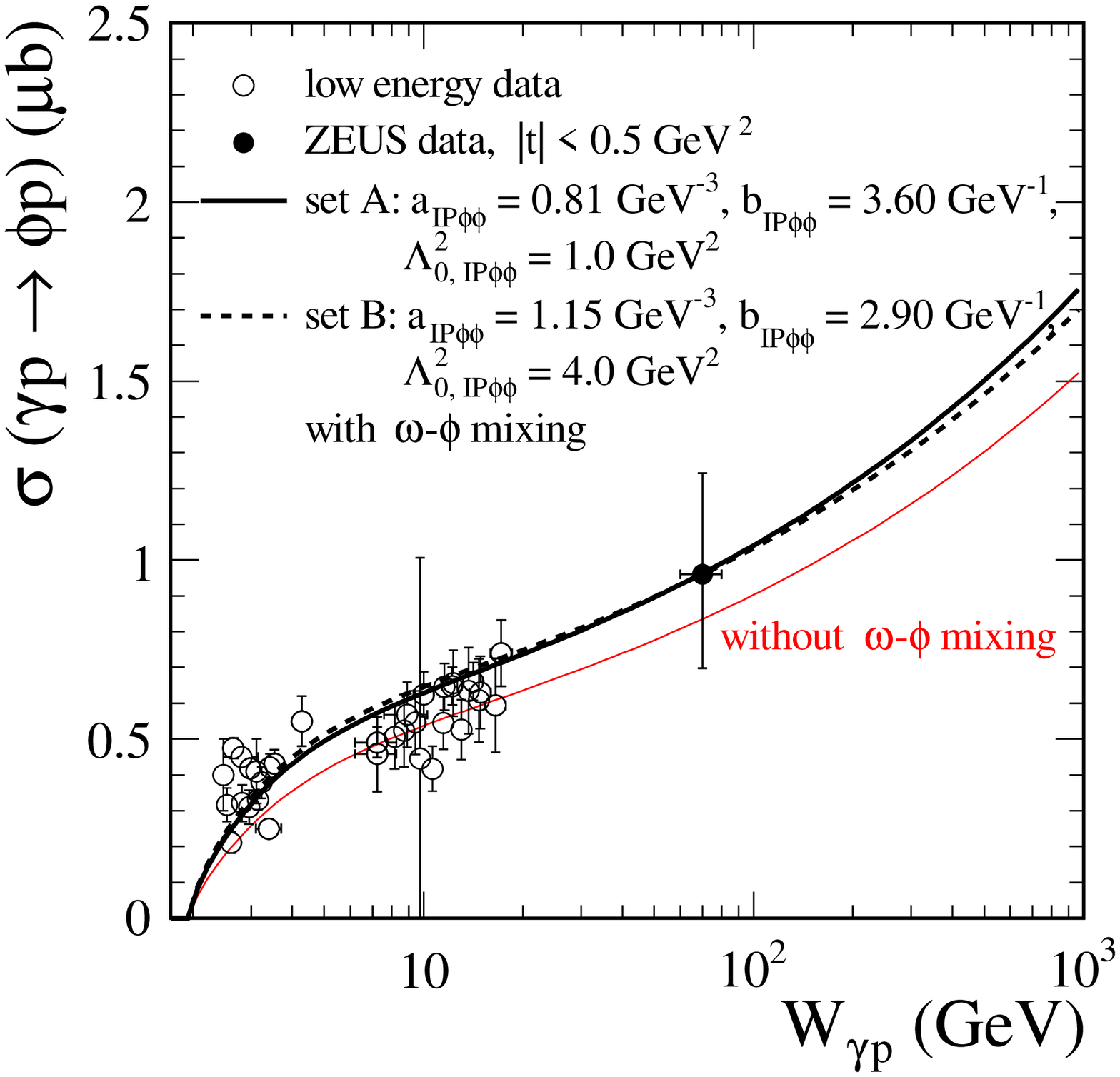}
\includegraphics[width=0.49\textwidth]{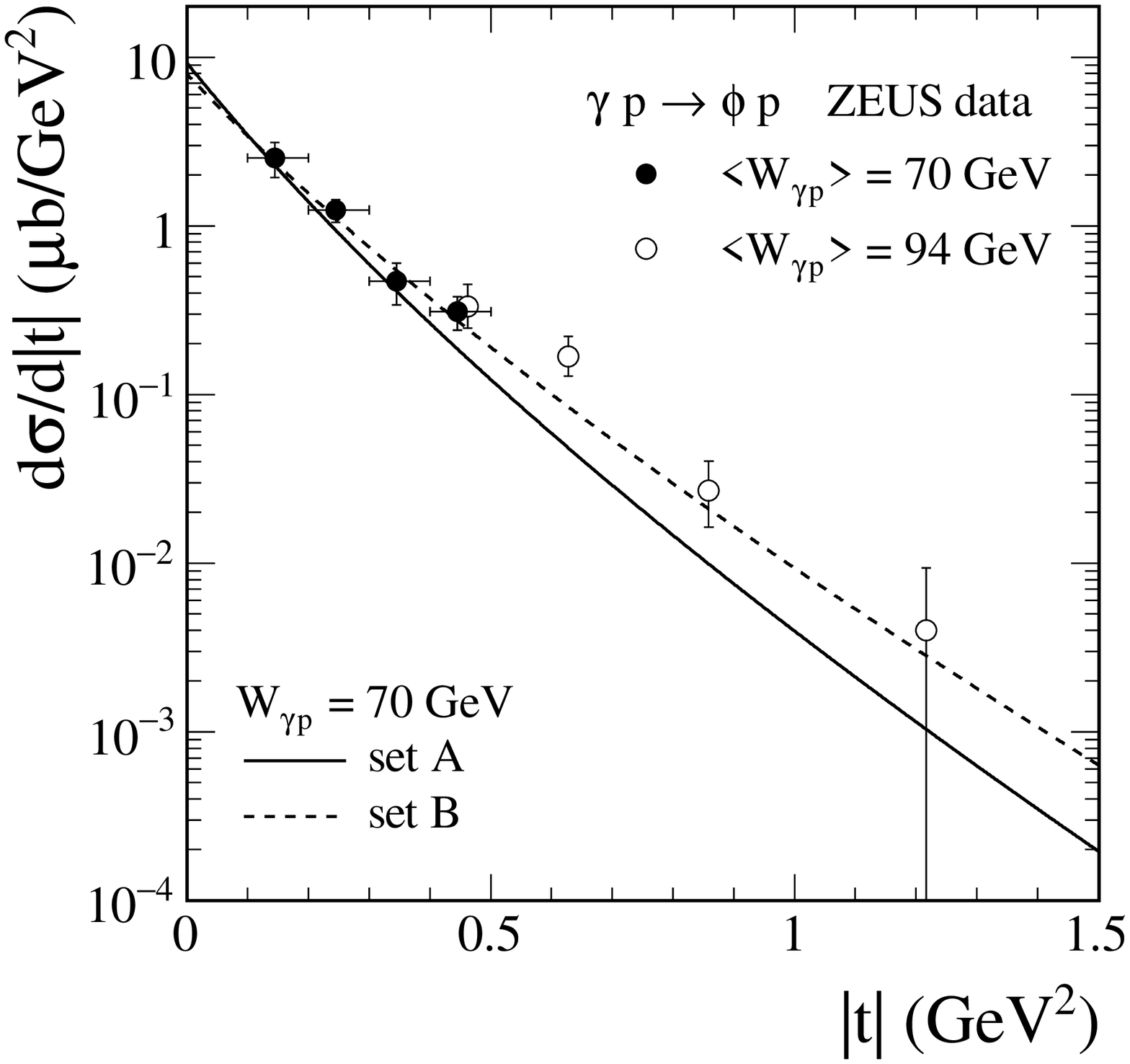}
\caption{\label{fig:gamp_phip_highW}
\small
Left panel: The elastic $\phi$ photoproduction cross section
as function of the center-of-mass energy ${\rm W}_{\gamma p}$.
Our results are compared with the HERA data \cite{Derrick:1996af}
at ${\rm W}_{\gamma p} = 70~{\rm GeV}$
and with a compilation of low-energy experimental data
(see the caption of Fig.~6 of \cite{Lebiedowicz:2018eui} for references).
The upper lines represent results for two parameter sets, set~A and set~B,
including the $\omega \to \phi$ transition terms with
(\ref{coupling_constant_pom_phi_ome}), (\ref{coupling_constant_f2_ome_phi}),
(\ref{coupling_constant_a2_rho_phi}).
Here we take in (\ref{Fpion}), in set A (\ref{photoproduction_setA}), 
$\Lambda_{0,\,\Pom\phi\phi}^{2} = 1.0$~GeV$^{2}$
and, in set B (\ref{photoproduction_setB}), 
$\Lambda_{0,\,\Pom\phi\phi}^{2} = 4.0$~GeV$^{2}$.
The lower red line represents the result 
for the diagram (a) of Fig.~\ref{fig:gamp_phip_diagrams} only 
with the parameter set~(\ref{photoproduction_setA}).
Right panel: The differential cross section $d\sigma/d|t|$ for the 
$\gamma p \to \phi p$ process.
We show the ZEUS data at low $|t|$ (at ${\rm W}_{\gamma p} = 70$~GeV 
and the squared photon virtuality $Q^{2} = 0$~GeV$^{2}$, 
solid marks, \cite{Derrick:1996af}) 
and at higher $|t|$ (at ${\rm W}_{\gamma p} = 94$~GeV 
and $Q^{2} < 0.01$~GeV$^{2}$, open circles, \cite{Breitweg:1999jy}).
Again, the results for the two parameter sets,
set~A~(\ref{photoproduction_setA}) and set~B~(\ref{photoproduction_setB}), are presented.}
\end{figure}
In Fig.~\ref{fig:gamp_phip_highW} we show the elastic 
$\phi$ photoproduction cross section
as a function of the center-of-mass energy ${\rm W}_{\gamma p}$ (left panel)
and the differential cross section $d\sigma/d|t|$ (right panel).
To estimate the $\Pom \phi \phi$ coupling constants
we use the relation [see Eq.~(4.20) of \cite{Lebiedowicz:2018eui}]
\begin{eqnarray}
2 m_{\phi}^{2} \, a_{\Pom \phi \phi} + b_{\Pom \phi \phi}
= 4 \,(2 \beta_{\Pom K K} - \beta_{\Pom \pi \pi})
= 5.28 \; \mathrm{GeV}^{-1} \,.
\label{rhop_tot_opt_aux_pom}
\end{eqnarray}
We show results for two parameter sets, set~A and set~B,
\begin{eqnarray}
&&{\rm set\,A}: \;
a_{\Pom \phi \phi} = 0.81 \;{\rm GeV^{-3}}\,,\quad
b_{\Pom \phi \phi} = 3.60 \;{\rm GeV^{-1}}\,,\quad
\Lambda_{0,\,\Pom\phi\phi}^{2} = 1.0\;{\rm GeV^{2}}\,, 
\label{photoproduction_setA}\\
&&{\rm set\,B}: \;
a_{\Pom \phi \phi} = 1.15 \;{\rm GeV^{-3}}\,,\quad
b_{\Pom \phi \phi} = 2.90 \;{\rm GeV^{-1}}\,,\quad
\Lambda_{0,\,\Pom\phi\phi}^{2} = 4.0\;{\rm GeV^{2}}\,,
\label{photoproduction_setB}
\end{eqnarray}
which were obtained 
based on the diagrams (a) and (b) of Fig.~\ref{fig:gamp_phip_diagrams}
including the diffractive $\omega$-$\phi$ transition terms 
with 
\begin{eqnarray}
a_{\Pom \widetilde{\omega} \phi} = 0\,, \quad 
b_{\Pom \widetilde{\omega} \phi} = -0.46\;{\rm GeV^{-1}}
\label{coupling_constant_pom_phi_ome}
\end{eqnarray}
using (\ref{AA10a}) and (\ref{couplings_omega_pomeron}).
Similarly we obtain from (\ref{AA10a}) and (\ref{couplings_omega_reggeon_f2}),
(\ref{couplings_omega_reggeon_a2})
\begin{eqnarray}
&&a_{f_{2 \Reg} \widetilde{\omega} \phi} = 
-\tan(\Delta \theta_V) \,a_{f_{2 \Reg} \omega \omega} = -0.19\;{\rm GeV^{-3}} \,, \nonumber\\
&&b_{f_{2 \Reg} \widetilde{\omega} \phi} = 
-\tan(\Delta \theta_V) \,b_{f_{2 \Reg} \omega \omega} = -0.33\;{\rm GeV^{-1}} \,;
\label{coupling_constant_f2_ome_phi} \\
&&a_{a_{2 \Reg} \widetilde{\rho} \phi} = 
-\tan(\Delta \theta_V) \,a_{a_{2 \Reg} \rho \omega} = -0.17\;{\rm GeV^{-3}} \,, \nonumber\\
&&b_{a_{2 \Reg} \widetilde{\rho} \phi} = 
-\tan(\Delta \theta_V) \,b_{a_{2 \Reg} \rho \omega} = -0.30\;{\rm GeV^{-1}} \,.
\label{coupling_constant_a2_rho_phi}
\end{eqnarray}
Note that the parameter set (\ref{photoproduction_setA})
for $\Lambda_{0,\,\Pom\phi\phi}^{2} = 1.0\;{\rm GeV^{2}}$
is different than found by us in Sec.~IV~B of \cite{Lebiedowicz:2018eui}
(see Fig.~6 there)
\begin{eqnarray}
a_{\Pom \phi \phi} = 0.49 \;{\rm GeV^{-3}}\,,\quad
b_{\Pom \phi \phi} = 4.27 \;{\rm GeV^{-1}}\,,\quad
\Lambda_{0,\,\Pom\phi\phi}^{2} = 1.0\;{\rm GeV^{2}}\,,
\label{photoproduction_old}
\end{eqnarray}
where the $\omega$-$\phi$ mixing effect was not included.
For comparison, the red lower line represents 
the result without the $\omega$-$\phi$ mixing,
i.e., it contains only the terms represented 
by the diagram (a) of Fig.~\ref{fig:gamp_phip_diagrams}.
We can see from Fig.~\ref{fig:gamp_phip_highW} (right panel)
that the parameter set~B (\ref{photoproduction_setB})
for $\Lambda_{0,\,\Pom\phi\phi}^{2} = 4.0\;{\rm GeV^{2}}$
with the relevant values of the coupling constants $a$ and $b$
describes more accurately the $t$ distribution.

In Fig.~\ref{fig:gamp_phip} we show the integrated cross section
for the $\gamma p \to \phi p$ reaction at low ${\rm W}_{\gamma p}$ energies.
We can see that the diffractive pomeron and reggeon exchanges,
even including the pseudoscalar and scalar meson exchange contributions,
are not sufficient to describe the low-energy data.
Here we want to examine the uncertainties of the photoproduction contribution
due to the meson exchanges in the $t$ channel.
In the left panel, for the meson exchanges, we use the values of the coupling constants 
and the cutoff parameters from \cite{Titov:1999eu} 
while in the right panel we choose
$\Lambda_{\widetilde{M}NN} = \Lambda_{\phi \gamma \widetilde{M}} = 1.2$~GeV 
in (\ref{ff_PSNN}) and (\ref{ff_phigamPS}) below.

Our extrapolations of the cross section, using the theory applicable
at high energies, represents the experimental data roughly on the average.
But the scatter of the experimental data is quite considerable.
Thus, it is impossible for us to draw any further conclusions
concerning these low-energy results at the moment.

\begin{figure}[!ht]
\includegraphics[width=0.49\textwidth]{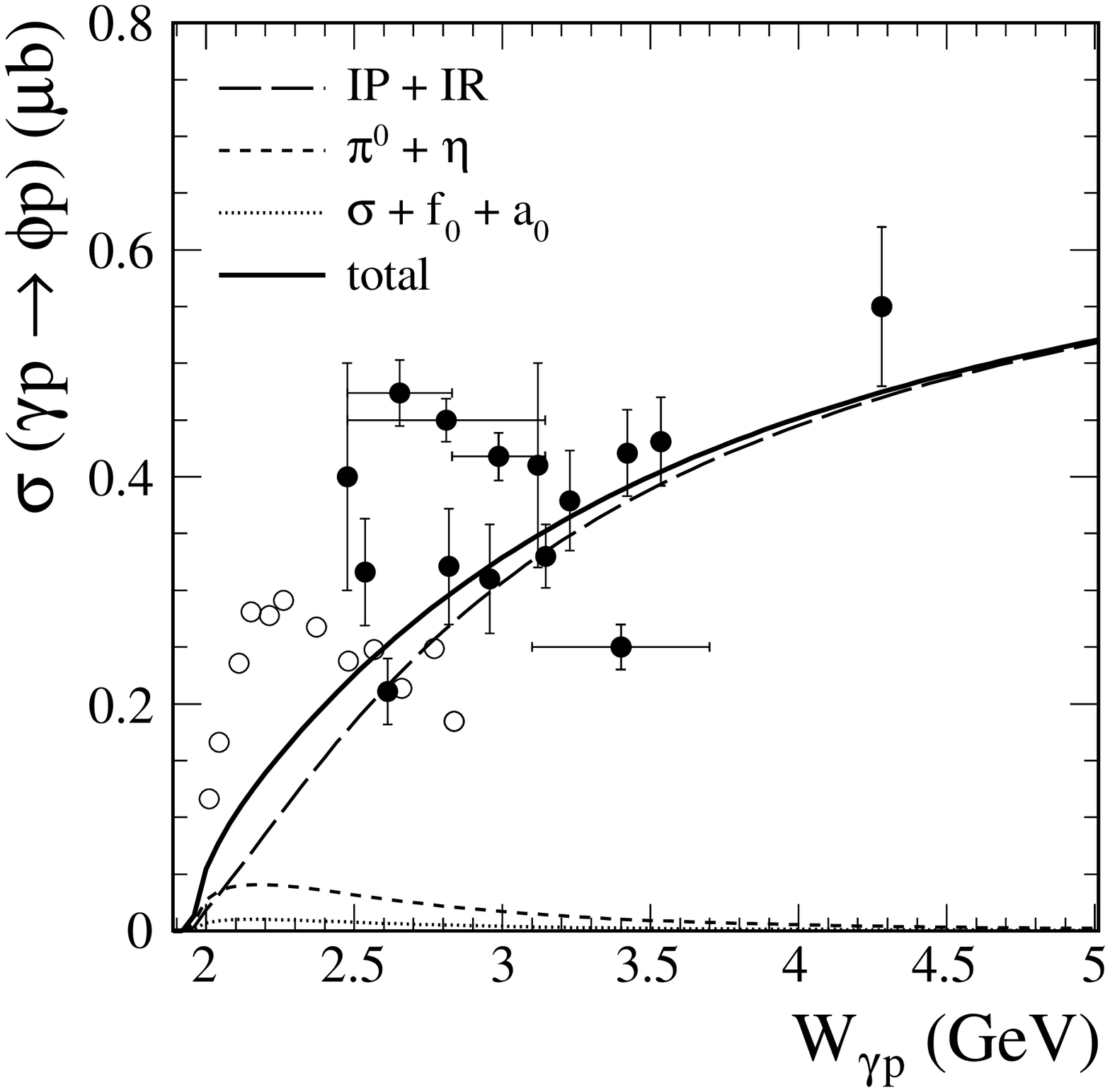}
\includegraphics[width=0.49\textwidth]{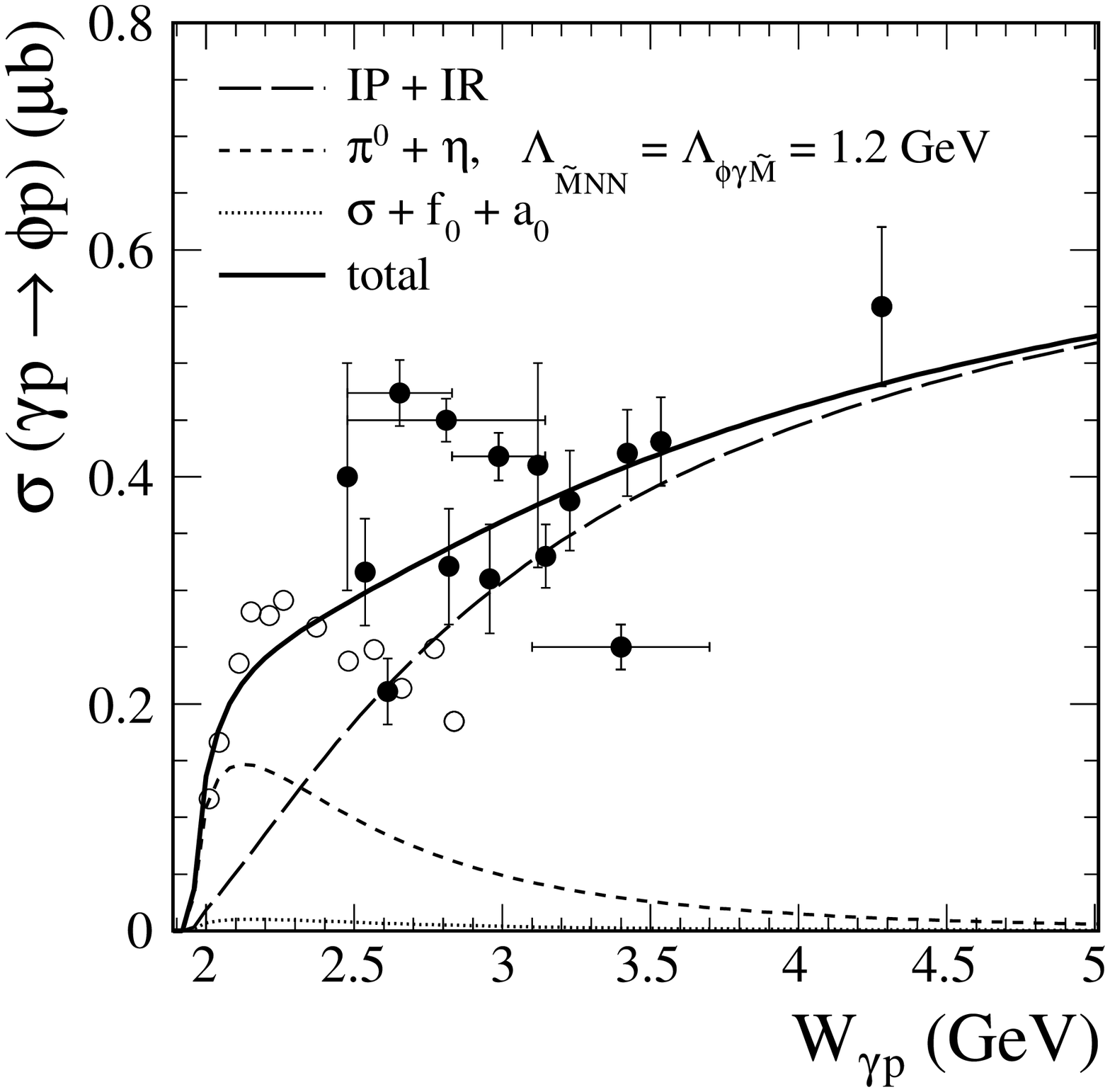}
\caption{\label{fig:gamp_phip}
\small
The elastic $\phi$ photoproduction cross section
as a function of ${\rm W}_{\gamma p}$
integrated over $t_{\rm min} < |t| < 1$~GeV$^{2}$.
The theoretical results are compared with a compilation of
low-energy experimental data from \cite{Ballam:1972eq,Fries:1978di,Barber:1981fj}, 
and \cite{Laget:2000gj}.
The open data points are taken from \cite{Kong:2016scm}
(data was obtained there by integrating over the differential cross sections given in \cite{Dey:2014tfa}).
The solid lines correspond to a coherent sum of pomeron, $f_{2 \Reg}$ reggeon, 
pseudoscalar, and scalar exchanges.
For the diffractive component ($\Pom + \Reg$) 
we take the set~A of parameters from Fig.~\ref{fig:gamp_phip_highW}.
The results for the pseudoscalar and scalar exchanges 
shown in the left panel were obtained with the parameters from \cite{Titov:1999eu};
see Appendix~\ref{sec:subleading}, Sec.~1.
In the right panel, for comparison, we show results obtained 
for different values of the cutoff parameters in the pseudoscalar term. 
Here we take
$\Lambda_{\widetilde{M}NN} = \Lambda_{\phi \gamma \widetilde{M}} = 1.2$~GeV
in (\ref{ff_PSNN}) and (\ref{ff_phigamPS}).}
\end{figure}

\section{Subleading contributions to $\phi$ CEP}
\label{sec:subleading}

In this section we discuss the following subleading processes
contributing to $pp \to pp \phi$.
The fusion processes
$\gamma$-$\pi^{0}$, $\gamma$-$\eta$, $\gamma$-$\eta'$, and $\gamma$-$f_{0}$, $\gamma$-$a_{0}$,
and fusion processes involving vector mesons
$\phi$-$\Pom$, $\omega$-$\Pom$, $\omega$-$f_{2 \Reg}$, 
$\rho$-$\pi^{0}$, $\omega$-$\eta$, and $\omega$-$\eta'$.
We can have also $\omega$-$f_{0}$ and $\omega$-$f_{2}'$ contributions.
But these contributions are expected to be very small since the $\phi$ is
nearly a pure $s \bar{s}$ state, the $\omega$ nearly a pure $u \bar{u} + d \bar{d}$ state.
In the following we shall, therefore, neglect such contributions.

Below we present formulas for $\phi$ production with
subsequent decay $\phi \to K^{+}K^{-}$.
The formulas for $\phi$ production are obtained from those
by the replacement (\ref{2to3_replacement}).

The discussions of the subleading processes for $\phi$ CEP
are very important for the comparison of our theory with
the WA102 experimental results. See in particular
Figs.~\ref{fig:WA102_ABS_noodd} and \ref{fig:WA102_ABS_phi12} of Sec.~\ref{sec:comparison_WA102}.
At LHC energies the subleading processes should be negligible
for mid-rapidity $\phi$ production.
In Secs.~1 and 2 of this Appendix we discuss
$\gamma$-pseudoscalar- and $\gamma$-scalar-fusion contributions
to $\phi$ CEP. The couplings which we find there can also be used
to calculate subleading contributions to
photoproduction of the $\phi$ meson.
The corresponding results are shown together with
the leading contributions in Fig.~\ref{fig:gamp_phip} of Appendix~\ref{sec:appendixB}.
\subsection{$\gamma$-pseudoscalar-meson contributions}
\label{sec:gamma_PS}

First we consider processes with pseudoscalar meson 
$\widetilde{M} = \pi^{0}, \eta, \eta'$ exchanges.
The generic diagrams for these contributions are shown 
in Fig.~\ref{diagrams_gamma_PS} (a), (b).
We have for the total $\gamma$-pseudoscalar-meson-fusion contribution
\begin{eqnarray}
^{(3)}{\cal M}_{pp \to pp K^{+}K^{-}}^{(\phi \to K^{+}K^{-})} = 
\sum_{\widetilde{M} = \pi^{0},\, \eta,\, \eta'} 
\left( {\cal M}^{(\gamma \widetilde{M})}_{pp \to pp K^{+}K^{-}} + 
       {\cal M}^{(\widetilde{M} \gamma)}_{pp \to pp K^{+}K^{-}} \right) \,.
\label{pp_ppphi_PS}
\end{eqnarray}
%

\begin{figure}
(a)\includegraphics[width=6.5cm]{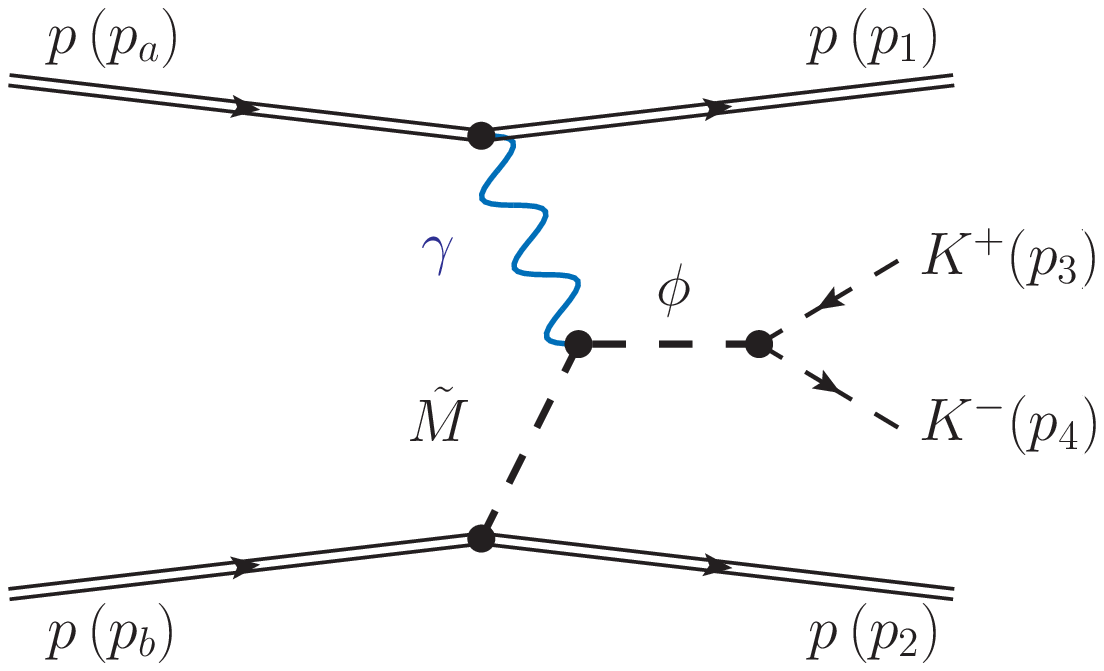}   
(b)\includegraphics[width=6.5cm]{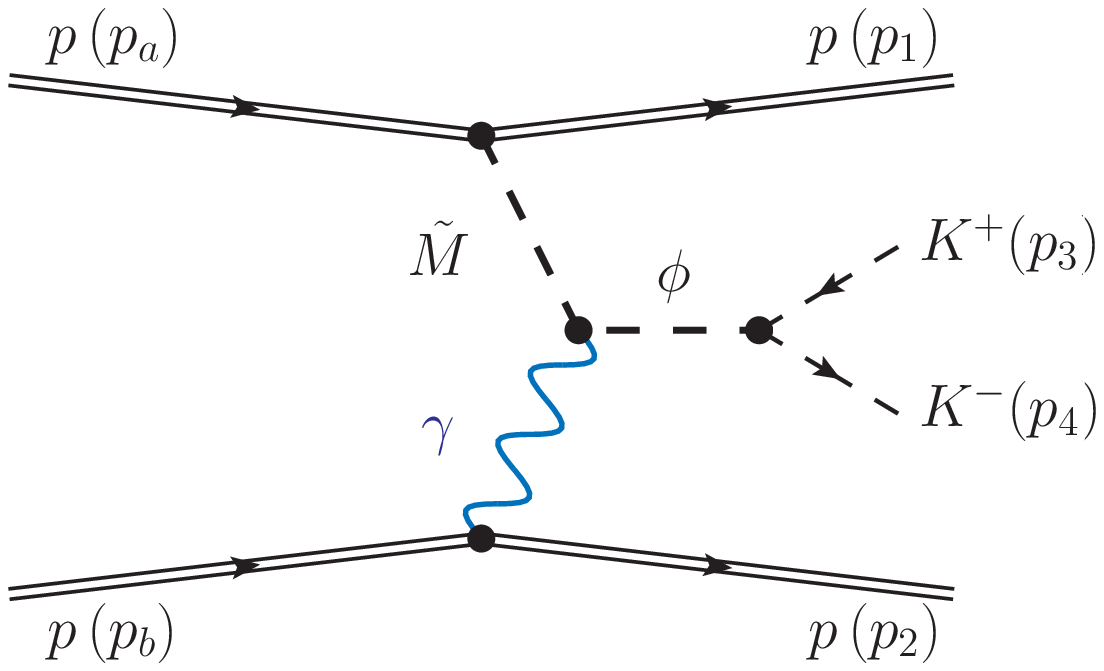}  
  \caption{\label{diagrams_gamma_PS}
The Born-level diagrams for central-exclusive production of $\phi$
decaying to $K^+ K^-$ in proton-proton collisions with
pseudoscalar meson $\widetilde{M}$ exchange:
(a) $\gamma$-$\widetilde{M}$ fusion; (b) $\widetilde{M}$-$\gamma$ fusion.}
\end{figure}

The $\gamma$-$\widetilde{M}$ amplitude can be written as
\begin{eqnarray}
{\cal M}^{(\gamma \widetilde{M})}_{pp \to pp K^{+}K^{-}} 
&=& (-i)
\bar{u}(p_{1}, \lambda_{1}) 
i\Gamma^{(\gamma pp)}_{\mu}(p_{1},p_{a}) 
u(p_{a}, \lambda_{a}) \nonumber \\
&&\times  
i\Delta^{(\gamma)\,\mu \rho_{1}}(q_{1})\, 
i\Gamma^{(\phi \gamma \widetilde{M})}_{\rho_{2} \rho_{1}}(p_{34},q_{1})\, 
i\Delta^{(\phi)\,\rho_{2} \kappa}(p_{34})\,
i\Gamma^{(\phi KK)}_{\kappa}(p_{3},p_{4})
\nonumber \\
&& \times 
i\Delta^{(\widetilde{M})}(t_{2}) \,
\bar{u}(p_{2}, \lambda_{2}) 
i\Gamma^{(\widetilde{M} pp)}(p_{2},p_{b}) 
u(p_{b}, \lambda_{b}) \,.
\label{amplitude_gamma_pion}
\end{eqnarray}
For the $\widetilde{M}$-proton vertex we have (see (3.4) of \cite{Lebiedowicz:2016ryp})
\begin{eqnarray}
i\Gamma^{(\widetilde{M} pp)}(p',p) = -\gamma_{5} g_{\widetilde{M} pp} 
F^{(\widetilde{M} pp)}((p'-p)^{2})\,.
\label{PSNN}
\end{eqnarray}
We take $g_{\pi pp} = \sqrt{4 \pi \times 14.0}$,
$g_{\eta pp} = \sqrt{4 \pi \times 0.99}$; see Eqs.~(28) and (29) of \cite{Titov:1999eu}.

An effective Lagrangian for the $\phi \gamma \widetilde{M}$ coupling
is given in (22) of \cite{Titov:1999eu}
\begin{eqnarray}
{\cal L}'_{\phi \gamma \widetilde{M}}
= \frac{e\,g_{\phi \gamma \widetilde{M}}}{m_{\phi}}\, 
\varepsilon^{\mu \nu \alpha \beta}\,
(\partial_{\mu} \phi_{\nu})\,
(\partial_{\alpha} A_{\beta})\,
\widetilde{M}\,
\label{phi_gam_PS_Lagrangian}
\end{eqnarray}
with $A_{\beta}$ the photon field and $g_{\phi \gamma \widetilde{M}}$ a dimensionless coupling constant.
From this we get the $\phi \gamma \widetilde{M}$ vertex,
including a form factor, as follows
\newline
\hspace*{1.8cm}\includegraphics[width=5.cm]{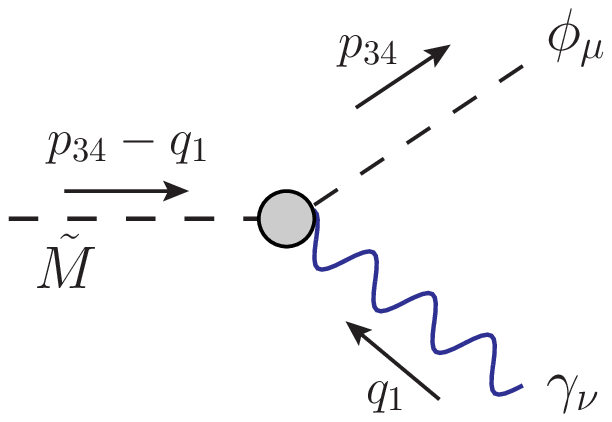}
\begin{equation}\label{phi_gam_PS}
\begin{split}
i\Gamma_{\mu \nu}^{(\phi \gamma \widetilde{M})}(p_{34},q_{1}) 
= -i e \frac{g_{\phi \gamma \widetilde{M}}}{m_{\phi}}\, 
\varepsilon_{\mu \nu \rho \sigma} p_{34}^{\rho} q_{1}^{\sigma}\,
\tilde{F}^{(\phi \gamma \widetilde{M})}(p_{34}^{2},q_{1}^{2},(p_{34}-q_{1})^{2})\,.
\end{split}
\end{equation}
We use a factorised ansatz for the $\phi \gamma \widetilde{M}$ form factor
\begin{eqnarray}
\tilde{F}^{(\phi \gamma \widetilde{M})}(p_{34}^{2},q_{1}^{2},(p_{34}-q_{1})^{2}) = 
\tilde{F}^{(\gamma)}(q_{1}^{2}) \,
\tilde{F}^{(\phi)}(p_{34}^{2})\,
F^{(\phi \gamma \widetilde{M})}((p_{34}-q_{1})^{2})\,.
\label{ff_phi_gam_PS}
\end{eqnarray}
Based on considerations of the vector-meson-dominance model (VMD)
we write the $\tilde{F}^{(\gamma)}$ form factor as
\begin{eqnarray}
\tilde{F}^{(\gamma)}(q_{1}^{2}) = 
\frac{m_{V}^{2}}{m_{V}^{2}-q_{1}^{2}}\, 
\tilde{F}^{(V)}(q_{1}^{2})
\label{ff_phi_gam_PS_VDM}
\end{eqnarray}
with $V = \rho^{0}$ for $\widetilde{M} = \pi^{0}$ and
$V = \omega$ for $\widetilde{M} = \eta, \eta'$.
For the form factors $\tilde{F}^{(V)}$ 
we choose the form as for $\tilde{F}^{(\phi)}$ in (\ref{ff_Nachtmann}) 
replacing $\phi$ by $V = \rho^{0}, \omega$.

The effective coupling constant $g_{\phi \gamma \widetilde{M}}$ 
is related to the decay width of $\phi \to \gamma \widetilde{M}$,
see (31) of \cite{Titov:1999eu},
\begin{eqnarray}
\Gamma(\phi \to \gamma \widetilde{M}) = 
\frac{\alpha}{24}\frac{(m_{\phi}^{2}-m_{\widetilde{M}}^{2})^{3}}{m_{\phi}^{5}}
|g_{\phi \gamma \widetilde{M}}|^{2}\,.
\label{coupling_phigamPS}
\end{eqnarray}
Using the most recent values from \cite{Tanabashi:2018oca}, and
taking the negative signs as in \cite{Titov:1999eu}, we have found
$g_{\phi \gamma \pi^{0}} = -0.137$, $g_{\phi \gamma \eta} = -0.705$,
and $|g_{\phi \gamma \eta'}| = 0.726$.
Note that $|g_{\phi \gamma \eta'}| > |g_{\phi \gamma \eta}|$.
But the contribution of $\eta'$ exchange is suppressed relative to the $\eta$ exchange
because of the heavier mass occurring in the propagator
and of the smaller value of $g_{\eta' pp} \simeq g_{\eta pp}/2$,
where we follow \cite{Titov:1999eu}.
However, we note that there is no consensus on this latter relation in the literature.
In \cite{Nakayama:1999jx} $g_{\eta' pp} \cong 6.1$ and $g_{\eta pp} = 6.14$ are given.

We follow \cite{Titov:1999eu,Titov:2003bk,Ryu:2012tw} 
and use monopole ans{\"a}tze for the form factors
$F^{(\widetilde{M} pp)}$ (\ref{PSNN})
and $F^{(\phi \gamma \widetilde{M})}$ (\ref{ff_phi_gam_PS})
\begin{eqnarray}
&&F^{(\widetilde{M} pp)}(t)= 
\frac{\Lambda_{\widetilde{M}NN}^{2}-m_{\widetilde{M}}^{2}}{\Lambda_{\widetilde{M}NN}^{2}-t}\,,
\label{ff_PSNN}\\
&&F^{(\phi \gamma \widetilde{M})}(t)= 
\frac{\Lambda_{\phi \gamma \widetilde{M}}^{2}-m_{\widetilde{M}}^{2}}{\Lambda_{\phi \gamma \widetilde{M}}^{2}-t}\,.
\label{ff_phigamPS}
\end{eqnarray}
The cutoff parameters 
$\Lambda_{\pi NN} = 0.7$~GeV, 
$\Lambda_{\phi \gamma \pi} = 0.77$~GeV,
$\Lambda_{\eta NN} = 1.0$~GeV,
$\Lambda_{\phi \gamma \eta} = 0.9$~GeV are taken from \cite{Titov:1999eu}.

To examine uncertainties of the photoproduction contribution 
in the $p p \to p p \phi$ reaction we intend to 
show also the result with 
$\Lambda_{\widetilde{M}NN} = 1.2$~GeV 
and $\Lambda_{\phi \gamma \widetilde{M}} = 1.2$~GeV 
in (\ref{ff_PSNN}) and (\ref{ff_phigamPS}), respectively,
which are slightly different from the values given in \cite{Titov:1999eu}.
This choice of parameters was used in \cite{Kiswandhi:2011cq}; see Sec.~II~B there.

In Appendix~\ref{sec:appendixB} we discuss the $\gamma p \to \phi p$ reaction.
There we compare our model calculations for different parameter sets
with the experimental data.

Inserting (\ref{PSNN})--(\ref{ff_phigamPS}) 
in (\ref{amplitude_gamma_pion}) we can write
the amplitude for the $\gamma \widetilde{M}$ exchange as follows
\begin{eqnarray}
{\cal M}^{(\gamma \widetilde{M})}_{pp \to pp K^{+}K^{-}} 
&=& i\,e^{2}\,
\bar{u}(p_{1}, \lambda_{1}) 
\left[ \gamma^{\alpha}F_{1}(t_{1}) + \frac{i}{2m_{p}} \sigma^{\alpha \alpha'} (p_{1}-p_{a})_{\alpha'} F_{2}(t_{1}) \right]
u(p_{a}, \lambda_{a})\nonumber \\
&&\times  
\frac{1}{t_{1}}\, 
\frac{g_{\phi \gamma \widetilde{M}}}{m_{\phi}}\, 
\varepsilon_{\beta \alpha \rho \sigma} p_{34}^{\rho} q_{1}^{\sigma}\,
\tilde{F}^{(\phi \gamma \widetilde{M})}(p_{34}^{2},q_{1}^{2},q_{2}^{2})\nonumber \\
&&\times  
\Delta_{T}^{(\phi)}(p_{34}^{2})\,
\frac{g_{\phi K^{+}K^{-}}}{2}\,(p_{3}-p_{4})^{\beta}\, F^{(\phi K K)}(p_{34}^{2})
\nonumber \\
&&\times 
\frac{1}{t_{2} - m_{\widetilde{M}}^{2}}\, 
g_{\widetilde{M} pp} \, F^{(\widetilde{M} pp)}(t_{2})\, 
\bar{u}(p_{2}, \lambda_{2}) \,\gamma_{5} \,u(p_{b}, \lambda_{b})\,.
\label{amplitude_gamma_pion_aux}
\end{eqnarray}
The amplitude ${\cal M}^{(\widetilde{M} \gamma)}_{pp \to pp K^{+}K^{-}}$
is obtained from (\ref{amplitude_gamma_pion_aux})
with the replacements (\ref{replace}).

\subsection{$\gamma$-scalar-meson contributions}
\label{sec:gamma_S}

Next we turn to the amplitudes for $\phi$ production through 
the fusion of $\gamma$ with scalar mesons
$S = f_{0}(500), f_{0}(980)$, and $a_{0}(980)$.
Their contribution is
\begin{eqnarray}
^{(4)}{\cal M}_{pp \to pp K^{+}K^{-}}^{(\phi \to K^{+}K^{-})} = 
\sum_{S = f_{0}(500),\, f_{0}(980),\, a_{0}(980)} 
\left( {\cal M}^{(\gamma S)}_{pp \to pp K^{+}K^{-}} + 
       {\cal M}^{(S \gamma)}_{pp \to pp K^{+}K^{-}} \right) \,.
\label{pp_ppphi_S}
\end{eqnarray}
The generic diagrams for these contributions are as in Fig.~\ref{diagrams_gamma_PS}
with $\widetilde{M}$ replaced by $S$.
The same applies to the analytic expressions.
We get ${\cal M}^{(\gamma S)}$ from ${\cal M}^{(\gamma \widetilde{M})}$
in (\ref{amplitude_gamma_pion}) replacing
$\Gamma_{\rho_{2} \rho_{1}}^{(\phi \gamma \widetilde{M})}$, $\Delta^{(\widetilde{M})}$, 
and $\Gamma^{(\widetilde{M} pp)}$ by 
$\Gamma_{\rho_{2} \rho_{1}}^{(\phi \gamma S)}$, $\Delta^{(S)}$, and $\Gamma^{(S pp)}$, respectively.
We use the following expressions 
for the $S$-proton and for the $\phi \gamma S$
effective coupling Lagrangians, see (34) and (35),
respectively, of \cite{Titov:1999eu},
\begin{eqnarray}
&&{\cal L}'_{Spp} = g_{S pp}\, \bar{p}\,p\,S\,,
\label{Spp_Lagrangian} \\
&&{\cal L}'_{\phi \gamma S}
= \frac{e\,g_{\phi \gamma S}}{m_{\phi}}\, 
(\partial^{\alpha} \phi^{\beta})\,
(\partial_{\alpha} A_{\beta} - \partial_{\beta} A_{\alpha})\,
S\,.
\label{phi_gam_S_Lagrangian}
\end{eqnarray}
From these we get the vertices including form factors, as follows,
where the momentum flow and the indices are chosen
as for the $\widetilde{M} pp$ and $\phi \gamma \widetilde{M}$ vertices,
respectively, see (\ref{PSNN}) and (\ref{phi_gam_PS}),
\begin{eqnarray}
&&i\Gamma^{(S pp)}(p',p) = i g_{S pp} F^{(S pp)}((p'-p)^{2})\,,
\label{SNN}\\
&&i\Gamma_{\mu \nu}^{(\phi \gamma S)}(p_{34},q_{1}) 
= -i e \frac{g_{\phi \gamma S}}{m_{\phi}}\, 
\left[ q_{1\,\mu} p_{34\,\nu} - (p_{34} \cdot q_{1}) g_{\mu \nu}  \right]\,
\tilde{F}^{(\phi \gamma S)}(p_{34}^{2},q_{1}^{2},(p_{34}-q_{1})^{2})\,. \qquad
\label{phi_gam_S}
\end{eqnarray}

For the contributions of scalar exchanges we take the parameters 
found in Appendix~C of \cite{Titov:1999eu}:
$g_{\phi \gamma f_{0}(500)} = 0.047$,
$g_{f_{0}(500) pp} = \sqrt{4 \pi \times 8.0}$, 
$g_{\phi \gamma f_{0}(980)} = -1.81$,
$g_{f_{0}(980) pp} = 0.56$, 
$g_{\phi \gamma a_{0}(980)} = -0.16$,
$g_{a_{0}(980) pp} = 21.7$.
For $f_{0}(500)$ the monopole form of the form factors 
as in (\ref{ff_PSNN}) and (\ref{ff_phigamPS}) with
$\widetilde{M}$ replaced by $f_{0}(500)$ and
$\Lambda_{f_{0}(500) NN} = \Lambda_{\phi \gamma f_{0}(500)} = 2$~GeV is used.
For the heavier mesons ($f_{0}(980)$ and $a_{0}(980)$) 
the following compact form is used \cite{Titov:1999eu}:
\begin{eqnarray}
F^{(S pp)}(t) F^{(\phi \gamma S)}(t)
= \frac{\Lambda_{S}^{4}}{\Lambda_{S}^{4} + (t-m_{S}^{2})^{2}}\,,
\quad \Lambda_{S} = 0.6\;{\rm GeV}\,.
\label{ff_product}
\end{eqnarray}

The final expression for the $\gamma S$-exchange amplitude in (\ref{pp_ppphi_S})
reads
\begin{eqnarray}
{\cal M}^{(\gamma S)}_{pp \to pp K^{+}K^{-}} 
&=& \,e^{2}\,
\bar{u}(p_{1}, \lambda_{1}) 
\left[ \gamma^{\alpha}F_{1}(t_{1}) + \frac{i}{2m_{p}} \sigma^{\alpha \alpha'} (p_{1}-p_{a})_{\alpha'} F_{2}(t_{1}) \right]
u(p_{a}, \lambda_{a})\nonumber \\
&&\times  
\frac{1}{t_{1}}\, 
\frac{g_{\phi \gamma S}}{m_{\phi}}\, 
\left[ q_{1 \,\beta} p_{34 \,\alpha} - (p_{34} \cdot q_{1}) g_{\beta \alpha}  \right]\,
\tilde{F}^{(\phi \gamma S)}(p_{34}^{2},q_{1}^{2},q_{2}^{2})\nonumber \\
&&\times 
\Delta_{T}^{(\phi)}(p_{34}^{2})\,
\frac{g_{\phi K^{+}K^{-}}}{2}\,(p_{3}-p_{4})^{\beta}\, F^{(\phi K K)}(p_{34}^{2})
\nonumber \\
&&\times 
\frac{1}{t_{2} - m_{S}^{2}}\, 
g_{S pp} \, F^{(S pp)}(t_{2})\, 
\bar{u}(p_{2}, \lambda_{2}) \,u(p_{b}, \lambda_{b})\,.
\label{amplitude_gamma_S_aux}
\end{eqnarray}
For ${\cal M}^{(S \gamma)}_{pp \to pp K^{+}K^{-}}$
we have to make the replacements (\ref{replace}).

\subsection{$\phi$-$\Pom$ and $\omega$-$\Pom$ contributions}
\label{sec:phi_pomeron}

Here we discuss two approaches,
reggeized-vector-meson-exchange approach (I) 
and reggeon-exchange approach (II).
For the second approach the corresponding diagrams are shown 
in Fig.~\ref{fig:RP}.
\begin{figure}
(a)\includegraphics[width=6.5cm]{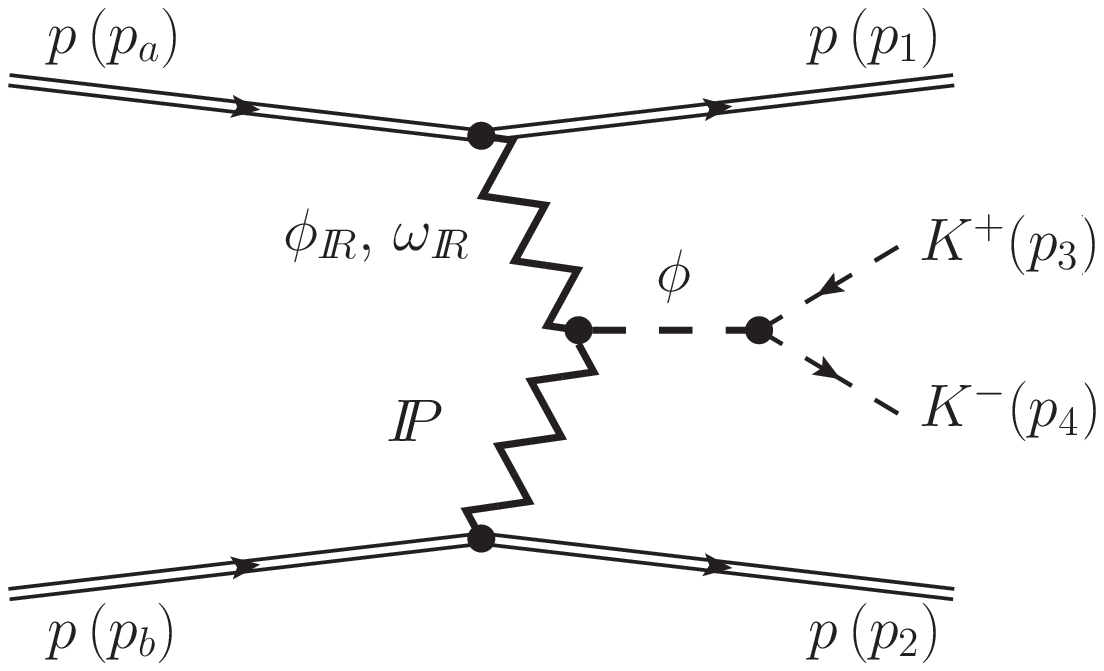}   
(b)\includegraphics[width=6.5cm]{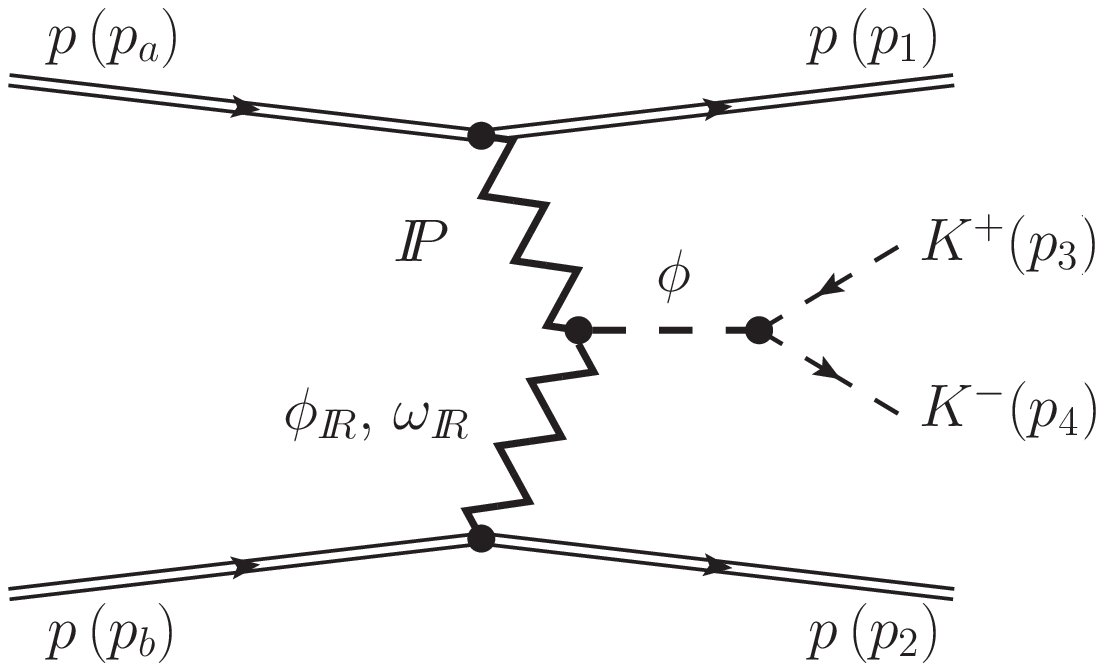}  
  \caption{\label{fig:RP}
The Born-level diagrams for diffractive production of a $\phi$ meson with the subsequent decay $\phi \to K^+ K^-$:
(a) reggeon-pomeron fusion;
(b) pomeron-reggeon fusion.}
\end{figure}

First we consider the contributions 
through the vector mesons $V = \phi$ and $\omega$:
\begin{eqnarray}
^{(5)}{\cal M}_{pp \to pp K^{+}K^{-}}^{(\phi \to K^{+}K^{-})} = 
\sum_{V = \phi,\, \omega} 
\left( {\cal M}^{(V \Pom)}_{pp \to pp K^{+}K^{-}} 
     + {\cal M}^{(\Pom V)}_{pp \to pp K^{+}K^{-}} \right) \,.
\label{pp_ppphi_VP}
\end{eqnarray}
The amplitude for the $V \Pom$-exchange can be written as
\begin{eqnarray}
{\cal M}^{(V \Pom)}_{pp \to pp K^{+}K^{-}} 
&=& (-i)
\bar{u}(p_{1}, \lambda_{1}) 
i\Gamma^{(V pp)}_{\mu}(p_{1},p_{a}) 
u(p_{a}, \lambda_{a}) \nonumber \\
&&\times  
i\Delta^{(V)\,\mu \rho_{1}}(q_{1}) \,
i\Gamma^{(\Pom V \phi)}_{\rho_{2} \rho_{1} \alpha \beta}(p_{34},q_{1})\, 
i\Delta^{(\phi)\,\rho_{2} \kappa}(p_{34})\,
i\Gamma^{(\phi KK)}_{\kappa}(p_{3},p_{4})
\nonumber \\
&&\times 
i\Delta^{(\Pom)\,\alpha \beta, \delta \eta}(s_{2},t_{2}) \,
\bar{u}(p_{2}, \lambda_{2}) 
i\Gamma^{(\Pom pp)}_{\delta \eta}(p_{2},p_{b}) 
u(p_{b}, \lambda_{b}) \,.
\label{amplitude_V_pomeron}
\end{eqnarray}

The $V$-proton vertex is
\begin{eqnarray}
i\Gamma_{\mu}^{(V pp)}(p',p)
=-i g_{V pp}\, F^{(V pp)}(t)
\left[\gamma_{\mu} - i\frac{\kappa_{V}}{2m_{p}} \sigma_{\mu \nu} (p'-p)^{\nu} 
\right],
\label{vertex_VNN}
\end{eqnarray}
with the tensor-to-vector coupling ratio, $\kappa_{V} = f_{V pp}/g_{V pp}$.
Following \cite{Nakayama:1999jx} 
we assume $\kappa_{\phi} = \kappa_{\omega}$ to be in the range $\simeq \pm 0.5$,
$g_{\phi pp} = -0.6$ and $g_{\omega pp} = 9.0$; see also \cite{Meissner:1997qt}.
Thus, the tensor term in (\ref{vertex_VNN}) is small and
in the calculation we take the vectorial term only 
with $g_{\phi pp} = -0.6$ and $g_{\omega pp} = 8.65$.
This latter value was determined in Sec.~6.3 of \cite{Ewerz:2013kda} 
and, as discussed there, we assume $g_{\omega pp} = g_{\omega_{\Reg} pp}$.

We also make the assumption that the $t$-dependence 
of the $V$-proton coupling can be parametrised
in a simple exponential form
\begin{eqnarray}
F^{(V pp)}(t)= \exp \left( \frac{t-m_{V}^{2}}{\Lambda_{VNN}^{2}} \right)\,,
\quad \Lambda_{VNN} = 1\;{\rm GeV}\,.
\label{F_VNN}
\end{eqnarray}
This form factor is normalized to unity when the vector meson $V$ is on its mass shell, 
i.e., when $t = m_{V}^{2}$.

The amplitude for the $V \Pom$-exchange can now be written as
\begin{eqnarray}
&&{\cal M}^{(V \Pom)}_{pp \to pp K^{+}K^{-}} =
-i\,g_{V pp}\, F^{(V pp)}(t_{1}) \, \bar{u}(p_{1}, \lambda_{1}) 
\gamma^{\alpha} 
u(p_{a}, \lambda_{a})\nonumber \\
&&\qquad \times  
\Delta_{T}^{(V)}(t_{1})\,
\Delta_{T}^{(\phi)}(p_{34}^{2})\,
\frac{g_{\phi K^{+}K^{-}}}{2}\,(p_{3}-p_{4})^{\beta}\, F^{(\phi K K)}(p_{34}^{2})
\nonumber \\
&& \qquad \times 
\left[ 2 a_{\Pom V \phi}\, \Gamma^{(0)}_{\beta \alpha \kappa \lambda}(p_{34},-q_{1})
- b_{\Pom V \phi}\,\Gamma^{(2)}_{\beta \alpha \kappa \lambda}(p_{34},-q_{1}) \right]
F_{M}(q_{2}^{2})\,\tilde{F}^{(V)}(q_{1}^{2})\,\tilde{F}^{(\phi)}(p_{34}^{2})
\nonumber \\
&& \qquad \times 
\frac{1}{2 s_{2}} \left( -i s_{2} \alpha'_{\Pom} \right)^{\alpha_{\Pom}(t_{2})-1}\,
3 \beta_{\Pom NN} \, F_{1}(t_{2})\, 
\bar{u}(p_{2}, \lambda_{2}) 
\left[ \gamma^{\kappa} (p_{2} + p_{b})^{\lambda} \right]
u(p_{b}, \lambda_{b})\,.
\label{amplitude_V_pomeron_aux}
\end{eqnarray}
For the $\Pom \phi \phi$ and $\Pom \omega \phi$ coupling vertices and constants 
see the discussion in the Appendices~\ref{sec:appendixA} and \ref{sec:appendixB}.

For small values of $s_{1} = (p_{1} + p_{34})^{2}$
the standard form of the vector-meson propagator factor
$\Delta_{T}^{(V)}(t_{1})$ in (\ref{amplitude_V_pomeron_aux}) 
should be adequate; see (\ref{phi_propagator}) for $V = \phi$.
For higher values of $s_{1}$ we must take into account the reggeization.
We do this, following (3.21), (3.24) of \cite{Lebiedowicz:2019jru},
by making in the amplitude ${\cal M}^{(V \Pom)}$ (\ref{amplitude_V_pomeron_aux})
the replacement 
\begin{eqnarray}
&&\Delta_{T}^{(V)}(t_{1}) \to \Delta_{T}^{(V)}(t_{1})
\left( \exp (i \phi(s_{1}))\,s_{1} \alpha'_{V}  \right)^{\alpha_{V}(t_{1})-1}\,,
\label{reggeization_2}\\
&&\phi(s_{1}) =
\frac{\pi}{2}\exp\left(\frac{s_{\rm thr}-s_{1}}{s_{{\rm thr}}}\right)-\frac{\pi}{2}\,;
\label{reggeization_aux}
\end{eqnarray}
where $s_{\rm thr}$ is the lowest value of $s_{1}$ (\ref{2to4_kinematic})
possible here:
\begin{eqnarray}
s_{\rm thr} = (m_p+2m_{K})^2\,. 
\label{sthr}
\end{eqnarray}
Note, that in (\ref{reggeization_2}) we take $s_{1}\alpha'_{V}$ 
instead of $s_{1}/s_{\rm thr}$ as in (3.21) of \cite{Lebiedowicz:2019jru}.
We assume for the Regge trajectories
\begin{eqnarray}
&&\alpha_{V}(t) = \alpha_{V}(0)+\alpha'_{V}\,t, \quad V = \phi,\, \omega,\\
&&\alpha_{\phi}(0) = 0.1\,, \;\;
\alpha'_{\phi} = 0.9 \; {\rm GeV}^{-2}\,, \\
&&\alpha_{\omega}(0) = 0.5\,, \;\;
\alpha'_{\omega} = 0.9 \; {\rm GeV}^{-2}\,;
\label{trajectory_phi}
\end{eqnarray}
see Eq.~(5.3.1) of \cite{Collins:1977}.

Alternatively, we shall consider the exchange of
the reggeons $\phi_{\Reg}$ and $\omega_{\Reg}$
instead of the mesons $\phi$ and $\omega$ as discussed above.
We recall that $C = -1$ exchanges ($\omega_{\Reg}$, $\phi_{\Reg}$)
are treated as effective vector exchanges in our model.
In order to obtain the $\omega_{\Reg} \Pom$-exchange amplitude
we make in (\ref{amplitude_V_pomeron})
the following replacements:
\begin{eqnarray}
&&\Gamma^{(V pp)}_{\mu}(p_{1},p_{a}) \to \Gamma^{(\omega_{\Reg} pp)}_{\mu}(p_{1},p_{a})\,,
\label{omeR_to_phi_aux1}\\
&&\Delta^{(V)\,\mu \rho_{1}}(q_{1}) \to \Delta^{(\omega_{\Reg})\,\mu \rho_{1}}(s_{1},t_{1})\,,
\label{omeR_to_phi_aux2}\\
&&\Gamma^{(\Pom V \phi)}_{\rho_{2} \rho_{1} \alpha \beta}(p_{34},q_{1}) \to 
\Gamma^{(\Pom \omega_{\Reg} \phi)}_{\rho_{2} \rho_{1} \alpha \beta}(p_{34},q_{1})\,.
\label{omeR_to_phi_aux3}
\end{eqnarray}
We take the corresponding terms 
(\ref{omeR_to_phi_aux1}) and (\ref{omeR_to_phi_aux2})
from (3.59)--(3.60) and (3.14)--(3.15) of \cite{Ewerz:2013kda}, respectively.
In (\ref{omeR_to_phi_aux3}) we use the relations (\ref{AA10a}) and
(\ref{coupling_constant_pom_phi_ome})
and we take the factorised form for the $\Pom \omega_{\Reg} \phi$ form factor
\begin{eqnarray}
F^{(\Pom \omega_{\Reg} \phi)}(q_{2}^{2},q_{1}^{2},p_{34}^{2}) = 
F_{M}(q_{2}^{2})\,F_{M}(q_{1}^{2})\,F^{(\phi)}(p_{34}^{2})
\label{FpomomeRphi}
\end{eqnarray}
with $F_{M}(q^{2})$ as in (\ref{Fpion}) but with $\Lambda_{0}^{2} = 0.5$~GeV$^{2}$
and $F^{(\phi)}(p_{34}^{2}) = F^{(\phi KK)}(p_{34}^{2})$; see (\ref{FphiKK_ff}).
Then, the $\omega_{\Reg} \Pom$-exchange amplitude can be written as
\begin{eqnarray}
&&{\cal M}^{(\omega_{\Reg} \Pom)}_{pp \to pp K^{+}K^{-}} =
i\,g_{\omega_{\Reg} pp}\, F_{1}(t_{1}) \, \bar{u}(p_{1}, \lambda_{1}) 
\gamma^{\alpha}
u(p_{a}, \lambda_{a})\nonumber \\
&&\qquad \times  
\frac{1}{M_{-}^{2}} \left( -i s_{1} \alpha'_{\omega_{\Reg}} \right)^{\alpha_{\omega_{\Reg}}(t_{1})-1}\,
\Delta_{T}^{(\phi)}(p_{34}^{2})\,
\frac{g_{\phi K^{+}K^{-}}}{2}\,(p_{3}-p_{4})^{\beta}\, F^{(\phi K K)}(p_{34}^{2})
\nonumber \\
&& \qquad \times 
\left[ 2 a_{\Pom \omega_{\Reg} \phi}\, \Gamma^{(0)}_{\beta \alpha \kappa \lambda}(p_{34},-q_{1})
- b_{\Pom \omega_{\Reg} \phi}\,\Gamma^{(2)}_{\beta \alpha \kappa \lambda}(p_{34},-q_{1}) \right]
F^{(\Pom \omega_{\Reg} \phi)}(q_{2}^{2},q_{1}^{2},p_{34}^{2})
\nonumber \\
&& \qquad \times 
\frac{1}{2 s_{2}} \left( -i s_{2} \alpha'_{\Pom} \right)^{\alpha_{\Pom}(t_{2})-1}\,
3 \beta_{\Pom NN} \, F_{1}(t_{2})\, 
\bar{u}(p_{2}, \lambda_{2}) 
\left[ \gamma^{\kappa} (p_{2} + p_{b})^{\lambda} \right]
u(p_{b}, \lambda_{b})\,. \quad
\label{amplitude_omeR_pomeron_aux}
\end{eqnarray}
We use for the parameter $M_{-}$ in the $\omega_{\Reg}$ propagator
the value found \mbox{in (3.14), (3.15) of \cite{Ewerz:2013kda}}
\begin{eqnarray}
M_{-} = 1.41\; \mathrm{GeV}\,. 
\label{M_minus}
\end{eqnarray}
In a similar way we obtain the $\phi_{\Reg} \Pom$-exchange amplitude.
We assume that $g_{\phi_{\Reg} pp} = g_{\phi pp}$.

The ${\cal M}^{(\Pom V)}$ and ${\cal M}^{(\Pom V_{\Reg})}$ amplitudes
are obtained from (\ref{amplitude_V_pomeron_aux}) and (\ref{amplitude_omeR_pomeron_aux}),
respectively, with the replacements (\ref{replace}).

For the WA102 energy, $\sqrt{s} = 29.1$~GeV,
also the secondary $f_{2 \Reg}$ exchange may play an important role.
Setting $\sqrt{s_{1}} \approx \sqrt{s_{2}}$
[$\sqrt{s_{1}}$ and $\sqrt{s_{2}}$ are the energies of the subprocesses
$p\,(p_{a})\, \Pom \,(q_{2}) \to p\,(p_{1})\, \phi \,(p_{34})$ and
$p\,(p_{b})\, \Pom \,(q_{1}) \to p\,(p_{2})\, \phi \,(p_{34})$, respectively]
and using the relation 
$s_{1} \,s_{2} \approx s\, m_{\phi}^{2}$ 
we obtain $\sqrt{s_{1}} \approx \sqrt{s_{2}} \approx 5.4$~GeV.
Therefore, in interpreting the WA102 data it is necessary
to take possible contributions from
$\omega$-$f_{2 \Reg}$ and $\omega_{\Reg}$-$f_{2 \Reg}$ exchanges into account,
in addition to the $\omega$-$\Pom$ and $\omega_{\Reg}$-$\Pom$ exchanges.

In a way similar to (\ref{amplitude_V_pomeron})--(\ref{amplitude_omeR_pomeron_aux})
we can write the amplitudes for the 
$\omega$-$f_{2 \Reg}$ and $\omega_{\Reg}$-$f_{2 \Reg}$ exchanges,
since both, $\Pom$ and $f_{2 \Reg}$ exchange, 
are treated as tensor exchanges in our model.
The effective $f_{2 \Reg}$-proton vertex function 
and the $f_{2 \Reg}$ propagator are given in \cite{Ewerz:2013kda} 
by Eqs.~(3.49) and (3.12), respectively.
As an example, the $\omega_{\Reg} f_{2 \Reg}$-exchange amplitude can be written 
as in (\ref{amplitude_omeR_pomeron_aux}) with the following replacements:
\begin{eqnarray}
&&\alpha_{\Pom}(t) \to \alpha_{\Reg_{+}}(t)\,, 
\label{omeR_f2R_1}\\
&&3 \beta_{\Pom NN} \to \frac{g_{f_{2 \Reg}pp}}{M_{0}}\,,
\label{omeR_f2R_2}\\
&&a_{\Pom \omega_{\Reg} \phi} \to a_{f_{2 \Reg} \omega_{\Reg} \phi}\,, \quad
b_{\Pom \omega_{\Reg} \phi} \to b_{f_{2 \Reg} \omega_{\Reg} \phi}\,,
\label{omeR_f2R_3}\\
&&F^{(\Pom \omega_{\Reg} \phi)} \to F^{(f_{2 \Reg} \omega_{\Reg} \phi)}\,.
\label{omeR_f2R_4}
\end{eqnarray}
We take 
$\alpha_{\Reg_{+}}(t) = \alpha_{\Reg_{+}}(0)+\alpha'_{\Reg_{+}}\,t$,
$\alpha_{\Reg_{+}}(0) = 0.5475$, $\alpha'_{\Reg_{+}} = 0.9$~GeV$^{-2}$
from (3.13) of \cite{Ewerz:2013kda} and
$g_{f_{2 \Reg} pp} = 11.04$, $M_{0} = 1$~GeV 
from (3.50) of \cite{Ewerz:2013kda}.
For the $f_{2 \Reg} \omega_{\Reg} \phi$ coupling parameters 
we assume that
$a_{f_{2 \Reg} \omega_{\Reg} \phi} = a_{f_{2 \Reg} \widetilde{\omega} \phi}$,
$b_{f_{2 \Reg} \omega_{\Reg} \phi} = b_{f_{2 \Reg} \widetilde{\omega} \phi}$
and use the relations~(\ref{coupling_constant_f2_ome_phi}).
We assume that $F^{(f_{2 \Reg} \omega_{\Reg} \phi)} = F^{(\Pom \omega_{\Reg} \phi)}$ (\ref{FpomomeRphi})
and take $\Lambda_{0}^{2} = 0.5$~GeV$^{2}$.

In addition, we could have also the $\rho$-$a_{2 \Reg}$ and $\rho_{\Reg}$-$a_{2 \Reg}$ exchanges,
but the couplings of $\rho_{\Reg}$ and $a_{2 \Reg}$ to the protons
are much smaller than those of $\omega_{\Reg}$ and $f_{2 \Reg}$;
see (3.62), (3.52), (3.60), and (3.50) of \cite{Ewerz:2013kda}.
Therefore, we neglect the $\rho$-$a_{2 \Reg}$ and $\rho_{\Reg}$-$a_{2 \Reg}$
terms in our considerations.

\subsection{$\rho$-$\pi^{0}$ contribution}
\label{sec:rho-pion}
Finally, we consider the contribution from $\rho \pi^{0}$,
respectively $\rho_{\Reg} \pi^{0}$, fusion.
\begin{eqnarray}
^{(6)}{\cal M}_{pp \to pp K^{+}K^{-}}^{(\phi \to K^{+}K^{-})} = 
  {\cal M}^{(\rho \pi^{0})}_{pp \to pp K^{+}K^{-}} 
+ {\cal M}^{(\pi^{0} \rho)}_{pp \to pp K^{+}K^{-}} \,.
\label{pp_ppphi_rhopi}
\end{eqnarray}
For the $\rho$-$\pi^{0}$ amplitude we have
\begin{eqnarray}
{\cal M}^{(\rho \pi^{0})}_{pp \to pp K^{+}K^{-}} 
&=& (-i)
\bar{u}(p_{1}, \lambda_{1}) 
i\Gamma^{(\rho pp)}_{\mu}(p_{1},p_{a}) 
u(p_{a}, \lambda_{a}) \nonumber \\
&&\times  
i\Delta^{(\rho)\,\mu \rho_{1}}(q_{1})\, 
i\Gamma^{(\phi \rho \pi^{0})}_{\rho_{2} \rho_{1}}(p_{34},q_{1})\, 
i\Delta^{(\phi)\,\rho_{2} \kappa}(p_{34})\,
i\Gamma^{(\phi KK)}_{\kappa}(p_{3},p_{4})
\nonumber \\
&&\times 
i\Delta^{(\pi^{0})}(t_{2}) \,
\bar{u}(p_{2}, \lambda_{2}) 
i\Gamma^{(\pi^{0} pp)}(p_{2},p_{b}) 
u(p_{b}, \lambda_{b}) \,.
\label{amplitude_rho_pion}
\end{eqnarray}

The $\rho$-proton vertex is given by (\ref{vertex_VNN}) 
and (\ref{F_VNN}) with $V = \rho$. 
The $\phi \rho \pi^{0}$ vertex is as the $\phi \gamma \widetilde{M}$ vertex in (\ref{phi_gam_PS})
with the replacements
\begin{eqnarray}
\gamma \to \rho\,, \quad \widetilde{M} \to \pi^{0} \,, \quad
e g_{\phi \gamma \widetilde{M}} \to g_{\phi \rho \pi^{0}}\,.
\label{phi_rho_pi_vertex}
\end{eqnarray}
The proton-$\pi^{0}$ vertex is given in (\ref{PSNN}).

Then the $\rho$-$\pi^{0}$ amplitude can be written as
\begin{eqnarray}
{\cal M}^{(\rho \pi^{0})}_{pp \to pp K^{+}K^{-}} &=&
i\,g_{\rho pp}\, F^{(\rho pp)}(t_{1}) \, \bar{u}(p_{1}, \lambda_{1}) 
\left[\gamma^{\alpha} - i\frac{\kappa_{\rho}}{2m_{p}} \sigma^{\alpha \alpha'} (p_{1}-p_{a})_{\alpha'} 
\right]
u(p_{a}, \lambda_{a})\nonumber \\
&&\times  
\Delta_{T}^{(\rho)}(t_{1})\,
\Delta_{T}^{(\phi)}(p_{34}^{2})\,
\frac{g_{\phi K^{+}K^{-}}}{2}\,(p_{3}-p_{4})^{\beta}\, F^{(\phi K K)}(p_{34}^{2})
\nonumber \\
&&\times  
\frac{g_{\phi \rho \pi}}{m_{\phi}}\, 
\varepsilon_{\beta \alpha \rho \sigma} p_{34}^{\rho} q_{1}^{\sigma}\,
\tilde{F}^{(\rho)}(q_{1}^{2})
\tilde{F}^{(\phi)}(p_{34}^{2})
F^{(\phi \rho \pi^{0})}(q_{2}^{2})
\nonumber \\
&&\times 
\frac{1}{t_{2} - m_{\pi^{0}}^{2}}\, 
g_{\pi^{0} pp} \, F^{(\pi^{0} pp)}(t_{2})\, 
\bar{u}(p_{2}, \lambda_{2}) \,\gamma_{5} \,u(p_{b}, \lambda_{b})\,.
\label{amplitude_rho_pion_aux}
\end{eqnarray}
We take $g_{\rho pp} = 3.72$, $\kappa_{\rho} = 6.1$, and
$g_{\phi \rho \pi} = -1.258$ from \cite{Ryu:2012tw}.
Here we choose monopole form factors (\ref{ff_PSNN}) and (\ref{ff_phigamPS})
with $\Lambda_{\pi^{0} pp} = 1.2$~GeV and $\Lambda_{\phi \rho \pi^{0}} = 1.2$~GeV, 
respectively.
However, in \cite{Nakayama:1999jx} smaller numerical values can be found,
$g_{\rho pp} = 2.63$--$3.36$ and $g_{\phi \rho \pi} = -0.65$, respectively.
Therefore, our result should be considered rather 
as an upper limit for the $\rho$-$\pi^{0}$ contribution.

The reggeization of the $\rho$-meson propagator in the $t$-channel
in ${\cal M}^{(\rho \pi^{0})}$
is taken into account here by the prescription (\ref{reggeization_2}) for $V = \rho$.
We assume for the $\rho$ trajectory 
\begin{eqnarray}
&&\alpha_{\rho}(t) = \alpha_{\rho}(0)+\alpha'_{\rho}\,t\,,\\ 
&&\alpha_{\rho}(0) = 0.5\,, \;\; 
\alpha'_{\rho} = 0.9 \; \mathrm{GeV}^{-2}\,.
\label{trajectory_rho}
\end{eqnarray}
The amplitude ${\cal M}^{(\pi^{0} \rho)}$ is obtained from
${\cal M}^{(\rho \pi^{0})}$ (\ref{amplitude_rho_pion})
by the replacements (\ref{replace}).

In principle we can also have $\omega$-$\eta$ and $\omega$-$\eta'$ fusion contributions.
$g_{\phi \omega \eta}$ and $g_{\phi \omega \eta'}$
cannot be obtained from mesonic decays.
Then one could rely only on models.
Due to these model uncertainties of the coupling constants
for the $\omega$-$\eta$ and $\omega$-$\eta'$ fusion processes
we neglect these contributions in our present study.

\section{The Collins-Soper frame}
\label{sec:appendixC}

To make our present article self contained we give
here the definition of the Collins-Soper (CS) frame used
in our paper; see \cite{Lebiedowicz:2019por} 
and for general remarks on various
reference frames of this type Appendix~A of \cite{Bolz:2014mya}.

We go to the $K^{+}K^{-}$ or $\mu^{+}\mu^{-}$ rest frame
for studying the reactions (\ref{2to4_reaction_KK}) 
or (\ref{2to4_reaction_mumu}), respectively.
Let $\bpa$, $\bpb$ be the three-momenta 
of the initial protons in this system.
We define the unit vectors
\begin{eqnarray}
\bhpa = \bpa / |\bpa|, \quad \bhpb = \bpb / |\bpb|\,.
\label{AC}
\end{eqnarray}
The CS frame is then defined by the coordinate-axes unit vectors
\begin{equation}
\begin{split}
& \bea_{,\,\rm CS} = \frac{\bhpa+\bhpb}{|\bhpa+\bhpb|}\,,\\
& \beb_{,\,\rm CS} = \frac{\bhpa \times \bhpb}{|\bhpa \times \bhpb|}\,,\\
& \bec_{,\,\rm CS} = \frac{\bhpa-\bhpb}{|\bhpa-\bhpb|}\,.
\end{split}
\label{CS}
\end{equation}

The angles $\theta_{K^{+}, \,{\rm CS}}$ 
and $\phi_{K^{+}, \,{\rm CS}}$,
respectively $\theta_{\mu^{+}, \,{\rm CS}}$ 
and $\phi_{\mu^{+}, \,{\rm CS}}$,
are the polar and azimuthal angles of the momentum vector $\bhpc$ in this system.
We have then, e.g.,
\begin{equation}
\cos\theta_{K^{+},\,{\rm CS}} = \bhpc \cdot \bec_{,\,\rm CS}\,.
\end{equation}
%

\acknowledgments

The authors are grateful to L. Adamczyk, C. Ewerz, S. Glazov,
L. G{\"o}rlich, R. McNulty, B. Rachwa{\l}, 
and T. Szumlak for useful discussions.
This work was partially supported by
the Polish National Science Centre Grant No. 2018/31/B/ST2/03537
and by the Center for Innovation and Transfer
of Natural Sciences and Engineering Knowledge
in Rzesz{\'o}w (Poland).

{
\begin{small}
\bibliography{refs}
\end{small}
}

\end{document}